\documentclass[prd,aps,twocolumn,a4paper,floatfix,nofootinbib]{revtex4-2}

\usepackage[utf8]{inputenc}
\usepackage{graphicx,psfrag}
\usepackage{microtype}
\usepackage{mathrsfs}
\usepackage{amsmath,amsfonts,amssymb}
\usepackage{multirow}
\usepackage{diagbox}
\usepackage{comment}
\usepackage{xcolor}
\usepackage{xspace} 
\usepackage{enumerate}
\usepackage{booktabs}
\usepackage[normalem]{ulem}
\usepackage{hyperref}
\usepackage{cleveref}
\usepackage{mathtools}
\usepackage{siunitx}
\usepackage{textgreek}
\usepackage[version=4]{mhchem}
\usepackage{chemformula}
\usepackage{orcidlink}
\hypersetup{
    colorlinks = true,
    linkcolor = {blue},
    citecolor = {blue},
    urlcolor = {blue},
    linkbordercolor = {white},
    }

\usepackage{color}
\definecolor{cyan}{rgb}{0,0.9,0.9}
\definecolor{orange}{rgb}{0.9,0.5,0}
\definecolor{magenta}{rgb}{1,0,1}
\definecolor{purple}{rgb}{0.8,0.4,0.8}
\definecolor{gray}{rgb}{0.8242,0.8242,0.8242}
\definecolor{green}{rgb}{0.,0.8,0.}

\def\bam{{\textsc{bam}}}
\def\sgrid{{\textsc{sgrid}}}
\def\winnet{{\textsc{WinNet}}}
\def\msun{M_\odot}
\def\abar{\Bar{A}}

\DeclareSIUnit{\year}{yr}

\definecolor{BerlinU1}{HTML}{62AD2D}
\definecolor{BerlinU2}{HTML}{E94D10}
\definecolor{BerlinU3}{HTML}{00A192}

\usepackage[normalem]{ulem}

\newcommand{\msol}{$M_\odot$\xspace}

\begin{document}

\title{General-relativistic radiation magnetohydrodynamics simulations of binary neutron star mergers: The influence of spin on the multi-messenger picture}

\author{Anna \surname{Neuweiler}$^{1}$\orcidlink{0000-0003-3205-8373}}
\author{Henrique \surname{Gieg}$^{1}$\orcidlink{0000-0002-1830-2694}}
\author{Henrik \surname{Rose}$^{1}$\orcidlink{0009-0009-2025-8256}}
\author{Hauke \surname{Koehn}$^{1}$\orcidlink{0009-0001-5350-7468}}
\author{Ivan \surname{Markin}$^{1}$\orcidlink{0000-0001-5731-1633}}
\author{Federico \surname{Schianchi}$^{2,1}$\orcidlink{0000-0001-7646-5988}}
\author{Liam \surname{Brodie}$^{3}$\orcidlink{0000-0001-7708-2073}}
\author{Alexander \surname{Haber}$^{3,4}$\orcidlink{0000-0002-5511-9565}}
\author{Vsevolod \surname{Nedora}$^{1,5}$\orcidlink{0000-0002-5196-2029}}
\author{Mattia \surname{Bulla}$^{6,7,8}$\orcidlink{0000-0002-8255-5127}}
\author{Tim \surname{Dietrich}$^{1,5}$\orcidlink{0000-0003-2374-307X}}

\affiliation{${}^1$ Institut f\"ur Physik und Astronomie, Universit\"at Potsdam, Haus 28, Karl-Liebknecht-Str. 24/25, 14476, Potsdam, Germany}
\affiliation{${}^2$ Departament de Física \& IAC3, Universitat de les Illes Balears, Palma de Mallorca, Baleares E-07122, Spain}
\affiliation{${}^3$ Department of Physics, Washington University in St.~Louis, St.~Louis, MO 63130, USA}
\affiliation{${}^4$Mathematical Sciences and STAG Research Centre, University of Southampton, Southampton SO17 1BJ, United Kingdom}
\affiliation{${}^5$ Max Planck Institute for Gravitational Physics (Albert Einstein Institute), Am M\"uhlenberg 1, Potsdam 14476, Germany}
\affiliation{$^{6}$Department of Physics and Earth Science, University of Ferrara, via Saragat 1, I-44122 Ferrara, Italy}
\affiliation{$^{7}$INFN, Sezione di Ferrara, via Saragat 1, I-44122 Ferrara, Italy}
\affiliation{$^{8}$INAF, Osservatorio Astronomico d’Abruzzo, via Mentore Maggini snc, 64100 Teramo, Italy}

\date{October 17, 2025}

\begin{abstract}
The rich phenomenology of binary neutron star mergers offers a unique opportunity to test general relativity, investigate matter at supranuclear densities, and learn more about the origin of heavy elements. As multi-messenger sources, they emit both gravitational waves and electromagnetic radiation across several frequency bands. The interpretation of these signals relies heavily on accurate numerical-relativity simulations that incorporate the relevant microphysical processes. 
Using the latest updates of the \bam\ code, we perform general-relativistic radiation magnetohydrodynamic simulations of binary neutron star mergers with two different spin configurations. We adopt a state-of-the-art equation of state based on relativistic mean-field theory developed for dense matter in neutron star mergers. 
To capture both dynamical ejecta and secular outflows from magnetic and neutrino-driven winds, we evolve the systems up to $\sim 100\ \rm ms$ after the merger at considerably high resolution with a grid spacing of $\Delta x \approx 93\ \rm m$ across the neutron stars. 
Our results show that the non-spinning configuration undergoes a more violent merger, producing more ejecta with lower electron fraction and higher velocities, while the spinning configuration forms a larger disk due to its higher angular momentum. 
Although the initial magnetic field amplification within $\lesssim 10\ \rm ms$ after merger is similar in both systems, the non-spinning system reaches stronger magnetic fields and higher energies at later times. 
For a detailed view of the multi-messenger observables, we extract the gravitational-wave signal and compute nucleosynthesis yields, the expected kilonova and afterglow light curves from our ejecta profiles.
\end{abstract}

\maketitle

\section{Introduction}
\label{sec:Intro}

\begin{figure*}[t]
    \centering
    \includegraphics[width=\linewidth]{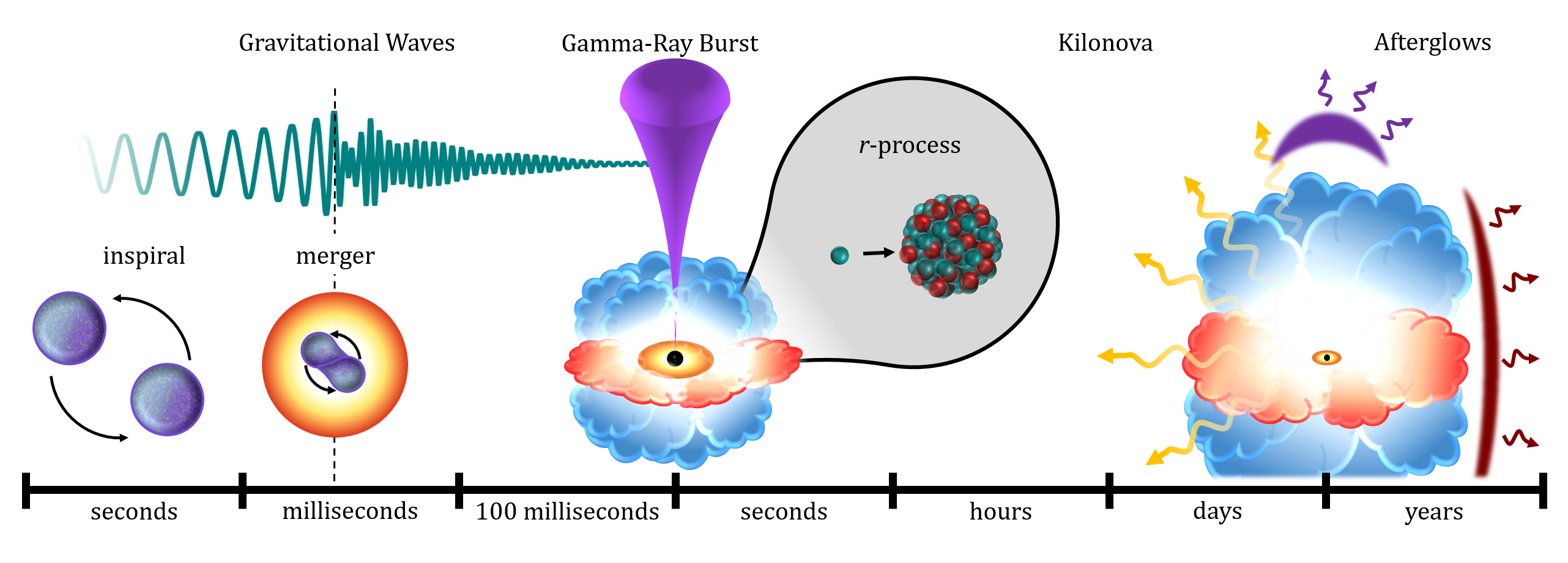}
    \caption{BNS merger phenomena for different timescales, ranging from seconds to milliseconds before the merger and from milliseconds to years after the merger. As simplified overview, we illustrate the physical phenomena together with the associated observable multi-messenger signals, consisting of GWs and EM signatures: inspiral and merger of the two neutron stars, ejection of lanthanide-rich (in red) and lanthanide-poor (in blue) material, formation of a black hole remnant with accretion disk, launch of a relativistic jet (in purple), formation of heavy elements via $r$-process, kilonova emission (in yellow), and non-thermal afterglows of the GRB (in purple) and the kilonova (in darkred).}
    \label{fig:overview}
\end{figure*}

Binary neutron star (BNS) mergers are high-energy multi-messenger events that emit gravitational waves (GWs) and electromagnetic (EM) radiation covering multiple frequency bands, e.g.,~\cite{Fernandez:2015use,Metzger:2016pju,Baiotti:2016qnr,Duez:2018jaf,Bernuzzi:2020tgt,Ruiz:2021gsv}. As these compact objects orbit each other and eventually merge, a GW signal is generated, whereas the EM emission originates from the merger remnant and ejecta. The neutron-rich material that is ejected in this process is an important site of rapid neutron-capture ($r$-process) nucleosynthesis~\cite{Lattimer:1974slx,Rosswog:1998hy,Korobkin:2012uy,Wanajo:2014wha,Cowan:2019pkx}. The radioactive decay of the formed nuclei powers an EM transient, called a kilonova. Furthermore, BNS merger remnants can launch a relativistic jet, most likely driven by a large-scale magnetic field, e.g.,~\cite{Rezzolla:2011da,Ruiz:2016rai,Mosta:2020hlh,Kiuchi:2023obe}, that can trigger a gamma-ray burst (GRB). Both the kilonova and the GRB produce an afterglow when the outflows interact with the surrounding interstellar medium. A simplified overview of the physical phenomena of the BNS merger and the associated observable GW and EM signatures is shown in Fig.~\ref{fig:overview}, highlighting the different timescales at which the signals are emitted, ranging from seconds to milliseconds before the merger to years after the merger.

Indeed, the first GW detection of a BNS merger in August 2017, associated with the event GW170817~\cite{LIGOScientific:2017vwq}, was accompanied by the kilonova signal AT2017gfo~\cite{LIGOScientific:2017pwl,Arcavi:2017xiz, Coulter:2017wya,Lipunov:2017dwd,Tanvir:2017pws,Valenti:2017ngx}, the short GRB GRB170817A~\cite{Goldstein:2017mmi,Savchenko:2017ffs}, and its non-thermal afterglow~\cite{Hajela:2019mjy,Hajela:2021faz,Balasubramanian:2022sie}. Among others, this joint observation has been used to probe general relativity~\cite{LIGOScientific:2017zic,LIGOScientific:2020tif,LIGOScientific:2018dkp,Sakstein:2017xjx,Ezquiaga:2017ekz,Creminelli:2017sry,Baker:2017hug} and to tighten constraints on the equation of state (EOS) of neutron stars, e.g., ~\cite{LIGOScientific:2018dkp,LIGOScientific:2017ync,LIGOScientific:2018cki,Bauswein:2017vtn,Margalit:2017dij,Radice:2017lry,Coughlin:2018miv,Most:2018hfd,Capano:2019eae,Raithel:2019uzi,Dietrich:2020efo}.

In addition to the intensively studied impact of the component masses, neutron star spins are essential to a proper modeling of BNS mergers and their observable signals. While observations confirm that isolated neutron stars can have dimensionless spins up to $\chi \sim 0.4$~\cite{Hessels:2006ze}, the fastest-spinning BNS systems known to merge within a Hubble time, i.e., PSR J0737-3039~\cite{Burgay:2003jj} and PSR J1946+2052~\cite{Stovall:2018ouw}, are expected to have spins of only $\chi \sim 0.04$ or $\chi \lesssim 0.05$ at merger. 
GW170817 itself provides only weak constraints on the binary components' spins, also affecting estimates on the component masses due to a degeneracy between the mass ratio $q$ and the aligned spin components in gravitational waveforms. For a high-spin prior ($|\chi| \leq 0.89$), the inferred component masses range from $0.86\ M_\odot$ to $2.26\ M_\odot$~\cite{LIGOScientific:2017vwq}. 
Numerous studies have shown that neutron star spins impact the lifetime of the remnant, the disk mass, and the amount of ejected material, e.g.,~\cite{Rosswog:2023rqa,East:2019lbk,Ruiz:2019ezy,Bernuzzi:2013rza,Dietrich:2016lyp,Kastaun:2016elu,Most:2019pac,Papenfort:2022ywx}. Consequently, the stars' spin might also affect signatures of the EM counterparts. 

In this work, we investigate the role of neutron-star spins in BNS mergers and assess the impact on nucleosynthesis yields and multi-messenger observables, i.e., the GW signal, the kilonova light curves, and the kilonova/GRB afterglow. For this purpose, we perform state-of-the-art numerical-relativity (NR) simulations that solve Einstein's field equations together with the equations for general-relativistic radiation magnetohydrodynamics (GRRMHD), including a proper treatment to model the underlying microphysics based on fundamental theories of the strong and weak interaction. In particular, we use the \bam\ code~\cite{Bruegmann:2006ulg,Thierfelder:2011yi} with recent updates to incorporate tabulated, nuclear-physics-informed EOSs~\cite{Gieg:2022mut}, a first-order multipolar (M1) neutrino-transfer scheme~\cite{Schianchi:2023uky}, and magnetic fields~\cite{Neuweiler:2024jae}. Several studies have highlighted the importance of incorporating magnetohydrodynamic effects, e.g.,~\cite{Kiuchi:2014hja,Dionysopoulou:2015tda,Ruiz:2016rai,Cook:2025frw,Gutierrez:2025gkx,Ciolfi:2019fie,Chabanov:2022twz,Aguilera-Miret:2021fre,Aguilera-Miret:2023qih}, and neutrino-radiation, e.g., \cite{Foucart:2015vpa,Sekiguchi:2016bjd,Vincent:2019kor,Radice:2018pdn,Radice:2021jtw,Espino:2023mda,Kawaguchi:2025con} to reliably model BNS mergers, their remnants, and outflowing material. Although the number of NR simulations that take both magnetic fields and neutrino radiation into account is growing, e.g.,~\cite{Palenzuela:2015dqa,Kiuchi:2023obe,Palenzuela:2022kqk,Musolino:2024sju,Curtis:2023zfo,Sun:2022vri,Hayashi:2022cdq,Combi:2023yav,Bamber:2025ggq}, the BNS systems considered are still limited and -- to our knowledge -- restricted to neutron stars without spin. We point out that spinning BNS configurations, either with a neutrino leakage scheme or with the inclusion of magnetohydrodynamic effects, have already been simulated and analyzed, e.g., in Refs.~\cite{Most:2019pac,Ruiz:2019ezy}. However, we include both magnetic fields and neutrino radiation with the M1 scheme to study spin effects in BNS mergers as comprehensively as possible.

We study equal-mass BNS systems in which both stars have a gravitational mass of $1.35\, M_\odot$. One configuration is initially non-rotating, while the stars in the other have a dimensionless spin of $\chi = 0.1$ aligned to the orbital angular momentum. Although this spin is higher than previously observed in merging BNS systems, it allows us to investigate the extent to which we can distinguish observable features between the two systems. We simulate these BNS systems with high resolution, using a grid spacing of $\Delta x \approx 93\ \rm m$ to cover the neutron stars. This is necessary to capture small-scale dynamics and resolve instabilities relevant for magnetic field amplification. Additionally, we perform the same simulations with reduced resolution to verify our results and distinguish artifacts caused by finite resolution. The simulations are performed for about $100\ \rm ms$ into the post-merger to also cover magnetic- and neutrino-driven winds. In total, the simulations consumed $29.88$~Mio. CPUh at the HPE Apollo (Hawk) at the High-Performance Computing Center Stuttgart (HLRS, see also Appendix~\ref{app:ressources}). We analyze the magnetic field amplification during and after the merger, and study the post-merger remnant, matter outflow, and its composition. Additionally, we co-evolve Lagrangian tracer particles that we use to study nucleosynthesis in post-processing. In order to connect the simulated BNS systems to observable signatures, we extract the GW strain and the ejecta from the simulation data. Based on the latter, we calculate the associated kilonova signal and its afterglow. 

The article is structured as follows: 
Section~\ref{sec:Methods} summarizes numerical methods of the employed codes, the simulated BNS configurations, and the EOS used. The results of the BNS simulations are discussed in Sec.~\ref{sec:Simulations}, focusing on the merger remnant, magnetic field amplification, and matter outflows, as well as the nucleosynthesis yields of the ejecta. The multi-messenger observables, including the GW signal, kilonova light curves, and afterglow emissions, are presented in Sec.~\ref{sec:MMApicture}. Finally, Sec.~\ref{sec:Conclusions} summarizes our key findings and results.
Unless stated otherwise, we use a $(-,+,+,+)$ signature for the metric and geometric units with $G=c=M_\odot=1$.

\section{Methods and setups}
\label{sec:Methods}

\subsection{Spacetime and matter evolution}
\label{subsec:evolution}

We perform the evolution of the BNS configurations using \bam~\cite{Bruegmann:2006ulg,Thierfelder:2011yi,Dietrich:2015iva, Bernuzzi:2016pie}, incorporating recent extensions for neutrino radiation~\cite{Gieg:2022mut,Schianchi:2023uky} and magnetic fields~\cite{Neuweiler:2024jae}. \bam\ solves Einstein's field equations numerically in the $3+1$ formulation. For the spacetime evolution, we use the Z4c reformulation with constraint damping terms \cite{Bernuzzi:2009ex,Hilditch:2012fp}, together with the $1+\log$ slicing~\cite{Bona:1994a} and $\Gamma$-driver shift~\cite{Alcubierre:2002kk} conditions.

For the matter evolution, our GRRMHD implementation employs the Valencia formulation~\cite{Marti:1991wi,Banyuls:1997zz,Anton:2005gi,Font:2008fka}. We adopt the ideal magnetohydrodynamics approximation, i.e., assuming infinite conductivity. To ensure the divergence-free constraint for the magnetic field and prevent the formation of unphysical magnetic monopoles, we apply the hyperbolic divergence-cleaning scheme~\cite{Liebling:2010bn,Mosta:2013gwu}\footnote{We note that our implementation had a mistake in the flux computation of the magnetic field in the damping term. We analyze the average relative error weighted by the flux and estimate it to be $< 0.1\ \%$. Hence, we assume the bias to be negligible in the simulations presented, but we remark that the same issue existed in previous studies, i.e., \cite{Neuweiler:2024jae,Neuweiler:2025lte}}. 
Neutrino-driven interactions are modeled using the M1 scheme~\cite{Thorn:1981,Shibata:2011kx,Foucart:2015gaa,Foucart:2016rxm,Weih:2020wpo,Radice:2021jtw,Musolino:2023pao}. 
Details about the implementation, as well as the set of weak reactions employed in this work, can be found in Ref.~\cite{Schianchi:2023uky}, except for elastic scattering on \textalpha-particles. In this work, we employ the elastic opacities of Ref.~\cite{Ruffert:1995fs} as implemented in Ref.~\cite{Gieg:2022mut}. We highlight that, in this manner, the opacities are computed using the same microscopic model as that of the EOS described in Sec.~\ref{subsec:eos}.
Moreover, we have implemented the computation of neutrino emissivities in the trapped regime based on black body functions at equilibrium as in Refs.~\cite{Radice:2021jtw,Perego:2019adq} (see Appendix~\ref{app:fluid-radiation}). This improves the stability under rapid variations of the fluid's temperature.

The computational grid in \bam\ is structured as a hierarchy of nested refinement levels $l=0,1,...,L-1$, where each of the $L$ levels contains one or more Cartesian boxes with fixed grid spacing. The resolution between successive levels increases by a factor of two, such that the grid spacing on level $l$ is given by $h_l = h_0 / 2^l$. To ensure that the neutron stars are always covered by the finest refinement level, the Cartesian boxes for the inner levels with $l \geq l_{\rm mv}$ move dynamically to track the neutron stars.

We use a fourth-order Runge-Kutta scheme for time integration with the Courant-Friedrichs-Lewy factor of $0.25$. \bam\ employs the Berger-Oliger scheme~\cite{Berger:1984zza} for local-time stepping (see Ref.~\cite{Bruegmann:2006ulg}), and the Berger-Colella scheme~\cite{Berger:1989a} to ensure the conservation of baryonic mass, energy, and momentum across refinement boundaries (see Ref.~\cite{Dietrich:2015iva}). 

We approximate the spatial derivatives of the metric using fourth-order finite-difference stencils. For the evolution of fluid variables, we use high-resolution shock-capturing methods. Specifically, we employ the fifth-order weighted-essentially-non-oscillatory WENOZ~\cite{Borges:2008} reconstruction and Harten-Lax-van-Leer (HLL) Riemann solver~\cite{Harten:1983} with a two-wave approximation.
We adopt a cold, static atmosphere to model the vacuum region surrounding the compact objects. In particular, the fluid quantities of a grid cell are set to atmosphere values once the rest-mass density $\rho$ falls below a density threshold $\rho_{\rm thr}$. The atmosphere density $\rho_{\rm atm}$ is chosen as the minimum rest-mass density $\rho_{\rm min}$ covered by the EOS table used (see Sec.~\ref{subsec:eos}). The temperature is set to $0.1\ \rm MeV$, and the fluid velocity is set to zero. The magnetic field remains unchanged. The density threshold is defined as $\rho_{\rm thr} = 10 \times \rho_{\rm atm}$ to avoid fluctuations around the density floor.
Furthermore, in regions with rest-mass densities below $10^{8} \times \rho_{\rm atm}$, we follow the procedure described in Ref.~\cite{Neuweiler:2024jae} to avoid enhanced oscillations when using high-order reconstruction methods. The oscillations of all reconstructed variables are examined, and if necessary, we fall back locally to a lower-order method, specifically the third-order convex-essential-non-oscillating (CENO3) scheme~\cite{Liu:1998,Zanna:2002qr}. In addition, physical validity is ensured by demanding a positive rest-mass density and a positive pressure. If these cannot be satisfied even with the lower-order scheme, we fall back to a linear reconstruction method.

\subsection{Codes used for post-processing}
\label{subsec:codes}

For analyzing the simulation data, we additionally use the nuclear reaction network \winnet~\cite{Reichert:2023xqy} to determine nuclear abundances, the radiative transfer code \textsc{possis}~\cite{Bulla:2019muo,Bulla:2022mwo} to calculate kilonova light curves, and the \textsc{pyblastafterglow} code~\cite{Nedora:2022kjv,Nedora:2023hiz, Nedora:2024vrv} to obtain the kilonova and GRB afterglows. The methods employed by each code are briefly outlined below, along with a description of how data from the NR simulations are incorporated.

\subsubsection{\textsc{WinNet}}
\label{subsubsec:WinNet}

To analyze and understand the properties of the outflowing material, we employ tracer particles that record the history of Lagrangian fluid trajectories (see Appendix~\ref{app:Tracers} for details). We use these thermodynamic trajectories as inputs to compute nucleosynthesis yields with the single-zone reaction network code \winnet~\cite{Reichert:2023xqy}.

Given that all trajectories reach temperatures above \qty{10}{GK}, we assume each tracer to evolve from nuclear-statistical equilibrium (NSE). We start to follow the NSE composition, only evolving weak reactions once temperatures drop below \qty{9}{GK} and switch to the full network evolution below \qty{7}{GK}. After the end of the trajectories, we assume homologous expansion and continue the network calculation until \qty{1}{\giga\year}.

Generally, we employ \winnet's default nuclear input. This comprises more than $6500$~isotopes up to \ce{^337Rg}. Their basic nuclear properties build on the FRDM(2012) model~\cite{Moller:2015fba}, which also underlies the parametrization of theoretically and experimentally determined reaction rates according to the JINA Reaclib reaction library~\cite{Cyburt:2010}.
While this provides the majority of reactions under consideration, specific reactions in certain regimes may be superseded by information from other sources, e.g., detailed rates for \textalpha- and \textbeta-decays and weak rates.

In addition, \winnet\ allows for different treatments of fission and neutrino reactions. While weak interactions play an important role in protonizing the ejecta and driving winds from the disk, the tracers' densities have already dropped significantly as we initiate the network evolution, such that neutrino reactions have a negligible impact on the later evolution. The fission prescription is more relevant for the final abundance pattern, though it only plays a significant role in tracers that produce the third peak. For performance reasons, we use the computationally more efficient treatment of Ref.~\cite{Panov:2001rus} for low-density trajectories, while using the spontaneous fission fragment distribution of Ref.~\cite{Kodama:1975aff} and the \textbeta-delayed fission prescription of Ref.~\cite{Mumpower:2019uid} otherwise.

\subsubsection{\textsc{possis}}
\label{subsubsec:possis}

We compute the kilonova signals associated to the simulated BNS systems by performing Monte Carlo radiative transfer simulations with \textsc{possis}~\cite{Bulla:2019muo,Bulla:2022mwo}. We use the same procedure as described in Ref.~\cite{Schianchi:2023uky} to extract ejecta data, i.e., the rest-mass density $\rho$ and electron fraction $Y_e$, from NR simulations. Specifically, we take a 3D snapshot at a time $t_{\rm cut}$, when the entire ejecta is still fully within the computational domain. Subsequently, we use the ejecta information from a detection sphere at $r \simeq 440\ \rm km$. Both data sets are rescaled assuming homologous expansion and combined to be used as input in \textsc{possis}, representing the ejecta at a reference time $t_0$.
While Refs.~\cite{Neuweiler:2022eum,Kawaguchi:2020vbf} demonstrated that the ejecta may not yet be fully homologously expanding at $\mathcal{O}\left(100\ \rm ms\right)$ after the merger, the resulting biases in light curve computation using \textsc{possis} have been shown to be negligible for $t> 80\ \rm ms$ after the merger~\cite{Neuweiler:2022eum} and this procedure has proven to give robust results.

For the computation of light curves, \textsc{possis} continues to expand the ejecta homologously, assuming a constant velocity $v^i$ of each fluid cell. The code generates photon packets at each time step and assigns them an energy, frequency, and direction of propagation.
\textsc{possis} employs heating rate libraries from Ref.~\cite{Rosswog:2022tus} and thermalization efficiencies from Refs.~\cite{Barnes:2016umi,Wollaeger:2017ahm}. Furthermore, it adopts wavelength- and time-dependent opacities as functions of local ejecta properties, i.e., rest-mass density, temperature, and electron fraction $\left(\rho, T, Y_e\right)$, from Ref.~\cite{Tanaka:2019iqp}.
The photon packets are propagated through the ejecta, accounting for interactions with matter through electron scattering and bound-bound absorption that alter their assigned propagation direction, frequency, and energy. Finally, synthetic observables for different observation angles $\iota$ are computed `on the fly' using Monte Carlo estimators with the event-based technique discussed in Ref.~\cite{Bulla:2019muo}.
We perform the radiative transfer simulations using $N_{\rm ph} = 10^6$~photon packets. 

\subsubsection{\textsc{pyblastafterglow}}
\label{subsubsec:PyBlastAfterglow}

The outflows from BNS mergers produce not only intrinsic radiation in the form of a kilonova, but also radiation through interactions with the ambient matter. Ejected matter with relativistic escape velocities hits the interstellar medium surrounding the merger site, producing a non-thermal EM signature. This signature, commonly referred to as an afterglow, arises from relativistic electrons gyrating around magnetic field lines in collisionless shocks.
Specifically, BNS mergers can produce two types of afterglows: a GRB afterglow from an ultra-relativistic, collimated jet and a kilonova afterglow from the mildly relativistic, circum-planar ejecta.

We determine the expected afterglow emissions from our simulations using \textsc{pyblastafterglow}~\cite{Nedora:2022kjv,Nedora:2023hiz,Nedora:2024vrv}, a modular code that evolves the dynamics of the discretized blast wave semi-analytically as it hits the cold, constant-density ambient interstellar medium. In a second step, it calculates the energy distribution of accelerated electrons and their synchrotron emission.

For the kilonova afterglow calculation, \textsc{pyblastafterglow} takes the angular ejecta velocity distributions from our NR simulations as input to initialize the blast waves. The code assumes azimuthal symmetry, i.e., the ejecta velocities only depend on the polar angle. Each blast wave is evolved independently by solving energy conservation equations~\cite{Nedora:2022kjv}. 
To determine the emitted radiation, it is important to note that the dynamical and secular ejecta from BNS mergers are at most mildly relativistic, with the velocity and Lorentz factor obeying $\beta\Gamma\lesssim 5$. Previous studies have shown that in these types of mildly relativistic collisionless shocks, both a thermal (Maxwellian) and a non-thermal (power-law) electron component can potentially deliver significant contributions to the produced synchrotron spectrum~\cite{Margalit:2021kuf, Margalit:2024asc}. This is in contrast to highly relativistic shocks with $\Gamma \gg 1$, where the contribution from the thermal population is negligible~\cite{Sironi:2013ri, Crumley:2018kvf}. Hence, \textsc{pyblastafterglow} includes an implementation of the thermal synchrotron model from Ref.~\cite{Margalit:2021kuf} that is used to determine the kilonova afterglow emission.
Thus, computing kilonova afterglows requires us to set five additional parameters, namely the ambient density $n_0$, the power law index of the non-thermal electron component $p$, as well as the fraction of the internal blast wave energy that is converted to magnetic fields $\epsilon_B$ and is used for particle acceleration $\epsilon_e$. Finally, $\epsilon_T$ is the fraction of the shock energy that is converted to thermal electrons. 

For the calculation of the GRB afterglow, we assume a Gaussian jet with the following distribution for the isotropic-energy equivalent
\begin{align}
    E_{\rm iso}(\theta) = \begin{cases}
        E_{0} \exp\left(-\frac{\theta^2}{2\theta_c^2}\right)\qquad \text{if $\theta\leq\theta_{\rm w}$} \\
        0 \qquad \qquad \qquad \qquad \text{else}
    \end{cases}\ .
\end{align}
This structure is inspired by numerical jet simulations~\cite{Xie:2018vya, Ryan:2019fhz, Gottlieb:2020raq}. Here, $E_0$ is the central isotropic kinetic energy-equivalent, $\theta_c$ denotes the core angle, and $\theta_w$ is the cut-off.
Since we cannot infer the jet structure from our NR simulations, we will resort to fiducial values taken from GRB170817A, as discussed below in Sec.~\ref{subsec:afterglow}.
The jet's blast wave elements are evolved according to the same shock-jump conditions and lateral spreading prescriptions of Ref.~\cite{Nedora:2024vrv}.
The non-thermal electron distribution is evolved according to a Fokker-Planck-type continuity equation and convolved with the classical synchrotron kernel to determine the comoving radiation~\cite{Aharonian:2010va}.
This step relies on the microphysical parameters $p$, $\epsilon_e$, $\epsilon_B$ with the same meaning as above for the kilonova afterglow, although they do not necessarily need to coincide with the values for the kilonova blast waves.

We further assume that GRB and kilonova blast waves evolve independently. This simplification is justified by the fact that only a small part of the kilonova blast wave travels through the circum-burst density of the collimated jet. These are mainly the polar ejecta, which in our case carry the lowest mass, and we confirmed by excising their blast waves from our kilonova afterglow computations that they do not affect the light curves.
Moreover, the investigation in Ref.~\cite{Nedora:2022kjv} shows that even if the blast waves are coupled to the pre-shocked density of the GRB jet, the effect on the dynamics of the kilonova ejecta is small.

\subsection{Configurations}
\label{subsec:configurations}

We present simulations for two equal-mass BNS systems with gravitational masses of $1.35\ M_\odot$ and baryonic masses $M_b = 1.48\ M_\odot$, corresponding to neutron stars in isolation with coordinate radius $R = 12.22\ \rm km$ and tidal deformability $\Lambda=478.6$. We employ the ABHT(QMC-RMF3) EOS~\cite{Alford:2023rgp} to describe neutron-star matter, satisfying constraints from terrestrial experiments, astrophysical observations, and first-principles calculations of neutron matter; cf. Sec.~\ref{subsec:eos}. 
One configuration has non-spinning neutron stars, and the other one has spinning neutron stars, aligned with the orbital angular momentum. For the latter, we chose the dimensionless spin of the two stars as $\chi_1 = \chi_2 = 0.1$. Both setups have an initial coordinate distance $d_0 \approx 41.2\ \rm km$.

We construct initial data using the pseudo-spectral code \sgrid~\cite{Tichy:2009yr,Tichy:2012rp,Dietrich:2015pxa,Tichy:2019ouu}, which uses the extended conformal thin sandwich formulation~\cite{York:1998hy,Pfeiffer:2002iy} to solve the Einstein Constraint Equations.
The magnetic field is subsequently superimposed for the evolution as a purely poloidal field inside each star. It is defined by a vector potential with
\begin{align}
    A_x &= -\left(y - y_{\rm NS}\right) A_b \max(p - p_{\rm cut},0)^2, \\
    A_y &=  \left(x - x_{\rm NS}\right) A_b \max(p - p_{\rm cut},0)^2, \\
    A_z &=  0,
\end{align}
where $x_{\rm NS}$ and $y_{\rm NS}$ refer to the coordinate center of each neutron star.
We set $A_b = 1510$ to obtain a maximum magnetic field strength on the order of $10^{15}\ \rm G$ inside the stars. The cut-off pressure $p_{\rm cut}$ is chosen as $0.004 \times p_{\rm max}$, where $p_{\rm max}$ is the initial maximum pressure in the neutron star at $t=0\ \rm ms$. This adopted magnetic field is several orders of magnitude stronger than observed in neutron-stars in binary systems, where the surface magnetic fields are typically between $\sim \left(10^8 - 10^{12}\right)\ \rm G$~\cite{Lorimer:2008se,Tauris:2017omb}.
It is nevertheless common in NR simulations to assume a strength up to $\sim 10^{15}\ \rm G$ just before the merger, e.g., \cite{Kiuchi:2022ninc,Hayashi:2022cdq,Kiuchi:2023obe,Most:2023sft,Combi:2022nhg,Mosta:2020hlh,Ruiz:2020via,Ciolfi:2019fie}. The idea of setting this strong magnetic field is to compensate for the unresolved small-scale dynamics, which lead to magnetic field amplification but are not captured by the limited resolution. 

Each system is simulated in a low-resolution (R1) and a high-resolution (R2) setup, respectively. The grid configuration in both cases consists of $L=8$ refinement levels, comprising three outer, non-moving, and five inner, moving levels. 
For the R1 setup, the inner and outer refinement boxes contain $n_{\rm mv}=128$ and $n=256$ points per direction, respectively, with a grid spacing of $h_L = 186\ \rm m$ on the finest level. We double the resolution in the R2 configuration, obtaining $n_{\rm mv}=256$ and $n=512$ points per direction for the inner and outer refinement boxes, respectively, with a grid spacing of $h_L = 93\ \rm m$ on the finest level.

\subsection{Equation of state}
\label{subsec:eos}

The ABHT(QMC-RMF3) EOS~\cite{Alford:2022bpp, Alford:2023rgp,qmc_rmf3} is specifically developed for dense matter in BNS mergers.
It is based on a relativistic mean-field theory (RMFT) that describes the interaction of nucleons via meson exchange. The free parameters of most RMFTs are fitted such that the model describes matter and nuclei that are nearly isospin-symmetric, i.e., have a proton fraction of roughly $50\, \%$, and then extrapolated to the neutron-rich matter in neutron stars. QMC-RMF3 is fitted to the binding energy per nucleon of pure neutron matter obtained by first-principle chiral effective field theory calculations~\cite{Tews:2018kmu}. Simultaneously, the nucleon-meson and meson-meson coupling constants are fitted for consistency with the zero-temperature properties of isospin-symmetric nuclear matter and observations of neutron star mass, radius, and tidal deformability. The EOS prescription includes neutrons, anti-neutrons, protons, anti-protons, electrons, positrons, and a non-interacting photon gas. At low densities, a thermodynamically consistent procedure is used to match the EOS in temperature, density, and proton-fraction space to a low-density RMFT called HS(IUF)~\cite{Hempel:2009mc,Fattoyev:2010mx} that is optimized for the more isospin-symmetric matter expected at low densities, and includes finite nuclei. Since the EOS is derived via RMFTs, it is always causal, and the extension of the EOS from zero temperature to finite temperatures is self-consistent. 

At nuclear saturation density $n_{\text{sat}}=\qty{0.1546}{\per\cubic\femto\metre}$, the ABHT(QMC-RMF3) EOS has a binding energy per nucleon $E_B=\qty{-16.39}{MeV}$, incompressibility $\kappa=\qty{231.3}{MeV}$, symmetry energy $J=\qty{31.29}{MeV}$, slope of the symmetry energy $L=\qty{47.20}{MeV}$, and effective nucleon mass divided its vacuum mass $m^*/m = 0.739$. These values are consistent with the white intersection region in Fig.~2 of Ref.~\cite{Drischler:2020hwi} and with Refs.~\cite{Shlomo:2006ole,Horowitz:2020evx}. The RMFT coupling constants' fitting further ensures consistency with the neutron matter binding energy derived from the chiral effective field theory calculation of Ref.~\cite{Tews:2018kmu} between $0.5\ n_0$ and $1.5\ n_0$ at zero temperature, where $n_0 = \qty{0.16}{\per\cubic\femto\metre}$. In addition, in Ref.~\cite{Alford:2023rgp} ABHT(QMC-RMF3) was shown to be consistent with the finite temperature chiral effective field theory calculation of Ref.~\cite{Keller:2020qhx} at $T=\qty{20}{MeV}$. The ABHT(QMC-RMF3) EOS predicts the following neutron star properties $R_{\qty{1.4}{\msun}} = \qty{12.21}{\kilo\meter}$, $M_{\text{max}} = \qty{2.15}{\msun}$, and $\Lambda_{\qty{1.4}{\msun}} = 387$, where $\Lambda$ is the dimensionless tidal deformability. These values are consistent with current astrophysical observations~\cite{LIGOScientific:2018cki,Miller:2019cac,Riley:2019yda,Miller:2021qha,Riley:2021pdl,Choudhury:2024xbk,Salmi:2022cgy,Salmi:2024aum,Dittmann:2024mbo,Fonseca:2021wxt,Vinciguerra:2023qxq}.

\section{Binary Neutron Star Merger Simulations}
\label{sec:Simulations}

\subsection{Qualitative overview}
\label{subsec:Overview}

As the main focus of our study lies on the merger and post-merger phase, our BNS simulations start with a relatively small initial separation and cover only the last few orbits of the inspiral phase. Both BNS configurations merge already after approximately $\sim 4.5$~orbits, emit GWs, and form a massive neutron star (MNS) remnant.
We define the merger time $t_{\rm merger}$ as the time when the amplitude of the GW strain in its dominant $(2,2)$ mode reaches its maximum. For the non-spinning and spinning BNS systems, the merger time is at $t_{\rm merger}=11.17\ \rm ms$ and $t_{\rm merger} = 12.05\ \rm ms$, respectively. As expected, the configuration with aligned spin $\chi=0.1$ merges slightly later because of the orbital hang-up effect~\cite{Campanelli:2006uy}. This is due to a repulsive spin-orbit interaction and has already been discussed in several studies of BNS mergers with rotating neutron stars, e.g., Refs.~\cite{Bernuzzi:2013rza,Dietrich:2015pxa,Ruiz:2019ezy,Kastaun:2016elu,Most:2019pac,Dudi:2021wcf}. 

\begin{figure*}[htp!]
    \centering
    \includegraphics[width=\linewidth]{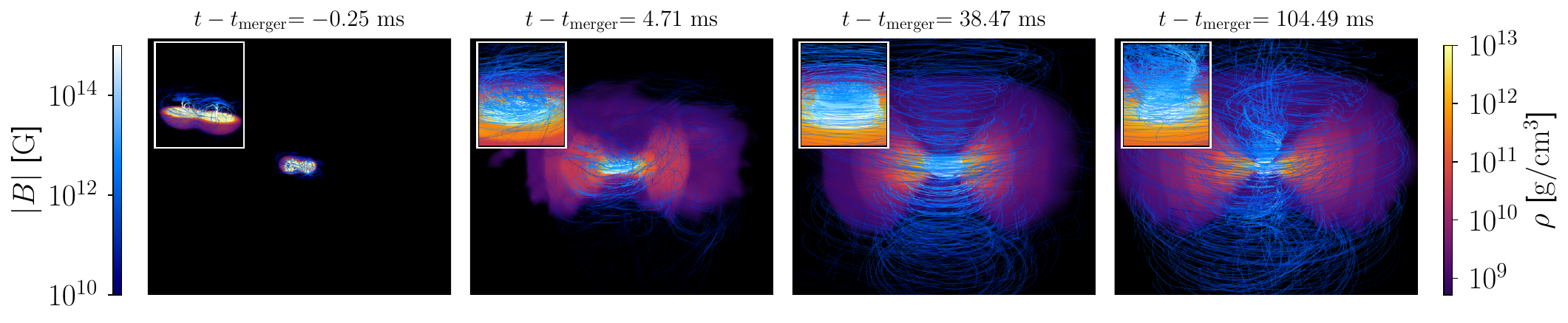}
    \caption{3D-Snapshots of the high-resolution simulation without spin. Each panel shows a rendering of the rest-mass density in purple to orange scales and magnetic field lines in blue. For visualization, the rest-mass density is sliced along the $x$-$z$ plane. Additionally, insets sliced along the $x$-$y$ plane are shown in the upper left corner of each panel. The snapshots are extracted from refinement level $l=3$ and represent different stages of the merger and post-merger phase, visualizing the evolution and winding of the magnetic field lines.}
    \label{fig:3D-magnetic}
\end{figure*}

Figure~\ref{fig:3D-magnetic} visualizes the matter dynamics, the ejecta expansion, and the magnetic field evolution as 3D snapshots of the BNS system without spin during and after the merger. 
Understanding the different mass ejection mechanisms, e.g., through magnetic and neutrino-driven winds, is particularly important for interpreting the associated EM counterparts.
The first snapshot presents the neutron stars at the onset of merger, exhibiting a purely poloidal magnetic field inside the stars as initialized. Subsequently, the magnetic field lines twist and wind up, leading to a strong toroidal field in the surrounding disk and forming a helicoidal structure at $\sim 40\ \rm ms$ after the merger. Later, at $\sim 100\ \rm ms$ after the merger, we observe a magnetic flux emanating from the remnant along the polar axis, forming a jet-like structure. 
In Sec.~\ref{subsec:MagneticField}, a more detailed analysis of the magnetic field amplification and mechanism is provided.

\begin{figure}[ht]
    \centering
    \includegraphics[width=\linewidth]{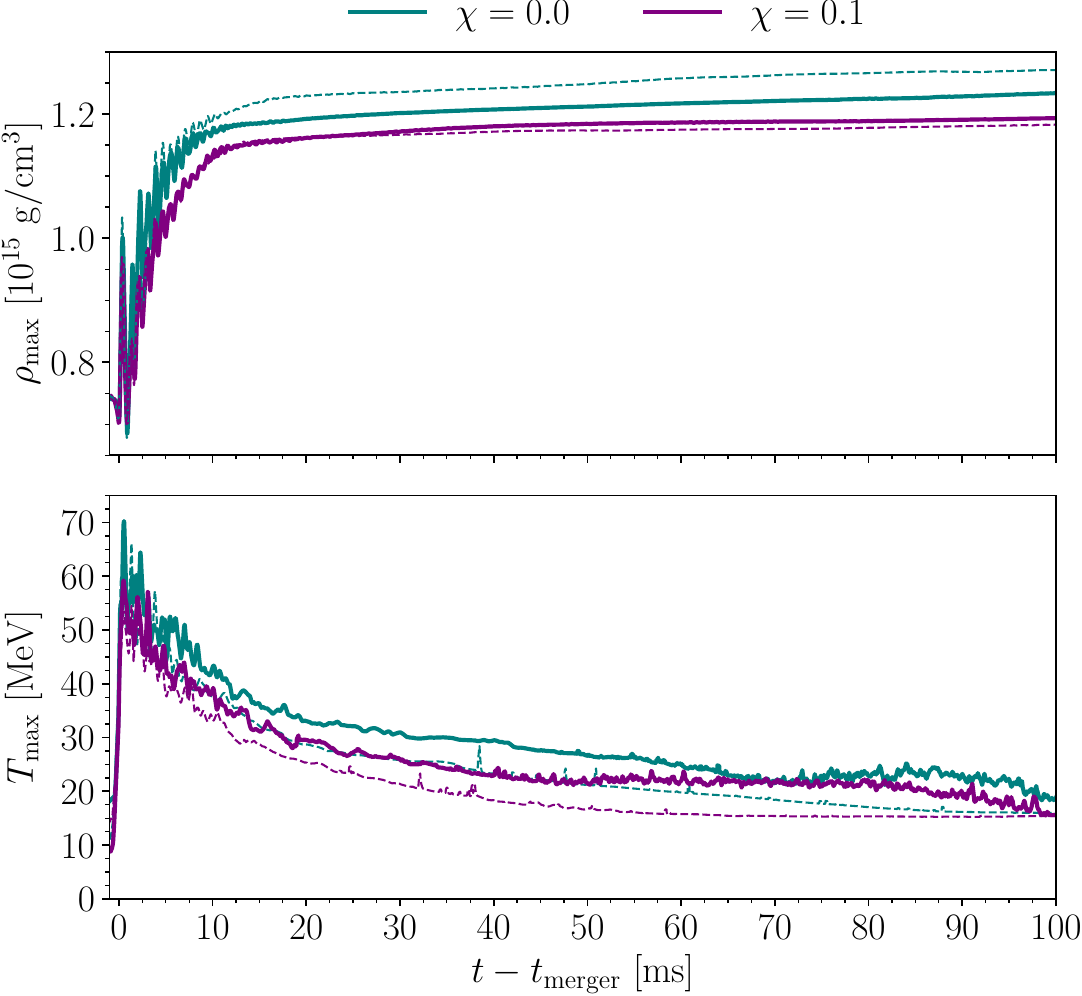}
    \caption{Evolution of maximum rest-mass density $\rho_{\rm max}$ (top panel) and maximum temperature $T_{\rm max}$ (bottom panel). Results of the R2 and R1 simulations are shown in solid and dashed lines, respectively, for the non-spinning (green) and spinning (purple) configurations extracted from refinement level $l=6$. For a smoother visualization, data is presented with a moving-average window of width $0.2\ \rm ms$.}
    \label{fig:rho-T}
\end{figure}

\begin{table}[tb]
\caption{Remnant properties of the BNS simulations. From left to right: initial dimensionless spin $\chi$ of the neutron stars, resolution of the simulation, remnant MNS mass $M_{\rm MNS}$, disk mass $M_{\rm disk}$, and ejecta properties, comprising its mass $M_{\rm eje}$, mass-averaged electron fraction $\left<Y_e^{\rm eje}\right>$, and mass-averaged asymptotic velocity $\left<v_{\infty}^{\rm eje}\right>$. The ejecta quantities are extracted on a sphere with radius $\sim 440\ \rm km$ on refinement level $l=2$. The remnant and disk mass are computed as the integral of the bound conserved rest-mass density $D$, defining the massive neutron star remnant in regions where $\rho > 10^{13}\ \rm g/cm^3$, at the end of the simulation.}
\label{tab:results}
\begin{tabular}{c c||c|c|c|c|c}
\toprule 
$\chi$  & res.  & $M_{\rm MNS}\ [M_{\odot}]$ & $M_{\rm disk}\ [M_{\odot}]$ & $M_{\rm eje}\ [M_{\odot}]$  & $\left<Y_e^{\rm eje}\right>$ & $\left<v_{\infty}^{\rm eje}\right>$ \\ 
\hline \hline
0.0 & R2 & 2.7910 & 0.1426 & 0.0137 & 0.407 & 0.165 \\
0.1 & R2 & 2.7480 & 0.1919 & 0.0098 & 0.438 & 0.148 \\
0.0 & R1 & 2.7992 & 0.1364 & 0.0233 & 0.406 & 0.192 \\
0.1 & R1 & 2.7371 & 0.2164 & 0.0124 & 0.397 & 0.184 \\
\bottomrule
\end{tabular}
\end{table}

Table~\ref{tab:results} provides a summary of the merger remnant properties, including remnant masses, disk masses, and key ejecta properties for all simulations performed. Each system produces an MNS remnant, which remains stable within the simulation time of $\sim 100\ \rm ms$ after the merger.
For the classification of ejected matter, both dynamical and secular ejecta, we use the geodesic criterion, according to which matter is considered unbound if the time-component of the four-velocity $u_t < -1$ and the radial component of the Eulerian velocity points outward, i.e., $v_r >0$; cf.~\cite{Hotokezaka:2012ze}. The gravitationally bound matter comprises the remnant system with the MNS and the surrounding disk. To obtain $M_{\rm MNS}$ and $M_{\rm disk}$ separately, we adopt the usual convention and define the disk where $\rho < 10^{13}\ \rm g/cm^3$ and the MNS where $\rho \geq 10^{13}\ \rm g/cm^3$~\cite{Shibata:2017xdx,Kiuchi:2019lls,Vincent:2019kor,Radice:2018pdn,Schianchi:2023uky}.

Our results show that the spin-aligned BNS configuration produces a remnant with a less massive MNS, but a more massive disk. We attribute this to the conservation of angular momentum, which leads to stronger centrifugal forces in the remnant system in the spinning BNS system. As a result, the remnant is more spatially extended, yielding a large amount of bound material, but at lower densities.
This is also confirmed in Fig.~\ref{fig:rho-T}, which shows the evolution of the maximum rest mass density (upper panel) and the maximum temperature (lower panel). The central rest-mass density and temperature of the remnant MNS are lower in the BNS systems with aligned spin, which is consistent with the results from Ref.~\cite{Most:2019pac}.
The ejecta mass is larger in the non-spinning BNS configuration with a higher mass-averaged asymptotic velocity $\left<v_\infty^{\rm eje}\right>$. We note that although there are stronger variations for the ejecta results at different resolutions, both resolutions predict the same trend for $M_{\rm eje}$ and $\left<v_\infty^{\rm eje}\right>$. However, the mass-averaged electron fraction $\left<Y_e^{\rm eje}\right>$ is larger for the $\chi=0.1$ BNS system at R2 resolution and smaller at R1 resolution. We discuss the ejecta properties in more detail in Sec.~\ref{subsec:Ejecta}.

Overall, the spin-aligned BNS configuration with repulsive spin-orbit interaction and increased gravitational binding energy results in a less violent merger scenario.
This is indicated in Fig.~\ref{fig:rho-T} by a smaller density variation between the late inspiral and the first density peak, when the neutron star cores first bounce and release substantial amounts of shocked ejecta. 
On the one hand, this also explains the previously mentioned larger ejecta mass and higher ejecta velocities for the non-spinning BNS configuration, where stronger core-bounces are regarded as the dominant ejection mechanism. While rotating BNS systems are typically expected to enhance the dynamical ejecta if tidal disruption plays a more significant role, e.g., in simulations with a different EOS or unequal component masses, aligned spin reduces the impact velocity and thus decreases shock-driven outflows~\cite{Dietrich:2016lyp,Dudi:2021wcf,East:2015vix}. For detailed studies on the ejection mechanisms in equal and unequal mass binaries, we refer the reader to Refs.~\cite{Dietrich:2015pxa,Radice:2018pdn} and, more recently, Ref.~\cite{Rosswog:2024vfe}. 
On the other hand, the higher maximum temperatures probed in the early post-merger corroborate our interpretation, given that more kinetic energy of the fluid elements may be converted to internal energy. Then, over longer timescales, radiative losses in the form of neutrino irradiation effectively cool the remnant. 

\subsection{Magnetic field amplification}
\label{subsec:MagneticField}

Various magnetohydrodynamic processes and instabilities operating across different spatial and temporal scales in BNS mergers influence the magnetic field evolution and amplify its strength and energy. One key mechanism is the Kelvin-Helmholtz instability (KHI), e.g.,~\cite{Price:2006fi,Kiuchi:2015sga,Giacomazzo:2014qba,Aguilera-Miret:2025nts}, which develops in the shear layer between the two colliding stars. Other magnetic instabilities are expected to arise in differentially rotating flows: Magnetic winding leads to a linear increase of the toroidal magnetic field component of the remnant on Alfvén timescales,e.g., ~\cite{Duez:2006qe,Sun:2018gcl,Kiuchi:2017zzg}. Distortions in the magnetic field in the radially decreasing angular velocity profile of the remnant disk can cause turbulence and induce magneto-rotational instability (MRI), e.g.,~\cite{Duez:2005cj,Siegel:2013nrw,Kiuchi:2023obe,Aguilera-Miret:2023qih}.
Furthermore, Refs.~\cite{Kiuchi:2023obe,Most:2023sme} demonstrated the presence of an $\alpha\Omega$-type dynamo in the post-merger remnant that contributes to magnetic field amplification over larger length scales, and Ref.~\cite{Reboul-Salze:2024jst} proposed a Taylor-Spruit dynamo to occur in the MNS remnant.

\begin{figure}[t]
    \centering
    \includegraphics[width=\linewidth]{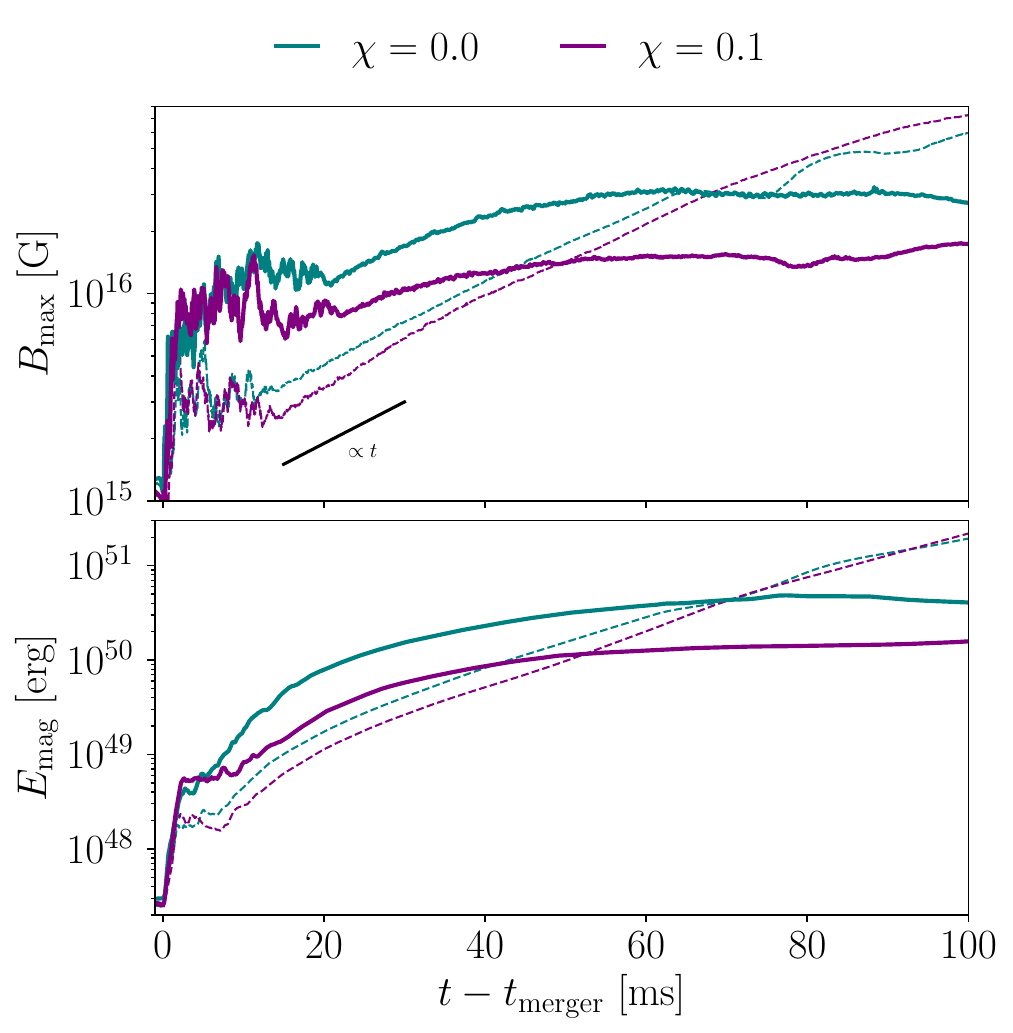}
    \caption{Evolution of maximum magnetic field strength $B_{\rm max}$ (top panel) and magnetic energy $E_{\rm mag}$ (bottom panel). Results of the R2 and R1 simulations are shown in solid and dashed lines, respectively, for the non-spinning (green) and spinning (purple) configurations extracted from refinement level $l=6$ for $B_{\rm max}$ and $l=1$ for $E_{\rm mag}$.}
    \label{fig:B-Emag}
\end{figure}

In Fig.~\ref{fig:B-Emag}, we show for both BNS configurations simulated with R1 and R2 resolutions the temporal evolution of the maximum magnetic field strength $B_{\rm max}$ and the magnetic energy $E_{\rm mag}$, defined as
\begin{equation}
    E_{\rm mag} = \frac{1}{2} \int u^t \sqrt{-g} b^2\, d^3x,
\end{equation}
with $b^2=b^\mu b_\mu$ and $b^\mu$ being the magnetic four-vector that describes the magnetic field for a comoving observer.
The initial amplification within the first milliseconds after the merger, typically ascribed to KHI, is considerably stronger at higher resolution since the shear layer is better resolved. In both R2 simulations, a similar magnetic field strength of $10^{16}\ \rm G$ is reached. Only later, at $\gtrsim 15\ \rm ms$ after the merger, as the magnetic energy and strength increase further (potentially through magnetic winding and/or MRI) deviations between the two systems become more pronounced. At R2 resolution, the non-spinning and spinning BNS configurations saturate at energies of about $5 \times 10^{50}\ \rm erg$ and $1 \times10^{50}\ \rm erg$, respectively. By contrast, the magnetic field strength continuously increases $\propto t$ throughout the entire simulation period for the R1 simulations.

The resolution dependence is not unexpected, since the length scales that must be resolved to capture the relevant magnetohydrodynamic processes and turbulence that trigger these instabilities are extremely small~\cite{Kiuchi:2015sga,Kiuchi:2017zzg,Kiuchi:2023obe}. 
In particular, KHI and MRI require very high resolutions, whereas magnetic winding and braking act on larger scales, which are easier to resolve. Consequently, the R1 resolution cannot resolve the same physical processes as R2. While the approximately linear growth of the magnetic field observed in the R1 simulations can be attributed to magnetic winding~\cite{Duez:2006qe}, magnetic braking does not set in within the simulated time, and neither the amplification through KHI nor the MRI is sufficiently resolved at this low resolution.
We therefore focus our analysis on results from our high-resolution simulations. Particularly, we discuss below the roles of KHI and MRI in magnetic field amplification.

\begin{figure*}[t]
    \centering
    \includegraphics[width=\linewidth]{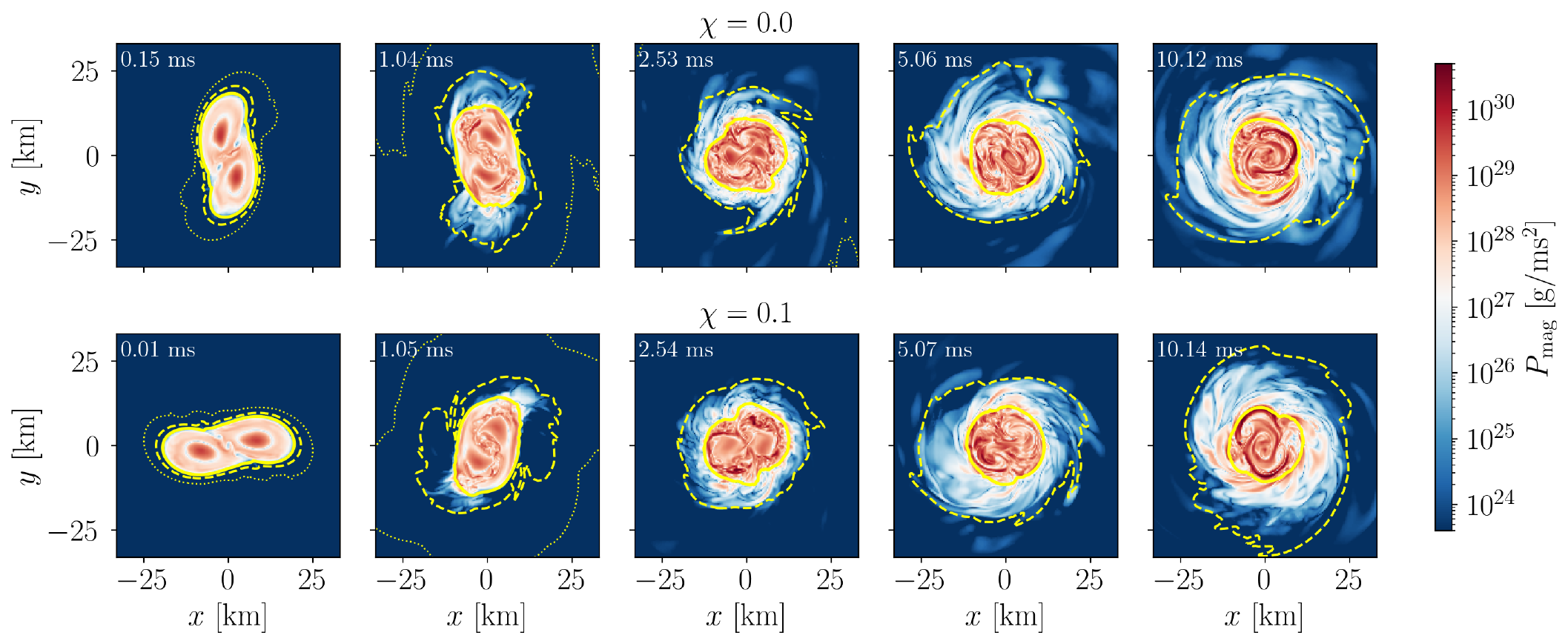}
    \caption{Shear layer of the non-spinning (upper panels) and spinning (lower panels) BNS systems at merger time for the simulations with R2 resolution. The snapshots show the magnetic pressure $P_{\rm mag} = 0.5 b^2$ in the $x$-$y$ plane at $\{0,1,2.5,5,10 \}\ \rm ms$ after the merger at refinement level $l=5$. The yellow lines are contours of the rest-mass density at $\rho = \{ 10^{10},10^{12},10^{14}\} \ {\rm g/cm^3}$, respectively as dotted, dashed, and solid lines.}
    \label{fig:KHI2D}
\end{figure*}

\subsubsection{Kelvin-Helmholtz instability}

Once the two neutron stars come into contact, KHI becomes prominent in the shear layer between the stars. To illustrate this process, Fig.~\ref{fig:KHI2D} shows the magnetic pressure $P_{\rm mag} = b^2/2$ for our R2 simulations in the orbital plane of the non-spinning (upper panels) and spinning (lower panels) BNS configurations at the merger and up to $\sim 10\ \rm ms$ afterwards. Additional contour lines for the rest-mass density at $\rho = \{10^{10},10^{12},10^{14}\}\ {\rm g/cm^3}$ mark the envelope, the bulk, and the inner core regions, respectively.

At the merger onset (first panel from left to right), most of the magnetic field is still found within the inner core region, as set by the initial data. The formation of small eddies in the shear layer between the two stars can be observed. The vortices during the merger further twist the magnetic field lines within the first few milliseconds, leading to an exponential growth. This is also evident in Fig.~\ref{fig:B-Emag} for the initial rise of the magnetic field strength and energy in the first $\lesssim 10\ \rm ms$ after the merger. The snapshots in Fig.~\ref{fig:KHI2D} suggest that KHI mainly amplifies the magnetic field within the inner core region and the bulk region of the remnant.
We note again that the energies and intensities of the magnetic field reached through the amplification by KHI within the first $\lesssim 10\ \rm ms$ are similar in both BNS systems for R2 resolution. 

For a more comprehensive analysis of the energy distribution over spatial scales, we compute the energy power spectrum density for the kinetic and magnetic energy.
For this purpose, we define $\epsilon_{\rm mag} = \sqrt{0.5 b^2}$ and $\epsilon_{\rm kin} = \sqrt{\rho v^2}$ and compute the three-dimensional Fourier transform
\begin{align}
    \hat{\epsilon}_{\rm mag}(\vec{k}) &= \frac{1}{\left(2 \pi \right)^3}\int \epsilon_{\rm mag}(\vec{x}) e^{i \vec{k}\cdot \vec{x}} d^3x,\\
    \hat{\epsilon}_{\rm kin}(\vec{k}) &= \frac{1}{\left(2 \pi \right)^3}\int \epsilon_{\rm kin}(\vec{x}) e^{i \vec{k} \cdot \vec{x}} d^3x,
\end{align}
with the wave vector $\vec{k}$ and the position of a fluid element $\vec{x}$. Then, we obtain the energy spectra as integrals over the solid angle $\Omega_k$ in $k$-space via
\begin{align}
    \varepsilon_{\rm mag}\left(k\right) &= \int  \hat{\epsilon}_{\rm mag}(\vec{k})  {\hat{\epsilon}^*_{\rm mag}}(\vec{k}) k^2 d\Omega,\\
    \varepsilon_{\rm kin}(k) &= \int  \hat{\epsilon}_{\rm kin}(\vec{k})  {\hat{\epsilon}^*_{\rm kin}}(\vec{k}) k^2 d\Omega,
\end{align}
with wave number $k=(\vec{k}\cdot\vec{k})^{1/2}$ defining the characteristic length of the system.

We perform this calculation on a cube with a side length $L$ of $\sim 100\ \rm km$ along the $x$, $y$, and $z$ directions for a sufficient coverage of the merger remnant, including the inner core, the bulk, and the envelope. This region matches the computational domain of refinement level $l=5$, and is thus partially covered by nested boxes of refinement levels $l=6$ and $l=7$.
To ensure that the analysis is performed using information from the finest grid available, we follow a similar approach to that described in Ref.~\cite{Aguilera-Miret:2020dhz}. In particular, we set for the cube the same grid spacing as on our finest refinement level $l=7$ and fill the cube with data from this level in the respective region. The rest of the domain is then interpolated using information from refinement levels $l=6$ and $l=5$, respectively. 
We note that this method can influence results on the smallest scales, and therefore at the highest wave numbers $k$, due to the interpolation from coarser grids to a finer grid in the outer refinement levels. To avoid fluctuations from low-density regions, we compute the spectra only in regions with a rest-mass density $\rho > 6 \times 10^{9}\ \rm g/cm^3$. Finally, we perform a discrete fast Fourier transform over our cube with $N^3$ equally spaced points, where the components of the wave vector are given by $k_d = n \frac{2 \pi}{L}$ with $n \in \left[ 0, N/2\right]$.

\begin{figure*}[t]
    \centering
    \includegraphics[width=\linewidth]{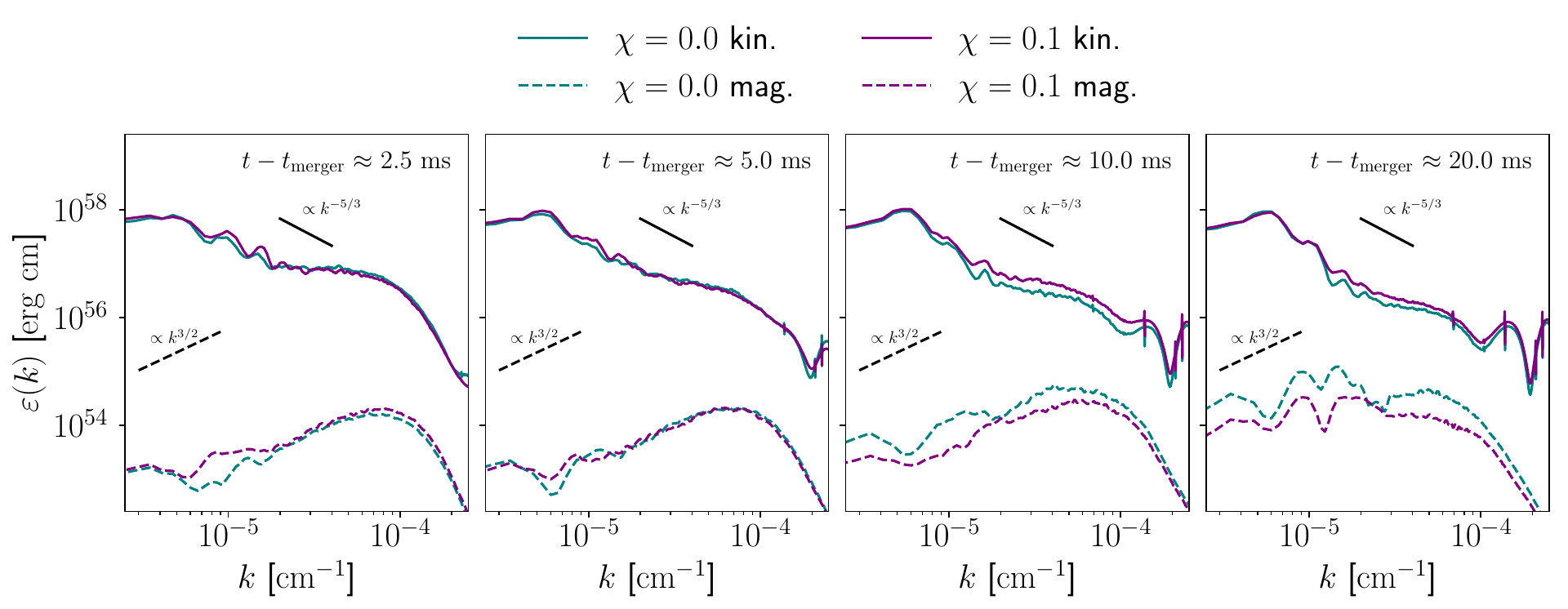}
    \caption{Evolution of the kinetic (solid) and magnetic (dashed) energy spectra as a function of the wave number. Results are shown for the non-spinning (green lines) and spinning (purple lines) high-resolution BNS simulations at $t=\{2.5, 5, 10, 20\}\ \rm ms$ after the merger. The black lines correspond to Kolmogorov (solid) and Kazantsev (dashed) slopes.}
    \label{fig:KHIspectra}
\end{figure*}

We present our resulting spectral distributions for the kinetic and magnetic energy in Fig.~\ref{fig:KHIspectra} for $t = \{2.5,5,10,20\}\ \rm ms$ after the merger.
The spectra exhibit the expected behavior for a turbulent magnetohydrodynamic scenario with a weak magnetic field, in which the magnetic field evolves passively and the turbulence is predominantly hydrodynamical. The shape of the kinetic energy spectra follows approximately a $k^{-5/3}$ slope, consistent with Kolmogorov's theory~\cite{Kolmogorov:1991}. Moreover, during the first $10\ \rm ms$ after the merger, the magnetic energy spectra can be approximated at longer wavelengths (smaller wave numbers), by a slope of $k^{3/2}$, in agreement with Kazantsev’s model~\cite{Kazantsev:1968}.
Accordingly, the kinetic energy is dominated by larger length scales and the magnetic energy by smaller ones. This also highlights the importance of numerical resolution for capturing KHI. Although resolution is expected to have less impact on the evolution of the kinetic energy on large scales, most magnetic energy is created on small scales. The drop in the magnetic energy spectra at high wave numbers $\gtrsim 10^{-4}\ \rm cm^{-1} $ corresponding to length scales of the order $\lesssim 100\ \rm m$ is likely due to numerical dissipation on these length scales, as our grid spacing is of similar order with $\Delta x \approx 93\ \rm m$.

Our results reproduce the typical dynamical stages of the KHI, as discussed in previous studies, e.g., Refs.~\cite{Zrake:2013mra,Aguilera-Miret:2020dhz}. Initially, within the first few milliseconds after the merger, the hydrodynamic cascade is fully developed, following the Kolmogorov spectral slope. Subsequently, during the kinematic phase, the magnetic field grows by a hydrodynamical turbulent mechanism, as in Kazantsev’s theory, until reaching saturation.
The spectra for the two different BNS systems during the first few milliseconds after the merger are largely identical. Thus, the initial magnetic field amplification through KHI seems independent of the star's intrinsic spin, as it behaves identically in both BNS systems.
Differences become noticeable only at later times of $\geq 10\ \rm ms$ after the merger. On these timescales, the magnetic energy spectra also increase at smaller wave numbers, indicating an inverse cascade. In agreement with Fig.~\ref{fig:B-Emag}, the magnetic energy amplifies more strongly for the BNS system without spin. We conclude that this difference for the BNS systems is related to the transfer of energy at larger scales, potentially through magnetic winding or MRI-driven dynamo mechanisms, which we discuss in the following section.

\begin{figure*}[t]
    \centering
    \includegraphics[width=\linewidth]{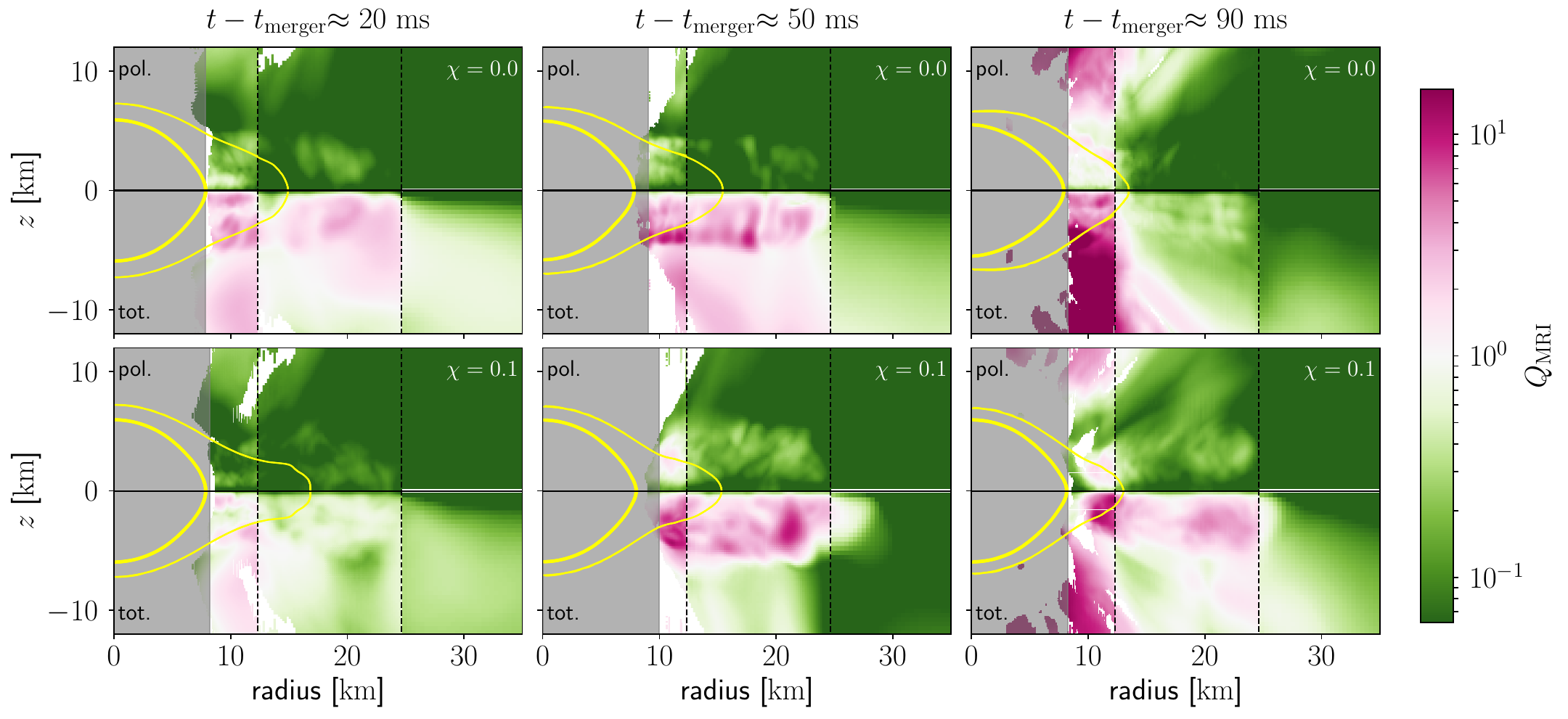}
    \caption{MRI quality factor $Q_{\rm MRI}$ of the non-spinning (upper panels) and spinning (lower panels) BNS simulations with R2 resolution. The snapshots show the azimuthal average of $Q_{\rm MRI}$ at $t=\{20,50,90\}\ \rm ms$ after the merger. Each panel shows, in the upper half, the quality factor considering only the poloidal magnetic field component and, in the lower half, considering the full magnetic field strength. The yellow lines are contours of the rest-mass density at $\rho = \{10^{13},10^{14.5}\} \ {\rm g/ cm^3}$, respectively as thick and thin lines. The black dashed lines mark the regions covered by the different refinement levels $l=\{5,6,7\}$. While $Q_{\rm MRI}$ is set to zero if the condition $\partial_R\Omega<0$ is not satisfied, the gray-colored region marks the area where MRI is inactive.}
    \label{fig:MRI2D}
\end{figure*}

\subsubsection{Magneto-rotational instability}
 
The interaction of the magnetic field with a sheared background flow, such as in an accretion disk, is expected to power MRI~\cite{Balbus:1998ja}. Several studies have investigated its role in BNS merger scenarios, e.g.,~\cite{Duez:2005cj,Siegel:2013nrw,Kiuchi:2023obe,Aguilera-Miret:2023qih}, suggesting that MRI could sustain turbulence in the remnant's envelope and drive a large-scale dynamo. 

MRI is only active for a radially decreasing angular velocity, $\partial_R \Omega <0$, where distortions of the magnetic field lines and the velocity field lead to turbulence and exponential growth. Its fastest-growing mode can be estimated as
\begin{equation}
    \lambda_{\rm MRI}^i \approx v_A^i \frac{2 \pi}{\Omega}
    \label{eq:MRI}
\end{equation}
where $v_A^i = \frac{B^i}{\sqrt{4 \pi \rho}}$ is the Alfvén velocity in $x^i$ direction and $\Omega$ is the angular velocity of the fluid, which we simplify by $\Omega = (xv_y - yv_x)/R^2 $ with cylindrical radius $R$. 
We note that this standard MRI criterion assumes a Keplerian rotation profile and neglects magnetic field gradients (see Ref.~\cite{Celora:2025fxm} for a derivation of extended MRI criteria). Furthermore, neutrino viscosity and drag may suppress MRI through diffusive and damping effects. Thus, Eq.~\eqref{eq:MRI} should be regarded as a rough approximation.

\begin{figure*}[htp!]
    \centering
    \includegraphics[width=\linewidth]{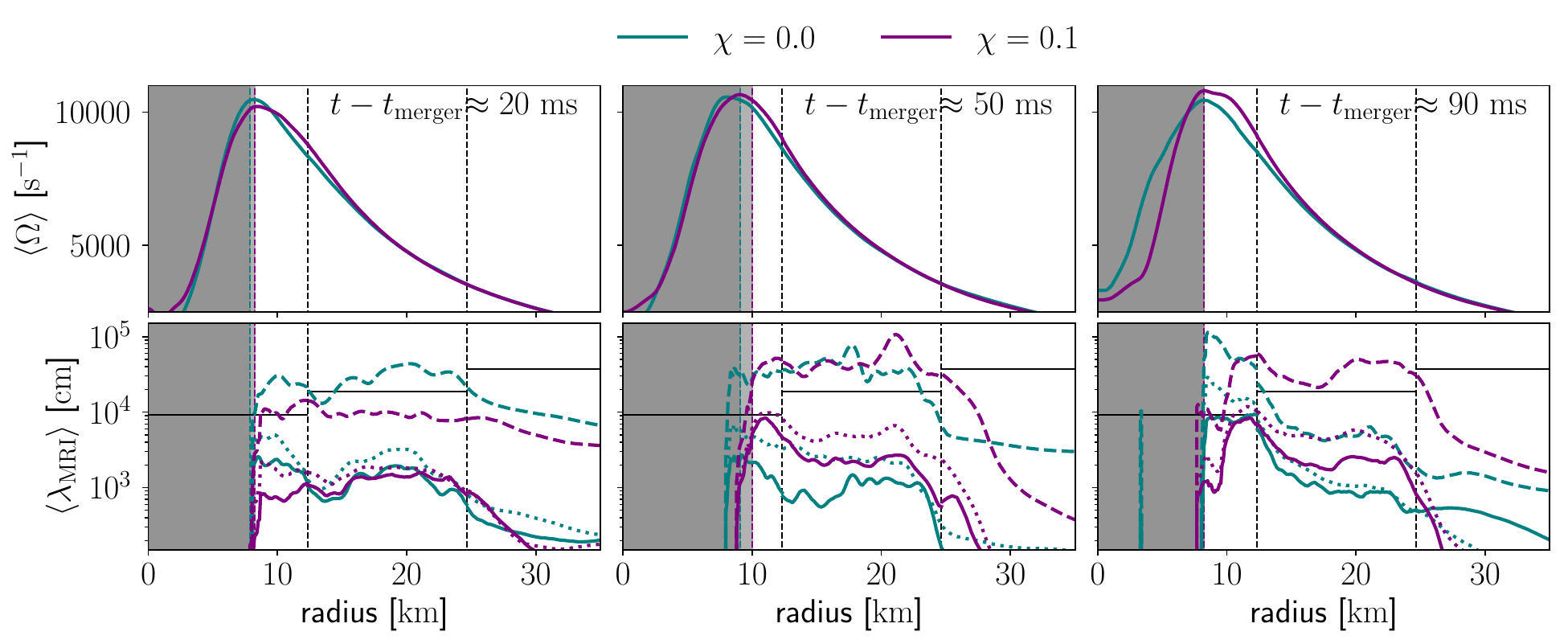}
    \caption{Profiles for the fluid angular velocity $\Omega$ in the top panels and the fastest-growing MRI mode $\lambda_{\rm MRI}$ in the bottom panels considering different components, i.e., poloidal $\lambda_{\rm MRI}^z$ in solid lines, toroidal $\lambda_{\rm MRI}^\Phi$ in dashed lines, and radial $\lambda_{\rm MRI}^R$ in dotted lines. In particular, the profiles show the averaged values over cylindrical radii $R$ for $|z|<5\ \rm km$ at $t=\{20,50,90\}\ \rm ms$ after the merger for both the non-spinning (green) and spinning (purple) BNS simulations at R2 resolution. The black vertical, dashed lines mark the regions covered by the different refinement levels, i.e., $l=\{5,6,7\}$, and the black horizontal lines in the bottom panel mark the respective grid spacing $\Delta x$. While the $\lambda_{\rm MRI}^i$ components are set to zero if the condition $\partial_R\Omega<0$ is not satisfied, the gray-colored region marks the area where MRI is inactive, respectively contoured by green and purple vertical, dashed lines for the non-spinning and spinning BNS systems.}
    \label{fig:MRI1D}
\end{figure*}

We compute azimuthal averages of the quality factor, defined as
\begin{equation}
    Q_{\rm MRI}=\lambda_{\rm MRI}/\Delta x,
\end{equation}
and present results for $t=\{20,50,90\}\ \rm ms$ after the merger for both BNS systems in Fig.~\ref{fig:MRI2D}. For each panel, we show the quality factor based on the poloidal magnetic field component only, in the upper half, and based on the total magnetic field strength, in the lower half.
Our results emphasize the challenges in resolving $\lambda_{\rm MRI}$. During the first $\lesssim 50\ \rm ms$ after the merger, our resolution can only partially capture MRI effects, considering the full magnetic field strength. However, our grid resolution is not suited to individually resolve the poloidal component. As the magnetic field strength increases, resolving $\lambda_{\rm MRI}$ becomes feasible, because $\lambda_{\rm MRI}^i \propto B^i$, which leads to a higher quality factor. 

On average, the quality factor $Q_{\rm MRI}$ seems slightly higher in the non-spinning than in the spinning configuration. This trend may explain the stronger magnetic field amplification observed in this BNS system, which indicates that MRI is better resolved.
For better comparison, we compute one-dimensional profiles of the fluid angular velocity and different $\lambda_{\rm MRI}^i$ components, shown in Fig.~\ref{fig:MRI1D}. 
In particular, we present the averaged values $\left< \Omega\right>$ and $\left<\lambda_{\rm MRI}^i\right>$ over cylindrical radii $R$ for a disk within $|z|<5\ \rm km$. This selection focuses on the orbital plane, where $Q_{\rm MRI}$ first begins to grow. As in Fig.~\ref{fig:MRI2D}, the profiles are shown at $t=\{20,50,90\}\ \rm ms$ after the merger. We also plot the grid spacing $\Delta x$ for the respective domain alongside the $\left<\lambda_{\rm MRI}^i\right>$ profiles: MRI features can only be resolved when $\left<\lambda_{\rm MRI}^i\right>$ lies above $\Delta x$, i.e., when the corresponding length scale exceeds the grid resolution.

Again, we find that the poloidal component $\lambda_{\rm MRI}^z$ is not resolved in our simulations, as the grid spacing $\Delta x$ on the corresponding refinement levels exceeds the required length scales. The same holds for the radial component $\lambda_{\rm MRI}^R$. Only the toroidal component $\lambda_{\rm MRI}^\Phi$ exceeds $\Delta x$ and may therefore be partially resolved.
Indeed, at $\sim 20\ \rm ms$ after the merger, $\lambda_{\rm MRI}^\Phi$ is larger for the non-spinning BNS configuration than for the spinning one. This could explain the faster growth of the magnetic energy and field strength for this BNS system, as the MRI, although not fully resolved, is still better captured. 
The higher $\lambda_{\rm MRI}$ values can easily be explained by the slightly lower angular velocities in the MRI-active region. Since $\lambda_{\rm MRI}$ is inversely proportional to $\Omega$, higher angular velocities reduce the length scale of the fastest-growing MRI mode. Consequently, it is more difficult to resolve MRI in the spinning BNS configuration.

\subsection{Ejecta dynamics and matter outflow}
\label{subsec:Ejecta}

BNS mergers are sources of massive outflows, whose kinematical, thermodynamical, and compositional properties are closely connected to EM counterparts arising from such events. Therefore, realistic modeling of the relevant physical processes in the ejecta is necessary in order to provide accurate theoretical predictions of observable signals and to interpret the available observational data.

The various ejection mechanisms operate on distinct timescales. It is hence instructive to divide the ejecta into a dynamical component and a secular component. The former is primarily driven by hydrodynamical processes and develops on timescales from a few milliseconds prior to the merger to tens of milliseconds after the merger. First, tidal torques act on deformed surface layers, leading to typically cold, neutron-rich outflows. Shocked matter streams follow these, characterized by high entropies and electron fractions $Y_e \gtrsim 0.25$. The shocked material is predominantly launched from the collision interface between the neutron stars when they first come into contact and then by successive core-bounces as the remnant evolves towards a quasi-equilibrium state. 

\begin{figure}[t]
    \centering
    \includegraphics[width=\linewidth]{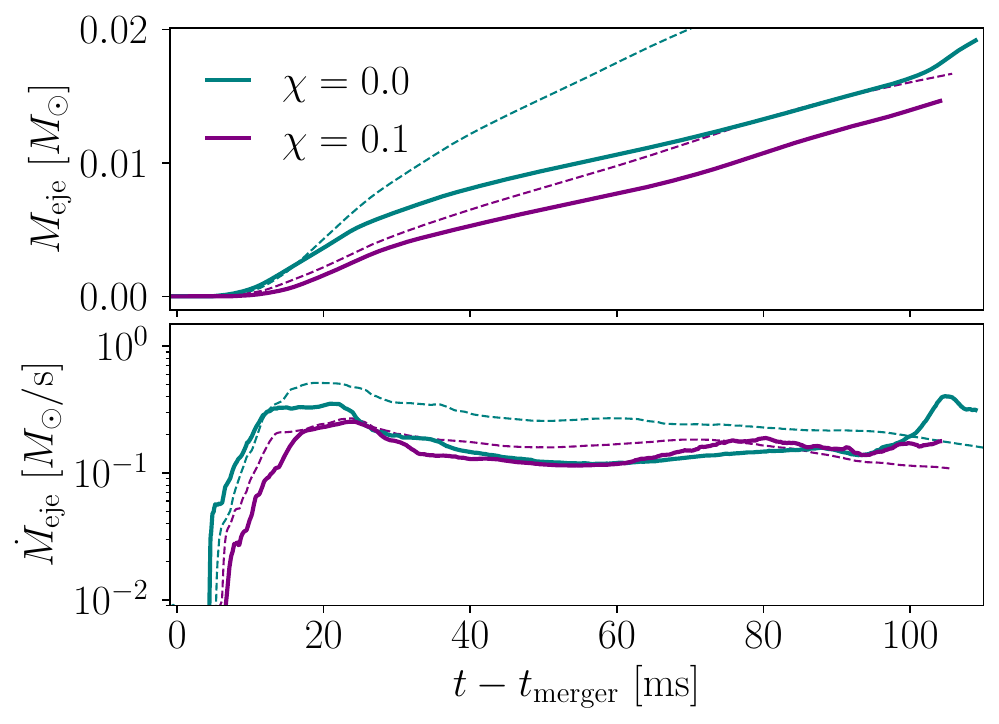}
    \caption{Matter outflow for the non-spinning (green) and spinning (purple) BNS simulations at R1 (dashed) and R2 resolution (solid). The ejecta mass (top panel) and ejected mass flux (bottom panel) are extracted on a coordinate sphere of radius $\sim 737\ \rm km$ on refinement level $l=2$.}
    \label{fig:ejectamass}
\end{figure}

Over longer timescales $t - t_{\rm merger} \gtrsim 10\ \rm ms$, where the internal motions of the remnant have subsided due to energy losses in the form of GWs, secular ejection mechanisms operate, resulting in a steady outflow of material from the disk and outer layers of the remnant, commonly interpreted as winds. At this stage, a component of the wind is induced by matter elements that acquire energy and momentum by absorption of neutrinos, most importantly $\nu_e$. This increases the electron fraction to $Y_e \gtrsim 0.4$. Another component of the wind is driven by magnetic fields as a consequence of the amplification mechanisms explored in Sec.~\ref{subsec:MagneticField}. Finally, although not considered in this manuscript, viscous effects might contribute to the production of secular ejecta by angular momentum transport, e.g., \cite{Fernandez:2013tya,Just:2014fka,Fujibayashi:2017puw,Radice:2018pdn,Nedora:2019jhl,Shibata:2019wef}.

The following analysis is based on the methods of Ref.~\cite{Schianchi:2023uky}, where properties of the ejecta are extracted on coordinate spheres of constant radii. 
The mass flux across a coordinate sphere of radius $r$ is
\begin{equation}
    \dot{M}_{\rm eje} = \oint [\sqrt{\gamma}D_{\rm u}(\alpha v^r - \beta^r)] r^2d\Omega,
\end{equation}
where $D_u$ is the Eulerian rest-mass density of unbound matter, $\gamma$ is the determinant of the spatial metric, $\alpha$ is the lapse function, $\beta^r$ ($v^r$) is the shift (Eulerian velocity) vector projected in the radial direction, and $d\Omega$ is the usual surface element over the sphere. Integrating the mass flux in time from $t' = 0$ to $t' = t$ then gives the ejecta mass $M_{\rm eje}$ that crossed a coordinate sphere until $t$.

\begin{figure*}[t]
    \centering
    \includegraphics[width=\linewidth]{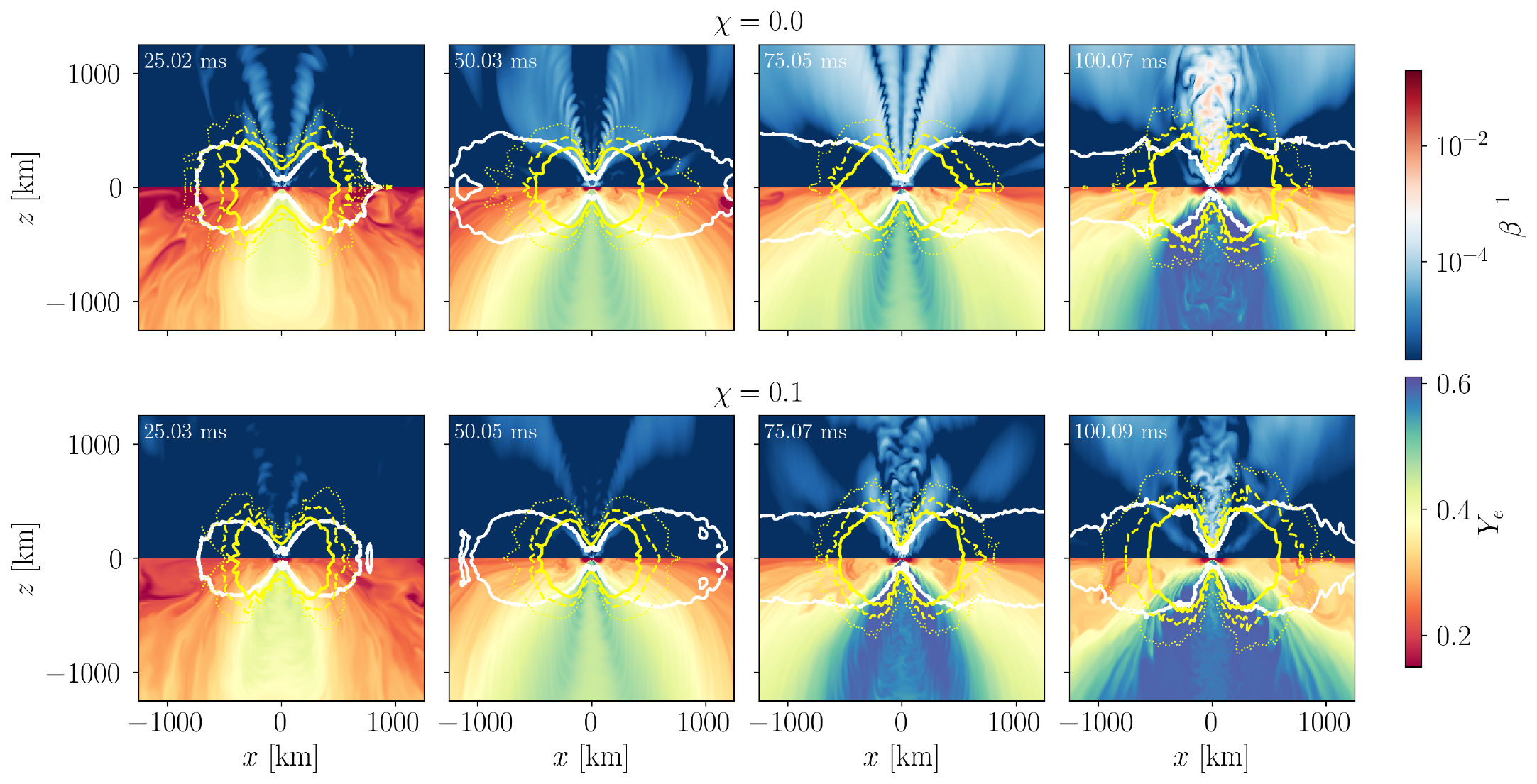}
    \caption{Snapshots of the inverse plasma parameter $\beta^{-1}$ ($z>0$) and the electron fraction $Y_e$ ($z<0$) of the non-spinning (upper panels) and spinning (lower panels) R2 simulations. The respective times after the merger are given in the upper left corner of each panel. The white line gives the contour of the gravitationally bound region. The snapshots are extracted from refinement level $l=1$. The yellow lines are contours of the rest-mass density at $\rho = \{ 0.25,0.5,1\} \times 10^{7} \ \rm g / cm^3 $, respectively as dotted, dashed, and solid lines.}
    \label{fig:BNS-remnant}
\end{figure*}

\begin{figure}[t!]
    \centering
    \includegraphics[width=\linewidth]{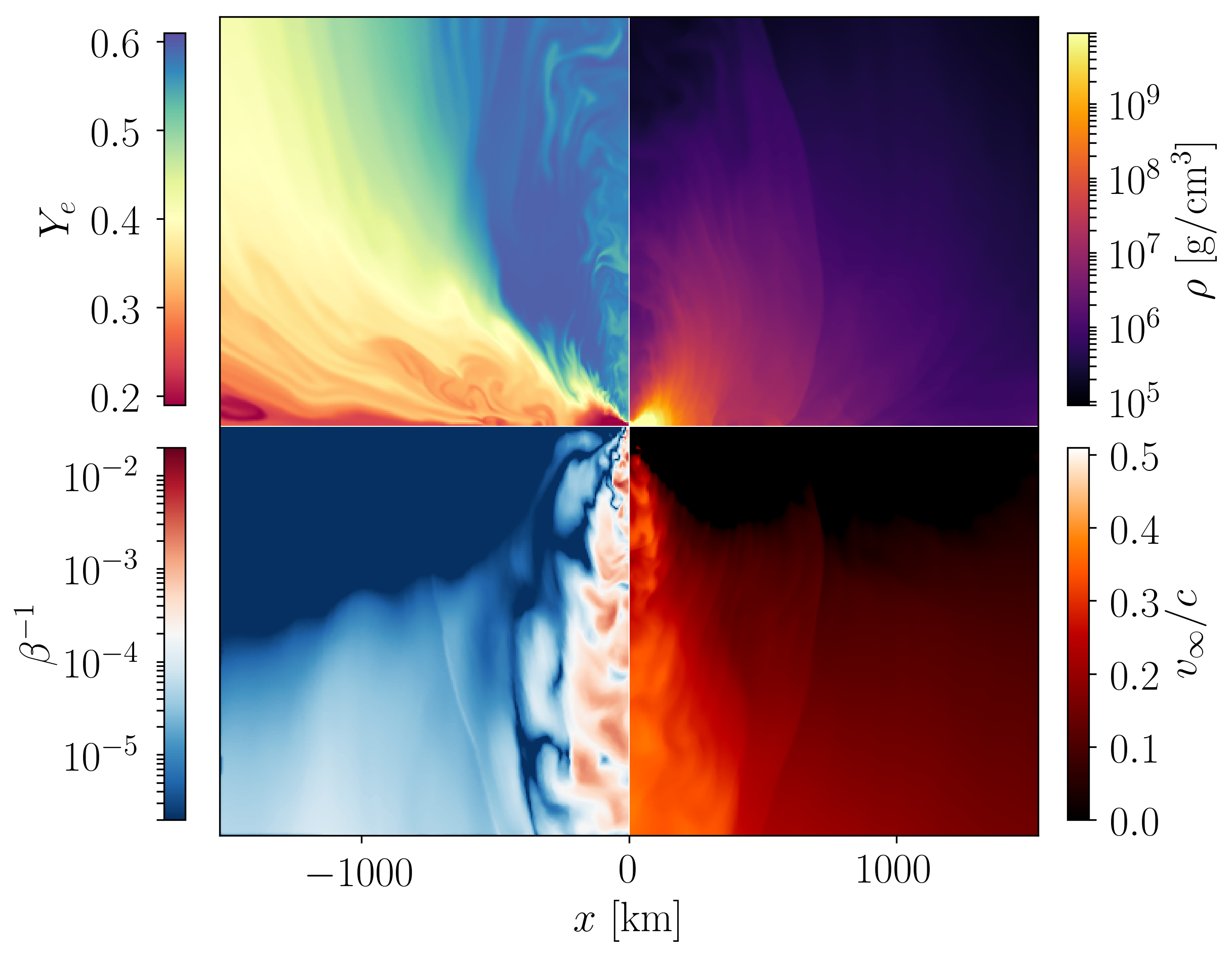}
    \caption{Snapshots of the electron fraction $Y_e$, the rest-mass density $\rho$, the inverse plasma parameter $\beta^{-1}$, and the asymptotic velocity $v_{\infty}$ in the $x$-$z$ plane for the high-resolution simulation of the $\chi = 0.0$ system at $\sim 103.5\ \rm ms$ after the merger.}
    \label{fig:2D-snapshots}
\end{figure}

Figure~\ref{fig:ejectamass} presents the evolution of the ejecta mass passing through a sphere with radius $\sim 737\ \rm km$ for the BNS simulations with and without spin at R1 and R2 resolutions. The top and bottom panels, respectively, show the integrated ejecta mass $M_{\rm eje}$ and the corresponding mass flux $\dot{M}_{\rm eje}$.
Due to the finite propagation time, the fastest ejecta reach the extraction sphere $\sim 5\ \rm ms$ after the merger. The mass flux then reaches values of $3 \times 10^{-1}\ M_\odot / \rm s$ and $2 \times 10^{-1}\ M_\odot / \rm s$ for the non-spinning and spinning BNS configurations, respectively. The R1 simulation of the BNS system without spin even shows peak values of $5 \times 10^{-1}\ M_\odot / \rm s$. Accordingly, the merger of the non-spinning BNS system produces considerably more dynamical ejecta. As discussed in Sec.~\ref{subsec:Overview}, the system without spin merges more violently than the spin-aligned configuration, resulting in more material being ejected during the collision.
Subsequently, as the remnant migrates to a more axis-symmetric state, the outflow rate decreases slightly in both systems. At high resolution, both systems exhibit a nearly constant mass flow of approximately $1 \times 10^{-1}\ M_\odot / \rm s$ from $20\ \rm ms$ to $60\ \rm ms$ after the merger, leading to an almost linear increase of the ejecta mass. 
This behavior is commonly observed in BNS merger simulations with M1 neutrino transport. Thus, we attribute this outflow component to neutrino-driven winds, as interactions of neutrinos emitted from the remnant steadily unbind material from the outer parts of the disk.
While the mass flux in the R1 simulations continue to decrease, we observe a slight increase in the R2 simulations at $\geq 60\ \rm ms$ after the merger. This time roughly coincides with the saturation of the magnetic field strength (cf. Fig.~\ref{fig:B-Emag}), which could indicate the onset of magnetically driven winds. 
Around $\geq 95\ \rm ms$, there is another increase in the mass flux for the non-rotating configuration, with peak values of $4 \times 10^{-1}\ M_\odot / \rm s$ at $\sim 105\ \rm ms$ after the merger. We analyze this outflow component in more detail below. \\

In order to better understand geometric features of the ejection mechanisms, Fig.~\ref{fig:BNS-remnant} depicts snapshots of the inverse plasma parameter $\beta^{-1} = P_{\rm mag}/p$ and the electron fraction $Y_e$ in the $x-z$ plane for the two high-resolution simulations, illustrating the evolution of the disk and the ejecta.
At $25\ \rm ms $ after the merger, the electron fraction ranges from $\sim 0.1$ in the orbital plane to $\sim 0.4$ in the polar region. The low $Y_e$ material in the orbital plane is primarily emitted by tidal torques during the merger and later partially reprocessed by the interaction with faster shocked ejecta.
The shock created by the collision between the two ejecta components causes a modest protonization of the matter due the heating and consequent emission of $\bar\nu_e$, which decouples from matter deeper in the remnant and disk than $\nu_e$~\cite{Endrizzi:2019trv, Perego:2019adq}. In the polar region, the electron fraction gradually increases due to absorption of $\nu_e$ produced in the remnant, which powers the steady secular outflow observed in Fig.~\ref{fig:ejectamass}.

While the magnetic field is amplified by the various instabilities discussed in Sec.~\ref{subsec:MagneticField}, the neutrino-driven wind reduces baryon pollution in the polar region, forming a funnel region with suitable conditions for the development of jet-like structures \cite{Sun:2022vri,Mosta:2020hlh}. In this region, the magnetic field forms a helicoidal structure (see Fig.~\ref{fig:3D-magnetic}) and $\beta^{-1}$ grows.
Indeed, we observe for our simulations the emergence of an outflow with considerably enhanced $\beta^{-1}$ value starting at $\sim 50\ \rm ms$ and $\sim 75\ \rm ms$ after the merger for $\chi=0.0$ and $\chi=0.1$, respectively. This supports the previous hypothesis that magnetic-driven winds contribute to the increase of mass flux observed in Fig.~\ref{fig:ejectamass} at this time.

\begin{figure}[t]
    \centering
    \includegraphics[width=\linewidth]{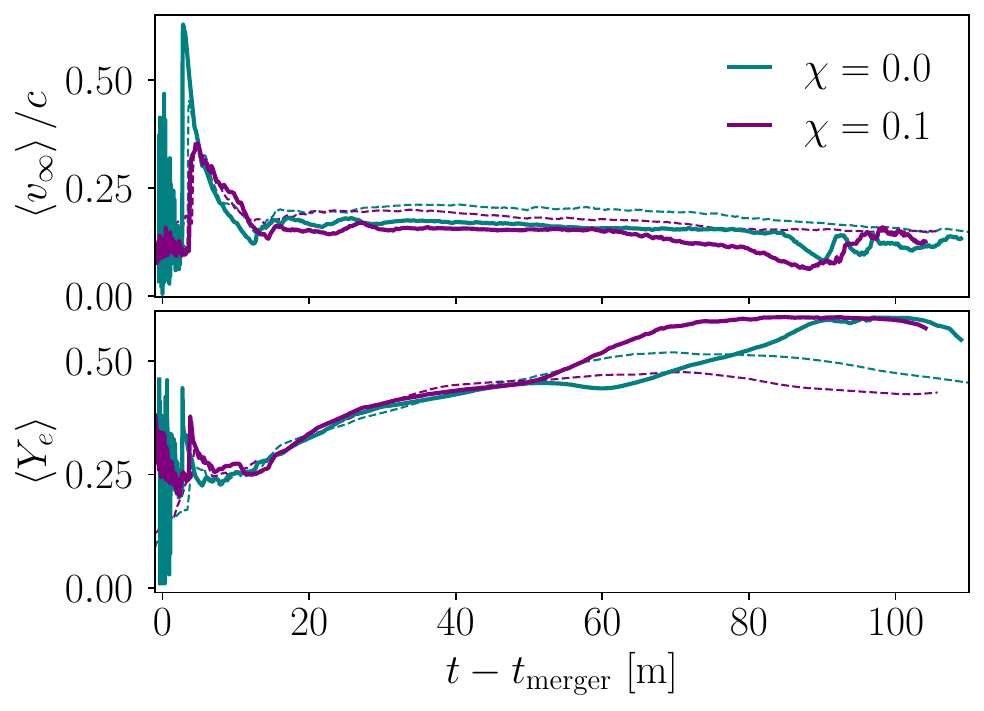}
    \caption{Averaged asymptotic velocity (top panel) and electron fraction (bottom panel) of ejected matter passing through the detection sphere of radius $\sim440\ \rm km$ for the non-spinning (green) and spinning (purple) BNS simulations at R1 (dashed) and R2 resolution (solid). }
    \label{fig:ejecta}
\end{figure}

For the non-spinning configuration, the morphology of $\beta^{-1}$ in the funnel around the polar cap at $\geq 75\ \rm ms$ after the merger suggests the occurrence of a buoyant instability. Hence, an outflow with considerably high $\beta^{-1}$, i.e., primarily magnetically-driven, is launched along the polar axis at $\sim 100\ \rm ms$ after the merger, coinciding with the time of the significant increase in mass flux in Fig.~\ref{fig:ejectamass} for this simulation. 
We provide additional snapshots for the rest-mass density $\rho$ and the asymptotic velocity $v_{\infty}$, defined as
\begin{equation}
    v_{\infty} = \sqrt{1 - 1/u_t^2},
\end{equation}
of this ejecta structure in Fig.~\ref{fig:2D-snapshots}. This collimated outflow component is consistent with the magnetic eruption scenario proposed by Ref.~\cite{Musolino:2024sju}, exhibiting a jet-like structure with mildly-relativistic velocities of $\lesssim 0.4\ c$ along the polar direction. \\

For a quantitative comparison of the ejecta between our simulations, we analyze relevant properties using information from the detection sphere at radius $\sim 440\ \rm km$. In Fig.~\ref{fig:ejecta}, we show mass-weighted averages of $Y_e$ and $v_\infty$ for the ejecta passing through this sphere as a function of time, where the mass-weighted average $\langle X \rangle$ of a quantity $X$ on the sphere of radius $r$ is defined as in Ref.~\cite{Schianchi:2023uky}
\begin{equation}
    \langle X\rangle = \frac{\oint f_u X r^2 d\Omega}{\oint f_u r^2 d\Omega}
\end{equation}
where $\oint$ is the integral on the detection sphere, and for a shorter notation, $f_u=\sqrt{\gamma}D_u\left(\alpha v^r - \beta^r\right)$ is the local radial mass flux of unbound material.

As seen in the upper panel of Fig.~\ref{fig:ejecta}, the early peak in the $\left<v_\infty\right>$ evolution corresponds to the faster material produced in the core-bounce, and is followed by progressively slower material. In the R2 simulations, the peak reaches higher values for $\chi=0.0$ , up to $\sim 0.6\ c$ as opposed to $\sim 0.4\ c$. Again, the slower ejecta from the spinning configurations trace back to the repulsive nature of the spin-orbit interaction and the lower impact velocities for these spin-aligned systems. 
Although the average velocities and electron fractions evolve in a rather similar way for both setups from $20\ \rm ms$ to $60\ \rm ms$, the spinning configuration develops a higher $\langle Y_e\rangle$ earlier than the non-spinning counterpart. Combined with the observation that $\beta^{-1}$ grows faster for $\chi=0.0$, we conclude that the secular ejecta for $\chi=0.1$ is predominantly driven by neutrino winds.

\begin{figure}[t]
    \centering
    \includegraphics[width=\linewidth]{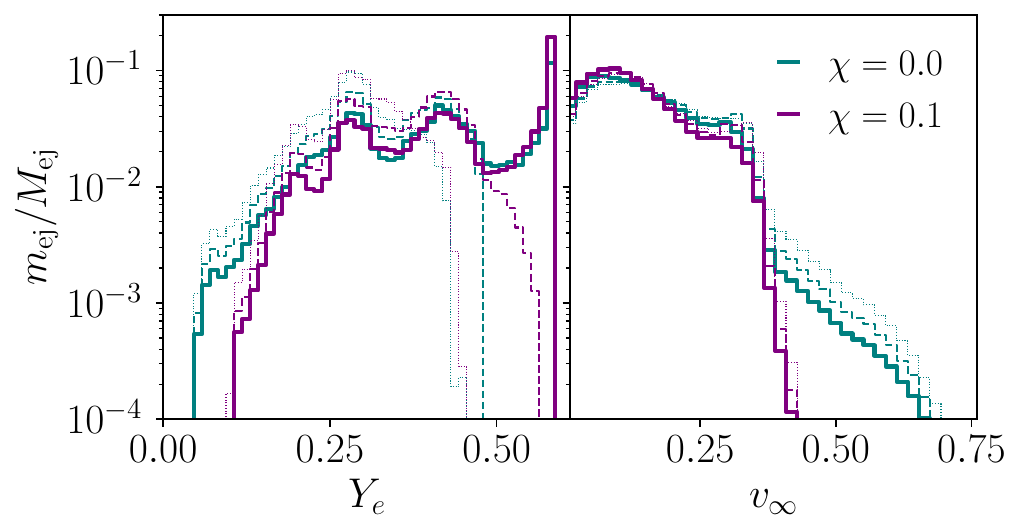}
    \caption{Histograms of the electron fraction (left panel) and asymptotic velocity (right panel) for the matter outflow of the non-spinning (green) and spinning (purple) BNS simulations at R2 resolution. Ejecta properties are extracted on the detection sphere of radius $\sim 440\ \rm km$. The dotted, dashed, and solid lines refer, respectively, to the ejected material passing through the sphere within the first  $20\ \rm ms$ after the merger, within the first $50\ \rm ms$ after the merger, and over the entire simulation period.}
    \label{fig:hist}
\end{figure}

This is in good agreement with the distributions depicted in the right panel of Fig.~\ref{fig:hist}, where mass-weighted histograms of the electron fraction (left panel) and asymptotic velocity (right panel) are presented for the R2 simulations at different timespans after the merger. For $\chi = 0.1$, we observe at most $v_\infty \sim 0.4\ c$. Consequently, material escapes the remnant more slowly and remains in its vicinity for longer in this configuration, only to reach the detection sphere after being strongly protonized due to neutrino irradiation. This results in the higher $\langle Y_e\rangle$ at $t-t_{\rm merger} \approx 50\ \rm ms$ and, correspondingly, in the higher-$Y_e$ tails of the distributions depicted in the left panel of Fig.~\ref{fig:hist}.
Over time, the ejecta's $Y_e$ progressively increases, as seen in the sequence of histograms for both configurations, until eventually reaching a sizable fraction of material with the maximum tabulated value $Y_e = 0.6$, well within expectations for such long-term simulations.

\begin{figure}[t]
    \centering
    \includegraphics[width=\linewidth]{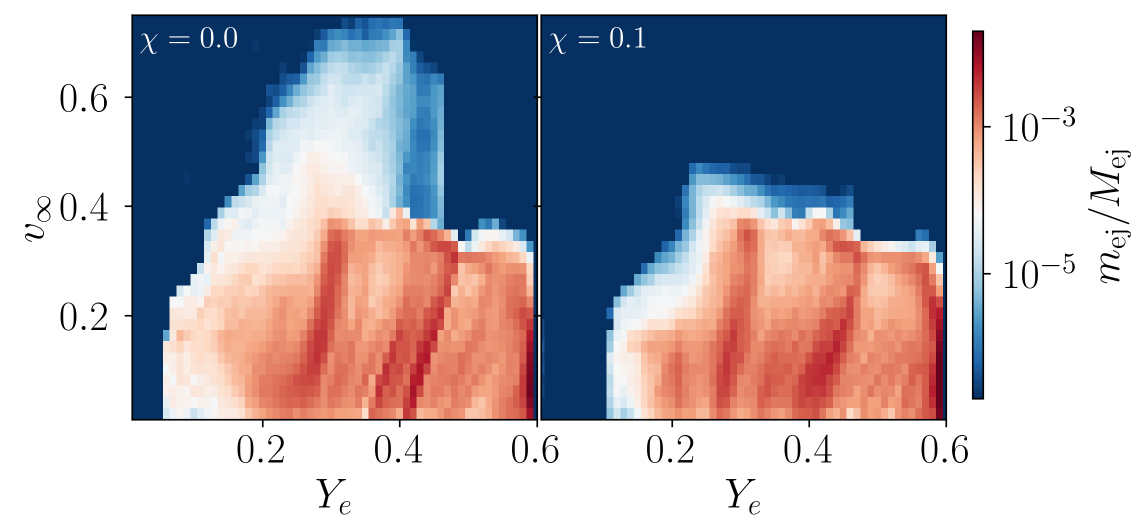}
    \caption{2D mass-weighted histograms of the ejecta's electron fraction and asymptotic velocity for the non-spinning (left panel) and spinning (right panel) BNS simulations at R2 resolution. The ejecta are extracted on the detection sphere of radius $\sim 440\ \rm km$.}
    \label{fig:hist2D}
\end{figure}

In general, the ejecta reaches lower $Y_e$ but also higher $v_\infty$ for $\chi=0.0$ than for $\chi=0.1$. 
For a better qualitative understanding of the ejection mechanisms at play in our simulations, we perform a combined analysis of the electron fraction and asymptotic velocity distributions and show two-dimensional histograms of the ejecta in Fig.~\ref{fig:hist2D}. We find that the bulk of the ejecta in the region with $Y_e \gtrsim 0.2$ and $v_\infty \lesssim 0.4$ is broadly similar between the two configurations. However, the non-spinning configuration has a pronounced ejecta component with high velocities $v_\infty\gtrsim0.4$ and additionally a low $Y_e$ tail with $v_\infty\lesssim0.2$. We think both components are consistent with the description of dynamical ejecta mechanisms proposed in Ref. ~\cite{Rosswog:2024vfe}.
We know that electron fraction is typically related to the region where the ejecta originates, whereas the velocity is correlated to details of the ejection episode. Thus, neutron-rich material $Y_e \lesssim 0.1$ is usually found in the interior of cold, catalyzed neutron stars, intermediate $Y_e \sim 0.2-0.4$ can be reached by overproduction of $\bar\nu_e$ with respect to $\nu_e$, usually taking place at higher temperatures, and even higher $Y_e$ can be reached through neutrino absorption over longer timescales. 
Ref.~\cite{Rosswog:2024vfe} identified that the bulk of the dynamical ejecta has two main origins: A predominantly equatorial `spray' component that is launched early on from the shear layer between the merging neutron stars with velocities $v \lesssim 0.6\ c$ and a more isotropic `bounce' component, where material from the interior of the merging neutron stars is released with $v \sim 0.8\ c$ during the first core bounce and with lower velocities between $0.2\ c$ and $0.4\ c$ during a second, weaker bounce. 
We interpret the high velocity ejecta with $v_\infty \gtrsim 0.4\ c$ for the non-spinning configuration as `spray' or `first bounce' component, since the velocities are consistent with this scenario and the electron fraction of $0.2 - 0.4$ is well within the expected range of matter that is protonized at the high temperatures developed in the shear layer and/or in the hot core-disk interface. 
Finally, we interpret the low $Y_e$ tail as matter from the (relatively cold) interior, launched by subsequent bounces. Naturally, the absence of both, the high $v_\infty$ and the low $Y_e$, ejecta components for $\chi=0.1$ can be traced to the overall weaker core bounces of this configuration.

\subsection{Nuclear abundances}
\label{subsec:yields}

\begin{figure}[t]
    \centering
    \includegraphics[width=\linewidth]{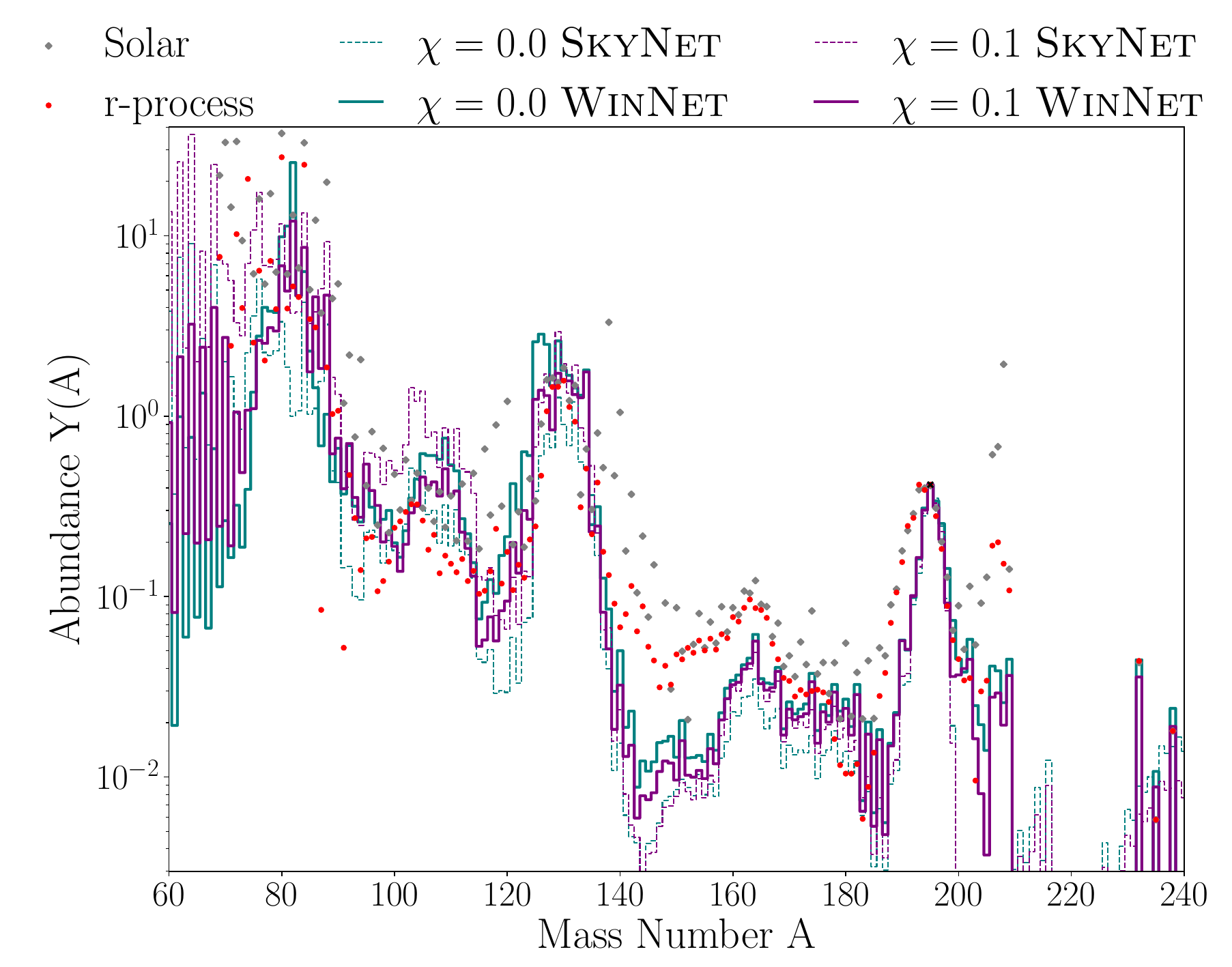}
    \caption{Nuclear abundances as computed with \winnet\ after \qty{1}{\giga\year} for both setups (solid lines). For comparison, we show solar-system abundances (gray diamonds,~\cite{Lodders:2025}) and $r$-process contributions (red dots,~\cite{Prantzos:2020}). Dashed lines indicate abundances with a simplified setup using \textsc{SkyNet}. All patterns are rescaled to match $r$-processed \ce{^195Pt}.}
    \label{fig:yields}
\end{figure}

The dynamical evolution of the ejecta properties corresponds to variations in nucleosynthesis yields and hence in the related EM signatures (see Sec.~\ref{subsec:kilonova}). To evaluate the abundances of newly synthesized heavy elements, we use information from tracer particles that are evolved along with the fluid in our BNS simulations. 

As detailed in Appendix~\ref{app:Tracers}, the advection scheme for the tracers leads to a high number of early, low-mass trajectories and cannot capture well the later wind ejecta.
The overall picture is thus subject to stochastic variations from the low number of high-mass wind trajectories, which is particularly problematic for the non-spinning case, where we need to ascribe almost $40\, \%$ of the total ejecta mass to a single trajectory.
Notwithstanding, the full abundance patterns in Fig.~\ref{fig:yields} are fairly consistent with other recent works~\cite{Bernuzzi:2024mfx, Ricigliano:2024lwf}.
Moreover, nucleosynthesis results obtained with \textsc{SkyNet}~\cite{Lippuner:2017tyn}, purely based on the ejecta extraction spheres in the simplified setup previously employed in Ref.~\cite{Schianchi:2023uky}, exhibit very similar trends.

In general, the nucleosynthesis features for both merger simulations share most characteristics. Abundances of nuclei in the iron-group and up to about $A=80$ are much lower than in the solar system~\cite{Prantzos:2020,Lodders:2025}. The $r$-process peaks are robustly produced, although the third peak is severely narrowed on its lower mass shoulder, associated with a deficit of cold, tidal ejecta in equal-mass mergers as presented~\cite{Marketin:2015gya}. While we observe a considerable excess of nuclei around $A\approx110$, corresponding roughly to the transition metals in the 5th period, lanthanides are partly deficient as compared to solar-system values.
The latter feature has also been found in other studies~\cite{Pfeiffer:1997,Cowan:2019pkx}, and is intricately associated with the choice of mass model, \textbeta-decay, and fission prescription. The underlying FRDM mass-model is known to lead to a `pile-up' near magic numbers, broadening the second peak while depleting low-mass lanthanides~\cite{Arcones:2010dz}. Similarly, the simplified fission prescription of Ref.~\cite{Panov:2001rus} deepens this trough, while the formulation of Ref.~\cite{Kodama:1975aff} alone tends to erase the peaks~\cite{Eichler:2014kma}.
The lower trough around $A=140-150$ in the spinning case, hence, correlates with a lower production of fissioning material. Consequently, lanthanides up to \ce{Eu} are deficient, while the relative abundances of \ce{Gd} - \ce{Lu} align well with those both in in old, \ce{Fe}-deficient stars~\cite{Roederer:2022exr} and in the solar system~\cite{Prantzos:2020}.

\begin{figure}[t]
    \centering
    \includegraphics[width=\linewidth]{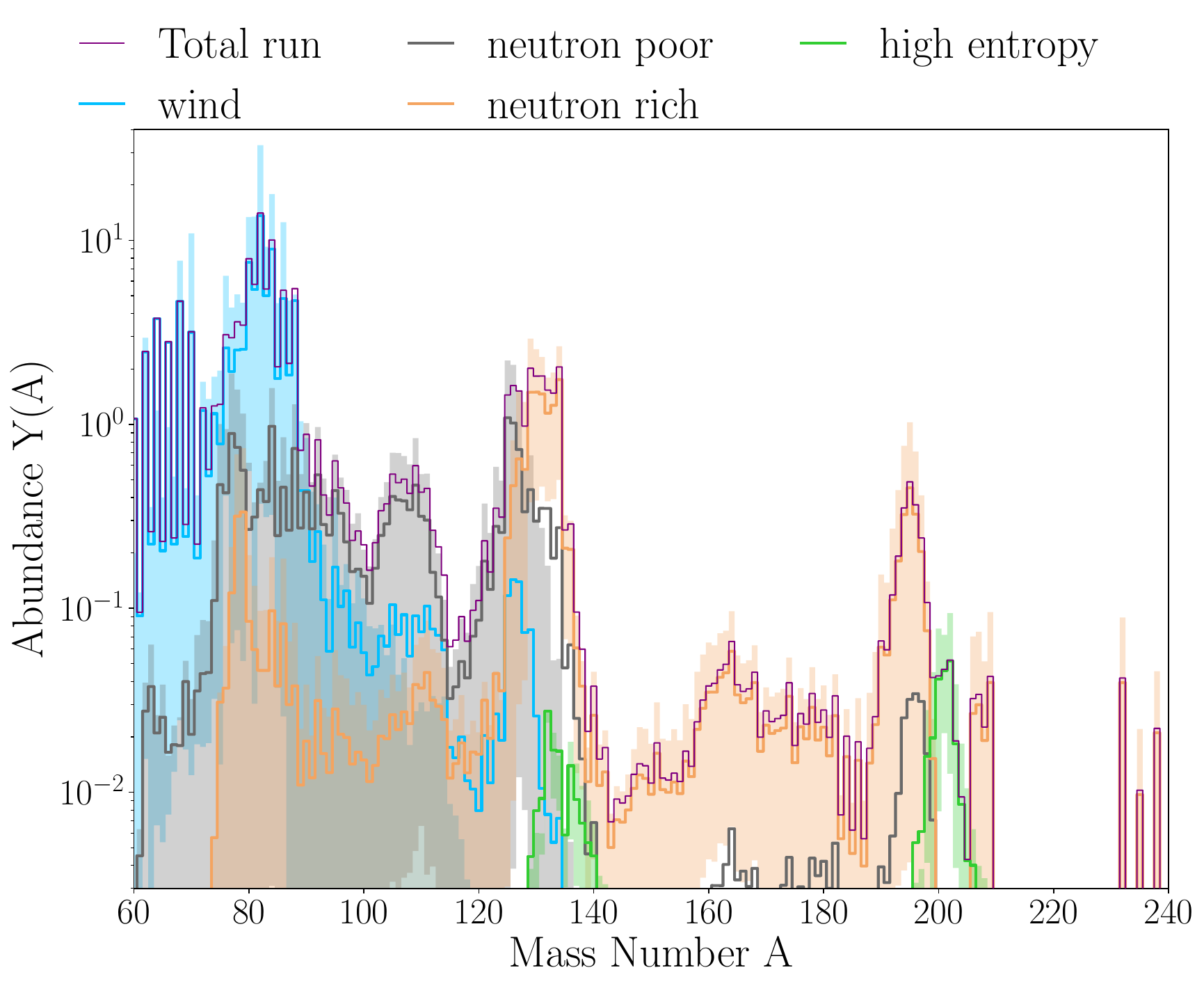}
    \caption{Contribution of different tracer subgroups to the full abundance pattern for the spinning system. The solid lines indicate the mass-weighted average, while the bands indicate the 16th-84th percentile in the group for each mass number.}
    \label{fig:yield_groups}
\end{figure}

To distinguish between different ejecta components and analyze their contribution to the total nuclear abundances, we identify the wind component as tracers crossing the extraction sphere at $t_{\rm ext}>\qty{30}{\milli\second}$.
From the remaining dynamical ejecta, we separate tracers with high entropy $s_0 \geq \qty{80}{k_B \per nuc}$, by and large coinciding with the fast, shock-ejected component, and split the rest into a neutron-rich component with $Y_e \leq 0.25$ and a neutron-poor component with $Y_e>0.25$.
\Cref{fig:yield_groups} shows the mass-weighted contributions of these groups to the total abundance in the spinning case, while Table~\ref{tab:yields} summarizes the total ejecta masses in different mass ranges for both configurations.

Winds account for about $60\, \%$ of the total ejecta mass in both setups. Despite comprising only about $150$~tracers each, they cast an extremely homogeneous picture:
They fall out of NSE only as they are ejected from the disk at velocities of about \qty{0.1}{c} with a fairly narrow density distribution around $\log(\rho_{\rm NSE}/\unit[per-mode=power]{\gram\per\centi\meter\cubed}) =\num{6.8(2)}$, corresponding to a similarly narrow distribution of entropies around $S\approx\qty{20}{k_B\per nuc}$.
Early in the post-merger, fast protonization occurs in the disk due to the overproduction of $\bar\nu_e$ via positron capture on free neutrons, which decouples from matter deeper in the remnant, and leads to relatively high electron fractions around $Y_e\gtrsim 0.35$. Hence, the final composition of the wind ejecta remains similar to the respective NSE-composition~\cite{Kuske:2025ffp}:
Most matter rests in nuclei up to \ce{^88Sr} in the first $r$-process peak and favors even mass numbers by roughly one order of magnitude. Only traces of second-peak elements form in trajectories with $Y_e<0.35$.
For $Y_e\gtrsim 0.45$, the mean mass number $\abar$ sharply drops and nucleosynthesis reaches only up to the iron group. While the low number of available tracers thus does not allow robust statements on the exact pattern, Fig.~\ref{fig:yield_groups} highlights that these poorly sampled tracers have no impact on the heavier elements.

\begin{table}[t]
    \caption{Mass of different ejecta components in \qty{e-3}{M_\odot}}

    \begin{tabular}{ccccc}
        \toprule
        Ejecta component & \multicolumn{2}{c}{Wind} & \multicolumn{2}{c}{Dynamical}  \\
        $\chi$& $0.0$ & $0.1$ & $0.0$ & $0.1$ \\
         \midrule
            \textalpha-particles $(A=4)$        & 0.00  & 0.08 & 0.16 & 0.06 \\
            Iron group $(50\leq A\leq 56)$      & 0.60  & 0.48 & 0.00 & 0.00 \\
            First peak $(73 \leq A \leq 91)$    & 5.74  & 3.66 & 0.65 & 0.53 \\
            Second peak $(121 \leq A \leq 138)$ & 0.13  & 0.06 & 2.72 & 1.50 \\
            Third peak $(186 \leq A \leq 203)$  & 0.00  & 0.00 & 0.40 & 0.29 \\
            Lanthanides $(139 \leq A \leq 176)$ & 0.00  & 0.00 & 0.15 & 0.10 \\
            Actinides $(210 \leq A \leq 260)$   & 0.00  & 0.00 & 0.02 & 0.01 \\
            Rest                                & 0.41  & 0.97 & 0.94 & 0.56 \\
            Total                               & 6.89  & 5.25 & 5.04 & 3.06 
    \end{tabular}
    \label{tab:yields}
\end{table}

\begin{figure*}[t]
    \centering
    \includegraphics[width=\linewidth]{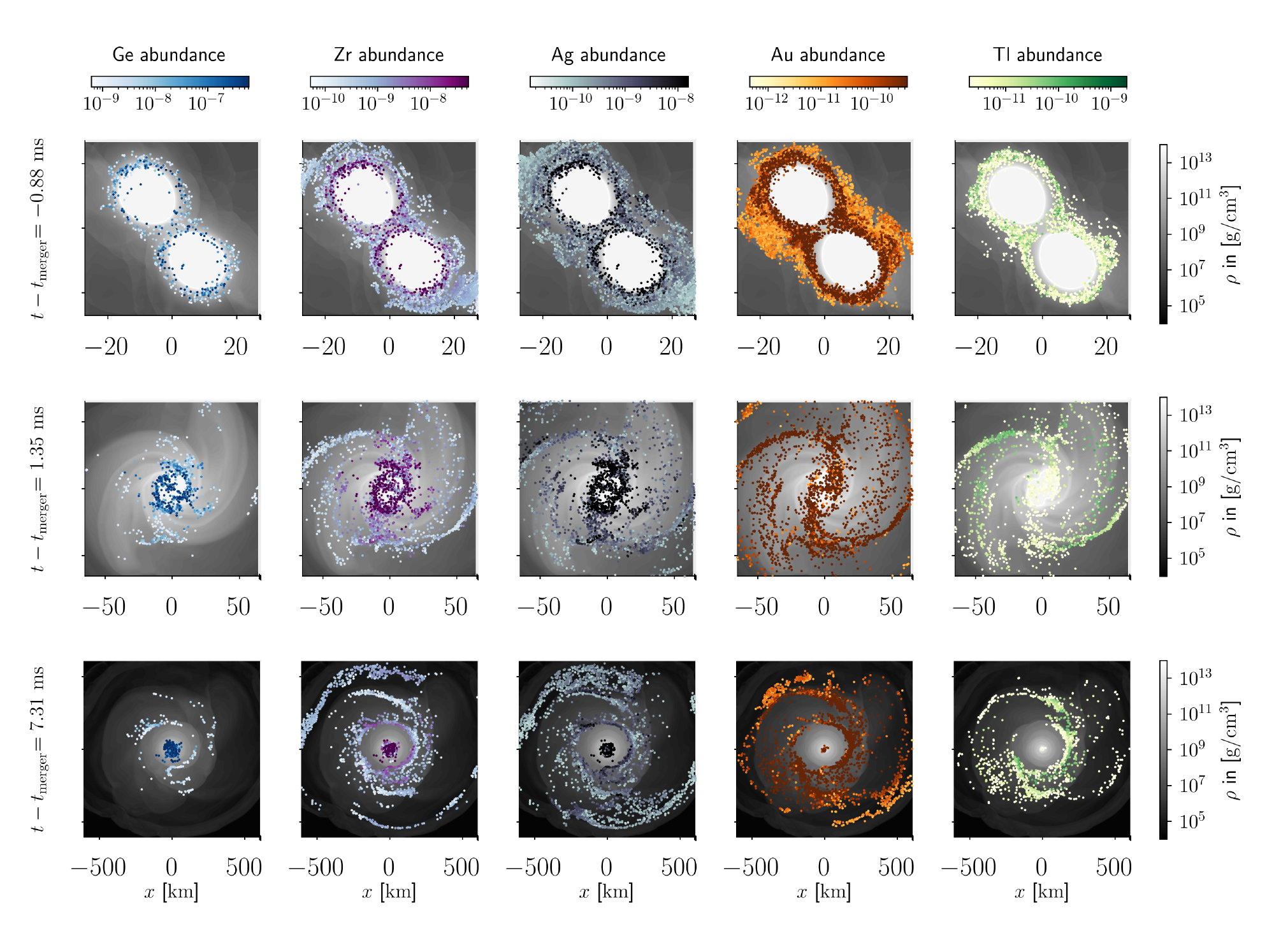}
    \caption{``Treasure map'' showing tracer particles projected onto the $x$-$y$ plane and their mass-weighted relative abundances of selected elements for the high-resolution spinning BNS simulation. The panels in the upper, middle, and lower rows respectively show the maps for $t=\{-0.88,1.25,7.31\}\ \rm ms$ after the merger and project tracer particles for $|z| < \{5,10,50\}\ \rm km$. For each element, the tracer particles shown combined account for $99.9\, \%$ to the total element abundance, while tracer particles with negligible contribution are not displayed for better visibility. In the background, we show the rest-mass density profile in grayscale.}
    \label{fig:treasuremaps}
\end{figure*}

Nuclei beyond the first peak result from the dynamical ejecta instead, which draw a more diverse picture:
The neutron-poor and neutron-rich components contribute each about a fifth of the total ejecta mass and have entropies comparable to the wind ejecta, but slightly higher velocities around \qty{0.15(5)}{c} and NSE densities $\log(\rho_{\rm NSE}/\unit[per-mode=power]{\gram\per\centi\meter\cubed})=\num{7.0(2)}$.

The neutron-poor ejecta contain some iron-group elements, but are dominated by ejecta up to the second peak. These account for the strong excess of transition metals with $A\approx 110$ and exhibit a broadened first peak beyond the $N=50$ shell closure, while the abundance of nuclei beyond tellurium ($Z=50$) isotopes declines rapidly. Only a few tracers close to $Y_e=0.25$ produce a subdominant yield of lanthanides and third-peak elements. These are primarily synthesized by the neutron-rich ejecta that account for almost all ejecta with $A>128$. As expected, the yield of heavy elements is hence strongly correlated with $Y_e$.
The highest actinide yields with mass fractions up to $X_{\rm act}\approx 0.07$ are found in tracers with the lowest $Y_e\approx 0.15$. These are also the most efficient producers of lanthanides, with $X_{\rm lanth}\geq 0.10$, while second-peak elements are relatively deficient.

There is a small (up to $3\, \%$ in the non-spinning case), yet noteworthy contribution to the total actinide mass from the fastest, shock-ejected tracers that encounter a `frustrated' $r$-process~\cite{deJesusMendoza-Temis:2014owk,Cowan:2019pkx}.
These have high-entropy ($s\sim \qty{120}{k_B\per nuc}$) and $Y_e\approx 0.3$. As they are flung out from the merger site, they encounter a \textalpha-rich freeze-out~\cite{Woosley:1992ek,Cowan:2019pkx}:
Their density drops too quickly for large amounts of \textalpha-particles to cross the \ce{^8Be} bottleneck and form heavier seeds. Correspondingly, they make a significant contribution to the final He-abundance.
Many neutrons also avoid capture in total, decaying to the remaining hydrogen on the order of hours.
Ref.~\cite{Metzger:2014yda} predicts this to be observable as a kilonova-precursor. The few massive seeds neutronize very efficiently, though, piling up along the $N={82,126}$ shell closures to produce close to $100\, \%$ of the final mercury and thallium yields.

\Cref{fig:treasuremaps} confirms these general trends based on ``treasure maps'' for a few characteristic elements at initialization, shortly after the merger, and around \qty{7}{\milli\second} after the merger. 
These project the location of tracers of our highest-resolution simulation of the spinning BNS configuration into the $x$-$y$ plane.
We can see that \ce{Ge} mostly originates from tracers in the outer layers of the star that remain longer in the disk. \ce{Zr} and \ce{Ag} have a significant wind contribution, but are increasingly abundant in the tidally ejected material. 
Gold is primarily formed in dense regions of the faster ejecta components, though there are some traces that leave the disk at later times. Ultimately, \ce{Tl} forms in the fastest shock-ejected tracers without any contribution from the disk.

\section{Multi-messenger picture}
\label{sec:MMApicture}

\subsection{Gravitational waves}
\label{subsec:GW}

\begin{figure}[t!]
    \centering
    \includegraphics[width=\linewidth]{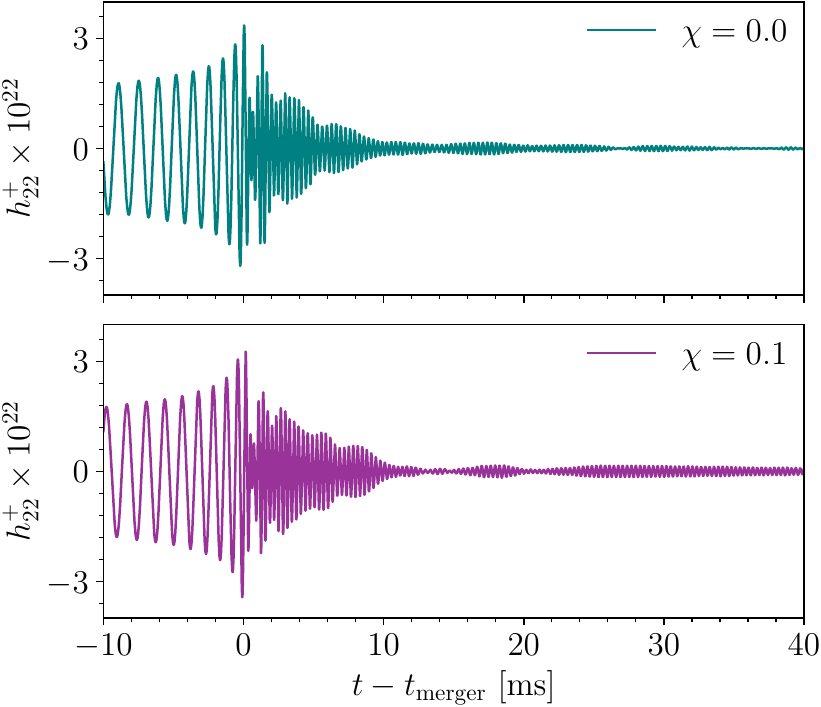}
    \caption{GW strain of the dominant $\left(2,2\right)$-mode with face-on view at $100\ {\rm Mpc}$ for the non-spinning (upper panel) and spinning (lower panel) configurations. The time is shifted by the merger times, where the amplitude of this mode is maximal.}
    \label{fig:GW-waveform}
\end{figure}

Throughout the inspiral phase, the continuous emission of GWs draws energy and angular momentum from the system. In response, the neutron stars coalesce and merge into a compact object surrounded by a disk of material with intermediate to low densities. 
The dynamics of the inspiral stage can be observed in the characteristic chirp signal, as depicted in the waveforms of Fig.~\ref{fig:GW-waveform} for $t-t_{\rm merger} < 0$, where the GW frequency and amplitude increase as the binary orbit decays, until the neutron stars come into contact and merge. As stated above, the merger time is defined as the instant at which the amplitude of the dominant $(2,2)$ mode is maximal. 

While the GW signal during the inspiral phase in BNS systems with spinning and non-spinning stars is well prescribed by various approximants, e.g., Ref.~\cite{Abac:2023ujg}, the post-merger phase is more ambiguous. Hydrodynamic instabilities in the remnant can excite inertial modes, which are potentially observable in planned third-generation GW detectors~\cite{DePietri:2018tpx,DePietri:2019mti} and could thus be used to investigate the rotational and thermal states of the remnant.
We therefore focus our analysis of the GW signal on the post-merger phase, where most differences between our configurations appear. The post-merger signal is characterized by a rapid decrease of the GW amplitude, as the system transitions to an approximately axis-symmetric state over a timescale of $\gtrsim 10\ {\rm ms}$. On this timescale, the remnant's internal motions are effectively dampened by radiative losses.

\begin{figure}[t!]
    \centering
    \includegraphics[width=\linewidth]{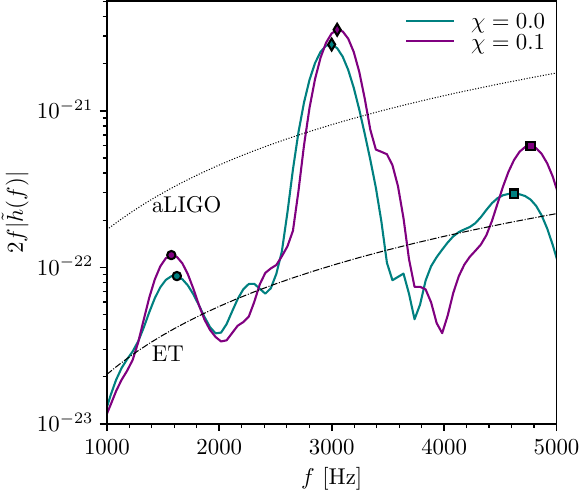}
    \caption{Characteristic frequency strain of the post-merger GW signal extracted from our R2 simulations at $100\ {\rm Mpc}$ for the non-spinning (teal) and spinning (purple) setups. Circle, diamond, and square markers correspond, respectively, to the $f_1,$ $f_2$, and $f_3$ peaks. Dotted and dash-dotted black lines correspond to design sensitivity curves of Advanced LIGO~\cite{Barsotti:2018} and Einstein Telescope~\cite{Hild:2010id}, respectively.}
    \label{fig:GW-spectra}
\end{figure}

In order to better understand different contributions to the post-merger signal, we show in Fig.~\ref{fig:GW-spectra} the characteristic frequency strain $h_c(f) = 2f|\tilde{h}(f)|$, windowed at $t - t_{\rm merger} \geq 1.5\ {\rm ms}$~\cite{Moore:2014lga}. For a smoother spectrum, we compute the power spectral density (PSD) of the strain $h(t)$ including all $l=2,3$ multipoles using Welch's method~\cite{Welch:1967}, with the signal being divided into segments with $25\, \%$ overlap. Additionally, each segment is tapered with a Hann window, and zero-padding ensures $4096$~points per segment. The peak frequencies are listed in Table~\ref{tab:peak-freq}. 

The dominant $f_2$ frequencies are related to quadrupolar oscillations of the remnant~\cite{Stergioulas:2011gd}. They are in general well described by fitting formulae depending on the combined tidal deformability (see, e.g., Ref.~\cite{Vretinaris:2019spn}). 
Furthermore, we find the $f_1$ peaks, mainly arising from the $l=2,~m=1$ mode, which roughly coincide with the expected $f_1 \approx f_2/2$. These peak frequencies are commonly associated with the one-arm instability, e.g., Refs.~\cite{Radice:2016gym,East:2015vix,East:2016zvv,Paschalidis:2015mla,Lehner:2016wjg,Radice:2023zlw}. The persistence of this mode in the post-merger suggests that spiral wind-waves might be active, contributing to the secular matter outflows~\cite{Nedora:2019jhl}. 
Finally, we remark that the physical origin of the $f_3$ peak, roughly corresponding to $f_3 \approx 3 f_2/2$ is still unclear~\cite{Stergioulas:2011gd, Takami:2014zpa, Baiotti:2016qnr}.

\begin{table}[t]
    \centering
        \caption{Peak post-merger GW frequencies. From left to right, the columns list simulation name, resolution, first peak frequency $f_1$, dominant peak frequency $f_2$, and third peak frequency $f_3$.}
    \begin{tabular}{ccccc}
    \toprule
    Simulation  &  Resolution & $f_1$ & $f_2$ &$f_3$ \\
     &  & $[{\rm kHz}]$ & $[{\rm kHz}]$ & $[{\rm kHz}]$\\
      \hline
     $\chi=0.0$    & R1 & $1.672$ & $3.048$ & $4.819$ \\
         & R2 & $1.622$ & $3.000$ & $4.622$ \\
    $\chi=0.1$  &R1  & $1.524$ & $3.000$ & $4.474$\\
      & R2 & $1.574$ & $3.049$ & $4.770$\\
    \bottomrule
    \end{tabular}
    \label{tab:peak-freq}
\end{table}

When comparing the peak frequencies of the $\chi=0.0$ and $\chi=0.1$ configurations, we note slight shifts. At R2 resolution, the frequencies $f_1,~f_2$ are only shifted by $\sim 50\ {\rm Hz}$, whereas $f_3$ is shifted by $\sim 150\ {\rm Hz}$. 
On the other hand, we quantify numerical uncertainties by the difference in frequency peaks for a given configuration between resolutions R1 and R2, being of $\sim 50\ {\rm Hz}$ for $f_1, ~f_2$, and $\sim \left(200 - 300\right) \ {\rm Hz}$ for $f_3$. 
Although the shift to higher frequencies at R2 resolution could easily be explained by the higher angular momentum and the less compact remnant in the spin-aligned BNS merger, we suspect that the observed differences in our simulations are dominated by grid resolution effects.

\subsection{Kilonova light curves}
\label{subsec:kilonova}

Kilonovae are thermal transients powered by the radioactive decay of $r$-process nuclei synthesized in the ejected merger material~\cite{Li:1998bw,Metzger:2016pju}. The emission of EM radiation therefore depends sensitively on ejecta properties (see Sec.~\ref{subsec:Ejecta}) with the corresponding nucleosynthesis yields (see Sec.~\ref{subsec:yields}). 
The observable signal is determined by several factors, including the ejecta geometry, viewing angle, energy released from the synthesized nuclei, and the interaction of the emitted radiation with the expanding matter, e.g.,~\cite{Wollaeger:2017ahm,Bulla:2022mwo,Darbha:2020lhz,Kawaguchi:2018ptg,Korobkin:2020spe,Collins:2023www,Groenewegen:2025ezj}.

To obtain the kilonova light curves associated with the two BNS mergers, we perform radiative transfer simulations with \textsc{possis}~\cite{Bulla:2019muo,Bulla:2022mwo} based on the data from our high-resolution NR simulations. 
As described in Sec.~\ref{subsec:codes}, we extract ejecta data at a time $t_{\rm cut}$, when the entire material is still contained within the computational domain, and supplement it with information from the detection sphere at $r \simeq 440\ \rm km$.
For the non-spinning and spinning BNS simulations, $t_{\rm cut}$ is at $12.50\ \rm ms$ and $15.05\ \rm ms$ after the merger, respectively.
Although the nuclear abundance calculations in Sec.~\ref{subsec:yields} also provide heating rates and thermalization efficiencies, which are key ingredients for kilonova modeling, we instead adopt the precomputed libraries from Refs.~\cite{Rosswog:2022tus,Barnes:2016umi,Wollaeger:2017ahm}, as implemented in \textsc{possis}~\cite{Bulla:2022mwo}. 
This choice avoids possible distortions caused by the coarse sampling of the late-time ejecta through the tracer particles used for the nuclear synthesis calculations, as discussed in Appendix~\ref{app:Tracers} and Sec.~\ref{subsec:yields}.

\begin{figure}[t!]
    \centering
    \includegraphics[width=\linewidth]{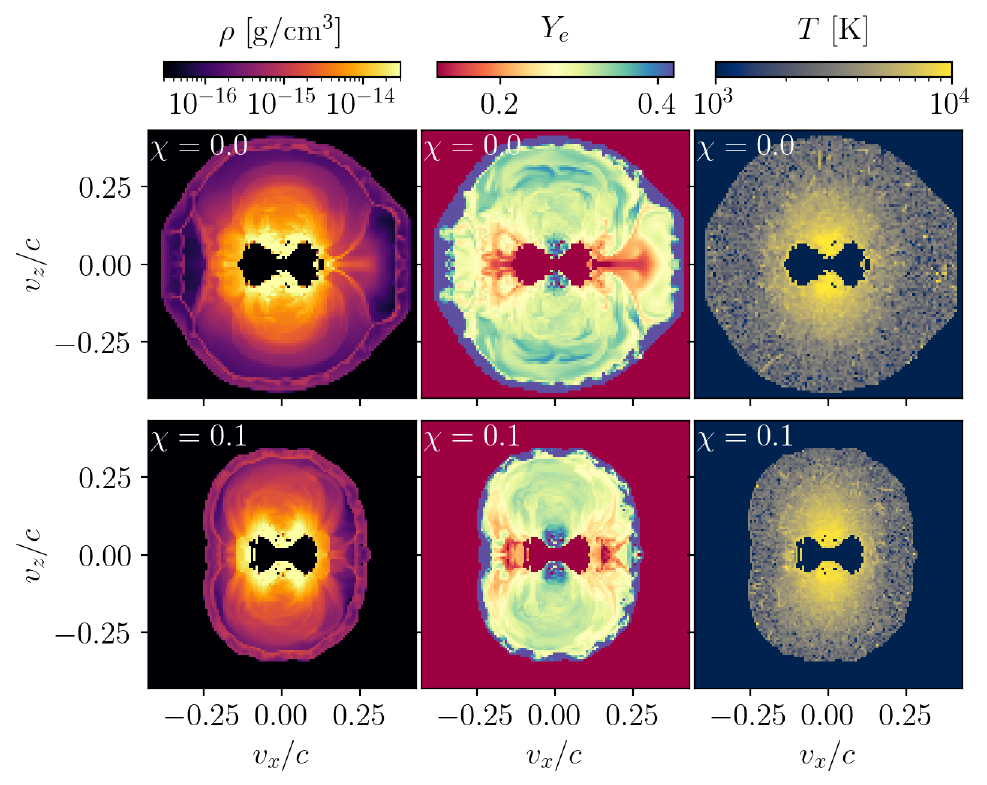}
    \caption{Ejecta data used as input for radiative transfer simulations with \textsc{possis}. The maps show the density $\rho$, the electron fraction $Y_e$, and the temperature $T$ in the $v_x$-$v_z$ plane at one day after the merger for the non-spinning (upper panels) and spinning (bottom panels) BNS systems.}
    \label{fig:KNmaps}
\end{figure}

The ejecta used as input for the radiative transfer simulations are presented in Fig.~\ref{fig:KNmaps}. The maps show the extracted rest-mass density and the electron fraction, as well as the temperatures calculated within \textsc{possis} in the $v_x$ - $v_z$ plane one day after the merger. The respective ejecta masses are $0.0137\ M_\odot$ and $0.0098\ M_\odot$ for the non-spinning and spinning BNS configuration. The maps summarize the main properties of ejecta discussed in Sec.~\ref{subsec:Ejecta}: The non-spinning BNS merger produces faster ejecta that are more extended in the $v_x$ - $v_z$ plane than in the spinning configuration. 
In both configurations, the electron fraction is higher in the polar regions than in the orbital plane, where the non-spinning configuration exhibits lower $Y_e$ values than the spinning one. Furthermore, both systems show a wind component with high $Y_e$ and slow velocities concentrated near the inner region of the ejecta.

\begin{figure}[t!]
    \centering
    \includegraphics[width=\linewidth]{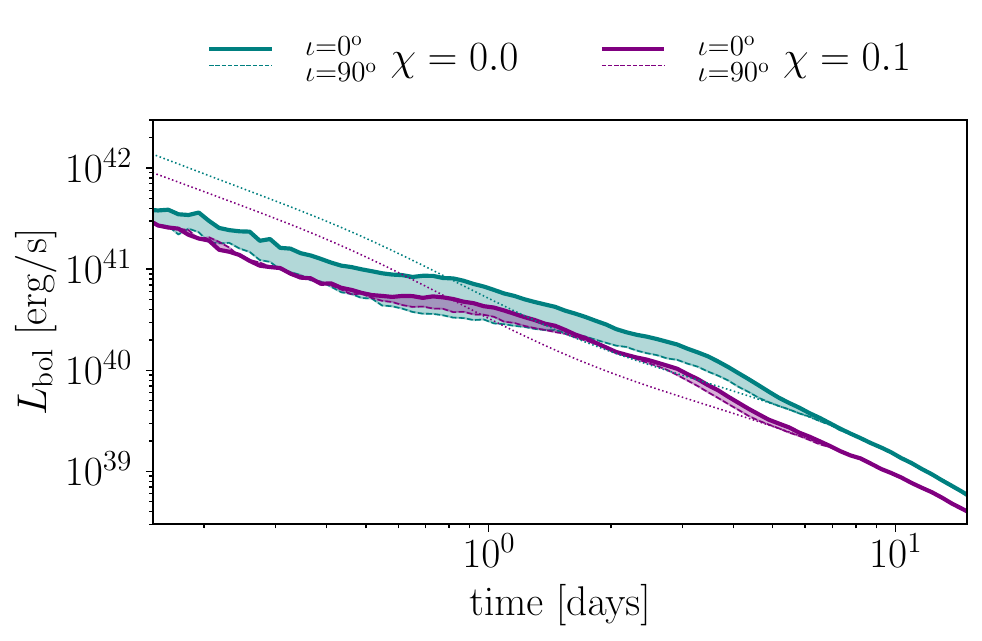}
    \caption{Bolometric light curves of the kilonova associated with the non-spinning (green) and spinning (purple) BNS configuration for observers looking along the polar axis $\iota = 0^\circ$ (solid line) and in the orbital plane with $\iota= 90^\circ$ (dashed line). The deposition curve is shown in dotted lines.}
    \label{fig:KNlbol}
\end{figure}

In Fig.~\ref{fig:KNlbol}, we present bolometric light curves for observers located at the pole with a viewing angle of $\iota=0^\circ$ and in the orbital plane with $\iota=90^\circ$, together with the deposition curve. The latter represents the total amount of luminosity injected into the system as a function of time. With the assumed axial symmetry of the ejecta, the analysis can be restricted to an azimuthal angle $\varphi=0^\circ$, i.e., we consider observers in the $x$-$z$ plane only.
The bolometric light curves are overall brighter in the non-spinning configuration due to the larger amount of ejecta contributing to the thermalized energy powering the kilonova emission.
As expected, the signals are brighter at a polar viewing angle and fainter for $\iota=90^\circ$, due to the lower electron fraction in the orbital plane. The material with low $Y_e$ produces more heavy elements (see Sec.~\ref{subsec:yields}) and thus higher opacities~\cite{Tanaka:2019iqp}, causing more absorption events of the photon packets. Consequently, photons propagate more freely in the polar region than in the orbital plane, which in turn leads to more radiation and brighter light curves observable from the pole. Because the non-spinning configuration has a greater variation in electron fractions than the spinning one, reaching lower $Y_e$ in the orbital plane, the variation for different viewing angles $\iota$ is also more significant.
The shapes of the light curves are otherwise similar: They initially lie below the deposition curve for up to $\sim 1\ \rm day$ because the inner ejecta are still optically thick, trapping the radiation at early times. As the ejecta expand and become more transparent, the previously trapped radiation escapes and produces an overshoot, exceeding the deposition curve $\sim 7\ \rm days$. Finally, once the ejecta is fully transparent, the light curves converge to the deposition curve.

\begin{figure}[t!]
    \centering
    \includegraphics[width=\linewidth]{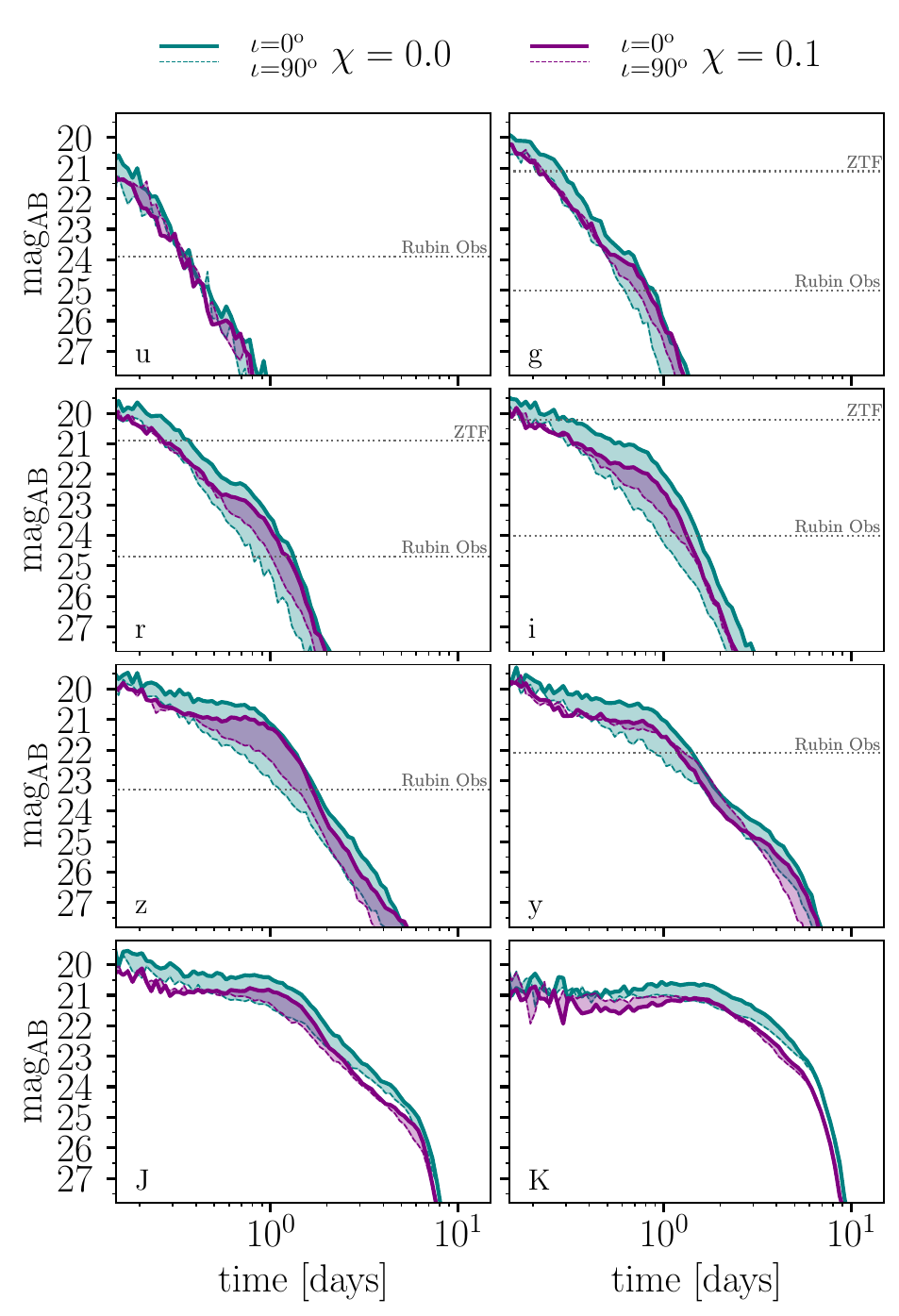}
    \caption{Kilonova light curves for the non-spinning (green) and spinning (purple) BNS configuration in different frequency bands, indicated in the bottom left corner of each panel. We show the apparent magnitude at a luminosity distance of $d_L=100\ \rm Mpc$. The different light curves correspond to observers looking along the polar axis $\iota = 0^\circ$ (solid line) and in the orbital plane with $\iota= 90^\circ$ (dashed line). The gray dotted lines indicate in the respective frequency bands $5\sigma$ limiting magnitudes for single exposures of the Zwicky Transient Facility (ZTF)~\cite{Dekany:2020tyb} and for design specifications of the Vera Rubin Observatory~\cite{LSST:2017}.}
    \label{fig:KN}
\end{figure}

Broad-band kilonova light curves are shown in Fig.~\ref{fig:KN} for different frequency bands in the ultraviolet, optical, and infrared ranges. The apparent magnitudes are scaled to a luminosity distance of $d_L=100\ \rm Mpc$.
As expected for kilonova signals, the light curves reach their peak earlier and decay faster in bluer filters than in redder ones. In fact, the light curves in the u-, g-, and r-bands merely peak and decrease immediately. In the i-, z-, and y-bands, the peak occurs at $\lesssim 1\ \rm day$, in the J-band at $\gtrsim 1\ \rm day$, and in the K-band only at $\sim 2\ \rm days$ after the merger.
This behavior is due to the fact that radiation is absorbed and re-emitted at longer wavelengths in the lanthanide-rich, low-$Y_e$ ejecta. Since radiation that undergoes this re-processing mechanism multiple times takes longer to diffuse and escape the ejecta, the kilonova signal shifts to redder wavelengths over time.
The non-spinning configuration is systematically brighter in the redder frequency bands, whereas the u- and g-bands show similar magnitudes for both configurations. We attribute this redder component to the lower $Y_e$ values reached in the material ejected in the non-spinning configuration, leading to a higher abundance of heavy elements with higher opacities. As a result, radiation is absorbed and re-emitted more frequently, enhancing the infrared component of the kilonova.
Again, we find a clear viewing angle dependence, most pronounced in the optical and near-infrared bands. In particular, in the g-, r-, i-, z-, and y-bands for the non-spinning BNS configuration, differences are in the order of $\sim 1-1.5\ \rm mag$, depending on filter and time. Similar to the bolometric light curves, the differences are smaller for the spinning BNS configuration.

To assess the observability of the computed kilonova signals, Fig.~\ref{fig:KN} also shows limiting magnitudes for single exposures of the Zwicky Transient Facility (ZTF) in the g-, r-, and i-bands~\cite{Dekany:2020tyb} and of the Vera Rubin Observatory in the u-, g-, r-, i-, z-, and y-bands~\cite{LSST:2017}.
At a luminosity distance of $100\ \rm Mpc$, the kilonova would be detectable with ZTF only within the first few hours after merger; depending on viewing angle and filter, within $\sim 0.2\ \rm days$ for the spinning configuration and slightly longer for the non-spinning one. In contrast, the signals in the infrared would be observable by the Vera Rubin Observatory for about one to two days after the merger. 
Although a detection by the Vera Rubin Observatory would be more likely, this result highlights the challenge of observing a kilonova signal as part of a multi-messenger detection of a BNS merger event at this luminosity distance.

\subsection{Afterglow of the kilonova and GRB}
\label{subsec:afterglow}

From an observer's perspective, at least two types of afterglows are expected to accompany BNS mergers from the interaction with the interstellar medium.
One type arises from the highly relativistic, collimated jet that is launched during or after the merger and leads to the emissions of a GRB through a largely unknown mechanism~\cite{Eichler:1989, Narayan:1992iy, Berger:2013jza, Goldstein:2017mmi, Savchenko:2017ffs} and is therefore referred to as GRB afterglow.
The second type emerges from the mildly relativistic material ejected during the inspiral, merger, and post-merger.
This material typically contributes to the kilonova, but it also contains a high-velocity component to create a kilonova afterglow as a long-term transient.

Our ejecta velocities range up to $0.8$~($0.65\ c$) in the non-spinning (spinning) case, with a total mass of $0.0137$~($0.0098$)~\msol. $5\times 10^{-5}\ (2\times10^{-9})$~\msol thereof have a velocity higher than $0.5\ c$.
We use the polar velocity distribution from the ejecta in our simulations as input in \textsc{pyblastafterglow} to compute the expected kilonova afterglow as described in Sec.~\ref{subsubsec:PyBlastAfterglow}.

The calculation of the GRB afterglow, however, requires knowledge of the jet structure and kinetic energy.
While various NR simulations have shown that jet-like structures can emerge from the post-merger remnant through a poloidal magnetic field~\cite{Rezzolla:2011da, Ruiz:2016rai} in combination with neutrino winds ~\cite{Mosta:2020hlh, Sun:2022vri, Kiuchi:2023obe, Combi:2023yav}, resolving the launch of an ultra-relativistic jet with $\Gamma \gtrsim 100$ remains currently infeasible, as temporal and spatial resolution is insufficient to accurately capture the magnetic field amplification and matter acceleration.
Thus, even though our simulations also display high-velocity outflows with $\beta\approx 0.4$ along the polar axes (see Fig.~\ref{fig:2D-snapshots}), it is not possible to establish robust estimates for the GRB jet parameters from our runs.
However, for comparison with the kilonova afterglow and to assess its observability, we calculate a GRB afterglow light curve from \textsc{pyblastafterglow} with fiducial values based on GRB170817A.
Specifically, we set $\theta_c=6.3\,^\circ$, $\theta_w=18.9\,^\circ$, and $\Gamma_0=500$ based on the posterior median from GRB170817A as inferred in Ref.~\cite{Koehn:2025zzb} with the \textsc{pyblastafterglow} model. Likewise, the microphysical parameters are set to $p=2.11$, $\epsilon_e=0.1$, $\epsilon_B=0.001$.
The posterior median for $E_0$ was $10^{52.2}\ \rm erg$, but since the isotropic kinetic energy-equivalent is arguably the parameter for which one could expect the most variability, we also show a case where the isotropic kinetic energy equivalent is two orders of magnitude lower.

\begin{figure}[t!]
    \centering
    \includegraphics[width=\linewidth]{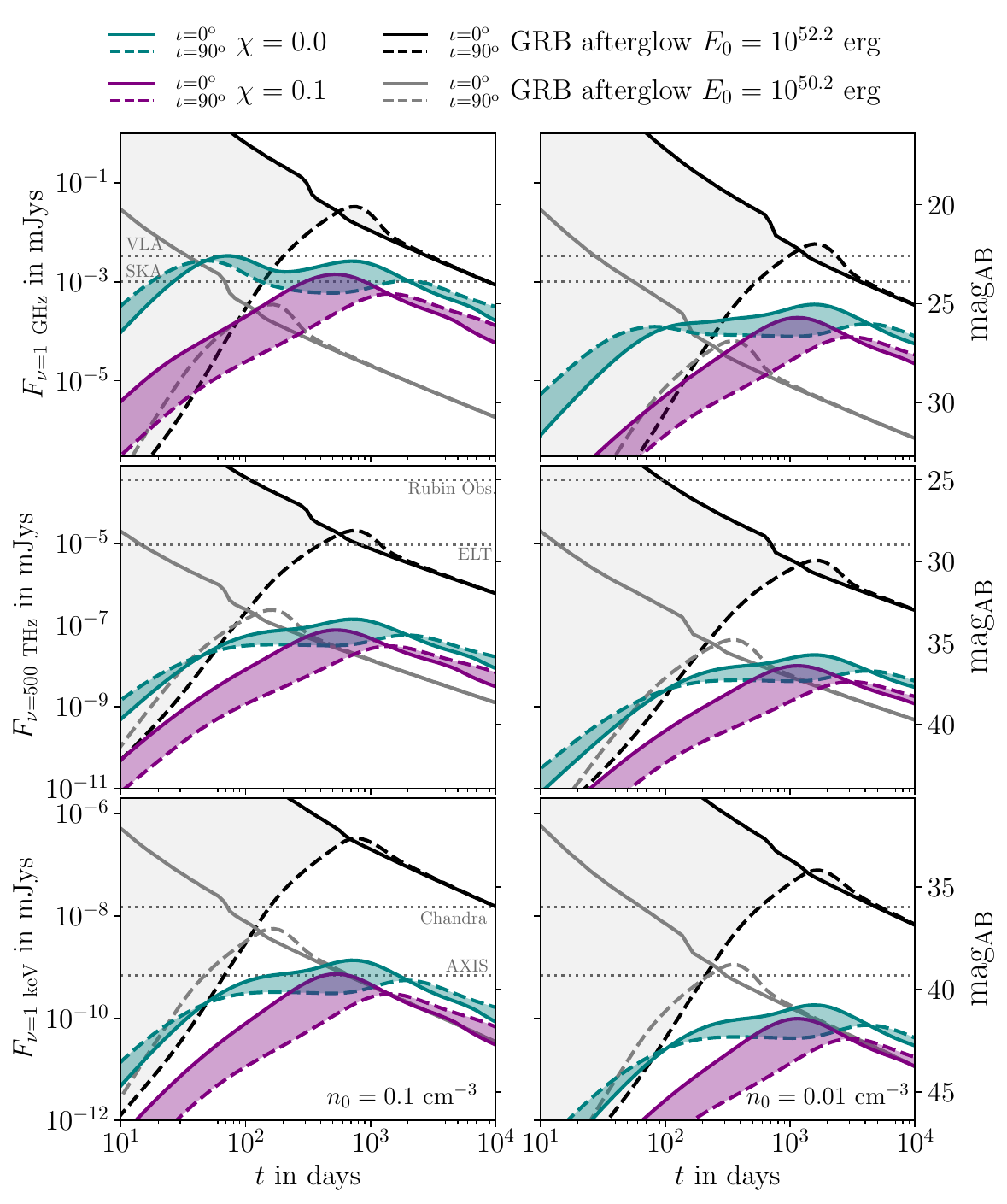}
    \caption{Kilonova afterglow light curves from the non-spinning (green) and spinning (purple) BNS simulations. Additionally, we show fiducial GRB afterglow light curves (black and gray) that are not directly related to the NR simulations. Solid lines mark observations with $\iota=0\,^\circ$, dashed lines with $\iota=90\,^\circ$. The panels on the left are for an interstellar medium density of $n_0=0.1\ \rm cm^{-3}$, the panels on the right are for $n_0=0.01\ \rm cm^{-3}$. From top to bottom, the panels show the emission at the frequencies of $1\ \rm GHz$, $500\ \rm THz$, and $1\ \rm keV$. All light curves are assumed at $d_L=100\ \rm Mpc$. Dotted lines indicate rough brightness limits for various observatories (see text).}
    \label{fig:afterglow}
\end{figure}

In Fig.~\ref{fig:afterglow}, we show the resulting kilonova afterglows from our BNS simulations and the fiducial GRB afterglow light curves. In particular, we show light curves in radio at $1\ \rm GHz$, optical at $500\ \rm THz$, and X-ray at $1\ \rm keV$ for two different interstellar medium densities, namely $n_0=0.1\ \rm cm^{-3}$ and $n_0=0.01\ \rm cm^{-3}$.

The kilonova afterglow in radio contains two separate contributions from the thermal electron population and the non-thermal population. 
The former is dominant at $t\leq100\ \rm days$, while the latter is the main contributor at $t\approx1000\ \rm days$.
Because of the strong dependence of the thermal emissivity on the shock velocity~\cite{Margalit:2021kuf}, the thermal component is more pronounced in the non-spinning case with higher ejecta masses and velocities. It there even gives rise to a double-peaked light curve, where the first peak is due to the thermal electrons.
At higher frequencies $\nu>1\ \rm GHz$, the emission is dominated by the non-thermal (or power-law) electron distribution.
The timescale of that non-thermal peak depends on the deceleration of the blast waves, where higher $n_0$ causes earlier and stronger deceleration. 
Hence, the peak is shifted from $t\approx550\ \rm days$ for $n_0=0.1\ \rm cm^{-3}$ to $t\approx1100\ \rm days$ for $n_0=0.01\ \rm cm^{-3}$.
Since the kilonova ejecta are not collimated, the observation angle $\iota$ only has a moderate effect on the observed light curve, though for an observer in the orbital plane at $\iota=90\,^\circ$, the light curve is generally dimmer and rises more slowly. 
This is because the kilonova afterglow is dominated by the material near the equatorial plane, where most of the ejecta mass resides. 
An observer located at the pole with a viewing angle of $\iota=0^\circ$ sees more of the equatorial ejecta, whereas an observer at $\iota=90\,^\circ$ receives emission mainly from one side.

For the kilonova afterglows shown in Fig.~\ref{fig:afterglow}, we have set $p=2.5$, $\epsilon_e=0.01$, $\epsilon_B=0.001$, and $\epsilon_T=1$.
These values are consistent with particle-in-cell simulations~\cite{Crumley:2018kvf, Marcowith:2020vho, Yuan:2025war}, though the possible ranges remain broad.
For this reason, we have conducted additional kilonova afterglow calculations varying the microphysical parameters.
For instance, when keeping the other parameters fixed, but setting $\epsilon_e=0.01$ and $\epsilon_B=10^{-5}$, the flux density is reduced by roughly a factor of $500$, while setting $\epsilon_e=0.2$ and $\epsilon_B=0.005$ increases the flux density by a factor of 10. 
Additionally, $\epsilon_e$ also influences the relative contribution of the thermal electron population, i.e., whether the radio light curve displays an early peak or not.
The electron power law index $p$ increases the non-thermal synchrotron flux and affects the power law decay of the late-time light curve.
The thermal part remains relatively unaffected.
The thermal efficiency $\epsilon_T$, in contrast, determines the energy budget for the thermal electrons and thus its choice can alter the early part of the radio light curve notably.
Recently, Ref.~\cite{Margalit:2024asc} suggested a more realistic value of $\epsilon_T=0.4$ based on the particle-in-cell simulations of Ref.~\cite{Vanthieghem:2024xov}.
Adopting this value while keeping the other parameters from Fig.~\ref{fig:afterglow} would reduce the thermal electron radiation notably and erase the first peak.

From Fig.~\ref{fig:afterglow} it is also apparent that, unless the BNS is observed far off-axis, the GRB afterglow dominates the early time light curve and practically prevents a distinct detection of the early kilonova afterglow.
However, if the isotropic jet energy is low, the thermal radio peak could rise above the GRB afterglow and cause a rebrightening of the light curve at $t\sim 100\ \rm days$.
This hinges on the uncertain amount of thermal electron energy and the interstellar density.
Rebrightening could also be caused in a similar way later at $t\sim1000\ \rm days$ from the non-thermal peak. 
At the higher jet energy of $10^{52.2}\ \rm erg$, the late-time GRB afterglow always outshines the kilonova afterglow by several orders of magnitude, rendering the latter effectively undetectable as a distinct source.
In general, at a luminosity distance of $100 \ \rm Mpc$, the absolute brightness of the kilonova afterglow remains low and does not exceed the typical brightness limits of current telescopes.
For $n_0=0.01\ \rm cm^{-3}$, the light curve peak remains below the 3$\sigma$ rms limit for a $12\times2.5\ \rm h$ observations with the VLA radio telescope~\cite{Balasubramanian:2022sie} or for a similar observation with the SKA~\cite{Braun:2019gdo, Bonaldi:2020ukl}.
However, at $n_0=0.1\ \rm cm^{-3}$ the thermal and non-thermal radio peak get close to the VLA detection limit and would be observable with SKA if the GRB afterglow is not too bright.
In near-optical frequency bands, however, detection of the kilonova afterglow might prove even more challenging.
The $5\sigma$ image depth of the Vera Rubin Observatory in its main survey~\cite{Bianco:2021ape, LSST:2017} and the estimated $5\sigma$ limiting magnitude of the ELT~\cite{ELT_HARMONI, ELT_MICADO} significantly exceed the expected non-thermal optical peak of the kilonova afterglow, regardless of $n_0$ or other microphysical parameters.
This also applies to Chandra X-ray detection, assuming a flux limit of $5.3\times10^{-17}\ \rm erg\ cm^{-2}\ s^{-1}$ at $200\ \rm ks$ exposure~\cite{Laird:2008va}.
For next-generation observatories like ATHENA~\cite{Nandra:2013jka} or AXIS~\cite{Reynolds:2023vvf}, the flux limit in this exposure time might be $2.5\times10^{-18}\ \rm erg\ cm^{-2}\ s^{-1}$~\cite{Marchesi:2020smf}, which would make the X-ray peak detectable for $n_0=0.1\ \rm cm^{-3}$.
We emphasize again that this assessment applies at $d_L=100\ \rm Mpc$ and assumes a typical exposure time for these telescopes.
Dedicated, ultra-deep follow-ups for a close BNS merger could enhance detectability prospects.
We also point out the possibility that a significant amount of the fast-tail ejecta is not resolved in our NR simulations due to artificial atmosphere settings and resolution issues~\cite{Rosswog:2024vfe}.
Thus, one could speculate whether the kilonova afterglow might be more powerful in certain cases than presented here.
Nevertheless, given our BNS ejecta profiles, a confident kilonova afterglow detection remains challenging, especially in a scenario with a powerful near-axis GRB afterglow.

\section{Conclusions}
\label{sec:Conclusions}

We perform state-of-the-art numerical-relativity simulations of equal-mass binary neutron star mergers in a spinning and a non-spinning configuration with \bam, incorporating both neutrino radiation and magneto-hydrodynamical effects. The neutron star matter is modeled using the ABHT(QMC-RMF) equation of state. We initialize a poloidal magnetic field with a maximum strength of $10^{15}\ \rm G$ inside the stars.
Each system is simulated at a high resolution with a grid spacing of $\Delta x \approx 93\ \rm m$ covering the neutron stars and a low resolution with $\Delta x \approx 186\ \rm m$, and is evolved up to $\sim100\ \rm ms$ after the merger, covering besides the dynamical ejecta also secular outflows driven by magnetic- and neutrino winds. 

With the same initial separation of $41.2\ \rm km $, the system with spin-aligned stars merges about $0.88\ \rm ms $ later than the non-spinning system due to repulsive spin–orbit interaction. The reduced impact velocity at the merger leads to a less violent collision in the spinning configuration, resulting in lower temperatures in the merger remnant and a smaller amount of shock-heated ejecta. Due to the higher angular momentum, the spinning system forms a more massive disk and a more extended remnant with lower central rest-mass densities. 

The initial magnetic-field amplification due to the Kelvin–Helmholtz instability, triggered in the shear layer during the merger, behaves similarly in both configurations and reaches similar field strengths and energies within the first $\lesssim 10\ \rm ms$ after merger. Subsequently, the magnetic energy increases further through magnetic winding and magneto-rotational instability, driven by the interaction of the magnetic field with the differentially rotating disk. Higher magnetic energies are reached in the non-spinning scenario. 

The amount of ejected material is larger in the non-spinning configuration. In the simulated equal-mass systems, the dynamical ejecta are mainly generated by shocks at the collision interface, with only minor contributions from tidal disruption. The more violent merger of the non-spinning system produces stronger shocks, leading to larger ejecta masses and higher outflow velocities. Furthermore, we find for this system a lower electron fraction in the orbital plane of $\lesssim 0.1$, and thus more neutron-rich ejecta. On later timescales of $> 10\ \rm ms$ after the merger, neutrino-driven winds power a slower outflow component with progressively higher electron fractions. 
Finally, we observe the launch of a mildly relativistic collimated outflow with velocities of up to $0.4\ c$ along the polar direction and high magnetization at $\sim100\ \rm ms$ after the merger in the non-spinning configuration. 

The thermodynamical conditions in the different ejecta components lead to characteristic abundance patterns from $r$-process nucleosynthesis. 
We confirm that the low-$Y_e$ components of the dynamical ejecta robustly reproduce the global features of the heavy-element abundances observed in the solar system, whereas the late winds cannot substantially surpass the first $r$-process peak. A substantial lack of low-mass lanthanides points towards our still incomplete understanding of various nuclear decay properties far from stability, while the relative paucity of nuclei around $A=190$ is linked to an underproduction of cold ejecta in equal-mass systems.\\

To obtain a comprehensive picture of the observable multi-messenger signatures, we extract the gravitational-wave signal and use the ejecta data to compute the light curves of the associated kilonova and its afterglow.

Since the description of the gravitational-wave signal during the inspiral is well-established by various approximants, we focus on the analysis of the post-merger frequencies, where we found slight shifts in the dominant peak frequencies between the two configurations. However, these differences are within numerical uncertainties of the simulations and are therefore likely artifacts of the grid resolution.

For the EM counterparts, we focus on the high-resolution simulations. 
We compute kilonova light curves using the Monte Carlo radiative transfer code \textsc{possis}. The emission is overall brighter in the non-spinning configuration, primarily due to the larger amount of ejecta. The kilonova is also slightly redder for this system, which we attribute to the lower electron fractions reached in the ejecta. Moreover, the light curves show a stronger dependence on the viewing angle in the non-spinning configuration than in the spinning one.

The afterglow emission are computed using the semi-analytic \textsc{pyblastafterglow} code. For the kilonova afterglow, we use ejecta profiles binned over different viewing angles $\iota$ and asymptotic velocities. While both a Maxwellian (thermal) and a power-law (non-thermal) electron component contribute to the afterglow, again the non-spinning configuration produces brighter light curves due to the larger ejecta mass and higher ejecta velocities.
Since our numerical-relativity simulations do not resolve an ultra-relativistic jet, we model the GRB afterglow assuming a fiducial Gaussian jet with parameters inferred from GRB170817A and explore different jet energies. When observed nearly on-axis, the GRB afterglow emission dominates over the kilonova afterglow. For an observer in the orbital plane, the GRB afterglow is initially suppressed. Nevertheless, the peak of the thermal emission of the kilonova afterglow is only observable for sufficiently low isotropic jet energies.\\

While our BNS simulations include a comprehensive treatment of the key microphysical and magnetohydrodynamic processes in mergers of binary neutron stars, the presence of muons and anti-muons, as well as resistive effects of the magnetic field are neglected. The former is expected to influence primarily the remnant lifetime and ejecta masses as well as their composition, thereby affecting the signatures of the kilonova~\cite{Gieg:2024jxs,Ng:2024zve}. The latter becomes significant in regions with low conductivity, such as in the magnetospheres of neutron stars and the polar cap above the remnant, e.g., \cite{Dionysopoulou:2015tda,Cheong:2024stz}. Accounting for finite conductivity would enable modeling magnetic reconnection and dissipation processes that can affect both jet dynamics and electromagnetic emission. In addition, simulations using more realistic initial magnetic field configurations consistent with astrophysical observations are needed. 
We plan to address these aspects in future work and extend our simulations to resistive magnetohydrodynamics with muons.

\section{Data availability}

Gravitational waveforms will be released as part of the CoRe database~\cite{Dietrich:2018phi,Gonzalez:2022mgo}.
Additionally, we provide an animation of the two high-resolution simulations~\cite{markin_2025_17303085}.
Further simulation data can be provided upon reasonable request.

\section*{Acknowledgments}

We thank B. Brügmann for valuable comments and discussions during this project. We further thank Rosa A. for her thoughtful supervision during the text review of this manuscript.
A.N., I.M., and T.D. gratefully acknowledge support from the Deutsche Forschungsgemeinschaft, DFG, project number 504148597 (DI 2553/7). Furthermore, T.D., H.K., H.R., and H.G. acknowledge funding from the EU Horizon under ERC Starting Grant, no.\ SMArt-101076369. L.B. is supported by the U.S. Department of Energy, Office of Science, Office of Nuclear Physics, under Award No. DE-FG02-05ER41375. A.H.~acknowledges support by the U.S. Department of Energy, Office of Science, Office of Nuclear Physics, under Award No. DE-FG02-05ER41375. A.H.~furthermore acknowledges financial support by the UKRI under the Horizon Europe Guarantee project EP/Z000939/1. M.B. acknowledges the Department of Physics and Earth Science of the University of Ferrara for the financial support through the FIRD 2024 grant. 
The simulations with \bam\ were performed on the national supercomputer HPE Apollo Hawk at the High Performance Computing (HPC) Center Stuttgart (HLRS) under the grant number mmapicture/44289, and on the DFG-funded research cluster Jarvis at the University of Potsdam (INST 336/173-1; project number: 502227537).
Views and opinions expressed are those of the authors only and do not necessarily reflect those of the European Union or the European Research Council. Neither the European Union nor the granting authority can be held responsible for them.

\appendix

\section{Fluid-radiation equilibration step}
\label{app:fluid-radiation}

The computation of the trapped equilibrium solution of Refs.~\cite{Radice:2021jtw,Perego:2019adq} is based on considering the fluid and radiation in the trapped regime as a single system in thermal equilibrium with a total energy density $e$ and lepton number $Y_l$. From Fermi--Dirac statistics, we know that the fluid is supposed to thermalize at a certain $Y_{e,{\rm eq}}$ and $T_{\rm eq}$ satisfying
\begin{align}
    e &= \rho \epsilon(\rho,T_{\rm eq},Y_{e,\rm eq}) + \frac{\rho}{m_b} \left[ Z_{\nu_e}(\rho,T_{\rm eq},Y_{e,\rm eq}) \right. \nonumber \\
    & \left.+ Z_{\overline{\nu}_e}(\rho,T_{\rm eq},Y_{e,\rm eq}) + 4Z_{\nu_x}(\rho,T_{\rm eq},Y_{e,\rm eq}) \right] \label{eq:T_eq}, \\
    Y_l &= Y_{e,\rm eq}  + Y_{\nu_e}(\rho,T_{\rm eq},Y_{e,\rm eq}) - Y_{\overline{\nu}_e}(\rho,T_{\rm eq},Y_{e,\rm eq}) \label{eq:Ye_eq},
\end{align}
with the neutrino number and energy fractions given by
\begin{align}
    Y_{\nu_i} &= \frac{4\pi}{(hc)^3} \frac{m_b}{\rho} T^3 F_2\left( \frac{\mu_{\nu_i}}{T} \right), \\
    Z_{\nu_i} &= \frac{4\pi}{(hc)^3} \frac{m_b}{\rho} T^4 F_3\left( \frac{\mu_{\nu_i}}{T} \right).
\end{align}

Ref.~\cite{Radice:2021jtw} solves this system of two equations in two variables on the fly at each computation of the right-hand side. In order to improve efficiency, we employ a tabulation strategy of a precomputed solution. However, tabulating the internal energy density $e$ is challenging given its wide range, covering several orders of magnitude, and its dependency on the EOS. Therefore, we implicitly define the auxiliary temperature $T^*$ by
\begin{equation}
    \rho \epsilon(\rho,T^*,Y_l) = e.
    \label{eq:T_star}
\end{equation}

We sample $\rho$, $T^*$, and $Y_l$ on the corresponding values of $\rho$, $T$ and $Y_e$ in the EOS table. For each triad $(\rho,T^*,Y_l)$, we compute the corresponding $e$ and solve Eqs.\eqref{eq:T_eq} and \eqref{eq:Ye_eq} for $T_{\rm eq}$ and $Y_{e,\rm eq}$.
The final result is a mapping 
\begin{equation}
   (\rho,e,Y_l)\mapsto (\rho,T^*,Y_l) \mapsto (T_{\rm eq}, Y_{e,\rm eq}),
\end{equation}
where the first mapping is performed solving Eq.~\eqref{eq:T_star} for $T^*$ using the standard temperature-recovering routine, and the second one using the same 3D interpolation routines used for computing EOS quantities. Finally, the emissivities computed using $(\rho,T,Y_e)$ and $(\rho,T_{\rm eq},Y_{e,\rm eq})$ are combined according to the weight function of Ref.~\cite{Radice:2021jtw}.

\section{Tracer particles}
\label{app:Tracers}

We extend \bam\ to evolve tracer particles in order to record the evolution history of Lagrangian fluid trajectories, allowing for a more detailed analysis of the outflowing material and its origin.
The tracers are advected via
\begin{equation}
    \partial_t x_{\rm tr}^i = \alpha v^i - \beta^i,
\end{equation}
where $x_{\rm tr}^i$ is the position of an individual tracer particle. 
The lapse $\alpha$, shift $\beta^i$, and fluid velocity $v^i$ as well as the recorded fluid quantities rest-mass density $\rho$, electron fraction $Y_e$, and temperature $T$ are interpolated to the position $x^i_{\rm tr}$ of the tracer.
The tracer particles are evolved during the BNS simulations using a simple forward Euler scheme, with the time step given by that of the finest refinement level covering the respective domain.

\begin{figure}[t]
    \centering
    \includegraphics[width=\linewidth]{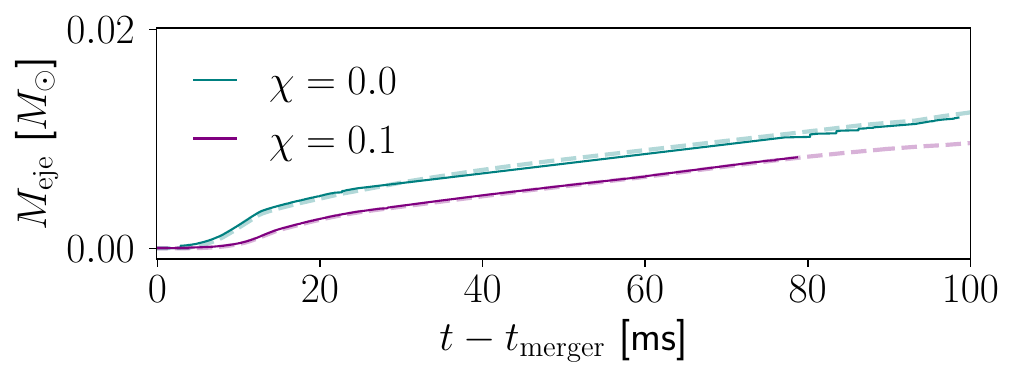}
    \caption{Comparison between ejecta passing through a sphere with a radius of $\sim 440\ \rm km$, represented by the tracer particles (solid lines), and the matter outflow extracted on the coordinate sphere (faint dashed lines) for the high-resolution non-spinning (green) and spinning (purple) BNS simulations.}
    \label{fig:tracermass}
\end{figure}

\begin{figure*}
    \centering
    \includegraphics[width=\linewidth]{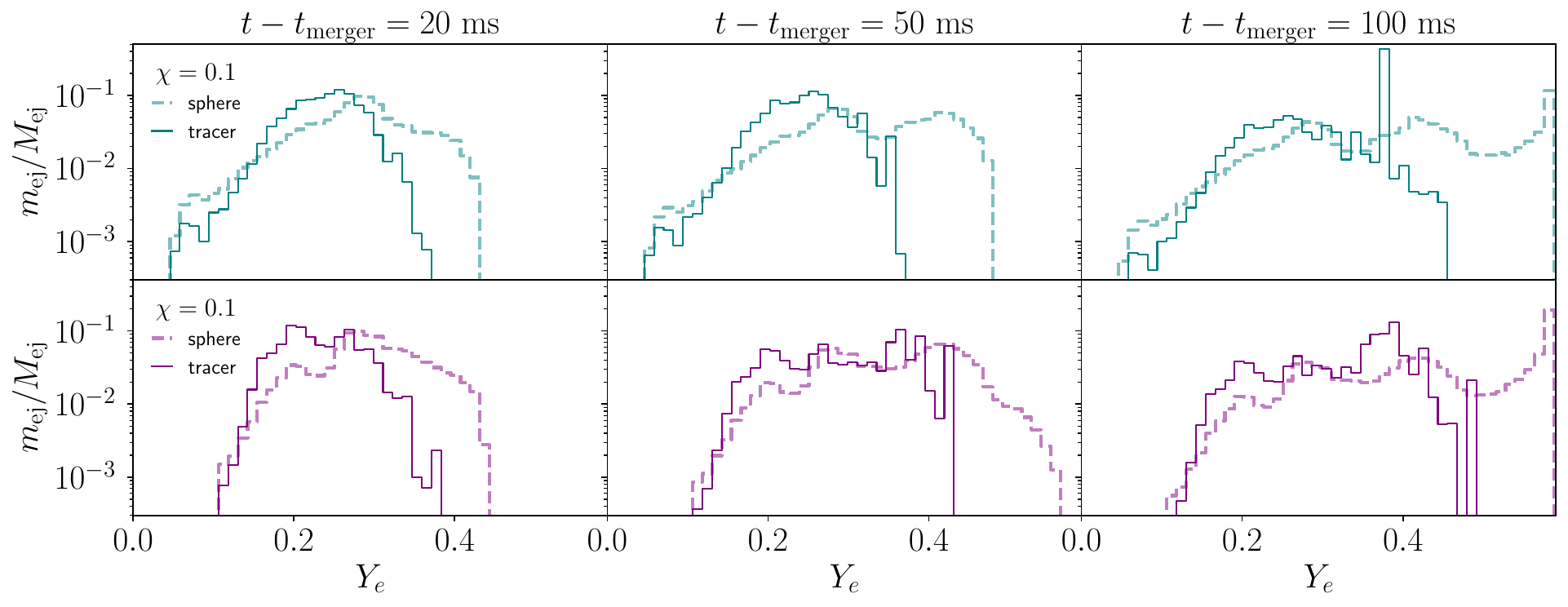}
    \caption{Comparison between the ejecta's electron fraction represented by the tracer particles (solid lines) and extracted on the coordinate sphere (faint dashed lines) for the high-resolution non-spinning (green) and spinning (purple) BNS simulations. The mass-weighted histograms are computed for ejecta passing through a sphere with a radius of $\sim 440\ \rm km$ until $\sim 20\ \rm ms$ (left panels), $\sim 50\ \rm ms$ (middle panels), $\sim 100\ \rm ms$ (right panels) after the merger.}
    \label{fig:tracerYe}
\end{figure*}

In total, we inject $\sim 10^5$ tracer particles into our BNS simulations at $\lesssim\qty{1}{\milli\second}$ before the merger. They are randomly distributed in neutron star matter for regions with densities $5000$~times above the artificial atmosphere value, corresponding to $\rho \geq \qty{6.176e7}{\gram\per\centi\meter\cubed}$.
The tracer particles are evolved until they cross a sphere of radius $\sim \qty{737}{\kilo\meter}$. Subsequently, we assume a homologous expansion.
For the nuclear reaction network, we consider only tracer particles that are unbound and traverse this sphere before the end of our simulation. 
Excluding $\approx 100$ tracers that are numerically unstable in the nucleosynthesis evolution, we ultimately obtain $58{,}264$ and $60{,}620$ tracers representing the ejecta for the high-resolution non-spinning and spinning BNS simulations, respectively. 

Although the tracers are massless by definition, nuclear abundance calculations require information on the mass represented by the individual tracer particles. For this purpose, we compute tracer masses on a sphere of radius $\sim 440\ \rm km$. We extract the ejected mass flux $\frac{d M_{\rm eje}}{d t}\left(t\right)$ on this detection sphere and impose
\begin{equation}
    \frac{d M_{\rm eje}}{d t}\left(t\right) = \sum_{\rm tr}^{N\left(t\right)} \frac{d M_{\rm tr}}{d t},
    \label{eq:massflux}
\end{equation}
where we sum over all $N(t)$ tracer particles that pass the sphere between $t$ and $t+\Delta t$. Here, $\Delta t$ is the output interval of the tracer data. We approximate the mass flux of the individual tracer particle by
\begin{equation}
    \frac{d M_{\rm tr}}{d t} \approx \Delta \Omega_{\rm tr} r_{\rm tr}^2 \rho_{\rm tr} v_{r,\rm tr},
\end{equation}
where $r_{\rm tr}$ is the distance of the tracer from the coordinate centre ($\sim 440\ \rm km$), and $\rho_{\rm tr}$ and $v_{r,{\rm tr}}$ denote the rest-mass density and radial velocity of the fluid element, respectively. 
$\Delta \Omega_{\rm tr}$ is the solid angle represented by the tracer particle. In the absence of more detailed information, we assume for simplicity that each particle passing through the sphere between $t$ and $t+\Delta t$ contributes to the total mass flux with the same $\Delta \Omega$. Thus, $\Delta \Omega$ can be scaled to ensure that Eq.~\eqref{eq:massflux} is satisfied. Finally, integration over time yields the mass $M_{\rm tr}$ ascribed to each tracer particle as it passes through the sphere. 

In Fig.~\ref{fig:tracermass}, we show the evolution of ejecta mass represented by the tracer particles, obtained by summing $M_{\rm tr}$ of all tracer crossing a sphere with a radius of $\sim 440\ \rm km$ for both high-resolution BNS simulations. For comparison, the ejecta masses extracted directly from the coordinate sphere are shown as faint dashed lines, demonstrating good agreement.
In Fig.~\ref{fig:tracerYe}, we further compare the mass-weighted histograms of the electron fraction $Y_e$ in the matter outflow at different times after the merger, computed once from the tracer particles and once from the detection sphere.
The histograms agree reasonably well for the low $Y_e$ material emitted at early times, i.e., $\lesssim 20\ \rm ms$ after the merger. Larger deviations occur at later times for the high $Y_e$ material because there are fewer tracer particles in the secular outflow with a higher electron fraction, and hence we sample this component more coarsely.

We assume this to be a consequence of the initial distribution of the tracer particles and the advection scheme: With a density-dependent distribution, the core of the neutron stars would be better sampled, and the tracer masses would possibly be more uniform. Our choice of uniform distribution, on the other hand, leads to better coverage of the outer layers of neutron stars with lower density. 
The vast majority of these tracers is ejected at early times and hence assigned to low masses. By contrast, only few tracers cross the extractions sphere at late times, coming from the accretion disk which represents an unstable advection state as opposed to infall to the black hole or ejection.
As a result, late ejecta have a much higher mass and dominate the full picture.
From each of the roughly $60000$~ejected tracers, only $62$~$(2228)$ and $29$~$(4084)$ tracers make up $50\, \%$ ($90\, \%$) of the total ejecta mass in the spinning and non-spinning case, respectively.
As discussed in Sec.~\ref{subsec:yields}, however, these have negligible impact on the massive nuclei yield. \\

We finally note that technical complications occurred in the R2 simulation for the non-spinning BNS configuration as the tracer particles were randomly redistributed in the neutron star material at $t_{\rm step} =7.81\ \rm ms$ and $t_{\rm step} =17.62\ \rm ms$. This error resulted from limited storage and failure of the checkpointing routine for the tracer particles, leading to a new tracer initialization after reaching the computing system's walltime. 
We resolve this by ``stacking'' matching trajectories across these time steps with the following procedure, where we refer to tracer particles before and after a restart as ``old'' and ``new'', respectively.
\begin{itemize}
    \item First, we identified tracer particles whose temperatures are still too high to be relevant for nuclear network calculation. 
    \item If $T>7\ \rm GK$, we stacked the trajectory of the ``old'' tracer particle with that of the closest ``new'' tracer particle that also satisfies $T>7\ \rm GK$, by looking for the minimum of $\left|\left(x_{\rm tr,old}^i+v^i_{\rm tr,old}\Delta t\right)-x_{\rm tr,new}^i\right|$.
    \item Otherwise, if $T \leq 7\ \rm GK$, we stacked the trajectory of the ``old'' tracer particle with that of a ``new'' tracer particle having similar fluid properties for $T$, $\rho$, and $Y_e$. Due to the assumed axial symmetry of the system, it is sufficient that the ``new'' tracer particle has a similar position in $z$ and $\sqrt{x^2 +y^2}$ as the ``old'' tracer particle. 
    The trajectories were only stacked, if the relative differences in rest-mass density and temperature were sufficiently small with $\left|\frac{\rho_{\rm tr,old}}{\rho_{\rm tr,new}} - 1\right| <1$ and $\left|\frac{T_{\rm tr,old}}{T_{\rm tr,new}} - 1\right| <10$, and the absolute difference in $Y_e$ was below $0.005$.
\end{itemize}

\begin{table}[h]
    \centering
    \caption{The consumed compute time and energy, and the induced amount of GHG emissions by our NR simulations.}
    \label{table:carbon_footprint}
    \begin{tabular}{l||c|c|c}
    \toprule
        & Compute time, & Energy, & Grid emissions, \\
        &  millions core-h & MWh & t\ch{CO2} \\
        \hline \hline
Evolution R1 & 2.00 & 8.18 & 1.57 \\
Evolution R2 & 27.89 & 114.37 & 21.96 \\
\hline
Total & 29.88 &  122.56 & 23.53 \\
    \bottomrule
    \end{tabular}
\end{table}

\section{Consumed resources and carbon footprint}
\label{app:ressources}

Performing high-resolution NR simulations with detailed physics is computationally expensive and thus requires a large amount of energy. Most of the time, computing facilities are attached to national grids, which acquire energy from a mix of different sources, including fossil fuels. This in turn induces a tangible amount of greenhouse gas (GHG) emissions, which are driving climate change~\cite{Oreskes_2004,Doran_2009,Cook_2013,Cook_2016,Lynas_2021,Myers_2021}.

We estimate the amount of GHG emissions produced by the NR simulations in this study. Starting from November 2024, HLRS began to provide detailed power consumption statistics. Using this data for the jobs after that date, we estimate the average power consumption of a single Hawk node used for our simulations to be $525\ \rm W$\footnote{Meanwhile, a simple estimate via the thermal design power (TDP) of the CPUs yields $450\ \rm W$, which is $\sim 15\, \%$ lower than the full node power consumption.}. Using this average node power, we estimate the energy consumed and the GHG emissions induced by our simulations, assuming the specific carbon intensity of the Universität Stuttgart at the Campus Vaihingen in 2023 with $192\ \rm g$\ch{CO2}$ \rm /kWh$~\cite{HLRS:2023}. In Table~\ref{table:carbon_footprint}, we list the consumed compute resources, energy, and the induced grid emissions.

\bibliography{ref.bib}

\begin{thebibliography}{263}%
\makeatletter
\providecommand \@ifxundefined [1]{%
 \@ifx{#1\undefined}
}%
\providecommand \@ifnum [1]{%
 \ifnum #1\expandafter \@firstoftwo
 \else \expandafter \@secondoftwo
 \fi
}%
\providecommand \@ifx [1]{%
 \ifx #1\expandafter \@firstoftwo
 \else \expandafter \@secondoftwo
 \fi
}%
\providecommand \natexlab [1]{#1}%
\providecommand \enquote  [1]{``#1''}%
\providecommand \bibnamefont  [1]{#1}%
\providecommand \bibfnamefont [1]{#1}%
\providecommand \citenamefont [1]{#1}%
\providecommand \href@noop [0]{\@secondoftwo}%
\providecommand \href [0]{\begingroup \@sanitize@url \@href}%
\providecommand \@href[1]{\@@startlink{#1}\@@href}%
\providecommand \@@href[1]{\endgroup#1\@@endlink}%
\providecommand \@sanitize@url [0]{\catcode `\\12\catcode `\$12\catcode
  `\&12\catcode `\#12\catcode `\^12\catcode `\_12\catcode `\%12\relax}%
\providecommand \@@startlink[1]{}%
\providecommand \@@endlink[0]{}%
\providecommand \url  [0]{\begingroup\@sanitize@url \@url }%
\providecommand \@url [1]{\endgroup\@href {#1}{\urlprefix }}%
\providecommand \urlprefix  [0]{URL }%
\providecommand \Eprint [0]{\href }%
\providecommand \doibase [0]{https://doi.org/}%
\providecommand \selectlanguage [0]{\@gobble}%
\providecommand \bibinfo  [0]{\@secondoftwo}%
\providecommand \bibfield  [0]{\@secondoftwo}%
\providecommand \translation [1]{[#1]}%
\providecommand \BibitemOpen [0]{}%
\providecommand \bibitemStop [0]{}%
\providecommand \bibitemNoStop [0]{.\EOS\space}%
\providecommand \EOS [0]{\spacefactor3000\relax}%
\providecommand \BibitemShut  [1]{\csname bibitem#1\endcsname}%
\let\auto@bib@innerbib\@empty
\bibitem [{\citenamefont {Fern{\'a}ndez}\ and\ \citenamefont
  {Metzger}(2016)}]{Fernandez:2015use}%
  \BibitemOpen
  \bibfield  {author} {\bibinfo {author} {\bibfnamefont {R.}~\bibnamefont
  {Fern{\'a}ndez}}\ and\ \bibinfo {author} {\bibfnamefont {B.~D.}\ \bibnamefont
  {Metzger}},\ }\href {https://doi.org/10.1146/annurev-nucl-102115-044819}
  {\bibfield  {journal} {\bibinfo  {journal} {Ann. Rev. Nucl. Part. Sci.}\
  }\textbf {\bibinfo {volume} {66}},\ \bibinfo {pages} {23} (\bibinfo {year}
  {2016})},\ \Eprint {https://arxiv.org/abs/1512.05435} {arXiv:1512.05435
  [astro-ph.HE]} \BibitemShut {NoStop}%
\bibitem [{\citenamefont {Metzger}(2017)}]{Metzger:2016pju}%
  \BibitemOpen
  \bibfield  {author} {\bibinfo {author} {\bibfnamefont {B.~D.}\ \bibnamefont
  {Metzger}},\ }\href {https://doi.org/10.1007/s41114-017-0006-z} {\bibfield
  {journal} {\bibinfo  {journal} {Living Rev. Rel.}\ }\textbf {\bibinfo
  {volume} {20}},\ \bibinfo {pages} {3} (\bibinfo {year} {2017})},\ \Eprint
  {https://arxiv.org/abs/1610.09381} {arXiv:1610.09381 [astro-ph.HE]}
  \BibitemShut {NoStop}%
\bibitem [{\citenamefont {Baiotti}\ and\ \citenamefont
  {Rezzolla}(2017)}]{Baiotti:2016qnr}%
  \BibitemOpen
  \bibfield  {author} {\bibinfo {author} {\bibfnamefont {L.}~\bibnamefont
  {Baiotti}}\ and\ \bibinfo {author} {\bibfnamefont {L.}~\bibnamefont
  {Rezzolla}},\ }\href {https://doi.org/10.1088/1361-6633/aa67bb} {\bibfield
  {journal} {\bibinfo  {journal} {Rept. Prog. Phys.}\ }\textbf {\bibinfo
  {volume} {80}},\ \bibinfo {pages} {096901} (\bibinfo {year} {2017})},\
  \Eprint {https://arxiv.org/abs/1607.03540} {arXiv:1607.03540 [gr-qc]}
  \BibitemShut {NoStop}%
\bibitem [{\citenamefont {Duez}\ and\ \citenamefont
  {Zlochower}(2019)}]{Duez:2018jaf}%
  \BibitemOpen
  \bibfield  {author} {\bibinfo {author} {\bibfnamefont {M.~D.}\ \bibnamefont
  {Duez}}\ and\ \bibinfo {author} {\bibfnamefont {Y.}~\bibnamefont
  {Zlochower}},\ }\href {https://doi.org/10.1088/1361-6633/aadb16} {\bibfield
  {journal} {\bibinfo  {journal} {Rept. Prog. Phys.}\ }\textbf {\bibinfo
  {volume} {82}},\ \bibinfo {pages} {016902} (\bibinfo {year} {2019})},\
  \Eprint {https://arxiv.org/abs/1808.06011} {arXiv:1808.06011 [gr-qc]}
  \BibitemShut {NoStop}%
\bibitem [{\citenamefont {Bernuzzi}(2020)}]{Bernuzzi:2020tgt}%
  \BibitemOpen
  \bibfield  {author} {\bibinfo {author} {\bibfnamefont {S.}~\bibnamefont
  {Bernuzzi}},\ }\href {https://doi.org/10.1007/s10714-024-03291-z} {\bibfield
  {journal} {\bibinfo  {journal} {Gen. Rel. Grav.}\ }\textbf {\bibinfo {volume}
  {52}},\ \bibinfo {pages} {108} (\bibinfo {year} {2020})},\ \Eprint
  {https://arxiv.org/abs/2004.06419} {arXiv:2004.06419 [astro-ph.HE]}
  \BibitemShut {NoStop}%
\bibitem [{\citenamefont {Ruiz}\ \emph {et~al.}(2021)\citenamefont {Ruiz},
  \citenamefont {Shapiro},\ and\ \citenamefont {Tsokaros}}]{Ruiz:2021gsv}%
  \BibitemOpen
  \bibfield  {author} {\bibinfo {author} {\bibfnamefont {M.}~\bibnamefont
  {Ruiz}}, \bibinfo {author} {\bibfnamefont {S.~L.}\ \bibnamefont {Shapiro}},\
  and\ \bibinfo {author} {\bibfnamefont {A.}~\bibnamefont {Tsokaros}},\ }\href
  {https://doi.org/10.3389/fspas.2021.656907} {\bibfield  {journal} {\bibinfo
  {journal} {Front. Astron. Space Sci.}\ }\textbf {\bibinfo {volume} {8}},\
  \bibinfo {pages} {39} (\bibinfo {year} {2021})},\ \Eprint
  {https://arxiv.org/abs/2102.03366} {arXiv:2102.03366 [astro-ph.HE]}
  \BibitemShut {NoStop}%
\bibitem [{\citenamefont {Lattimer}\ and\ \citenamefont
  {Schramm}(1974)}]{Lattimer:1974slx}%
  \BibitemOpen
  \bibfield  {author} {\bibinfo {author} {\bibfnamefont {J.~M.}\ \bibnamefont
  {Lattimer}}\ and\ \bibinfo {author} {\bibfnamefont {D.~N.}\ \bibnamefont
  {Schramm}},\ }\href {https://doi.org/10.1086/181612} {\bibfield  {journal}
  {\bibinfo  {journal} {Astrophys. J. Lett.}\ }\textbf {\bibinfo {volume}
  {192}},\ \bibinfo {pages} {L145} (\bibinfo {year} {1974})}\BibitemShut
  {NoStop}%
\bibitem [{\citenamefont {{Rosswog}}\ \emph {et~al.}(1999)\citenamefont
  {{Rosswog}}, \citenamefont {{Liebend{\"o}rfer}}, \citenamefont
  {{Thielemann}}, \citenamefont {{Davies}}, \citenamefont {{Benz}},\ and\
  \citenamefont {{Piran}}}]{Rosswog:1998hy}%
  \BibitemOpen
  \bibfield  {author} {\bibinfo {author} {\bibfnamefont {S.}~\bibnamefont
  {{Rosswog}}}, \bibinfo {author} {\bibfnamefont {M.}~\bibnamefont
  {{Liebend{\"o}rfer}}}, \bibinfo {author} {\bibfnamefont {F.~K.}\ \bibnamefont
  {{Thielemann}}}, \bibinfo {author} {\bibfnamefont {M.~B.}\ \bibnamefont
  {{Davies}}}, \bibinfo {author} {\bibfnamefont {W.}~\bibnamefont {{Benz}}},\
  and\ \bibinfo {author} {\bibfnamefont {T.}~\bibnamefont {{Piran}}},\ }\href
  {https://doi.org/10.48550/arXiv.astro-ph/9811367} {\bibfield  {journal}
  {\bibinfo  {journal} {Astron. Astrophys.}\ }\textbf {\bibinfo {volume}
  {341}},\ \bibinfo {pages} {499} (\bibinfo {year} {1999})},\ \Eprint
  {https://arxiv.org/abs/astro-ph/9811367} {arXiv:astro-ph/9811367 [astro-ph]}
  \BibitemShut {NoStop}%
\bibitem [{\citenamefont {Korobkin}\ \emph {et~al.}(2012)\citenamefont
  {Korobkin}, \citenamefont {Rosswog}, \citenamefont {Arcones},\ and\
  \citenamefont {Winteler}}]{Korobkin:2012uy}%
  \BibitemOpen
  \bibfield  {author} {\bibinfo {author} {\bibfnamefont {O.}~\bibnamefont
  {Korobkin}}, \bibinfo {author} {\bibfnamefont {S.}~\bibnamefont {Rosswog}},
  \bibinfo {author} {\bibfnamefont {A.}~\bibnamefont {Arcones}},\ and\ \bibinfo
  {author} {\bibfnamefont {C.}~\bibnamefont {Winteler}},\ }\href
  {https://doi.org/10.1111/j.1365-2966.2012.21859.x} {\bibfield  {journal}
  {\bibinfo  {journal} {Mon. Not. Roy. Astron. Soc.}\ }\textbf {\bibinfo
  {volume} {426}},\ \bibinfo {pages} {1940} (\bibinfo {year} {2012})},\ \Eprint
  {https://arxiv.org/abs/1206.2379} {arXiv:1206.2379 [astro-ph.SR]}
  \BibitemShut {NoStop}%
\bibitem [{\citenamefont {Wanajo}\ \emph {et~al.}(2014)\citenamefont {Wanajo},
  \citenamefont {Sekiguchi}, \citenamefont {Nishimura}, \citenamefont {Kiuchi},
  \citenamefont {Kyutoku},\ and\ \citenamefont {Shibata}}]{Wanajo:2014wha}%
  \BibitemOpen
  \bibfield  {author} {\bibinfo {author} {\bibfnamefont {S.}~\bibnamefont
  {Wanajo}}, \bibinfo {author} {\bibfnamefont {Y.}~\bibnamefont {Sekiguchi}},
  \bibinfo {author} {\bibfnamefont {N.}~\bibnamefont {Nishimura}}, \bibinfo
  {author} {\bibfnamefont {K.}~\bibnamefont {Kiuchi}}, \bibinfo {author}
  {\bibfnamefont {K.}~\bibnamefont {Kyutoku}},\ and\ \bibinfo {author}
  {\bibfnamefont {M.}~\bibnamefont {Shibata}},\ }\href
  {https://doi.org/10.1088/2041-8205/789/2/L39} {\bibfield  {journal} {\bibinfo
   {journal} {Astrophys. J. Lett.}\ }\textbf {\bibinfo {volume} {789}},\
  \bibinfo {pages} {L39} (\bibinfo {year} {2014})},\ \Eprint
  {https://arxiv.org/abs/1402.7317} {arXiv:1402.7317 [astro-ph.SR]}
  \BibitemShut {NoStop}%
\bibitem [{\citenamefont {Cowan}\ \emph {et~al.}(2021)\citenamefont {Cowan},
  \citenamefont {Sneden}, \citenamefont {Lawler}, \citenamefont {Aprahamian},
  \citenamefont {Wiescher}, \citenamefont {Langanke}, \citenamefont
  {Mart{\'\i}nez-Pinedo},\ and\ \citenamefont {Thielemann}}]{Cowan:2019pkx}%
  \BibitemOpen
  \bibfield  {author} {\bibinfo {author} {\bibfnamefont {J.~J.}\ \bibnamefont
  {Cowan}}, \bibinfo {author} {\bibfnamefont {C.}~\bibnamefont {Sneden}},
  \bibinfo {author} {\bibfnamefont {J.~E.}\ \bibnamefont {Lawler}}, \bibinfo
  {author} {\bibfnamefont {A.}~\bibnamefont {Aprahamian}}, \bibinfo {author}
  {\bibfnamefont {M.}~\bibnamefont {Wiescher}}, \bibinfo {author}
  {\bibfnamefont {K.}~\bibnamefont {Langanke}}, \bibinfo {author}
  {\bibfnamefont {G.}~\bibnamefont {Mart{\'\i}nez-Pinedo}},\ and\ \bibinfo
  {author} {\bibfnamefont {F.-K.}\ \bibnamefont {Thielemann}},\ }\href
  {https://doi.org/10.1103/RevModPhys.93.015002} {\bibfield  {journal}
  {\bibinfo  {journal} {Rev. Mod. Phys.}\ }\textbf {\bibinfo {volume} {93}},\
  \bibinfo {pages} {15002} (\bibinfo {year} {2021})},\ \Eprint
  {https://arxiv.org/abs/1901.01410} {arXiv:1901.01410 [astro-ph.HE]}
  \BibitemShut {NoStop}%
\bibitem [{\citenamefont {Rezzolla}\ \emph {et~al.}(2011)\citenamefont
  {Rezzolla}, \citenamefont {Giacomazzo}, \citenamefont {Baiotti},
  \citenamefont {Granot}, \citenamefont {Kouveliotou},\ and\ \citenamefont
  {Aloy}}]{Rezzolla:2011da}%
  \BibitemOpen
  \bibfield  {author} {\bibinfo {author} {\bibfnamefont {L.}~\bibnamefont
  {Rezzolla}}, \bibinfo {author} {\bibfnamefont {B.}~\bibnamefont
  {Giacomazzo}}, \bibinfo {author} {\bibfnamefont {L.}~\bibnamefont {Baiotti}},
  \bibinfo {author} {\bibfnamefont {J.}~\bibnamefont {Granot}}, \bibinfo
  {author} {\bibfnamefont {C.}~\bibnamefont {Kouveliotou}},\ and\ \bibinfo
  {author} {\bibfnamefont {M.~A.}\ \bibnamefont {Aloy}},\ }\href
  {https://doi.org/10.1088/2041-8205/732/1/L6} {\bibfield  {journal} {\bibinfo
  {journal} {Astrophys. J. Lett.}\ }\textbf {\bibinfo {volume} {732}},\
  \bibinfo {pages} {L6} (\bibinfo {year} {2011})},\ \Eprint
  {https://arxiv.org/abs/1101.4298} {arXiv:1101.4298 [astro-ph.HE]}
  \BibitemShut {NoStop}%
\bibitem [{\citenamefont {Ruiz}\ \emph {et~al.}(2016)\citenamefont {Ruiz},
  \citenamefont {Lang}, \citenamefont {Paschalidis},\ and\ \citenamefont
  {Shapiro}}]{Ruiz:2016rai}%
  \BibitemOpen
  \bibfield  {author} {\bibinfo {author} {\bibfnamefont {M.}~\bibnamefont
  {Ruiz}}, \bibinfo {author} {\bibfnamefont {R.~N.}\ \bibnamefont {Lang}},
  \bibinfo {author} {\bibfnamefont {V.}~\bibnamefont {Paschalidis}},\ and\
  \bibinfo {author} {\bibfnamefont {S.~L.}\ \bibnamefont {Shapiro}},\ }\href
  {https://doi.org/10.3847/2041-8205/824/1/L6} {\bibfield  {journal} {\bibinfo
  {journal} {Astrophys. J. Lett.}\ }\textbf {\bibinfo {volume} {824}},\
  \bibinfo {pages} {L6} (\bibinfo {year} {2016})},\ \Eprint
  {https://arxiv.org/abs/1604.02455} {arXiv:1604.02455 [astro-ph.HE]}
  \BibitemShut {NoStop}%
\bibitem [{\citenamefont {M\"osta}\ \emph {et~al.}(2020)\citenamefont
  {M\"osta}, \citenamefont {Radice}, \citenamefont {Haas}, \citenamefont
  {Schnetter},\ and\ \citenamefont {Bernuzzi}}]{Mosta:2020hlh}%
  \BibitemOpen
  \bibfield  {author} {\bibinfo {author} {\bibfnamefont {P.}~\bibnamefont
  {M\"osta}}, \bibinfo {author} {\bibfnamefont {D.}~\bibnamefont {Radice}},
  \bibinfo {author} {\bibfnamefont {R.}~\bibnamefont {Haas}}, \bibinfo {author}
  {\bibfnamefont {E.}~\bibnamefont {Schnetter}},\ and\ \bibinfo {author}
  {\bibfnamefont {S.}~\bibnamefont {Bernuzzi}},\ }\href
  {https://doi.org/10.3847/2041-8213/abb6ef} {\bibfield  {journal} {\bibinfo
  {journal} {Astrophys. J. Lett.}\ }\textbf {\bibinfo {volume} {901}},\
  \bibinfo {pages} {L37} (\bibinfo {year} {2020})},\ \Eprint
  {https://arxiv.org/abs/2003.06043} {arXiv:2003.06043 [astro-ph.HE]}
  \BibitemShut {NoStop}%
\bibitem [{\citenamefont {Kiuchi}\ \emph {et~al.}(2024)\citenamefont {Kiuchi},
  \citenamefont {Reboul-Salze}, \citenamefont {Shibata},\ and\ \citenamefont
  {Sekiguchi}}]{Kiuchi:2023obe}%
  \BibitemOpen
  \bibfield  {author} {\bibinfo {author} {\bibfnamefont {K.}~\bibnamefont
  {Kiuchi}}, \bibinfo {author} {\bibfnamefont {A.}~\bibnamefont
  {Reboul-Salze}}, \bibinfo {author} {\bibfnamefont {M.}~\bibnamefont
  {Shibata}},\ and\ \bibinfo {author} {\bibfnamefont {Y.}~\bibnamefont
  {Sekiguchi}},\ }\href {https://doi.org/10.1038/s41550-024-02194-y} {\bibfield
   {journal} {\bibinfo  {journal} {Nature Astron.}\ }\textbf {\bibinfo {volume}
  {8}},\ \bibinfo {pages} {298} (\bibinfo {year} {2024})},\ \Eprint
  {https://arxiv.org/abs/2306.15721} {arXiv:2306.15721 [astro-ph.HE]}
  \BibitemShut {NoStop}%
\bibitem [{\citenamefont {Abbott}\ \emph
  {et~al.}(2017{\natexlab{a}})\citenamefont {Abbott} \emph
  {et~al.}}]{LIGOScientific:2017vwq}%
  \BibitemOpen
  \bibfield  {author} {\bibinfo {author} {\bibfnamefont {B.~P.}\ \bibnamefont
  {Abbott}} \emph {et~al.} (\bibinfo {collaboration} {LIGO Scientific,
  Virgo}),\ }\href {https://doi.org/10.1103/PhysRevLett.119.161101} {\bibfield
  {journal} {\bibinfo  {journal} {Phys. Rev. Lett.}\ }\textbf {\bibinfo
  {volume} {119}},\ \bibinfo {pages} {161101} (\bibinfo {year}
  {2017}{\natexlab{a}})},\ \Eprint {https://arxiv.org/abs/1710.05832}
  {arXiv:1710.05832 [gr-qc]} \BibitemShut {NoStop}%
\bibitem [{\citenamefont {Abbott}\ \emph
  {et~al.}(2017{\natexlab{b}})\citenamefont {Abbott} \emph
  {et~al.}}]{LIGOScientific:2017pwl}%
  \BibitemOpen
  \bibfield  {author} {\bibinfo {author} {\bibfnamefont {B.~P.}\ \bibnamefont
  {Abbott}} \emph {et~al.} (\bibinfo {collaboration} {LIGO Scientific,
  Virgo}),\ }\href {https://doi.org/10.3847/2041-8213/aa9478} {\bibfield
  {journal} {\bibinfo  {journal} {Astrophys. J. Lett.}\ }\textbf {\bibinfo
  {volume} {850}},\ \bibinfo {pages} {L39} (\bibinfo {year}
  {2017}{\natexlab{b}})},\ \Eprint {https://arxiv.org/abs/1710.05836}
  {arXiv:1710.05836 [astro-ph.HE]} \BibitemShut {NoStop}%
\bibitem [{\citenamefont {Arcavi}\ \emph {et~al.}(2017)\citenamefont {Arcavi}
  \emph {et~al.}}]{Arcavi:2017xiz}%
  \BibitemOpen
  \bibfield  {author} {\bibinfo {author} {\bibfnamefont {I.}~\bibnamefont
  {Arcavi}} \emph {et~al.},\ }\href {https://doi.org/10.1038/nature24291}
  {\bibfield  {journal} {\bibinfo  {journal} {Nature}\ }\textbf {\bibinfo
  {volume} {551}},\ \bibinfo {pages} {64} (\bibinfo {year} {2017})},\ \Eprint
  {https://arxiv.org/abs/1710.05843} {arXiv:1710.05843 [astro-ph.HE]}
  \BibitemShut {NoStop}%
\bibitem [{\citenamefont {Coulter}\ \emph {et~al.}(2017)\citenamefont {Coulter}
  \emph {et~al.}}]{Coulter:2017wya}%
  \BibitemOpen
  \bibfield  {author} {\bibinfo {author} {\bibfnamefont {D.~A.}\ \bibnamefont
  {Coulter}} \emph {et~al.},\ }\href {https://doi.org/10.1126/science.aap9811}
  {\bibfield  {journal} {\bibinfo  {journal} {Science}\ }\textbf {\bibinfo
  {volume} {358}},\ \bibinfo {pages} {1556} (\bibinfo {year} {2017})},\ \Eprint
  {https://arxiv.org/abs/1710.05452} {arXiv:1710.05452 [astro-ph.HE]}
  \BibitemShut {NoStop}%
\bibitem [{\citenamefont {Lipunov}\ \emph {et~al.}(2017)\citenamefont {Lipunov}
  \emph {et~al.}}]{Lipunov:2017dwd}%
  \BibitemOpen
  \bibfield  {author} {\bibinfo {author} {\bibfnamefont {V.~M.}\ \bibnamefont
  {Lipunov}} \emph {et~al.},\ }\href {https://doi.org/10.3847/2041-8213/aa92c0}
  {\bibfield  {journal} {\bibinfo  {journal} {Astrophys. J. Lett.}\ }\textbf
  {\bibinfo {volume} {850}},\ \bibinfo {pages} {L1} (\bibinfo {year} {2017})},\
  \Eprint {https://arxiv.org/abs/1710.05461} {arXiv:1710.05461 [astro-ph.HE]}
  \BibitemShut {NoStop}%
\bibitem [{\citenamefont {Tanvir}\ \emph {et~al.}(2017)\citenamefont {Tanvir}
  \emph {et~al.}}]{Tanvir:2017pws}%
  \BibitemOpen
  \bibfield  {author} {\bibinfo {author} {\bibfnamefont {N.~R.}\ \bibnamefont
  {Tanvir}} \emph {et~al.},\ }\href {https://doi.org/10.3847/2041-8213/aa90b6}
  {\bibfield  {journal} {\bibinfo  {journal} {Astrophys. J. Lett.}\ }\textbf
  {\bibinfo {volume} {848}},\ \bibinfo {pages} {L27} (\bibinfo {year}
  {2017})},\ \Eprint {https://arxiv.org/abs/1710.05455} {arXiv:1710.05455
  [astro-ph.HE]} \BibitemShut {NoStop}%
\bibitem [{\citenamefont {Valenti}\ \emph {et~al.}(2017)\citenamefont
  {Valenti}, \citenamefont {Sand}, \citenamefont {Yang}, \citenamefont
  {Cappellaro}, \citenamefont {Tartaglia}, \citenamefont {Corsi}, \citenamefont
  {Jha}, \citenamefont {Reichart}, \citenamefont {Haislip},\ and\ \citenamefont
  {Kouprianov}}]{Valenti:2017ngx}%
  \BibitemOpen
  \bibfield  {author} {\bibinfo {author} {\bibfnamefont {S.}~\bibnamefont
  {Valenti}}, \bibinfo {author} {\bibfnamefont {D.~J.}\ \bibnamefont {Sand}},
  \bibinfo {author} {\bibfnamefont {S.}~\bibnamefont {Yang}}, \bibinfo {author}
  {\bibfnamefont {E.}~\bibnamefont {Cappellaro}}, \bibinfo {author}
  {\bibfnamefont {L.}~\bibnamefont {Tartaglia}}, \bibinfo {author}
  {\bibfnamefont {A.}~\bibnamefont {Corsi}}, \bibinfo {author} {\bibfnamefont
  {S.~W.}\ \bibnamefont {Jha}}, \bibinfo {author} {\bibfnamefont {D.~E.}\
  \bibnamefont {Reichart}}, \bibinfo {author} {\bibfnamefont {J.}~\bibnamefont
  {Haislip}},\ and\ \bibinfo {author} {\bibfnamefont {V.}~\bibnamefont
  {Kouprianov}},\ }\href {https://doi.org/10.3847/2041-8213/aa8edf} {\bibfield
  {journal} {\bibinfo  {journal} {Astrophys. J. Lett.}\ }\textbf {\bibinfo
  {volume} {848}},\ \bibinfo {pages} {L24} (\bibinfo {year} {2017})},\ \Eprint
  {https://arxiv.org/abs/1710.05854} {arXiv:1710.05854 [astro-ph.HE]}
  \BibitemShut {NoStop}%
\bibitem [{\citenamefont {Goldstein}\ \emph {et~al.}(2017)\citenamefont
  {Goldstein} \emph {et~al.}}]{Goldstein:2017mmi}%
  \BibitemOpen
  \bibfield  {author} {\bibinfo {author} {\bibfnamefont {A.}~\bibnamefont
  {Goldstein}} \emph {et~al.},\ }\href
  {https://doi.org/10.3847/2041-8213/aa8f41} {\bibfield  {journal} {\bibinfo
  {journal} {Astrophys. J. Lett.}\ }\textbf {\bibinfo {volume} {848}},\
  \bibinfo {pages} {L14} (\bibinfo {year} {2017})},\ \Eprint
  {https://arxiv.org/abs/1710.05446} {arXiv:1710.05446 [astro-ph.HE]}
  \BibitemShut {NoStop}%
\bibitem [{\citenamefont {Savchenko}\ \emph {et~al.}(2017)\citenamefont
  {Savchenko} \emph {et~al.}}]{Savchenko:2017ffs}%
  \BibitemOpen
  \bibfield  {author} {\bibinfo {author} {\bibfnamefont {V.}~\bibnamefont
  {Savchenko}} \emph {et~al.},\ }\href
  {https://doi.org/10.3847/2041-8213/aa8f94} {\bibfield  {journal} {\bibinfo
  {journal} {Astrophys. J. Lett.}\ }\textbf {\bibinfo {volume} {848}},\
  \bibinfo {pages} {L15} (\bibinfo {year} {2017})},\ \Eprint
  {https://arxiv.org/abs/1710.05449} {arXiv:1710.05449 [astro-ph.HE]}
  \BibitemShut {NoStop}%
\bibitem [{\citenamefont {Hajela}\ \emph {et~al.}(2019)\citenamefont {Hajela}
  \emph {et~al.}}]{Hajela:2019mjy}%
  \BibitemOpen
  \bibfield  {author} {\bibinfo {author} {\bibfnamefont {A.}~\bibnamefont
  {Hajela}} \emph {et~al.},\ }\href {https://doi.org/10.3847/2041-8213/ab5226}
  {\bibfield  {journal} {\bibinfo  {journal} {Astrophys. J. Lett.}\ }\textbf
  {\bibinfo {volume} {886}},\ \bibinfo {pages} {L17} (\bibinfo {year}
  {2019})},\ \Eprint {https://arxiv.org/abs/1909.06393} {arXiv:1909.06393
  [astro-ph.HE]} \BibitemShut {NoStop}%
\bibitem [{\citenamefont {Hajela}\ \emph {et~al.}(2022)\citenamefont {Hajela}
  \emph {et~al.}}]{Hajela:2021faz}%
  \BibitemOpen
  \bibfield  {author} {\bibinfo {author} {\bibfnamefont {A.}~\bibnamefont
  {Hajela}} \emph {et~al.},\ }\href {https://doi.org/10.3847/2041-8213/ac504a}
  {\bibfield  {journal} {\bibinfo  {journal} {Astrophys. J. Lett.}\ }\textbf
  {\bibinfo {volume} {927}},\ \bibinfo {pages} {L17} (\bibinfo {year}
  {2022})},\ \Eprint {https://arxiv.org/abs/2104.02070} {arXiv:2104.02070
  [astro-ph.HE]} \BibitemShut {NoStop}%
\bibitem [{\citenamefont {Balasubramanian}\ \emph {et~al.}(2022)\citenamefont
  {Balasubramanian}, \citenamefont {Corsi}, \citenamefont {Mooley},
  \citenamefont {Hotokezaka}, \citenamefont {Kaplan}, \citenamefont {Frail},
  \citenamefont {Hallinan}, \citenamefont {Lazzati},\ and\ \citenamefont
  {Murphy}}]{Balasubramanian:2022sie}%
  \BibitemOpen
  \bibfield  {author} {\bibinfo {author} {\bibfnamefont {A.}~\bibnamefont
  {Balasubramanian}}, \bibinfo {author} {\bibfnamefont {A.}~\bibnamefont
  {Corsi}}, \bibinfo {author} {\bibfnamefont {K.~P.}\ \bibnamefont {Mooley}},
  \bibinfo {author} {\bibfnamefont {K.}~\bibnamefont {Hotokezaka}}, \bibinfo
  {author} {\bibfnamefont {D.~L.}\ \bibnamefont {Kaplan}}, \bibinfo {author}
  {\bibfnamefont {D.~A.}\ \bibnamefont {Frail}}, \bibinfo {author}
  {\bibfnamefont {G.}~\bibnamefont {Hallinan}}, \bibinfo {author}
  {\bibfnamefont {D.}~\bibnamefont {Lazzati}},\ and\ \bibinfo {author}
  {\bibfnamefont {E.~J.}\ \bibnamefont {Murphy}},\ }\href
  {https://doi.org/10.3847/1538-4357/ac9133} {\bibfield  {journal} {\bibinfo
  {journal} {Astrophys. J.}\ }\textbf {\bibinfo {volume} {938}},\ \bibinfo
  {pages} {12} (\bibinfo {year} {2022})},\ \Eprint
  {https://arxiv.org/abs/2205.14788} {arXiv:2205.14788 [astro-ph.HE]}
  \BibitemShut {NoStop}%
\bibitem [{\citenamefont {Abbott}\ \emph
  {et~al.}(2017{\natexlab{c}})\citenamefont {Abbott} \emph
  {et~al.}}]{LIGOScientific:2017zic}%
  \BibitemOpen
  \bibfield  {author} {\bibinfo {author} {\bibfnamefont {B.~P.}\ \bibnamefont
  {Abbott}} \emph {et~al.} (\bibinfo {collaboration} {LIGO Scientific, Virgo,
  Fermi-GBM, INTEGRAL}),\ }\href {https://doi.org/10.3847/2041-8213/aa920c}
  {\bibfield  {journal} {\bibinfo  {journal} {Astrophys. J. Lett.}\ }\textbf
  {\bibinfo {volume} {848}},\ \bibinfo {pages} {L13} (\bibinfo {year}
  {2017}{\natexlab{c}})},\ \Eprint {https://arxiv.org/abs/1710.05834}
  {arXiv:1710.05834 [astro-ph.HE]} \BibitemShut {NoStop}%
\bibitem [{\citenamefont {Abbott}\ \emph {et~al.}(2021)\citenamefont {Abbott}
  \emph {et~al.}}]{LIGOScientific:2020tif}%
  \BibitemOpen
  \bibfield  {author} {\bibinfo {author} {\bibfnamefont {R.}~\bibnamefont
  {Abbott}} \emph {et~al.} (\bibinfo {collaboration} {LIGO Scientific,
  Virgo}),\ }\href {https://doi.org/10.1103/PhysRevD.103.122002} {\bibfield
  {journal} {\bibinfo  {journal} {Phys. Rev. D}\ }\textbf {\bibinfo {volume}
  {103}},\ \bibinfo {pages} {122002} (\bibinfo {year} {2021})},\ \Eprint
  {https://arxiv.org/abs/2010.14529} {arXiv:2010.14529 [gr-qc]} \BibitemShut
  {NoStop}%
\bibitem [{\citenamefont {Abbott}\ \emph {et~al.}(2019)\citenamefont {Abbott}
  \emph {et~al.}}]{LIGOScientific:2018dkp}%
  \BibitemOpen
  \bibfield  {author} {\bibinfo {author} {\bibfnamefont {B.~P.}\ \bibnamefont
  {Abbott}} \emph {et~al.} (\bibinfo {collaboration} {LIGO Scientific,
  Virgo}),\ }\href {https://doi.org/10.1103/PhysRevLett.123.011102} {\bibfield
  {journal} {\bibinfo  {journal} {Phys. Rev. Lett.}\ }\textbf {\bibinfo
  {volume} {123}},\ \bibinfo {pages} {011102} (\bibinfo {year} {2019})},\
  \Eprint {https://arxiv.org/abs/1811.00364} {arXiv:1811.00364 [gr-qc]}
  \BibitemShut {NoStop}%
\bibitem [{\citenamefont {Sakstein}\ and\ \citenamefont
  {Jain}(2017)}]{Sakstein:2017xjx}%
  \BibitemOpen
  \bibfield  {author} {\bibinfo {author} {\bibfnamefont {J.}~\bibnamefont
  {Sakstein}}\ and\ \bibinfo {author} {\bibfnamefont {B.}~\bibnamefont
  {Jain}},\ }\href {https://doi.org/10.1103/PhysRevLett.119.251303} {\bibfield
  {journal} {\bibinfo  {journal} {Phys. Rev. Lett.}\ }\textbf {\bibinfo
  {volume} {119}},\ \bibinfo {pages} {251303} (\bibinfo {year} {2017})},\
  \Eprint {https://arxiv.org/abs/1710.05893} {arXiv:1710.05893 [astro-ph.CO]}
  \BibitemShut {NoStop}%
\bibitem [{\citenamefont {Ezquiaga}\ and\ \citenamefont
  {Zumalac{\'a}rregui}(2017)}]{Ezquiaga:2017ekz}%
  \BibitemOpen
  \bibfield  {author} {\bibinfo {author} {\bibfnamefont {J.~M.}\ \bibnamefont
  {Ezquiaga}}\ and\ \bibinfo {author} {\bibfnamefont {M.}~\bibnamefont
  {Zumalac{\'a}rregui}},\ }\href
  {https://doi.org/10.1103/PhysRevLett.119.251304} {\bibfield  {journal}
  {\bibinfo  {journal} {Phys. Rev. Lett.}\ }\textbf {\bibinfo {volume} {119}},\
  \bibinfo {pages} {251304} (\bibinfo {year} {2017})},\ \Eprint
  {https://arxiv.org/abs/1710.05901} {arXiv:1710.05901 [astro-ph.CO]}
  \BibitemShut {NoStop}%
\bibitem [{\citenamefont {Creminelli}\ and\ \citenamefont
  {Vernizzi}(2017)}]{Creminelli:2017sry}%
  \BibitemOpen
  \bibfield  {author} {\bibinfo {author} {\bibfnamefont {P.}~\bibnamefont
  {Creminelli}}\ and\ \bibinfo {author} {\bibfnamefont {F.}~\bibnamefont
  {Vernizzi}},\ }\href {https://doi.org/10.1103/PhysRevLett.119.251302}
  {\bibfield  {journal} {\bibinfo  {journal} {Phys. Rev. Lett.}\ }\textbf
  {\bibinfo {volume} {119}},\ \bibinfo {pages} {251302} (\bibinfo {year}
  {2017})},\ \Eprint {https://arxiv.org/abs/1710.05877} {arXiv:1710.05877
  [astro-ph.CO]} \BibitemShut {NoStop}%
\bibitem [{\citenamefont {Baker}\ \emph {et~al.}(2017)\citenamefont {Baker},
  \citenamefont {Bellini}, \citenamefont {Ferreira}, \citenamefont {Lagos},
  \citenamefont {Noller},\ and\ \citenamefont {Sawicki}}]{Baker:2017hug}%
  \BibitemOpen
  \bibfield  {author} {\bibinfo {author} {\bibfnamefont {T.}~\bibnamefont
  {Baker}}, \bibinfo {author} {\bibfnamefont {E.}~\bibnamefont {Bellini}},
  \bibinfo {author} {\bibfnamefont {P.~G.}\ \bibnamefont {Ferreira}}, \bibinfo
  {author} {\bibfnamefont {M.}~\bibnamefont {Lagos}}, \bibinfo {author}
  {\bibfnamefont {J.}~\bibnamefont {Noller}},\ and\ \bibinfo {author}
  {\bibfnamefont {I.}~\bibnamefont {Sawicki}},\ }\href
  {https://doi.org/10.1103/PhysRevLett.119.251301} {\bibfield  {journal}
  {\bibinfo  {journal} {Phys. Rev. Lett.}\ }\textbf {\bibinfo {volume} {119}},\
  \bibinfo {pages} {251301} (\bibinfo {year} {2017})},\ \Eprint
  {https://arxiv.org/abs/1710.06394} {arXiv:1710.06394 [astro-ph.CO]}
  \BibitemShut {NoStop}%
\bibitem [{\citenamefont {Abbott}\ \emph
  {et~al.}(2017{\natexlab{d}})\citenamefont {Abbott} \emph
  {et~al.}}]{LIGOScientific:2017ync}%
  \BibitemOpen
  \bibfield  {author} {\bibinfo {author} {\bibfnamefont {B.~P.}\ \bibnamefont
  {Abbott}} \emph {et~al.} (\bibinfo {collaboration} {LIGO Scientific, Virgo,
  Fermi GBM, INTEGRAL, IceCube, AstroSat Cadmium Zinc Telluride Imager Team,
  IPN, Insight-Hxmt, ANTARES, Swift, AGILE Team, 1M2H Team, Dark Energy Camera
  GW-EM, DES, DLT40, GRAWITA, Fermi-LAT, ATCA, ASKAP, Las Cumbres Observatory
  Group, OzGrav, DWF (Deeper Wider Faster Program), AST3, CAASTRO, VINROUGE,
  MASTER, J-GEM, GROWTH, JAGWAR, CaltechNRAO, TTU-NRAO, NuSTAR, Pan-STARRS,
  MAXI Team, TZAC Consortium, KU, Nordic Optical Telescope, ePESSTO, GROND,
  Texas Tech University, SALT Group, TOROS, BOOTES, MWA, CALET, IKI-GW
  Follow-up, H.E.S.S., LOFAR, LWA, HAWC, Pierre Auger, ALMA, Euro VLBI Team, Pi
  of Sky, Chandra Team at McGill University, DFN, ATLAS Telescopes, High Time
  Resolution Universe Survey, RIMAS, RATIR, SKA South Africa/MeerKAT}),\ }\href
  {https://doi.org/10.3847/2041-8213/aa91c9} {\bibfield  {journal} {\bibinfo
  {journal} {Astrophys. J. Lett.}\ }\textbf {\bibinfo {volume} {848}},\
  \bibinfo {pages} {L12} (\bibinfo {year} {2017}{\natexlab{d}})},\ \Eprint
  {https://arxiv.org/abs/1710.05833} {arXiv:1710.05833 [astro-ph.HE]}
  \BibitemShut {NoStop}%
\bibitem [{\citenamefont {Abbott}\ \emph {et~al.}(2018)\citenamefont {Abbott}
  \emph {et~al.}}]{LIGOScientific:2018cki}%
  \BibitemOpen
  \bibfield  {author} {\bibinfo {author} {\bibfnamefont {B.~P.}\ \bibnamefont
  {Abbott}} \emph {et~al.} (\bibinfo {collaboration} {LIGO Scientific,
  Virgo}),\ }\href {https://doi.org/10.1103/PhysRevLett.121.161101} {\bibfield
  {journal} {\bibinfo  {journal} {Phys. Rev. Lett.}\ }\textbf {\bibinfo
  {volume} {121}},\ \bibinfo {pages} {161101} (\bibinfo {year} {2018})},\
  \Eprint {https://arxiv.org/abs/1805.11581} {arXiv:1805.11581 [gr-qc]}
  \BibitemShut {NoStop}%
\bibitem [{\citenamefont {Bauswein}\ \emph {et~al.}(2017)\citenamefont
  {Bauswein}, \citenamefont {Just}, \citenamefont {Janka},\ and\ \citenamefont
  {Stergioulas}}]{Bauswein:2017vtn}%
  \BibitemOpen
  \bibfield  {author} {\bibinfo {author} {\bibfnamefont {A.}~\bibnamefont
  {Bauswein}}, \bibinfo {author} {\bibfnamefont {O.}~\bibnamefont {Just}},
  \bibinfo {author} {\bibfnamefont {H.-T.}\ \bibnamefont {Janka}},\ and\
  \bibinfo {author} {\bibfnamefont {N.}~\bibnamefont {Stergioulas}},\ }\href
  {https://doi.org/10.3847/2041-8213/aa9994} {\bibfield  {journal} {\bibinfo
  {journal} {Astrophys. J. Lett.}\ }\textbf {\bibinfo {volume} {850}},\
  \bibinfo {pages} {L34} (\bibinfo {year} {2017})},\ \Eprint
  {https://arxiv.org/abs/1710.06843} {arXiv:1710.06843 [astro-ph.HE]}
  \BibitemShut {NoStop}%
\bibitem [{\citenamefont {Margalit}\ and\ \citenamefont
  {Metzger}(2017)}]{Margalit:2017dij}%
  \BibitemOpen
  \bibfield  {author} {\bibinfo {author} {\bibfnamefont {B.}~\bibnamefont
  {Margalit}}\ and\ \bibinfo {author} {\bibfnamefont {B.~D.}\ \bibnamefont
  {Metzger}},\ }\href {https://doi.org/10.3847/2041-8213/aa991c} {\bibfield
  {journal} {\bibinfo  {journal} {Astrophys. J. Lett.}\ }\textbf {\bibinfo
  {volume} {850}},\ \bibinfo {pages} {L19} (\bibinfo {year} {2017})},\ \Eprint
  {https://arxiv.org/abs/1710.05938} {arXiv:1710.05938 [astro-ph.HE]}
  \BibitemShut {NoStop}%
\bibitem [{\citenamefont {Radice}\ \emph
  {et~al.}(2018{\natexlab{a}})\citenamefont {Radice}, \citenamefont {Perego},
  \citenamefont {Zappa},\ and\ \citenamefont {Bernuzzi}}]{Radice:2017lry}%
  \BibitemOpen
  \bibfield  {author} {\bibinfo {author} {\bibfnamefont {D.}~\bibnamefont
  {Radice}}, \bibinfo {author} {\bibfnamefont {A.}~\bibnamefont {Perego}},
  \bibinfo {author} {\bibfnamefont {F.}~\bibnamefont {Zappa}},\ and\ \bibinfo
  {author} {\bibfnamefont {S.}~\bibnamefont {Bernuzzi}},\ }\href
  {https://doi.org/10.3847/2041-8213/aaa402} {\bibfield  {journal} {\bibinfo
  {journal} {Astrophys. J. Lett.}\ }\textbf {\bibinfo {volume} {852}},\
  \bibinfo {pages} {L29} (\bibinfo {year} {2018}{\natexlab{a}})},\ \Eprint
  {https://arxiv.org/abs/1711.03647} {arXiv:1711.03647 [astro-ph.HE]}
  \BibitemShut {NoStop}%
\bibitem [{\citenamefont {Coughlin}\ \emph {et~al.}(2018)\citenamefont
  {Coughlin} \emph {et~al.}}]{Coughlin:2018miv}%
  \BibitemOpen
  \bibfield  {author} {\bibinfo {author} {\bibfnamefont {M.~W.}\ \bibnamefont
  {Coughlin}} \emph {et~al.},\ }\href {https://doi.org/10.1093/mnras/sty2174}
  {\bibfield  {journal} {\bibinfo  {journal} {Mon. Not. Roy. Astron. Soc.}\
  }\textbf {\bibinfo {volume} {480}},\ \bibinfo {pages} {3871} (\bibinfo {year}
  {2018})},\ \Eprint {https://arxiv.org/abs/1805.09371} {arXiv:1805.09371
  [astro-ph.HE]} \BibitemShut {NoStop}%
\bibitem [{\citenamefont {Most}\ \emph {et~al.}(2018)\citenamefont {Most},
  \citenamefont {Weih}, \citenamefont {Rezzolla},\ and\ \citenamefont
  {Schaffner-Bielich}}]{Most:2018hfd}%
  \BibitemOpen
  \bibfield  {author} {\bibinfo {author} {\bibfnamefont {E.~R.}\ \bibnamefont
  {Most}}, \bibinfo {author} {\bibfnamefont {L.~R.}\ \bibnamefont {Weih}},
  \bibinfo {author} {\bibfnamefont {L.}~\bibnamefont {Rezzolla}},\ and\
  \bibinfo {author} {\bibfnamefont {J.}~\bibnamefont {Schaffner-Bielich}},\
  }\href {https://doi.org/10.1103/PhysRevLett.120.261103} {\bibfield  {journal}
  {\bibinfo  {journal} {Phys. Rev. Lett.}\ }\textbf {\bibinfo {volume} {120}},\
  \bibinfo {pages} {261103} (\bibinfo {year} {2018})},\ \Eprint
  {https://arxiv.org/abs/1803.00549} {arXiv:1803.00549 [gr-qc]} \BibitemShut
  {NoStop}%
\bibitem [{\citenamefont {Capano}\ \emph {et~al.}(2020)\citenamefont {Capano},
  \citenamefont {Tews}, \citenamefont {Brown}, \citenamefont {Margalit},
  \citenamefont {De}, \citenamefont {Kumar}, \citenamefont {Brown},
  \citenamefont {Krishnan},\ and\ \citenamefont {Reddy}}]{Capano:2019eae}%
  \BibitemOpen
  \bibfield  {author} {\bibinfo {author} {\bibfnamefont {C.~D.}\ \bibnamefont
  {Capano}}, \bibinfo {author} {\bibfnamefont {I.}~\bibnamefont {Tews}},
  \bibinfo {author} {\bibfnamefont {S.~M.}\ \bibnamefont {Brown}}, \bibinfo
  {author} {\bibfnamefont {B.}~\bibnamefont {Margalit}}, \bibinfo {author}
  {\bibfnamefont {S.}~\bibnamefont {De}}, \bibinfo {author} {\bibfnamefont
  {S.}~\bibnamefont {Kumar}}, \bibinfo {author} {\bibfnamefont {D.~A.}\
  \bibnamefont {Brown}}, \bibinfo {author} {\bibfnamefont {B.}~\bibnamefont
  {Krishnan}},\ and\ \bibinfo {author} {\bibfnamefont {S.}~\bibnamefont
  {Reddy}},\ }\href {https://doi.org/10.1038/s41550-020-1014-6} {\bibfield
  {journal} {\bibinfo  {journal} {Nature Astron.}\ }\textbf {\bibinfo {volume}
  {4}},\ \bibinfo {pages} {625} (\bibinfo {year} {2020})},\ \Eprint
  {https://arxiv.org/abs/1908.10352} {arXiv:1908.10352 [astro-ph.HE]}
  \BibitemShut {NoStop}%
\bibitem [{\citenamefont {Raithel}(2019)}]{Raithel:2019uzi}%
  \BibitemOpen
  \bibfield  {author} {\bibinfo {author} {\bibfnamefont {C.~A.}\ \bibnamefont
  {Raithel}},\ }\href {https://doi.org/10.1140/epja/i2019-12759-5} {\bibfield
  {journal} {\bibinfo  {journal} {Eur. Phys. J. A}\ }\textbf {\bibinfo {volume}
  {55}},\ \bibinfo {pages} {80} (\bibinfo {year} {2019})},\ \Eprint
  {https://arxiv.org/abs/1904.10002} {arXiv:1904.10002 [astro-ph.HE]}
  \BibitemShut {NoStop}%
\bibitem [{\citenamefont {Dietrich}\ \emph {et~al.}(2020)\citenamefont
  {Dietrich}, \citenamefont {Coughlin}, \citenamefont {Pang}, \citenamefont
  {Bulla}, \citenamefont {Heinzel}, \citenamefont {Issa}, \citenamefont
  {Tews},\ and\ \citenamefont {Antier}}]{Dietrich:2020efo}%
  \BibitemOpen
  \bibfield  {author} {\bibinfo {author} {\bibfnamefont {T.}~\bibnamefont
  {Dietrich}}, \bibinfo {author} {\bibfnamefont {M.~W.}\ \bibnamefont
  {Coughlin}}, \bibinfo {author} {\bibfnamefont {P.~T.~H.}\ \bibnamefont
  {Pang}}, \bibinfo {author} {\bibfnamefont {M.}~\bibnamefont {Bulla}},
  \bibinfo {author} {\bibfnamefont {J.}~\bibnamefont {Heinzel}}, \bibinfo
  {author} {\bibfnamefont {L.}~\bibnamefont {Issa}}, \bibinfo {author}
  {\bibfnamefont {I.}~\bibnamefont {Tews}},\ and\ \bibinfo {author}
  {\bibfnamefont {S.}~\bibnamefont {Antier}},\ }\href
  {https://doi.org/10.1126/science.abb4317} {\bibfield  {journal} {\bibinfo
  {journal} {Science}\ }\textbf {\bibinfo {volume} {370}},\ \bibinfo {pages}
  {1450} (\bibinfo {year} {2020})},\ \Eprint {https://arxiv.org/abs/2002.11355}
  {arXiv:2002.11355 [astro-ph.HE]} \BibitemShut {NoStop}%
\bibitem [{\citenamefont {Hessels}\ \emph {et~al.}(2006)\citenamefont
  {Hessels}, \citenamefont {Ransom}, \citenamefont {Stairs}, \citenamefont
  {Freire}, \citenamefont {Kaspi},\ and\ \citenamefont
  {Camilo}}]{Hessels:2006ze}%
  \BibitemOpen
  \bibfield  {author} {\bibinfo {author} {\bibfnamefont {J.~W.~T.}\
  \bibnamefont {Hessels}}, \bibinfo {author} {\bibfnamefont {S.~M.}\
  \bibnamefont {Ransom}}, \bibinfo {author} {\bibfnamefont {I.~H.}\
  \bibnamefont {Stairs}}, \bibinfo {author} {\bibfnamefont {P.~C.~C.}\
  \bibnamefont {Freire}}, \bibinfo {author} {\bibfnamefont {V.~M.}\
  \bibnamefont {Kaspi}},\ and\ \bibinfo {author} {\bibfnamefont
  {F.}~\bibnamefont {Camilo}},\ }\href
  {https://doi.org/10.1126/science.1123430} {\bibfield  {journal} {\bibinfo
  {journal} {Science}\ }\textbf {\bibinfo {volume} {311}},\ \bibinfo {pages}
  {1901} (\bibinfo {year} {2006})},\ \Eprint
  {https://arxiv.org/abs/astro-ph/0601337} {arXiv:astro-ph/0601337}
  \BibitemShut {NoStop}%
\bibitem [{\citenamefont {Burgay}\ \emph {et~al.}(2003)\citenamefont {Burgay}
  \emph {et~al.}}]{Burgay:2003jj}%
  \BibitemOpen
  \bibfield  {author} {\bibinfo {author} {\bibfnamefont {M.}~\bibnamefont
  {Burgay}} \emph {et~al.},\ }\href {https://doi.org/10.1038/nature02124}
  {\bibfield  {journal} {\bibinfo  {journal} {Nature}\ }\textbf {\bibinfo
  {volume} {426}},\ \bibinfo {pages} {531} (\bibinfo {year} {2003})},\ \Eprint
  {https://arxiv.org/abs/astro-ph/0312071} {arXiv:astro-ph/0312071}
  \BibitemShut {NoStop}%
\bibitem [{\citenamefont {Stovall}\ \emph {et~al.}(2018)\citenamefont {Stovall}
  \emph {et~al.}}]{Stovall:2018ouw}%
  \BibitemOpen
  \bibfield  {author} {\bibinfo {author} {\bibfnamefont {K.}~\bibnamefont
  {Stovall}} \emph {et~al.},\ }\href {https://doi.org/10.3847/2041-8213/aaad06}
  {\bibfield  {journal} {\bibinfo  {journal} {Astrophys. J. Lett.}\ }\textbf
  {\bibinfo {volume} {854}},\ \bibinfo {pages} {L22} (\bibinfo {year}
  {2018})},\ \Eprint {https://arxiv.org/abs/1802.01707} {arXiv:1802.01707
  [astro-ph.HE]} \BibitemShut {NoStop}%
\bibitem [{\citenamefont {Rosswog}\ \emph {et~al.}(2024)\citenamefont
  {Rosswog}, \citenamefont {Diener}, \citenamefont {Torsello}, \citenamefont
  {Tauris},\ and\ \citenamefont {Sarin}}]{Rosswog:2023rqa}%
  \BibitemOpen
  \bibfield  {author} {\bibinfo {author} {\bibfnamefont {S.}~\bibnamefont
  {Rosswog}}, \bibinfo {author} {\bibfnamefont {P.}~\bibnamefont {Diener}},
  \bibinfo {author} {\bibfnamefont {F.}~\bibnamefont {Torsello}}, \bibinfo
  {author} {\bibfnamefont {T.~M.}\ \bibnamefont {Tauris}},\ and\ \bibinfo
  {author} {\bibfnamefont {N.}~\bibnamefont {Sarin}},\ }\href
  {https://doi.org/10.1093/mnras/stae454} {\bibfield  {journal} {\bibinfo
  {journal} {Mon. Not. Roy. Astron. Soc.}\ }\textbf {\bibinfo {volume} {530}},\
  \bibinfo {pages} {2336} (\bibinfo {year} {2024})},\ \Eprint
  {https://arxiv.org/abs/2310.15920} {arXiv:2310.15920 [astro-ph.HE]}
  \BibitemShut {NoStop}%
\bibitem [{\citenamefont {East}\ \emph {et~al.}(2019)\citenamefont {East},
  \citenamefont {Paschalidis}, \citenamefont {Pretorius},\ and\ \citenamefont
  {Tsokaros}}]{East:2019lbk}%
  \BibitemOpen
  \bibfield  {author} {\bibinfo {author} {\bibfnamefont {W.~E.}\ \bibnamefont
  {East}}, \bibinfo {author} {\bibfnamefont {V.}~\bibnamefont {Paschalidis}},
  \bibinfo {author} {\bibfnamefont {F.}~\bibnamefont {Pretorius}},\ and\
  \bibinfo {author} {\bibfnamefont {A.}~\bibnamefont {Tsokaros}},\ }\href
  {https://doi.org/10.1103/PhysRevD.100.124042} {\bibfield  {journal} {\bibinfo
   {journal} {Phys. Rev. D}\ }\textbf {\bibinfo {volume} {100}},\ \bibinfo
  {pages} {124042} (\bibinfo {year} {2019})},\ \Eprint
  {https://arxiv.org/abs/1906.05288} {arXiv:1906.05288 [astro-ph.HE]}
  \BibitemShut {NoStop}%
\bibitem [{\citenamefont {Ruiz}\ \emph {et~al.}(2019)\citenamefont {Ruiz},
  \citenamefont {Tsokaros}, \citenamefont {Paschalidis},\ and\ \citenamefont
  {Shapiro}}]{Ruiz:2019ezy}%
  \BibitemOpen
  \bibfield  {author} {\bibinfo {author} {\bibfnamefont {M.}~\bibnamefont
  {Ruiz}}, \bibinfo {author} {\bibfnamefont {A.}~\bibnamefont {Tsokaros}},
  \bibinfo {author} {\bibfnamefont {V.}~\bibnamefont {Paschalidis}},\ and\
  \bibinfo {author} {\bibfnamefont {S.~L.}\ \bibnamefont {Shapiro}},\ }\href
  {https://doi.org/10.1103/PhysRevD.99.084032} {\bibfield  {journal} {\bibinfo
  {journal} {Phys. Rev. D}\ }\textbf {\bibinfo {volume} {99}},\ \bibinfo
  {pages} {084032} (\bibinfo {year} {2019})},\ \Eprint
  {https://arxiv.org/abs/1902.08636} {arXiv:1902.08636 [astro-ph.HE]}
  \BibitemShut {NoStop}%
\bibitem [{\citenamefont {Bernuzzi}\ \emph {et~al.}(2014)\citenamefont
  {Bernuzzi}, \citenamefont {Dietrich}, \citenamefont {Tichy},\ and\
  \citenamefont {Br{\"u}gmann}}]{Bernuzzi:2013rza}%
  \BibitemOpen
  \bibfield  {author} {\bibinfo {author} {\bibfnamefont {S.}~\bibnamefont
  {Bernuzzi}}, \bibinfo {author} {\bibfnamefont {T.}~\bibnamefont {Dietrich}},
  \bibinfo {author} {\bibfnamefont {W.}~\bibnamefont {Tichy}},\ and\ \bibinfo
  {author} {\bibfnamefont {B.}~\bibnamefont {Br{\"u}gmann}},\ }\href
  {https://doi.org/10.1103/PhysRevD.89.104021} {\bibfield  {journal} {\bibinfo
  {journal} {Phys. Rev. D}\ }\textbf {\bibinfo {volume} {89}},\ \bibinfo
  {pages} {104021} (\bibinfo {year} {2014})},\ \Eprint
  {https://arxiv.org/abs/1311.4443} {arXiv:1311.4443 [gr-qc]} \BibitemShut
  {NoStop}%
\bibitem [{\citenamefont {Dietrich}\ \emph {et~al.}(2017)\citenamefont
  {Dietrich}, \citenamefont {Bernuzzi}, \citenamefont {Ujevic},\ and\
  \citenamefont {Tichy}}]{Dietrich:2016lyp}%
  \BibitemOpen
  \bibfield  {author} {\bibinfo {author} {\bibfnamefont {T.}~\bibnamefont
  {Dietrich}}, \bibinfo {author} {\bibfnamefont {S.}~\bibnamefont {Bernuzzi}},
  \bibinfo {author} {\bibfnamefont {M.}~\bibnamefont {Ujevic}},\ and\ \bibinfo
  {author} {\bibfnamefont {W.}~\bibnamefont {Tichy}},\ }\href
  {https://doi.org/10.1103/PhysRevD.95.044045} {\bibfield  {journal} {\bibinfo
  {journal} {Phys. Rev. D}\ }\textbf {\bibinfo {volume} {95}},\ \bibinfo
  {pages} {044045} (\bibinfo {year} {2017})},\ \Eprint
  {https://arxiv.org/abs/1611.07367} {arXiv:1611.07367 [gr-qc]} \BibitemShut
  {NoStop}%
\bibitem [{\citenamefont {Kastaun}\ \emph {et~al.}(2017)\citenamefont
  {Kastaun}, \citenamefont {Ciolfi}, \citenamefont {Endrizzi},\ and\
  \citenamefont {Giacomazzo}}]{Kastaun:2016elu}%
  \BibitemOpen
  \bibfield  {author} {\bibinfo {author} {\bibfnamefont {W.}~\bibnamefont
  {Kastaun}}, \bibinfo {author} {\bibfnamefont {R.}~\bibnamefont {Ciolfi}},
  \bibinfo {author} {\bibfnamefont {A.}~\bibnamefont {Endrizzi}},\ and\
  \bibinfo {author} {\bibfnamefont {B.}~\bibnamefont {Giacomazzo}},\ }\href
  {https://doi.org/10.1103/PhysRevD.96.043019} {\bibfield  {journal} {\bibinfo
  {journal} {Phys. Rev. D}\ }\textbf {\bibinfo {volume} {96}},\ \bibinfo
  {pages} {043019} (\bibinfo {year} {2017})},\ \Eprint
  {https://arxiv.org/abs/1612.03671} {arXiv:1612.03671 [astro-ph.HE]}
  \BibitemShut {NoStop}%
\bibitem [{\citenamefont {Most}\ \emph {et~al.}(2019)\citenamefont {Most},
  \citenamefont {Papenfort}, \citenamefont {Tsokaros},\ and\ \citenamefont
  {Rezzolla}}]{Most:2019pac}%
  \BibitemOpen
  \bibfield  {author} {\bibinfo {author} {\bibfnamefont {E.~R.}\ \bibnamefont
  {Most}}, \bibinfo {author} {\bibfnamefont {L.~J.}\ \bibnamefont {Papenfort}},
  \bibinfo {author} {\bibfnamefont {A.}~\bibnamefont {Tsokaros}},\ and\
  \bibinfo {author} {\bibfnamefont {L.}~\bibnamefont {Rezzolla}},\ }\href
  {https://doi.org/10.3847/1538-4357/ab3ebb} {\bibfield  {journal} {\bibinfo
  {journal} {Astrophys. J.}\ }\textbf {\bibinfo {volume} {884}},\ \bibinfo
  {pages} {40} (\bibinfo {year} {2019})},\ \Eprint
  {https://arxiv.org/abs/1904.04220} {arXiv:1904.04220 [astro-ph.HE]}
  \BibitemShut {NoStop}%
\bibitem [{\citenamefont {Papenfort}\ \emph {et~al.}(2022)\citenamefont
  {Papenfort}, \citenamefont {Most}, \citenamefont {Tootle},\ and\
  \citenamefont {Rezzolla}}]{Papenfort:2022ywx}%
  \BibitemOpen
  \bibfield  {author} {\bibinfo {author} {\bibfnamefont {L.~J.}\ \bibnamefont
  {Papenfort}}, \bibinfo {author} {\bibfnamefont {E.~R.}\ \bibnamefont {Most}},
  \bibinfo {author} {\bibfnamefont {S.}~\bibnamefont {Tootle}},\ and\ \bibinfo
  {author} {\bibfnamefont {L.}~\bibnamefont {Rezzolla}},\ }\href
  {https://doi.org/10.1093/mnras/stac964} {\bibfield  {journal} {\bibinfo
  {journal} {Mon. Not. Roy. Astron. Soc.}\ }\textbf {\bibinfo {volume} {513}},\
  \bibinfo {pages} {3646} (\bibinfo {year} {2022})},\ \Eprint
  {https://arxiv.org/abs/2201.03632} {arXiv:2201.03632 [astro-ph.HE]}
  \BibitemShut {NoStop}%
\bibitem [{\citenamefont {Brügmann}\ \emph {et~al.}(2008)\citenamefont
  {Brügmann}, \citenamefont {Gonzalez}, \citenamefont {Hannam}, \citenamefont
  {Husa}, \citenamefont {Sperhake},\ and\ \citenamefont
  {Tichy}}]{Bruegmann:2006ulg}%
  \BibitemOpen
  \bibfield  {author} {\bibinfo {author} {\bibfnamefont {B.}~\bibnamefont
  {Brügmann}}, \bibinfo {author} {\bibfnamefont {J.~A.}\ \bibnamefont
  {Gonzalez}}, \bibinfo {author} {\bibfnamefont {M.}~\bibnamefont {Hannam}},
  \bibinfo {author} {\bibfnamefont {S.}~\bibnamefont {Husa}}, \bibinfo {author}
  {\bibfnamefont {U.}~\bibnamefont {Sperhake}},\ and\ \bibinfo {author}
  {\bibfnamefont {W.}~\bibnamefont {Tichy}},\ }\href
  {https://doi.org/10.1103/PhysRevD.77.024027} {\bibfield  {journal} {\bibinfo
  {journal} {Phys. Rev. D}\ }\textbf {\bibinfo {volume} {77}},\ \bibinfo
  {pages} {024027} (\bibinfo {year} {2008})},\ \Eprint
  {https://arxiv.org/abs/gr-qc/0610128} {arXiv:gr-qc/0610128} \BibitemShut
  {NoStop}%
\bibitem [{\citenamefont {Thierfelder}\ \emph {et~al.}(2011)\citenamefont
  {Thierfelder}, \citenamefont {Bernuzzi},\ and\ \citenamefont
  {Brügmann}}]{Thierfelder:2011yi}%
  \BibitemOpen
  \bibfield  {author} {\bibinfo {author} {\bibfnamefont {M.}~\bibnamefont
  {Thierfelder}}, \bibinfo {author} {\bibfnamefont {S.}~\bibnamefont
  {Bernuzzi}},\ and\ \bibinfo {author} {\bibfnamefont {B.}~\bibnamefont
  {Brügmann}},\ }\href {https://doi.org/10.1103/PhysRevD.84.044012} {\bibfield
   {journal} {\bibinfo  {journal} {Phys. Rev. D}\ }\textbf {\bibinfo {volume}
  {84}},\ \bibinfo {pages} {044012} (\bibinfo {year} {2011})},\ \Eprint
  {https://arxiv.org/abs/1104.4751} {arXiv:1104.4751 [gr-qc]} \BibitemShut
  {NoStop}%
\bibitem [{\citenamefont {Gieg}\ \emph {et~al.}(2022)\citenamefont {Gieg},
  \citenamefont {Schianchi}, \citenamefont {Dietrich},\ and\ \citenamefont
  {Ujevic}}]{Gieg:2022mut}%
  \BibitemOpen
  \bibfield  {author} {\bibinfo {author} {\bibfnamefont {H.}~\bibnamefont
  {Gieg}}, \bibinfo {author} {\bibfnamefont {F.}~\bibnamefont {Schianchi}},
  \bibinfo {author} {\bibfnamefont {T.}~\bibnamefont {Dietrich}},\ and\
  \bibinfo {author} {\bibfnamefont {M.}~\bibnamefont {Ujevic}},\ }\href
  {https://doi.org/10.3390/universe8070370} {\bibfield  {journal} {\bibinfo
  {journal} {Universe}\ }\textbf {\bibinfo {volume} {8}},\ \bibinfo {pages}
  {370} (\bibinfo {year} {2022})},\ \Eprint {https://arxiv.org/abs/2206.01337}
  {arXiv:2206.01337 [gr-qc]} \BibitemShut {NoStop}%
\bibitem [{\citenamefont {Schianchi}\ \emph {et~al.}(2024)\citenamefont
  {Schianchi}, \citenamefont {Gieg}, \citenamefont {Nedora}, \citenamefont
  {Neuweiler}, \citenamefont {Ujevic}, \citenamefont {Bulla},\ and\
  \citenamefont {Dietrich}}]{Schianchi:2023uky}%
  \BibitemOpen
  \bibfield  {author} {\bibinfo {author} {\bibfnamefont {F.}~\bibnamefont
  {Schianchi}}, \bibinfo {author} {\bibfnamefont {H.}~\bibnamefont {Gieg}},
  \bibinfo {author} {\bibfnamefont {V.}~\bibnamefont {Nedora}}, \bibinfo
  {author} {\bibfnamefont {A.}~\bibnamefont {Neuweiler}}, \bibinfo {author}
  {\bibfnamefont {M.}~\bibnamefont {Ujevic}}, \bibinfo {author} {\bibfnamefont
  {M.}~\bibnamefont {Bulla}},\ and\ \bibinfo {author} {\bibfnamefont
  {T.}~\bibnamefont {Dietrich}},\ }\href
  {https://doi.org/10.1103/PhysRevD.109.044012} {\bibfield  {journal} {\bibinfo
   {journal} {Phys. Rev. D}\ }\textbf {\bibinfo {volume} {109}},\ \bibinfo
  {pages} {044012} (\bibinfo {year} {2024})},\ \Eprint
  {https://arxiv.org/abs/2307.04572} {arXiv:2307.04572 [gr-qc]} \BibitemShut
  {NoStop}%
\bibitem [{\citenamefont {Neuweiler}\ \emph {et~al.}(2024)\citenamefont
  {Neuweiler}, \citenamefont {Dietrich}, \citenamefont {Br\"ugmann},
  \citenamefont {Giangrandi}, \citenamefont {Kiuchi}, \citenamefont
  {Schianchi}, \citenamefont {M\"osta}, \citenamefont {Shankar}, \citenamefont
  {Giacomazzo},\ and\ \citenamefont {Shibata}}]{Neuweiler:2024jae}%
  \BibitemOpen
  \bibfield  {author} {\bibinfo {author} {\bibfnamefont {A.}~\bibnamefont
  {Neuweiler}}, \bibinfo {author} {\bibfnamefont {T.}~\bibnamefont {Dietrich}},
  \bibinfo {author} {\bibfnamefont {B.}~\bibnamefont {Br\"ugmann}}, \bibinfo
  {author} {\bibfnamefont {E.}~\bibnamefont {Giangrandi}}, \bibinfo {author}
  {\bibfnamefont {K.}~\bibnamefont {Kiuchi}}, \bibinfo {author} {\bibfnamefont
  {F.}~\bibnamefont {Schianchi}}, \bibinfo {author} {\bibfnamefont
  {P.}~\bibnamefont {M\"osta}}, \bibinfo {author} {\bibfnamefont
  {S.}~\bibnamefont {Shankar}}, \bibinfo {author} {\bibfnamefont
  {B.}~\bibnamefont {Giacomazzo}},\ and\ \bibinfo {author} {\bibfnamefont
  {M.}~\bibnamefont {Shibata}},\ }\href
  {https://doi.org/10.1103/PhysRevD.110.084046} {\bibfield  {journal} {\bibinfo
   {journal} {Phys. Rev. D}\ }\textbf {\bibinfo {volume} {110}},\ \bibinfo
  {pages} {084046} (\bibinfo {year} {2024})},\ \Eprint
  {https://arxiv.org/abs/2407.20946} {arXiv:2407.20946 [gr-qc]} \BibitemShut
  {NoStop}%
\bibitem [{\citenamefont {Kiuchi}\ \emph {et~al.}(2014)\citenamefont {Kiuchi},
  \citenamefont {Kyutoku}, \citenamefont {Sekiguchi}, \citenamefont {Shibata},\
  and\ \citenamefont {Wada}}]{Kiuchi:2014hja}%
  \BibitemOpen
  \bibfield  {author} {\bibinfo {author} {\bibfnamefont {K.}~\bibnamefont
  {Kiuchi}}, \bibinfo {author} {\bibfnamefont {K.}~\bibnamefont {Kyutoku}},
  \bibinfo {author} {\bibfnamefont {Y.}~\bibnamefont {Sekiguchi}}, \bibinfo
  {author} {\bibfnamefont {M.}~\bibnamefont {Shibata}},\ and\ \bibinfo {author}
  {\bibfnamefont {T.}~\bibnamefont {Wada}},\ }\href
  {https://doi.org/10.1103/PhysRevD.90.041502} {\bibfield  {journal} {\bibinfo
  {journal} {Phys. Rev. D}\ }\textbf {\bibinfo {volume} {90}},\ \bibinfo
  {pages} {041502} (\bibinfo {year} {2014})},\ \Eprint
  {https://arxiv.org/abs/1407.2660} {arXiv:1407.2660 [astro-ph.HE]}
  \BibitemShut {NoStop}%
\bibitem [{\citenamefont {Dionysopoulou}\ \emph {et~al.}(2015)\citenamefont
  {Dionysopoulou}, \citenamefont {Alic},\ and\ \citenamefont
  {Rezzolla}}]{Dionysopoulou:2015tda}%
  \BibitemOpen
  \bibfield  {author} {\bibinfo {author} {\bibfnamefont {K.}~\bibnamefont
  {Dionysopoulou}}, \bibinfo {author} {\bibfnamefont {D.}~\bibnamefont
  {Alic}},\ and\ \bibinfo {author} {\bibfnamefont {L.}~\bibnamefont
  {Rezzolla}},\ }\href {https://doi.org/10.1103/PhysRevD.92.084064} {\bibfield
  {journal} {\bibinfo  {journal} {Phys. Rev. D}\ }\textbf {\bibinfo {volume}
  {92}},\ \bibinfo {pages} {084064} (\bibinfo {year} {2015})},\ \Eprint
  {https://arxiv.org/abs/1502.02021} {arXiv:1502.02021 [gr-qc]} \BibitemShut
  {NoStop}%
\bibitem [{\citenamefont {Cook}\ \emph {et~al.}(2025)\citenamefont {Cook},
  \citenamefont {Guti{\'e}rrez}, \citenamefont {Bernuzzi}, \citenamefont
  {Radice}, \citenamefont {Daszuta}, \citenamefont {Fields}, \citenamefont
  {Hammond}, \citenamefont {Bandyopadhyay},\ and\ \citenamefont
  {Jacobi}}]{Cook:2025frw}%
  \BibitemOpen
  \bibfield  {author} {\bibinfo {author} {\bibfnamefont {W.}~\bibnamefont
  {Cook}}, \bibinfo {author} {\bibfnamefont {E.~M.}\ \bibnamefont
  {Guti{\'e}rrez}}, \bibinfo {author} {\bibfnamefont {S.}~\bibnamefont
  {Bernuzzi}}, \bibinfo {author} {\bibfnamefont {D.}~\bibnamefont {Radice}},
  \bibinfo {author} {\bibfnamefont {B.}~\bibnamefont {Daszuta}}, \bibinfo
  {author} {\bibfnamefont {J.}~\bibnamefont {Fields}}, \bibinfo {author}
  {\bibfnamefont {P.}~\bibnamefont {Hammond}}, \bibinfo {author} {\bibfnamefont
  {H.}~\bibnamefont {Bandyopadhyay}},\ and\ \bibinfo {author} {\bibfnamefont
  {M.}~\bibnamefont {Jacobi}},\ }\Eprint {https://arxiv.org/abs/2508.19342}
  {arXiv:2508.19342 [astro-ph.HE]}  (\bibinfo {year} {2025})\BibitemShut
  {NoStop}%
\bibitem [{\citenamefont {Guti{\'e}rrez}\ \emph {et~al.}(2025)\citenamefont
  {Guti{\'e}rrez}, \citenamefont {Cook}, \citenamefont {Radice}, \citenamefont
  {Bernuzzi}, \citenamefont {Fields}, \citenamefont {Hammond}, \citenamefont
  {Daszuta}, \citenamefont {Bandyopadhyay},\ and\ \citenamefont
  {Jacobi}}]{Gutierrez:2025gkx}%
  \BibitemOpen
  \bibfield  {author} {\bibinfo {author} {\bibfnamefont {E.~M.}\ \bibnamefont
  {Guti{\'e}rrez}}, \bibinfo {author} {\bibfnamefont {W.}~\bibnamefont {Cook}},
  \bibinfo {author} {\bibfnamefont {D.}~\bibnamefont {Radice}}, \bibinfo
  {author} {\bibfnamefont {S.}~\bibnamefont {Bernuzzi}}, \bibinfo {author}
  {\bibfnamefont {J.}~\bibnamefont {Fields}}, \bibinfo {author} {\bibfnamefont
  {P.}~\bibnamefont {Hammond}}, \bibinfo {author} {\bibfnamefont
  {B.}~\bibnamefont {Daszuta}}, \bibinfo {author} {\bibfnamefont
  {H.}~\bibnamefont {Bandyopadhyay}},\ and\ \bibinfo {author} {\bibfnamefont
  {M.}~\bibnamefont {Jacobi}},\ }\Eprint {https://arxiv.org/abs/2506.18995}
  {arXiv:2506.18995 [astro-ph.HE]}  (\bibinfo {year} {2025})\BibitemShut
  {NoStop}%
\bibitem [{\citenamefont {Ciolfi}\ \emph {et~al.}(2019)\citenamefont {Ciolfi},
  \citenamefont {Kastaun}, \citenamefont {Kalinani},\ and\ \citenamefont
  {Giacomazzo}}]{Ciolfi:2019fie}%
  \BibitemOpen
  \bibfield  {author} {\bibinfo {author} {\bibfnamefont {R.}~\bibnamefont
  {Ciolfi}}, \bibinfo {author} {\bibfnamefont {W.}~\bibnamefont {Kastaun}},
  \bibinfo {author} {\bibfnamefont {J.~V.}\ \bibnamefont {Kalinani}},\ and\
  \bibinfo {author} {\bibfnamefont {B.}~\bibnamefont {Giacomazzo}},\ }\href
  {https://doi.org/10.1103/PhysRevD.100.023005} {\bibfield  {journal} {\bibinfo
   {journal} {Phys. Rev. D}\ }\textbf {\bibinfo {volume} {100}},\ \bibinfo
  {pages} {023005} (\bibinfo {year} {2019})},\ \Eprint
  {https://arxiv.org/abs/1904.10222} {arXiv:1904.10222 [astro-ph.HE]}
  \BibitemShut {NoStop}%
\bibitem [{\citenamefont {Chabanov}\ \emph {et~al.}(2023)\citenamefont
  {Chabanov}, \citenamefont {Tootle}, \citenamefont {Most},\ and\ \citenamefont
  {Rezzolla}}]{Chabanov:2022twz}%
  \BibitemOpen
  \bibfield  {author} {\bibinfo {author} {\bibfnamefont {M.}~\bibnamefont
  {Chabanov}}, \bibinfo {author} {\bibfnamefont {S.~D.}\ \bibnamefont
  {Tootle}}, \bibinfo {author} {\bibfnamefont {E.~R.}\ \bibnamefont {Most}},\
  and\ \bibinfo {author} {\bibfnamefont {L.}~\bibnamefont {Rezzolla}},\ }\href
  {https://doi.org/10.3847/2041-8213/acbbc5} {\bibfield  {journal} {\bibinfo
  {journal} {Astrophys. J. Lett.}\ }\textbf {\bibinfo {volume} {945}},\
  \bibinfo {pages} {L14} (\bibinfo {year} {2023})},\ \Eprint
  {https://arxiv.org/abs/2211.13661} {arXiv:2211.13661 [astro-ph.HE]}
  \BibitemShut {NoStop}%
\bibitem [{\citenamefont {Aguilera-Miret}\ \emph {et~al.}(2022)\citenamefont
  {Aguilera-Miret}, \citenamefont {Vigan{\`o}},\ and\ \citenamefont
  {Palenzuela}}]{Aguilera-Miret:2021fre}%
  \BibitemOpen
  \bibfield  {author} {\bibinfo {author} {\bibfnamefont {R.}~\bibnamefont
  {Aguilera-Miret}}, \bibinfo {author} {\bibfnamefont {D.}~\bibnamefont
  {Vigan{\`o}}},\ and\ \bibinfo {author} {\bibfnamefont {C.}~\bibnamefont
  {Palenzuela}},\ }\href {https://doi.org/10.3847/2041-8213/ac50a7} {\bibfield
  {journal} {\bibinfo  {journal} {Astrophys. J. Lett.}\ }\textbf {\bibinfo
  {volume} {926}},\ \bibinfo {pages} {L31} (\bibinfo {year} {2022})},\ \Eprint
  {https://arxiv.org/abs/2112.08406} {arXiv:2112.08406 [gr-qc]} \BibitemShut
  {NoStop}%
\bibitem [{\citenamefont {Aguilera-Miret}\ \emph {et~al.}(2023)\citenamefont
  {Aguilera-Miret}, \citenamefont {Palenzuela}, \citenamefont {Carrasco},\ and\
  \citenamefont {Vigan{\`o}}}]{Aguilera-Miret:2023qih}%
  \BibitemOpen
  \bibfield  {author} {\bibinfo {author} {\bibfnamefont {R.}~\bibnamefont
  {Aguilera-Miret}}, \bibinfo {author} {\bibfnamefont {C.}~\bibnamefont
  {Palenzuela}}, \bibinfo {author} {\bibfnamefont {F.}~\bibnamefont
  {Carrasco}},\ and\ \bibinfo {author} {\bibfnamefont {D.}~\bibnamefont
  {Vigan{\`o}}},\ }\href {https://doi.org/10.1103/PhysRevD.108.103001}
  {\bibfield  {journal} {\bibinfo  {journal} {Phys. Rev. D}\ }\textbf {\bibinfo
  {volume} {108}},\ \bibinfo {pages} {103001} (\bibinfo {year} {2023})},\
  \Eprint {https://arxiv.org/abs/2307.04837} {arXiv:2307.04837 [astro-ph.HE]}
  \BibitemShut {NoStop}%
\bibitem [{\citenamefont {Foucart}\ \emph {et~al.}(2015)\citenamefont
  {Foucart}, \citenamefont {O'Connor}, \citenamefont {Roberts}, \citenamefont
  {Duez}, \citenamefont {Haas}, \citenamefont {Kidder}, \citenamefont {Ott},
  \citenamefont {Pfeiffer}, \citenamefont {Scheel},\ and\ \citenamefont
  {Szilagyi}}]{Foucart:2015vpa}%
  \BibitemOpen
  \bibfield  {author} {\bibinfo {author} {\bibfnamefont {F.}~\bibnamefont
  {Foucart}}, \bibinfo {author} {\bibfnamefont {E.}~\bibnamefont {O'Connor}},
  \bibinfo {author} {\bibfnamefont {L.}~\bibnamefont {Roberts}}, \bibinfo
  {author} {\bibfnamefont {M.~D.}\ \bibnamefont {Duez}}, \bibinfo {author}
  {\bibfnamefont {R.}~\bibnamefont {Haas}}, \bibinfo {author} {\bibfnamefont
  {L.~E.}\ \bibnamefont {Kidder}}, \bibinfo {author} {\bibfnamefont {C.~D.}\
  \bibnamefont {Ott}}, \bibinfo {author} {\bibfnamefont {H.~P.}\ \bibnamefont
  {Pfeiffer}}, \bibinfo {author} {\bibfnamefont {M.~A.}\ \bibnamefont
  {Scheel}},\ and\ \bibinfo {author} {\bibfnamefont {B.}~\bibnamefont
  {Szilagyi}},\ }\href {https://doi.org/10.1103/PhysRevD.91.124021} {\bibfield
  {journal} {\bibinfo  {journal} {Phys. Rev. D}\ }\textbf {\bibinfo {volume}
  {91}},\ \bibinfo {pages} {124021} (\bibinfo {year} {2015})},\ \Eprint
  {https://arxiv.org/abs/1502.04146} {arXiv:1502.04146 [astro-ph.HE]}
  \BibitemShut {NoStop}%
\bibitem [{\citenamefont {Sekiguchi}\ \emph {et~al.}(2016)\citenamefont
  {Sekiguchi}, \citenamefont {Kiuchi}, \citenamefont {Kyutoku}, \citenamefont
  {Shibata},\ and\ \citenamefont {Taniguchi}}]{Sekiguchi:2016bjd}%
  \BibitemOpen
  \bibfield  {author} {\bibinfo {author} {\bibfnamefont {Y.}~\bibnamefont
  {Sekiguchi}}, \bibinfo {author} {\bibfnamefont {K.}~\bibnamefont {Kiuchi}},
  \bibinfo {author} {\bibfnamefont {K.}~\bibnamefont {Kyutoku}}, \bibinfo
  {author} {\bibfnamefont {M.}~\bibnamefont {Shibata}},\ and\ \bibinfo {author}
  {\bibfnamefont {K.}~\bibnamefont {Taniguchi}},\ }\href
  {https://doi.org/10.1103/PhysRevD.93.124046} {\bibfield  {journal} {\bibinfo
  {journal} {Phys. Rev. D}\ }\textbf {\bibinfo {volume} {93}},\ \bibinfo
  {pages} {124046} (\bibinfo {year} {2016})},\ \Eprint
  {https://arxiv.org/abs/1603.01918} {arXiv:1603.01918 [astro-ph.HE]}
  \BibitemShut {NoStop}%
\bibitem [{\citenamefont {Vincent}\ \emph {et~al.}(2020)\citenamefont
  {Vincent}, \citenamefont {Foucart}, \citenamefont {Duez}, \citenamefont
  {Haas}, \citenamefont {Kidder}, \citenamefont {Pfeiffer},\ and\ \citenamefont
  {Scheel}}]{Vincent:2019kor}%
  \BibitemOpen
  \bibfield  {author} {\bibinfo {author} {\bibfnamefont {T.}~\bibnamefont
  {Vincent}}, \bibinfo {author} {\bibfnamefont {F.}~\bibnamefont {Foucart}},
  \bibinfo {author} {\bibfnamefont {M.~D.}\ \bibnamefont {Duez}}, \bibinfo
  {author} {\bibfnamefont {R.}~\bibnamefont {Haas}}, \bibinfo {author}
  {\bibfnamefont {L.~E.}\ \bibnamefont {Kidder}}, \bibinfo {author}
  {\bibfnamefont {H.~P.}\ \bibnamefont {Pfeiffer}},\ and\ \bibinfo {author}
  {\bibfnamefont {M.~A.}\ \bibnamefont {Scheel}},\ }\href
  {https://doi.org/10.1103/PhysRevD.101.044053} {\bibfield  {journal} {\bibinfo
   {journal} {Phys. Rev. D}\ }\textbf {\bibinfo {volume} {101}},\ \bibinfo
  {eid} {044053} (\bibinfo {year} {2020})},\ \Eprint
  {https://arxiv.org/abs/1908.00655} {1908.00655 [gr-qc]} \BibitemShut
  {NoStop}%
\bibitem [{\citenamefont {Radice}\ \emph
  {et~al.}(2018{\natexlab{b}})\citenamefont {Radice}, \citenamefont {Perego},
  \citenamefont {Hotokezaka}, \citenamefont {Fromm}, \citenamefont {Bernuzzi},\
  and\ \citenamefont {Roberts}}]{Radice:2018pdn}%
  \BibitemOpen
  \bibfield  {author} {\bibinfo {author} {\bibfnamefont {D.}~\bibnamefont
  {Radice}}, \bibinfo {author} {\bibfnamefont {A.}~\bibnamefont {Perego}},
  \bibinfo {author} {\bibfnamefont {K.}~\bibnamefont {Hotokezaka}}, \bibinfo
  {author} {\bibfnamefont {S.~A.}\ \bibnamefont {Fromm}}, \bibinfo {author}
  {\bibfnamefont {S.}~\bibnamefont {Bernuzzi}},\ and\ \bibinfo {author}
  {\bibfnamefont {L.~F.}\ \bibnamefont {Roberts}},\ }\href
  {https://doi.org/10.3847/1538-4357/aaf054} {\bibfield  {journal} {\bibinfo
  {journal} {Astrophys. J.}\ }\textbf {\bibinfo {volume} {869}},\ \bibinfo
  {pages} {130} (\bibinfo {year} {2018}{\natexlab{b}})},\ \Eprint
  {https://arxiv.org/abs/1809.11161} {arXiv:1809.11161 [astro-ph.HE]}
  \BibitemShut {NoStop}%
\bibitem [{\citenamefont {Radice}\ \emph {et~al.}(2022)\citenamefont {Radice},
  \citenamefont {Bernuzzi}, \citenamefont {Perego},\ and\ \citenamefont
  {Haas}}]{Radice:2021jtw}%
  \BibitemOpen
  \bibfield  {author} {\bibinfo {author} {\bibfnamefont {D.}~\bibnamefont
  {Radice}}, \bibinfo {author} {\bibfnamefont {S.}~\bibnamefont {Bernuzzi}},
  \bibinfo {author} {\bibfnamefont {A.}~\bibnamefont {Perego}},\ and\ \bibinfo
  {author} {\bibfnamefont {R.}~\bibnamefont {Haas}},\ }\href
  {https://doi.org/10.1093/mnras/stac589} {\bibfield  {journal} {\bibinfo
  {journal} {Mon. Not. Roy. Astron. Soc.}\ }\textbf {\bibinfo {volume} {512}},\
  \bibinfo {pages} {1499} (\bibinfo {year} {2022})},\ \Eprint
  {https://arxiv.org/abs/2111.14858} {arXiv:2111.14858 [astro-ph.HE]}
  \BibitemShut {NoStop}%
\bibitem [{\citenamefont {Espino}\ \emph {et~al.}(2024)\citenamefont {Espino},
  \citenamefont {Radice}, \citenamefont {Zappa}, \citenamefont {Gamba},\ and\
  \citenamefont {Bernuzzi}}]{Espino:2023mda}%
  \BibitemOpen
  \bibfield  {author} {\bibinfo {author} {\bibfnamefont {P.~L.}\ \bibnamefont
  {Espino}}, \bibinfo {author} {\bibfnamefont {D.}~\bibnamefont {Radice}},
  \bibinfo {author} {\bibfnamefont {F.}~\bibnamefont {Zappa}}, \bibinfo
  {author} {\bibfnamefont {R.}~\bibnamefont {Gamba}},\ and\ \bibinfo {author}
  {\bibfnamefont {S.}~\bibnamefont {Bernuzzi}},\ }\href
  {https://doi.org/10.1103/PhysRevD.109.103027} {\bibfield  {journal} {\bibinfo
   {journal} {Phys. Rev. D}\ }\textbf {\bibinfo {volume} {109}},\ \bibinfo
  {pages} {103027} (\bibinfo {year} {2024})},\ \Eprint
  {https://arxiv.org/abs/2311.12923} {arXiv:2311.12923 [astro-ph.HE]}
  \BibitemShut {NoStop}%
\bibitem [{\citenamefont {Kawaguchi}\ \emph {et~al.}(2025)\citenamefont
  {Kawaguchi}, \citenamefont {Fujibayashi},\ and\ \citenamefont
  {Shibata}}]{Kawaguchi:2025con}%
  \BibitemOpen
  \bibfield  {author} {\bibinfo {author} {\bibfnamefont {K.}~\bibnamefont
  {Kawaguchi}}, \bibinfo {author} {\bibfnamefont {S.}~\bibnamefont
  {Fujibayashi}},\ and\ \bibinfo {author} {\bibfnamefont {M.}~\bibnamefont
  {Shibata}},\ }\href {https://doi.org/10.1103/yygv-6268} {\bibfield  {journal}
  {\bibinfo  {journal} {Phys. Rev. D}\ }\textbf {\bibinfo {volume} {112}},\
  \bibinfo {pages} {043001} (\bibinfo {year} {2025})},\ \Eprint
  {https://arxiv.org/abs/2506.01679} {arXiv:2506.01679 [astro-ph.HE]}
  \BibitemShut {NoStop}%
\bibitem [{\citenamefont {Palenzuela}\ \emph {et~al.}(2015)\citenamefont
  {Palenzuela}, \citenamefont {Liebling}, \citenamefont {Neilsen},
  \citenamefont {Lehner}, \citenamefont {Caballero}, \citenamefont {O'Connor},\
  and\ \citenamefont {Anderson}}]{Palenzuela:2015dqa}%
  \BibitemOpen
  \bibfield  {author} {\bibinfo {author} {\bibfnamefont {C.}~\bibnamefont
  {Palenzuela}}, \bibinfo {author} {\bibfnamefont {S.~L.}\ \bibnamefont
  {Liebling}}, \bibinfo {author} {\bibfnamefont {D.}~\bibnamefont {Neilsen}},
  \bibinfo {author} {\bibfnamefont {L.}~\bibnamefont {Lehner}}, \bibinfo
  {author} {\bibfnamefont {O.~L.}\ \bibnamefont {Caballero}}, \bibinfo {author}
  {\bibfnamefont {E.}~\bibnamefont {O'Connor}},\ and\ \bibinfo {author}
  {\bibfnamefont {M.}~\bibnamefont {Anderson}},\ }\href
  {https://doi.org/10.1103/PhysRevD.92.044045} {\bibfield  {journal} {\bibinfo
  {journal} {Phys. Rev. D}\ }\textbf {\bibinfo {volume} {92}},\ \bibinfo
  {pages} {044045} (\bibinfo {year} {2015})},\ \Eprint
  {https://arxiv.org/abs/1505.01607} {arXiv:1505.01607 [gr-qc]} \BibitemShut
  {NoStop}%
\bibitem [{\citenamefont {Palenzuela}\ \emph {et~al.}(2022)\citenamefont
  {Palenzuela}, \citenamefont {Liebling},\ and\ \citenamefont
  {Mi{\~n}ano}}]{Palenzuela:2022kqk}%
  \BibitemOpen
  \bibfield  {author} {\bibinfo {author} {\bibfnamefont {C.}~\bibnamefont
  {Palenzuela}}, \bibinfo {author} {\bibfnamefont {S.}~\bibnamefont
  {Liebling}},\ and\ \bibinfo {author} {\bibfnamefont {B.}~\bibnamefont
  {Mi{\~n}ano}},\ }\href {https://doi.org/10.1103/PhysRevD.105.103020}
  {\bibfield  {journal} {\bibinfo  {journal} {Phys. Rev. D}\ }\textbf {\bibinfo
  {volume} {105}},\ \bibinfo {pages} {103020} (\bibinfo {year} {2022})},\
  \Eprint {https://arxiv.org/abs/2204.02721} {arXiv:2204.02721 [gr-qc]}
  \BibitemShut {NoStop}%
\bibitem [{\citenamefont {Musolino}\ \emph {et~al.}(2025)\citenamefont
  {Musolino}, \citenamefont {Rezzolla},\ and\ \citenamefont
  {Most}}]{Musolino:2024sju}%
  \BibitemOpen
  \bibfield  {author} {\bibinfo {author} {\bibfnamefont {C.}~\bibnamefont
  {Musolino}}, \bibinfo {author} {\bibfnamefont {L.}~\bibnamefont {Rezzolla}},\
  and\ \bibinfo {author} {\bibfnamefont {E.~R.}\ \bibnamefont {Most}},\ }\href
  {https://doi.org/10.3847/2041-8213/adcd6d} {\bibfield  {journal} {\bibinfo
  {journal} {Astrophys. J. Lett.}\ }\textbf {\bibinfo {volume} {984}},\
  \bibinfo {pages} {L61} (\bibinfo {year} {2025})},\ \Eprint
  {https://arxiv.org/abs/2410.06253} {arXiv:2410.06253 [astro-ph.HE]}
  \BibitemShut {NoStop}%
\bibitem [{\citenamefont {Curtis}\ \emph {et~al.}(2024)\citenamefont {Curtis},
  \citenamefont {Bosch}, \citenamefont {M{\"o}sta}, \citenamefont {Radice},
  \citenamefont {Bernuzzi}, \citenamefont {Perego}, \citenamefont {Haas},\ and\
  \citenamefont {Schnetter}}]{Curtis:2023zfo}%
  \BibitemOpen
  \bibfield  {author} {\bibinfo {author} {\bibfnamefont {S.}~\bibnamefont
  {Curtis}}, \bibinfo {author} {\bibfnamefont {P.}~\bibnamefont {Bosch}},
  \bibinfo {author} {\bibfnamefont {P.}~\bibnamefont {M{\"o}sta}}, \bibinfo
  {author} {\bibfnamefont {D.}~\bibnamefont {Radice}}, \bibinfo {author}
  {\bibfnamefont {S.}~\bibnamefont {Bernuzzi}}, \bibinfo {author}
  {\bibfnamefont {A.}~\bibnamefont {Perego}}, \bibinfo {author} {\bibfnamefont
  {R.}~\bibnamefont {Haas}},\ and\ \bibinfo {author} {\bibfnamefont
  {E.}~\bibnamefont {Schnetter}},\ }\href
  {https://doi.org/10.3847/2041-8213/ad0fe1} {\bibfield  {journal} {\bibinfo
  {journal} {Astrophys. J. Lett.}\ }\textbf {\bibinfo {volume} {961}},\
  \bibinfo {pages} {L26} (\bibinfo {year} {2024})},\ \Eprint
  {https://arxiv.org/abs/2305.07738} {arXiv:2305.07738 [astro-ph.HE]}
  \BibitemShut {NoStop}%
\bibitem [{\citenamefont {Sun}\ \emph {et~al.}(2022)\citenamefont {Sun},
  \citenamefont {Ruiz}, \citenamefont {Shapiro},\ and\ \citenamefont
  {Tsokaros}}]{Sun:2022vri}%
  \BibitemOpen
  \bibfield  {author} {\bibinfo {author} {\bibfnamefont {L.}~\bibnamefont
  {Sun}}, \bibinfo {author} {\bibfnamefont {M.}~\bibnamefont {Ruiz}}, \bibinfo
  {author} {\bibfnamefont {S.~L.}\ \bibnamefont {Shapiro}},\ and\ \bibinfo
  {author} {\bibfnamefont {A.}~\bibnamefont {Tsokaros}},\ }\href
  {https://doi.org/10.1103/PhysRevD.105.104028} {\bibfield  {journal} {\bibinfo
   {journal} {Phys. Rev. D}\ }\textbf {\bibinfo {volume} {105}},\ \bibinfo
  {pages} {104028} (\bibinfo {year} {2022})},\ \Eprint
  {https://arxiv.org/abs/2202.12901} {arXiv:2202.12901 [astro-ph.HE]}
  \BibitemShut {NoStop}%
\bibitem [{\citenamefont {Hayashi}\ \emph {et~al.}(2023)\citenamefont
  {Hayashi}, \citenamefont {Kiuchi}, \citenamefont {Kyutoku}, \citenamefont
  {Sekiguchi},\ and\ \citenamefont {Shibata}}]{Hayashi:2022cdq}%
  \BibitemOpen
  \bibfield  {author} {\bibinfo {author} {\bibfnamefont {K.}~\bibnamefont
  {Hayashi}}, \bibinfo {author} {\bibfnamefont {K.}~\bibnamefont {Kiuchi}},
  \bibinfo {author} {\bibfnamefont {K.}~\bibnamefont {Kyutoku}}, \bibinfo
  {author} {\bibfnamefont {Y.}~\bibnamefont {Sekiguchi}},\ and\ \bibinfo
  {author} {\bibfnamefont {M.}~\bibnamefont {Shibata}},\ }\href
  {https://doi.org/10.1103/PhysRevD.107.123001} {\bibfield  {journal} {\bibinfo
   {journal} {Phys. Rev. D}\ }\textbf {\bibinfo {volume} {107}},\ \bibinfo
  {pages} {123001} (\bibinfo {year} {2023})},\ \Eprint
  {https://arxiv.org/abs/2211.07158} {arXiv:2211.07158 [astro-ph.HE]}
  \BibitemShut {NoStop}%
\bibitem [{\citenamefont {Combi}\ and\ \citenamefont
  {Siegel}(2023{\natexlab{a}})}]{Combi:2023yav}%
  \BibitemOpen
  \bibfield  {author} {\bibinfo {author} {\bibfnamefont {L.}~\bibnamefont
  {Combi}}\ and\ \bibinfo {author} {\bibfnamefont {D.~M.}\ \bibnamefont
  {Siegel}},\ }\href {https://doi.org/10.1103/PhysRevLett.131.231402}
  {\bibfield  {journal} {\bibinfo  {journal} {Phys. Rev. Lett.}\ }\textbf
  {\bibinfo {volume} {131}},\ \bibinfo {pages} {231402} (\bibinfo {year}
  {2023}{\natexlab{a}})},\ \Eprint {https://arxiv.org/abs/2303.12284}
  {arXiv:2303.12284 [astro-ph.HE]} \BibitemShut {NoStop}%
\bibitem [{\citenamefont {Bamber}\ \emph {et~al.}(2025)\citenamefont {Bamber},
  \citenamefont {Tsokaros}, \citenamefont {Ruiz}, \citenamefont {Shapiro},
  \citenamefont {Favata}, \citenamefont {Karlson},\ and\ \citenamefont
  {Pi{\~n}as}}]{Bamber:2025ggq}%
  \BibitemOpen
  \bibfield  {author} {\bibinfo {author} {\bibfnamefont {J.}~\bibnamefont
  {Bamber}}, \bibinfo {author} {\bibfnamefont {A.}~\bibnamefont {Tsokaros}},
  \bibinfo {author} {\bibfnamefont {M.}~\bibnamefont {Ruiz}}, \bibinfo {author}
  {\bibfnamefont {S.~L.}\ \bibnamefont {Shapiro}}, \bibinfo {author}
  {\bibfnamefont {M.}~\bibnamefont {Favata}}, \bibinfo {author} {\bibfnamefont
  {M.}~\bibnamefont {Karlson}},\ and\ \bibinfo {author} {\bibfnamefont {F.~V.}\
  \bibnamefont {Pi{\~n}as}},\ }\Eprint {https://arxiv.org/abs/2510.09742}
  {arXiv:2510.09742 [gr-qc]}  (\bibinfo {year} {2025})\BibitemShut {NoStop}%
\bibitem [{\citenamefont {Dietrich}\ \emph
  {et~al.}(2015{\natexlab{a}})\citenamefont {Dietrich}, \citenamefont
  {Bernuzzi}, \citenamefont {Ujevic},\ and\ \citenamefont
  {Brügmann}}]{Dietrich:2015iva}%
  \BibitemOpen
  \bibfield  {author} {\bibinfo {author} {\bibfnamefont {T.}~\bibnamefont
  {Dietrich}}, \bibinfo {author} {\bibfnamefont {S.}~\bibnamefont {Bernuzzi}},
  \bibinfo {author} {\bibfnamefont {M.}~\bibnamefont {Ujevic}},\ and\ \bibinfo
  {author} {\bibfnamefont {B.}~\bibnamefont {Brügmann}},\ }\href
  {https://doi.org/10.1103/PhysRevD.91.124041} {\bibfield  {journal} {\bibinfo
  {journal} {Phys. Rev. D}\ }\textbf {\bibinfo {volume} {91}},\ \bibinfo
  {pages} {124041} (\bibinfo {year} {2015}{\natexlab{a}})},\ \Eprint
  {https://arxiv.org/abs/1504.01266} {arXiv:1504.01266 [gr-qc]} \BibitemShut
  {NoStop}%
\bibitem [{\citenamefont {Bernuzzi}\ and\ \citenamefont
  {Dietrich}(2016)}]{Bernuzzi:2016pie}%
  \BibitemOpen
  \bibfield  {author} {\bibinfo {author} {\bibfnamefont {S.}~\bibnamefont
  {Bernuzzi}}\ and\ \bibinfo {author} {\bibfnamefont {T.}~\bibnamefont
  {Dietrich}},\ }\href {https://doi.org/10.1103/PhysRevD.94.064062} {\bibfield
  {journal} {\bibinfo  {journal} {Phys. Rev. D}\ }\textbf {\bibinfo {volume}
  {94}},\ \bibinfo {pages} {064062} (\bibinfo {year} {2016})},\ \Eprint
  {https://arxiv.org/abs/1604.07999} {arXiv:1604.07999 [gr-qc]} \BibitemShut
  {NoStop}%
\bibitem [{\citenamefont {Bernuzzi}\ and\ \citenamefont
  {Hilditch}(2010)}]{Bernuzzi:2009ex}%
  \BibitemOpen
  \bibfield  {author} {\bibinfo {author} {\bibfnamefont {S.}~\bibnamefont
  {Bernuzzi}}\ and\ \bibinfo {author} {\bibfnamefont {D.}~\bibnamefont
  {Hilditch}},\ }\href {https://doi.org/10.1103/PhysRevD.81.084003} {\bibfield
  {journal} {\bibinfo  {journal} {Phys. Rev. D}\ }\textbf {\bibinfo {volume}
  {81}},\ \bibinfo {pages} {084003} (\bibinfo {year} {2010})},\ \Eprint
  {https://arxiv.org/abs/0912.2920} {arXiv:0912.2920 [gr-qc]} \BibitemShut
  {NoStop}%
\bibitem [{\citenamefont {Hilditch}\ \emph {et~al.}(2013)\citenamefont
  {Hilditch}, \citenamefont {Bernuzzi}, \citenamefont {Thierfelder},
  \citenamefont {Cao}, \citenamefont {Tichy},\ and\ \citenamefont
  {Brügmann}}]{Hilditch:2012fp}%
  \BibitemOpen
  \bibfield  {author} {\bibinfo {author} {\bibfnamefont {D.}~\bibnamefont
  {Hilditch}}, \bibinfo {author} {\bibfnamefont {S.}~\bibnamefont {Bernuzzi}},
  \bibinfo {author} {\bibfnamefont {M.}~\bibnamefont {Thierfelder}}, \bibinfo
  {author} {\bibfnamefont {Z.}~\bibnamefont {Cao}}, \bibinfo {author}
  {\bibfnamefont {W.}~\bibnamefont {Tichy}},\ and\ \bibinfo {author}
  {\bibfnamefont {B.}~\bibnamefont {Brügmann}},\ }\href
  {https://doi.org/10.1103/PhysRevD.88.084057} {\bibfield  {journal} {\bibinfo
  {journal} {Phys. Rev. D}\ }\textbf {\bibinfo {volume} {88}},\ \bibinfo
  {pages} {084057} (\bibinfo {year} {2013})},\ \Eprint
  {https://arxiv.org/abs/1212.2901} {arXiv:1212.2901 [gr-qc]} \BibitemShut
  {NoStop}%
\bibitem [{\citenamefont {Bona}\ \emph {et~al.}(1996)\citenamefont {Bona},
  \citenamefont {Mass{\'o}}, \citenamefont {Stela},\ and\ \citenamefont
  {Seidel}}]{Bona:1994a}%
  \BibitemOpen
  \bibfield  {author} {\bibinfo {author} {\bibfnamefont {C.}~\bibnamefont
  {Bona}}, \bibinfo {author} {\bibfnamefont {J.}~\bibnamefont {Mass{\'o}}},
  \bibinfo {author} {\bibfnamefont {J.}~\bibnamefont {Stela}},\ and\ \bibinfo
  {author} {\bibfnamefont {E.}~\bibnamefont {Seidel}},\ }in\ \href@noop {}
  {\emph {\bibinfo {booktitle} {The Seventh {M}arcel {G}rossmann Meeting: On
  Recent Developments in Theoretical and Experimental General Relativity,
  Gravitation, and Relativistic Field Theories}}},\ \bibinfo {editor} {edited
  by\ \bibinfo {editor} {\bibfnamefont {R.~T.}\ \bibnamefont {Jantzen}},
  \bibinfo {editor} {\bibfnamefont {G.~M.}\ \bibnamefont {Keiser}},\ and\
  \bibinfo {editor} {\bibfnamefont {R.}~\bibnamefont {Ruffini}}}\ (\bibinfo
  {publisher} {World {S}cientific},\ \bibinfo {address} {Singapore},\ \bibinfo
  {year} {1996})\BibitemShut {NoStop}%
\bibitem [{\citenamefont {Alcubierre}\ \emph {et~al.}(2003)\citenamefont
  {Alcubierre}, \citenamefont {Brügmann}, \citenamefont {Diener},
  \citenamefont {Koppitz}, \citenamefont {Pollney}, \citenamefont {Seidel},\
  and\ \citenamefont {Takahashi}}]{Alcubierre:2002kk}%
  \BibitemOpen
  \bibfield  {author} {\bibinfo {author} {\bibfnamefont {M.}~\bibnamefont
  {Alcubierre}}, \bibinfo {author} {\bibfnamefont {B.}~\bibnamefont
  {Brügmann}}, \bibinfo {author} {\bibfnamefont {P.}~\bibnamefont {Diener}},
  \bibinfo {author} {\bibfnamefont {M.}~\bibnamefont {Koppitz}}, \bibinfo
  {author} {\bibfnamefont {D.}~\bibnamefont {Pollney}}, \bibinfo {author}
  {\bibfnamefont {E.}~\bibnamefont {Seidel}},\ and\ \bibinfo {author}
  {\bibfnamefont {R.}~\bibnamefont {Takahashi}},\ }\href
  {https://doi.org/10.1103/PhysRevD.67.084023} {\bibfield  {journal} {\bibinfo
  {journal} {Phys. Rev. D}\ }\textbf {\bibinfo {volume} {67}},\ \bibinfo
  {pages} {084023} (\bibinfo {year} {2003})},\ \Eprint
  {https://arxiv.org/abs/gr-qc/0206072} {arXiv:gr-qc/0206072} \BibitemShut
  {NoStop}%
\bibitem [{\citenamefont {Marti}\ \emph {et~al.}(1991)\citenamefont {Marti},
  \citenamefont {Ibanez},\ and\ \citenamefont {Miralles}}]{Marti:1991wi}%
  \BibitemOpen
  \bibfield  {author} {\bibinfo {author} {\bibfnamefont {J.~M.}\ \bibnamefont
  {Marti}}, \bibinfo {author} {\bibfnamefont {J.~M.}\ \bibnamefont {Ibanez}},\
  and\ \bibinfo {author} {\bibfnamefont {J.~A.}\ \bibnamefont {Miralles}},\
  }\href {https://doi.org/10.1103/PhysRevD.43.3794} {\bibfield  {journal}
  {\bibinfo  {journal} {Phys. Rev. D}\ }\textbf {\bibinfo {volume} {43}},\
  \bibinfo {pages} {3794} (\bibinfo {year} {1991})}\BibitemShut {NoStop}%
\bibitem [{\citenamefont {Banyuls}\ \emph {et~al.}(1997)\citenamefont
  {Banyuls}, \citenamefont {Font}, \citenamefont {Ibanez}, \citenamefont
  {Marti},\ and\ \citenamefont {Miralles}}]{Banyuls:1997zz}%
  \BibitemOpen
  \bibfield  {author} {\bibinfo {author} {\bibfnamefont {F.}~\bibnamefont
  {Banyuls}}, \bibinfo {author} {\bibfnamefont {J.~A.}\ \bibnamefont {Font}},
  \bibinfo {author} {\bibfnamefont {J.~M.~A.}\ \bibnamefont {Ibanez}}, \bibinfo
  {author} {\bibfnamefont {J.~M.~A.}\ \bibnamefont {Marti}},\ and\ \bibinfo
  {author} {\bibfnamefont {J.~A.}\ \bibnamefont {Miralles}},\ }\href
  {https://doi.org/10.1086/303604} {\bibfield  {journal} {\bibinfo  {journal}
  {Astrophys. J.}\ }\textbf {\bibinfo {volume} {476}},\ \bibinfo {pages} {221}
  (\bibinfo {year} {1997})}\BibitemShut {NoStop}%
\bibitem [{\citenamefont {Anton}\ \emph {et~al.}(2006)\citenamefont {Anton},
  \citenamefont {Zanotti}, \citenamefont {Miralles}, \citenamefont {Marti},
  \citenamefont {Ibanez}, \citenamefont {Font},\ and\ \citenamefont
  {Pons}}]{Anton:2005gi}%
  \BibitemOpen
  \bibfield  {author} {\bibinfo {author} {\bibfnamefont {L.}~\bibnamefont
  {Anton}}, \bibinfo {author} {\bibfnamefont {O.}~\bibnamefont {Zanotti}},
  \bibinfo {author} {\bibfnamefont {J.~A.}\ \bibnamefont {Miralles}}, \bibinfo
  {author} {\bibfnamefont {J.~M.}\ \bibnamefont {Marti}}, \bibinfo {author}
  {\bibfnamefont {J.~M.}\ \bibnamefont {Ibanez}}, \bibinfo {author}
  {\bibfnamefont {J.~A.}\ \bibnamefont {Font}},\ and\ \bibinfo {author}
  {\bibfnamefont {J.~A.}\ \bibnamefont {Pons}},\ }\href
  {https://doi.org/10.1086/498238} {\bibfield  {journal} {\bibinfo  {journal}
  {Astrophys. J.}\ }\textbf {\bibinfo {volume} {637}},\ \bibinfo {pages} {296}
  (\bibinfo {year} {2006})},\ \Eprint {https://arxiv.org/abs/astro-ph/0506063}
  {arXiv:astro-ph/0506063} \BibitemShut {NoStop}%
\bibitem [{\citenamefont {Font}(2008)}]{Font:2008fka}%
  \BibitemOpen
  \bibfield  {author} {\bibinfo {author} {\bibfnamefont {J.~A.}\ \bibnamefont
  {Font}},\ }\href {https://doi.org/10.12942/lrr-2008-7} {\bibfield  {journal}
  {\bibinfo  {journal} {Living Rev. Rel.}\ }\textbf {\bibinfo {volume} {11}},\
  \bibinfo {pages} {7} (\bibinfo {year} {2008})}\BibitemShut {NoStop}%
\bibitem [{\citenamefont {Liebling}\ \emph {et~al.}(2010)\citenamefont
  {Liebling}, \citenamefont {Lehner}, \citenamefont {Neilsen},\ and\
  \citenamefont {Palenzuela}}]{Liebling:2010bn}%
  \BibitemOpen
  \bibfield  {author} {\bibinfo {author} {\bibfnamefont {S.~L.}\ \bibnamefont
  {Liebling}}, \bibinfo {author} {\bibfnamefont {L.}~\bibnamefont {Lehner}},
  \bibinfo {author} {\bibfnamefont {D.}~\bibnamefont {Neilsen}},\ and\ \bibinfo
  {author} {\bibfnamefont {C.}~\bibnamefont {Palenzuela}},\ }\href
  {https://doi.org/10.1103/PhysRevD.81.124023} {\bibfield  {journal} {\bibinfo
  {journal} {Phys. Rev. D}\ }\textbf {\bibinfo {volume} {81}},\ \bibinfo
  {pages} {124023} (\bibinfo {year} {2010})},\ \Eprint
  {https://arxiv.org/abs/1001.0575} {arXiv:1001.0575 [gr-qc]} \BibitemShut
  {NoStop}%
\bibitem [{\citenamefont {M\"osta}\ \emph {et~al.}(2014)\citenamefont
  {M\"osta}, \citenamefont {Mundim}, \citenamefont {Faber}, \citenamefont
  {Haas}, \citenamefont {Noble}, \citenamefont {Bode}, \citenamefont
  {L\"offler}, \citenamefont {Ott}, \citenamefont {Reisswig},\ and\
  \citenamefont {Schnetter}}]{Mosta:2013gwu}%
  \BibitemOpen
  \bibfield  {author} {\bibinfo {author} {\bibfnamefont {P.}~\bibnamefont
  {M\"osta}}, \bibinfo {author} {\bibfnamefont {B.~C.}\ \bibnamefont {Mundim}},
  \bibinfo {author} {\bibfnamefont {J.~A.}\ \bibnamefont {Faber}}, \bibinfo
  {author} {\bibfnamefont {R.}~\bibnamefont {Haas}}, \bibinfo {author}
  {\bibfnamefont {S.~C.}\ \bibnamefont {Noble}}, \bibinfo {author}
  {\bibfnamefont {T.}~\bibnamefont {Bode}}, \bibinfo {author} {\bibfnamefont
  {F.}~\bibnamefont {L\"offler}}, \bibinfo {author} {\bibfnamefont {C.~D.}\
  \bibnamefont {Ott}}, \bibinfo {author} {\bibfnamefont {C.}~\bibnamefont
  {Reisswig}},\ and\ \bibinfo {author} {\bibfnamefont {E.}~\bibnamefont
  {Schnetter}},\ }\href {https://doi.org/10.1088/0264-9381/31/1/015005}
  {\bibfield  {journal} {\bibinfo  {journal} {Class. Quant. Grav.}\ }\textbf
  {\bibinfo {volume} {31}},\ \bibinfo {pages} {015005} (\bibinfo {year}
  {2014})},\ \Eprint {https://arxiv.org/abs/1304.5544} {arXiv:1304.5544
  [gr-qc]} \BibitemShut {NoStop}%
\bibitem [{\citenamefont {Neuweiler}\ \emph {et~al.}(2025)\citenamefont
  {Neuweiler}, \citenamefont {Dietrich},\ and\ \citenamefont
  {Br{\"u}gmann}}]{Neuweiler:2025lte}%
  \BibitemOpen
  \bibfield  {author} {\bibinfo {author} {\bibfnamefont {A.}~\bibnamefont
  {Neuweiler}}, \bibinfo {author} {\bibfnamefont {T.}~\bibnamefont
  {Dietrich}},\ and\ \bibinfo {author} {\bibfnamefont {B.}~\bibnamefont
  {Br{\"u}gmann}},\ }\href {https://doi.org/10.1103/tsg2-bp7l} {\bibfield
  {journal} {\bibinfo  {journal} {Phys. Rev. D}\ }\textbf {\bibinfo {volume}
  {112}},\ \bibinfo {pages} {023033} (\bibinfo {year} {2025})},\ \Eprint
  {https://arxiv.org/abs/2504.10228} {arXiv:2504.10228 [gr-qc]} \BibitemShut
  {NoStop}%
\bibitem [{\citenamefont {Thorne}(1981)}]{Thorn:1981}%
  \BibitemOpen
  \bibfield  {author} {\bibinfo {author} {\bibfnamefont {K.~S.}\ \bibnamefont
  {Thorne}},\ }\href {https://doi.org/10.1093/mnras/194.2.439} {\bibfield
  {journal} {\bibinfo  {journal} {Mon. Not. Roy. Astron. Soc.}\ }\textbf
  {\bibinfo {volume} {194}},\ \bibinfo {pages} {439} (\bibinfo {year}
  {1981})}\BibitemShut {NoStop}%
\bibitem [{\citenamefont {Shibata}\ \emph {et~al.}(2011)\citenamefont
  {Shibata}, \citenamefont {Kiuchi}, \citenamefont {Sekiguchi},\ and\
  \citenamefont {Suwa}}]{Shibata:2011kx}%
  \BibitemOpen
  \bibfield  {author} {\bibinfo {author} {\bibfnamefont {M.}~\bibnamefont
  {Shibata}}, \bibinfo {author} {\bibfnamefont {K.}~\bibnamefont {Kiuchi}},
  \bibinfo {author} {\bibfnamefont {Y.-i.}\ \bibnamefont {Sekiguchi}},\ and\
  \bibinfo {author} {\bibfnamefont {Y.}~\bibnamefont {Suwa}},\ }\href
  {https://doi.org/10.1143/PTP.125.1255} {\bibfield  {journal} {\bibinfo
  {journal} {Prog. Theor. Phys.}\ }\textbf {\bibinfo {volume} {125}},\ \bibinfo
  {pages} {1255} (\bibinfo {year} {2011})},\ \Eprint
  {https://arxiv.org/abs/1104.3937} {arXiv:1104.3937 [astro-ph.HE]}
  \BibitemShut {NoStop}%
\bibitem [{\citenamefont {Foucart}\ \emph
  {et~al.}(2016{\natexlab{a}})\citenamefont {Foucart}, \citenamefont {Haas},
  \citenamefont {Duez}, \citenamefont {O'Connor}, \citenamefont {Ott},
  \citenamefont {Roberts}, \citenamefont {Kidder}, \citenamefont {Lippuner},
  \citenamefont {Pfeiffer},\ and\ \citenamefont {Scheel}}]{Foucart:2015gaa}%
  \BibitemOpen
  \bibfield  {author} {\bibinfo {author} {\bibfnamefont {F.}~\bibnamefont
  {Foucart}}, \bibinfo {author} {\bibfnamefont {R.}~\bibnamefont {Haas}},
  \bibinfo {author} {\bibfnamefont {M.~D.}\ \bibnamefont {Duez}}, \bibinfo
  {author} {\bibfnamefont {E.}~\bibnamefont {O'Connor}}, \bibinfo {author}
  {\bibfnamefont {C.~D.}\ \bibnamefont {Ott}}, \bibinfo {author} {\bibfnamefont
  {L.}~\bibnamefont {Roberts}}, \bibinfo {author} {\bibfnamefont {L.~E.}\
  \bibnamefont {Kidder}}, \bibinfo {author} {\bibfnamefont {J.}~\bibnamefont
  {Lippuner}}, \bibinfo {author} {\bibfnamefont {H.~P.}\ \bibnamefont
  {Pfeiffer}},\ and\ \bibinfo {author} {\bibfnamefont {M.~A.}\ \bibnamefont
  {Scheel}},\ }\href {https://doi.org/10.1103/PhysRevD.93.044019} {\bibfield
  {journal} {\bibinfo  {journal} {Phys. Rev.}\ }\textbf {\bibinfo {volume}
  {D93}},\ \bibinfo {pages} {044019} (\bibinfo {year} {2016}{\natexlab{a}})},\
  \Eprint {https://arxiv.org/abs/1510.06398} {arXiv:1510.06398 [astro-ph.HE]}
  \BibitemShut {NoStop}%
\bibitem [{\citenamefont {Foucart}\ \emph
  {et~al.}(2016{\natexlab{b}})\citenamefont {Foucart}, \citenamefont
  {O'Connor}, \citenamefont {Roberts}, \citenamefont {Kidder}, \citenamefont
  {Pfeiffer},\ and\ \citenamefont {Scheel}}]{Foucart:2016rxm}%
  \BibitemOpen
  \bibfield  {author} {\bibinfo {author} {\bibfnamefont {F.}~\bibnamefont
  {Foucart}}, \bibinfo {author} {\bibfnamefont {E.}~\bibnamefont {O'Connor}},
  \bibinfo {author} {\bibfnamefont {L.}~\bibnamefont {Roberts}}, \bibinfo
  {author} {\bibfnamefont {L.~E.}\ \bibnamefont {Kidder}}, \bibinfo {author}
  {\bibfnamefont {H.~P.}\ \bibnamefont {Pfeiffer}},\ and\ \bibinfo {author}
  {\bibfnamefont {M.~A.}\ \bibnamefont {Scheel}},\ }\href
  {https://doi.org/10.1103/PhysRevD.94.123016} {\bibfield  {journal} {\bibinfo
  {journal} {Phys. Rev.}\ }\textbf {\bibinfo {volume} {D94}},\ \bibinfo {pages}
  {123016} (\bibinfo {year} {2016}{\natexlab{b}})},\ \Eprint
  {https://arxiv.org/abs/1607.07450} {arXiv:1607.07450 [astro-ph.HE]}
  \BibitemShut {NoStop}%
\bibitem [{\citenamefont {Weih}\ \emph {et~al.}(2020)\citenamefont {Weih},
  \citenamefont {Olivares},\ and\ \citenamefont {Rezzolla}}]{Weih:2020wpo}%
  \BibitemOpen
  \bibfield  {author} {\bibinfo {author} {\bibfnamefont {L.~R.}\ \bibnamefont
  {Weih}}, \bibinfo {author} {\bibfnamefont {H.}~\bibnamefont {Olivares}},\
  and\ \bibinfo {author} {\bibfnamefont {L.}~\bibnamefont {Rezzolla}},\ }\href
  {https://doi.org/10.1093/mnras/staa1297} {\bibfield  {journal} {\bibinfo
  {journal} {Mon. Not. Roy. Astron. Soc.}\ }\textbf {\bibinfo {volume} {495}},\
  \bibinfo {pages} {2285} (\bibinfo {year} {2020})},\ \Eprint
  {https://arxiv.org/abs/2003.13580} {arXiv:2003.13580 [gr-qc]} \BibitemShut
  {NoStop}%
\bibitem [{\citenamefont {Musolino}\ and\ \citenamefont
  {Rezzolla}(2024)}]{Musolino:2023pao}%
  \BibitemOpen
  \bibfield  {author} {\bibinfo {author} {\bibfnamefont {C.}~\bibnamefont
  {Musolino}}\ and\ \bibinfo {author} {\bibfnamefont {L.}~\bibnamefont
  {Rezzolla}},\ }\href {https://doi.org/10.1093/mnras/stae224} {\bibfield
  {journal} {\bibinfo  {journal} {Mon. Not. Roy. Astron. Soc.}\ }\textbf
  {\bibinfo {volume} {528}},\ \bibinfo {pages} {5952} (\bibinfo {year}
  {2024})},\ \Eprint {https://arxiv.org/abs/2304.09168} {arXiv:2304.09168
  [gr-qc]} \BibitemShut {NoStop}%
\bibitem [{\citenamefont {Ruffert}\ \emph {et~al.}(1996)\citenamefont
  {Ruffert}, \citenamefont {Janka},\ and\ \citenamefont
  {Schaefer}}]{Ruffert:1995fs}%
  \BibitemOpen
  \bibfield  {author} {\bibinfo {author} {\bibfnamefont {M.~H.}\ \bibnamefont
  {Ruffert}}, \bibinfo {author} {\bibfnamefont {H.~T.}\ \bibnamefont {Janka}},\
  and\ \bibinfo {author} {\bibfnamefont {G.}~\bibnamefont {Schaefer}},\ }\href
  {https://articles.adsabs.harvard.edu/pdf/1996A%26A...311..532R} {\bibfield
  {journal} {\bibinfo  {journal} {Astron. Astrophys.}\ }\textbf {\bibinfo
  {volume} {311}},\ \bibinfo {pages} {532} (\bibinfo {year} {1996})},\ \Eprint
  {https://arxiv.org/abs/astro-ph/9509006} {arXiv:astro-ph/9509006}
  \BibitemShut {NoStop}%
\bibitem [{\citenamefont {Perego}\ \emph {et~al.}(2019)\citenamefont {Perego},
  \citenamefont {Bernuzzi},\ and\ \citenamefont {Radice}}]{Perego:2019adq}%
  \BibitemOpen
  \bibfield  {author} {\bibinfo {author} {\bibfnamefont {A.}~\bibnamefont
  {Perego}}, \bibinfo {author} {\bibfnamefont {S.}~\bibnamefont {Bernuzzi}},\
  and\ \bibinfo {author} {\bibfnamefont {D.}~\bibnamefont {Radice}},\ }\href
  {https://doi.org/10.1140/epja/i2019-12810-7} {\bibfield  {journal} {\bibinfo
  {journal} {Eur. Phys. J. A}\ }\textbf {\bibinfo {volume} {55}},\ \bibinfo
  {pages} {124} (\bibinfo {year} {2019})},\ \Eprint
  {https://arxiv.org/abs/1903.07898} {arXiv:1903.07898 [gr-qc]} \BibitemShut
  {NoStop}%
\bibitem [{\citenamefont {Berger}\ and\ \citenamefont
  {Oliger}(1984)}]{Berger:1984zza}%
  \BibitemOpen
  \bibfield  {author} {\bibinfo {author} {\bibfnamefont {M.~J.}\ \bibnamefont
  {Berger}}\ and\ \bibinfo {author} {\bibfnamefont {J.}~\bibnamefont
  {Oliger}},\ }\href {https://doi.org/10.1016/0021-9991(84)90073-1} {\bibfield
  {journal} {\bibinfo  {journal} {Journal of Computational Physics}\ }\textbf
  {\bibinfo {volume} {53}},\ \bibinfo {pages} {484} (\bibinfo {year}
  {1984})}\BibitemShut {NoStop}%
\bibitem [{\citenamefont {{Berger}}\ and\ \citenamefont
  {{Colella}}(1989)}]{Berger:1989a}%
  \BibitemOpen
  \bibfield  {author} {\bibinfo {author} {\bibfnamefont {M.~J.}\ \bibnamefont
  {{Berger}}}\ and\ \bibinfo {author} {\bibfnamefont {P.}~\bibnamefont
  {{Colella}}},\ }\href {https://doi.org/10.1016/0021-9991(89)90035-1}
  {\bibfield  {journal} {\bibinfo  {journal} {Journal of Computational
  Physics}\ }\textbf {\bibinfo {volume} {82}},\ \bibinfo {pages} {64} (\bibinfo
  {year} {1989})}\BibitemShut {NoStop}%
\bibitem [{\citenamefont {Borges}\ \emph {et~al.}(2008)\citenamefont {Borges},
  \citenamefont {Carmona}, \citenamefont {Costa},\ and\ \citenamefont
  {Don}}]{Borges:2008}%
  \BibitemOpen
  \bibfield  {author} {\bibinfo {author} {\bibfnamefont {R.}~\bibnamefont
  {Borges}}, \bibinfo {author} {\bibfnamefont {M.}~\bibnamefont {Carmona}},
  \bibinfo {author} {\bibfnamefont {B.}~\bibnamefont {Costa}},\ and\ \bibinfo
  {author} {\bibfnamefont {W.~S.}\ \bibnamefont {Don}},\ }\href
  {https://doi.org/10.1016/j.jcp.2007.11.038} {\bibfield  {journal} {\bibinfo
  {journal} {Journal of Computational Physics}\ }\textbf {\bibinfo {volume}
  {227}},\ \bibinfo {pages} {3191} (\bibinfo {year} {2008})}\BibitemShut
  {NoStop}%
\bibitem [{\citenamefont {Harten}\ \emph {et~al.}(1983)\citenamefont {Harten},
  \citenamefont {Lax},\ and\ \citenamefont {Leer}}]{Harten:1983}%
  \BibitemOpen
  \bibfield  {author} {\bibinfo {author} {\bibfnamefont {A.}~\bibnamefont
  {Harten}}, \bibinfo {author} {\bibfnamefont {P.~D.}\ \bibnamefont {Lax}},\
  and\ \bibinfo {author} {\bibfnamefont {B.~v.}\ \bibnamefont {Leer}},\ }\href
  {https://doi.org/10.1137/1025002} {\bibfield  {journal} {\bibinfo  {journal}
  {SIAM Review}\ }\textbf {\bibinfo {volume} {25}},\ \bibinfo {pages} {35}
  (\bibinfo {year} {1983})}\BibitemShut {NoStop}%
\bibitem [{\citenamefont {{Liu}}\ and\ \citenamefont
  {{Osher}}(1998)}]{Liu:1998}%
  \BibitemOpen
  \bibfield  {author} {\bibinfo {author} {\bibfnamefont {X.-D.}\ \bibnamefont
  {{Liu}}}\ and\ \bibinfo {author} {\bibfnamefont {S.}~\bibnamefont
  {{Osher}}},\ }\href {https://doi.org/10.1006/jcph.1998.5937} {\bibfield
  {journal} {\bibinfo  {journal} {Journal of Computational Physics}\ }\textbf
  {\bibinfo {volume} {142}},\ \bibinfo {pages} {304} (\bibinfo {year}
  {1998})}\BibitemShut {NoStop}%
\bibitem [{\citenamefont {Zanna}\ and\ \citenamefont
  {Bucciantini}(2002)}]{Zanna:2002qr}%
  \BibitemOpen
  \bibfield  {author} {\bibinfo {author} {\bibfnamefont {L.~D.}\ \bibnamefont
  {Zanna}}\ and\ \bibinfo {author} {\bibfnamefont {N.}~\bibnamefont
  {Bucciantini}},\ }\href {https://doi.org/10.1051/0004-6361:20020776}
  {\bibfield  {journal} {\bibinfo  {journal} {Astron. Astrophys.}\ }\textbf
  {\bibinfo {volume} {390}},\ \bibinfo {pages} {1177} (\bibinfo {year}
  {2002})},\ \Eprint {https://arxiv.org/abs/astro-ph/0205290}
  {arXiv:astro-ph/0205290} \BibitemShut {NoStop}%
\bibitem [{\citenamefont {Reichert}\ \emph {et~al.}(2023)\citenamefont
  {Reichert} \emph {et~al.}}]{Reichert:2023xqy}%
  \BibitemOpen
  \bibfield  {author} {\bibinfo {author} {\bibfnamefont {M.}~\bibnamefont
  {Reichert}} \emph {et~al.},\ }\href
  {https://doi.org/10.3847/1538-4365/acf033} {\bibfield  {journal} {\bibinfo
  {journal} {Astrophys. J. Suppl.}\ }\textbf {\bibinfo {volume} {268}},\
  \bibinfo {pages} {66} (\bibinfo {year} {2023})},\ \Eprint
  {https://arxiv.org/abs/2305.07048} {arXiv:2305.07048 [astro-ph.IM]}
  \BibitemShut {NoStop}%
\bibitem [{\citenamefont {Bulla}(2019)}]{Bulla:2019muo}%
  \BibitemOpen
  \bibfield  {author} {\bibinfo {author} {\bibfnamefont {M.}~\bibnamefont
  {Bulla}},\ }\href {https://doi.org/10.1093/mnras/stz2495} {\bibfield
  {journal} {\bibinfo  {journal} {Mon. Not. Roy. Astron. Soc.}\ }\textbf
  {\bibinfo {volume} {489}},\ \bibinfo {pages} {5037} (\bibinfo {year}
  {2019})},\ \Eprint {https://arxiv.org/abs/1906.04205} {arXiv:1906.04205
  [astro-ph.HE]} \BibitemShut {NoStop}%
\bibitem [{\citenamefont {Bulla}(2023)}]{Bulla:2022mwo}%
  \BibitemOpen
  \bibfield  {author} {\bibinfo {author} {\bibfnamefont {M.}~\bibnamefont
  {Bulla}},\ }\href {https://doi.org/10.1093/mnras/stad232} {\bibfield
  {journal} {\bibinfo  {journal} {Mon. Not. Roy. Astron. Soc.}\ }\textbf
  {\bibinfo {volume} {520}},\ \bibinfo {pages} {2558} (\bibinfo {year}
  {2023})},\ \Eprint {https://arxiv.org/abs/2211.14348} {arXiv:2211.14348
  [astro-ph.HE]} \BibitemShut {NoStop}%
\bibitem [{\citenamefont {Nedora}\ \emph
  {et~al.}(2023{\natexlab{a}})\citenamefont {Nedora}, \citenamefont {Dietrich},
  \citenamefont {Shibata}, \citenamefont {Pohl},\ and\ \citenamefont
  {Menegazzi}}]{Nedora:2022kjv}%
  \BibitemOpen
  \bibfield  {author} {\bibinfo {author} {\bibfnamefont {V.}~\bibnamefont
  {Nedora}}, \bibinfo {author} {\bibfnamefont {T.}~\bibnamefont {Dietrich}},
  \bibinfo {author} {\bibfnamefont {M.}~\bibnamefont {Shibata}}, \bibinfo
  {author} {\bibfnamefont {M.}~\bibnamefont {Pohl}},\ and\ \bibinfo {author}
  {\bibfnamefont {L.~C.}\ \bibnamefont {Menegazzi}},\ }\href
  {https://doi.org/10.1093/mnras/stad175} {\bibfield  {journal} {\bibinfo
  {journal} {Mon. Not. Roy. Astron. Soc.}\ }\textbf {\bibinfo {volume} {520}},\
  \bibinfo {pages} {2727} (\bibinfo {year} {2023}{\natexlab{a}})},\ \Eprint
  {https://arxiv.org/abs/2208.01558} {arXiv:2208.01558 [astro-ph.HE]}
  \BibitemShut {NoStop}%
\bibitem [{\citenamefont {Nedora}\ \emph
  {et~al.}(2023{\natexlab{b}})\citenamefont {Nedora}, \citenamefont
  {Dietrich},\ and\ \citenamefont {Shibata}}]{Nedora:2023hiz}%
  \BibitemOpen
  \bibfield  {author} {\bibinfo {author} {\bibfnamefont {V.}~\bibnamefont
  {Nedora}}, \bibinfo {author} {\bibfnamefont {T.}~\bibnamefont {Dietrich}},\
  and\ \bibinfo {author} {\bibfnamefont {M.}~\bibnamefont {Shibata}},\ }\href
  {https://doi.org/10.1093/mnras/stad2128} {\bibfield  {journal} {\bibinfo
  {journal} {Mon. Not. Roy. Astron. Soc.}\ }\textbf {\bibinfo {volume} {524}},\
  \bibinfo {pages} {5514} (\bibinfo {year} {2023}{\natexlab{b}})},\ \Eprint
  {https://arxiv.org/abs/2302.12850} {arXiv:2302.12850 [astro-ph.HE]}
  \BibitemShut {NoStop}%
\bibitem [{\citenamefont {Nedora}\ \emph {et~al.}(2025)\citenamefont {Nedora},
  \citenamefont {Menegazzi}, \citenamefont {Peretti}, \citenamefont
  {Dietrich},\ and\ \citenamefont {Shibata}}]{Nedora:2024vrv}%
  \BibitemOpen
  \bibfield  {author} {\bibinfo {author} {\bibfnamefont {V.}~\bibnamefont
  {Nedora}}, \bibinfo {author} {\bibfnamefont {L.~C.}\ \bibnamefont
  {Menegazzi}}, \bibinfo {author} {\bibfnamefont {E.}~\bibnamefont {Peretti}},
  \bibinfo {author} {\bibfnamefont {T.}~\bibnamefont {Dietrich}},\ and\
  \bibinfo {author} {\bibfnamefont {M.}~\bibnamefont {Shibata}},\ }\href
  {https://doi.org/10.1093/mnras/staf302} {\bibfield  {journal} {\bibinfo
  {journal} {Mon. Not. Roy. Astron. Soc.}\ }\textbf {\bibinfo {volume} {538}},\
  \bibinfo {pages} {2089} (\bibinfo {year} {2025})},\ \Eprint
  {https://arxiv.org/abs/2409.16852} {2409.16852 [astro-ph.HE]} \BibitemShut
  {NoStop}%
\bibitem [{\citenamefont {M{\"o}ller}\ \emph {et~al.}(2016)\citenamefont
  {M{\"o}ller}, \citenamefont {Sierk}, \citenamefont {Ichikawa},\ and\
  \citenamefont {Sagawa}}]{Moller:2015fba}%
  \BibitemOpen
  \bibfield  {author} {\bibinfo {author} {\bibfnamefont {P.}~\bibnamefont
  {M{\"o}ller}}, \bibinfo {author} {\bibfnamefont {A.~J.}\ \bibnamefont
  {Sierk}}, \bibinfo {author} {\bibfnamefont {T.}~\bibnamefont {Ichikawa}},\
  and\ \bibinfo {author} {\bibfnamefont {H.}~\bibnamefont {Sagawa}},\ }\href
  {https://doi.org/10.1016/j.adt.2015.10.002} {\bibfield  {journal} {\bibinfo
  {journal} {Atom. Data Nucl. Data Tabl.}\ }\textbf {\bibinfo {volume}
  {109-110}},\ \bibinfo {pages} {1} (\bibinfo {year} {2016})},\ \Eprint
  {https://arxiv.org/abs/1508.06294} {arXiv:1508.06294 [nucl-th]} \BibitemShut
  {NoStop}%
\bibitem [{\citenamefont {{Cyburt}}\ \emph {et~al.}(2010)\citenamefont
  {{Cyburt}} \emph {et~al.}}]{Cyburt:2010}%
  \BibitemOpen
  \bibfield  {author} {\bibinfo {author} {\bibfnamefont {R.~H.}\ \bibnamefont
  {{Cyburt}}} \emph {et~al.},\ }\href
  {https://doi.org/10.1088/0067-0049/189/1/240} {\bibfield  {journal} {\bibinfo
   {journal} {Astrophys. J. Suppl.}\ }\textbf {\bibinfo {volume} {189}},\
  \bibinfo {pages} {240} (\bibinfo {year} {2010})}\BibitemShut {NoStop}%
\bibitem [{\citenamefont {Panov}\ \emph {et~al.}(2001)\citenamefont {Panov},
  \citenamefont {Freiburghaus},\ and\ \citenamefont
  {Thielemann}}]{Panov:2001rus}%
  \BibitemOpen
  \bibfield  {author} {\bibinfo {author} {\bibfnamefont {I.~V.}\ \bibnamefont
  {Panov}}, \bibinfo {author} {\bibfnamefont {C.}~\bibnamefont
  {Freiburghaus}},\ and\ \bibinfo {author} {\bibfnamefont {F.~K.}\ \bibnamefont
  {Thielemann}},\ }\href {https://doi.org/10.1016/S0375-9474(01)00797-7}
  {\bibfield  {journal} {\bibinfo  {journal} {Nucl. Phys. A}\ }\textbf
  {\bibinfo {volume} {688}},\ \bibinfo {pages} {587} (\bibinfo {year}
  {2001})}\BibitemShut {NoStop}%
\bibitem [{\citenamefont {Kodama}\ and\ \citenamefont
  {Takahashi}(1975)}]{Kodama:1975aff}%
  \BibitemOpen
  \bibfield  {author} {\bibinfo {author} {\bibfnamefont {T.}~\bibnamefont
  {Kodama}}\ and\ \bibinfo {author} {\bibfnamefont {K.}~\bibnamefont
  {Takahashi}},\ }\href {https://doi.org/10.1016/0375-9474(75)90381-4}
  {\bibfield  {journal} {\bibinfo  {journal} {Nucl. Phys. A}\ }\textbf
  {\bibinfo {volume} {239}},\ \bibinfo {pages} {489} (\bibinfo {year}
  {1975})}\BibitemShut {NoStop}%
\bibitem [{\citenamefont {Mumpower}\ \emph {et~al.}(2020)\citenamefont
  {Mumpower}, \citenamefont {Jaffke}, \citenamefont {Verriere},\ and\
  \citenamefont {Randrup}}]{Mumpower:2019uid}%
  \BibitemOpen
  \bibfield  {author} {\bibinfo {author} {\bibfnamefont {M.~R.}\ \bibnamefont
  {Mumpower}}, \bibinfo {author} {\bibfnamefont {P.}~\bibnamefont {Jaffke}},
  \bibinfo {author} {\bibfnamefont {M.}~\bibnamefont {Verriere}},\ and\
  \bibinfo {author} {\bibfnamefont {J.}~\bibnamefont {Randrup}},\ }\href
  {https://doi.org/10.1103/PhysRevC.101.054607} {\bibfield  {journal} {\bibinfo
   {journal} {Phys. Rev. C}\ }\textbf {\bibinfo {volume} {101}},\ \bibinfo
  {pages} {054607} (\bibinfo {year} {2020})},\ \Eprint
  {https://arxiv.org/abs/1911.06344} {arXiv:1911.06344 [nucl-th]} \BibitemShut
  {NoStop}%
\bibitem [{\citenamefont {Neuweiler}\ \emph {et~al.}(2023)\citenamefont
  {Neuweiler}, \citenamefont {Dietrich}, \citenamefont {Bulla}, \citenamefont
  {Chaurasia}, \citenamefont {Rosswog},\ and\ \citenamefont
  {Ujevic}}]{Neuweiler:2022eum}%
  \BibitemOpen
  \bibfield  {author} {\bibinfo {author} {\bibfnamefont {A.}~\bibnamefont
  {Neuweiler}}, \bibinfo {author} {\bibfnamefont {T.}~\bibnamefont {Dietrich}},
  \bibinfo {author} {\bibfnamefont {M.}~\bibnamefont {Bulla}}, \bibinfo
  {author} {\bibfnamefont {S.~V.}\ \bibnamefont {Chaurasia}}, \bibinfo {author}
  {\bibfnamefont {S.}~\bibnamefont {Rosswog}},\ and\ \bibinfo {author}
  {\bibfnamefont {M.}~\bibnamefont {Ujevic}},\ }\href
  {https://doi.org/10.1103/PhysRevD.107.023016} {\bibfield  {journal} {\bibinfo
   {journal} {Phys. Rev. D}\ }\textbf {\bibinfo {volume} {107}},\ \bibinfo
  {pages} {023016} (\bibinfo {year} {2023})},\ \Eprint
  {https://arxiv.org/abs/2208.13460} {arXiv:2208.13460 [astro-ph.HE]}
  \BibitemShut {NoStop}%
\bibitem [{\citenamefont {Kawaguchi}\ \emph {et~al.}(2021)\citenamefont
  {Kawaguchi}, \citenamefont {Fujibayashi}, \citenamefont {Shibata},
  \citenamefont {Tanaka},\ and\ \citenamefont {Wanajo}}]{Kawaguchi:2020vbf}%
  \BibitemOpen
  \bibfield  {author} {\bibinfo {author} {\bibfnamefont {K.}~\bibnamefont
  {Kawaguchi}}, \bibinfo {author} {\bibfnamefont {S.}~\bibnamefont
  {Fujibayashi}}, \bibinfo {author} {\bibfnamefont {M.}~\bibnamefont
  {Shibata}}, \bibinfo {author} {\bibfnamefont {M.}~\bibnamefont {Tanaka}},\
  and\ \bibinfo {author} {\bibfnamefont {S.}~\bibnamefont {Wanajo}},\ }\href
  {https://doi.org/10.3847/1538-4357/abf3bc} {\bibfield  {journal} {\bibinfo
  {journal} {Astrophys. J.}\ }\textbf {\bibinfo {volume} {913}},\ \bibinfo
  {pages} {100} (\bibinfo {year} {2021})},\ \Eprint
  {https://arxiv.org/abs/2012.14711} {arXiv:2012.14711 [astro-ph.HE]}
  \BibitemShut {NoStop}%
\bibitem [{\citenamefont {Rosswog}\ and\ \citenamefont
  {Korobkin}(2024)}]{Rosswog:2022tus}%
  \BibitemOpen
  \bibfield  {author} {\bibinfo {author} {\bibfnamefont {S.}~\bibnamefont
  {Rosswog}}\ and\ \bibinfo {author} {\bibfnamefont {O.}~\bibnamefont
  {Korobkin}},\ }\href {https://doi.org/10.1002/andp.202200306} {\bibfield
  {journal} {\bibinfo  {journal} {Annalen Phys.}\ }\textbf {\bibinfo {volume}
  {536}},\ \bibinfo {pages} {2200306} (\bibinfo {year} {2024})},\ \Eprint
  {https://arxiv.org/abs/2208.14026} {arXiv:2208.14026 [astro-ph.HE]}
  \BibitemShut {NoStop}%
\bibitem [{\citenamefont {Barnes}\ \emph {et~al.}(2016)\citenamefont {Barnes},
  \citenamefont {Kasen}, \citenamefont {Wu},\ and\ \citenamefont
  {Mart{\'\i}nez-Pinedo}}]{Barnes:2016umi}%
  \BibitemOpen
  \bibfield  {author} {\bibinfo {author} {\bibfnamefont {J.}~\bibnamefont
  {Barnes}}, \bibinfo {author} {\bibfnamefont {D.}~\bibnamefont {Kasen}},
  \bibinfo {author} {\bibfnamefont {M.-R.}\ \bibnamefont {Wu}},\ and\ \bibinfo
  {author} {\bibfnamefont {G.}~\bibnamefont {Mart{\'\i}nez-Pinedo}},\ }\href
  {https://doi.org/10.3847/0004-637X/829/2/110} {\bibfield  {journal} {\bibinfo
   {journal} {Astrophys. J.}\ }\textbf {\bibinfo {volume} {829}},\ \bibinfo
  {pages} {110} (\bibinfo {year} {2016})},\ \Eprint
  {https://arxiv.org/abs/1605.07218} {arXiv:1605.07218 [astro-ph.HE]}
  \BibitemShut {NoStop}%
\bibitem [{\citenamefont {Wollaeger}\ \emph {et~al.}(2018)\citenamefont
  {Wollaeger}, \citenamefont {Korobkin}, \citenamefont {Fontes}, \citenamefont
  {Rosswog}, \citenamefont {Even}, \citenamefont {Fryer}, \citenamefont
  {Sollerman}, \citenamefont {Hungerford}, \citenamefont {van Rossum},\ and\
  \citenamefont {Wollaber}}]{Wollaeger:2017ahm}%
  \BibitemOpen
  \bibfield  {author} {\bibinfo {author} {\bibfnamefont {R.~T.}\ \bibnamefont
  {Wollaeger}}, \bibinfo {author} {\bibfnamefont {O.}~\bibnamefont {Korobkin}},
  \bibinfo {author} {\bibfnamefont {C.~J.}\ \bibnamefont {Fontes}}, \bibinfo
  {author} {\bibfnamefont {S.~K.}\ \bibnamefont {Rosswog}}, \bibinfo {author}
  {\bibfnamefont {W.~P.}\ \bibnamefont {Even}}, \bibinfo {author}
  {\bibfnamefont {C.~L.}\ \bibnamefont {Fryer}}, \bibinfo {author}
  {\bibfnamefont {J.}~\bibnamefont {Sollerman}}, \bibinfo {author}
  {\bibfnamefont {A.~L.}\ \bibnamefont {Hungerford}}, \bibinfo {author}
  {\bibfnamefont {D.~R.}\ \bibnamefont {van Rossum}},\ and\ \bibinfo {author}
  {\bibfnamefont {A.~B.}\ \bibnamefont {Wollaber}},\ }\href
  {https://doi.org/10.1093/mnras/sty1018} {\bibfield  {journal} {\bibinfo
  {journal} {Mon. Not. Roy. Astron. Soc.}\ }\textbf {\bibinfo {volume} {478}},\
  \bibinfo {pages} {3298} (\bibinfo {year} {2018})},\ \Eprint
  {https://arxiv.org/abs/1705.07084} {arXiv:1705.07084 [astro-ph.HE]}
  \BibitemShut {NoStop}%
\bibitem [{\citenamefont {Tanaka}\ \emph {et~al.}(2020)\citenamefont {Tanaka},
  \citenamefont {Kato}, \citenamefont {Gaigalas},\ and\ \citenamefont
  {Kawaguchi}}]{Tanaka:2019iqp}%
  \BibitemOpen
  \bibfield  {author} {\bibinfo {author} {\bibfnamefont {M.}~\bibnamefont
  {Tanaka}}, \bibinfo {author} {\bibfnamefont {D.}~\bibnamefont {Kato}},
  \bibinfo {author} {\bibfnamefont {G.}~\bibnamefont {Gaigalas}},\ and\
  \bibinfo {author} {\bibfnamefont {K.}~\bibnamefont {Kawaguchi}},\ }\href
  {https://doi.org/10.1093/mnras/staa1576} {\bibfield  {journal} {\bibinfo
  {journal} {Mon. Not. Roy. Astron. Soc.}\ }\textbf {\bibinfo {volume} {496}},\
  \bibinfo {pages} {1369} (\bibinfo {year} {2020})},\ \Eprint
  {https://arxiv.org/abs/1906.08914} {arXiv:1906.08914 [astro-ph.HE]}
  \BibitemShut {NoStop}%
\bibitem [{\citenamefont {Margalit}\ and\ \citenamefont
  {Quataert}(2021)}]{Margalit:2021kuf}%
  \BibitemOpen
  \bibfield  {author} {\bibinfo {author} {\bibfnamefont {B.}~\bibnamefont
  {Margalit}}\ and\ \bibinfo {author} {\bibfnamefont {E.}~\bibnamefont
  {Quataert}},\ }\href {https://doi.org/10.3847/2041-8213/ac3d97} {\bibfield
  {journal} {\bibinfo  {journal} {Astrophys. J. Lett.}\ }\textbf {\bibinfo
  {volume} {923}},\ \bibinfo {pages} {L14} (\bibinfo {year} {2021})},\ \Eprint
  {https://arxiv.org/abs/2111.00012} {arXiv:2111.00012 [astro-ph.HE]}
  \BibitemShut {NoStop}%
\bibitem [{\citenamefont {Margalit}\ and\ \citenamefont
  {Quataert}(2024)}]{Margalit:2024asc}%
  \BibitemOpen
  \bibfield  {author} {\bibinfo {author} {\bibfnamefont {B.}~\bibnamefont
  {Margalit}}\ and\ \bibinfo {author} {\bibfnamefont {E.}~\bibnamefont
  {Quataert}},\ }\href {https://doi.org/10.3847/1538-4357/ad8b47} {\bibfield
  {journal} {\bibinfo  {journal} {Astrophys. J.}\ }\textbf {\bibinfo {volume}
  {977}},\ \bibinfo {pages} {134} (\bibinfo {year} {2024})},\ \Eprint
  {https://arxiv.org/abs/2403.07048} {arXiv:2403.07048 [astro-ph.HE]}
  \BibitemShut {NoStop}%
\bibitem [{\citenamefont {Sironi}\ \emph {et~al.}(2013)\citenamefont {Sironi},
  \citenamefont {Spitkovsky},\ and\ \citenamefont {Arons}}]{Sironi:2013ri}%
  \BibitemOpen
  \bibfield  {author} {\bibinfo {author} {\bibfnamefont {L.}~\bibnamefont
  {Sironi}}, \bibinfo {author} {\bibfnamefont {A.}~\bibnamefont {Spitkovsky}},\
  and\ \bibinfo {author} {\bibfnamefont {J.}~\bibnamefont {Arons}},\ }\href
  {https://doi.org/10.1088/0004-637X/771/1/54} {\bibfield  {journal} {\bibinfo
  {journal} {Astrophys. J.}\ }\textbf {\bibinfo {volume} {771}},\ \bibinfo
  {pages} {54} (\bibinfo {year} {2013})},\ \Eprint
  {https://arxiv.org/abs/1301.5333} {arXiv:1301.5333 [astro-ph.HE]}
  \BibitemShut {NoStop}%
\bibitem [{\citenamefont {Crumley}\ \emph {et~al.}(2019)\citenamefont
  {Crumley}, \citenamefont {Caprioli}, \citenamefont {Markoff},\ and\
  \citenamefont {Spitkovsky}}]{Crumley:2018kvf}%
  \BibitemOpen
  \bibfield  {author} {\bibinfo {author} {\bibfnamefont {P.}~\bibnamefont
  {Crumley}}, \bibinfo {author} {\bibfnamefont {D.}~\bibnamefont {Caprioli}},
  \bibinfo {author} {\bibfnamefont {S.}~\bibnamefont {Markoff}},\ and\ \bibinfo
  {author} {\bibfnamefont {A.}~\bibnamefont {Spitkovsky}},\ }\href
  {https://doi.org/10.1093/mnras/stz232} {\bibfield  {journal} {\bibinfo
  {journal} {Mon. Not. Roy. Astron. Soc.}\ }\textbf {\bibinfo {volume} {485}},\
  \bibinfo {pages} {5105} (\bibinfo {year} {2019})},\ \Eprint
  {https://arxiv.org/abs/1809.10809} {arXiv:1809.10809 [astro-ph.HE]}
  \BibitemShut {NoStop}%
\bibitem [{\citenamefont {Xie}\ \emph {et~al.}(2018)\citenamefont {Xie},
  \citenamefont {Zrake},\ and\ \citenamefont {MacFadyen}}]{Xie:2018vya}%
  \BibitemOpen
  \bibfield  {author} {\bibinfo {author} {\bibfnamefont {X.}~\bibnamefont
  {Xie}}, \bibinfo {author} {\bibfnamefont {J.}~\bibnamefont {Zrake}},\ and\
  \bibinfo {author} {\bibfnamefont {A.}~\bibnamefont {MacFadyen}},\ }\href
  {https://doi.org/10.3847/1538-4357/aacf9c} {\bibfield  {journal} {\bibinfo
  {journal} {Astrophys. J.}\ }\textbf {\bibinfo {volume} {863}},\ \bibinfo
  {pages} {58} (\bibinfo {year} {2018})},\ \Eprint
  {https://arxiv.org/abs/1804.09345} {arXiv:1804.09345 [astro-ph.HE]}
  \BibitemShut {NoStop}%
\bibitem [{\citenamefont {Ryan}\ \emph {et~al.}(2020)\citenamefont {Ryan},
  \citenamefont {van Eerten}, \citenamefont {Piro},\ and\ \citenamefont
  {Troja}}]{Ryan:2019fhz}%
  \BibitemOpen
  \bibfield  {author} {\bibinfo {author} {\bibfnamefont {G.}~\bibnamefont
  {Ryan}}, \bibinfo {author} {\bibfnamefont {H.}~\bibnamefont {van Eerten}},
  \bibinfo {author} {\bibfnamefont {L.}~\bibnamefont {Piro}},\ and\ \bibinfo
  {author} {\bibfnamefont {E.}~\bibnamefont {Troja}},\ }\href
  {https://doi.org/10.3847/1538-4357/ab93cf} {\bibfield  {journal} {\bibinfo
  {journal} {Astrophys. J.}\ }\textbf {\bibinfo {volume} {896}},\ \bibinfo
  {pages} {166} (\bibinfo {year} {2020})},\ \Eprint
  {https://arxiv.org/abs/1909.11691} {arXiv:1909.11691 [astro-ph.HE]}
  \BibitemShut {NoStop}%
\bibitem [{\citenamefont {Gottlieb}\ \emph {et~al.}(2020)\citenamefont
  {Gottlieb}, \citenamefont {Nakar},\ and\ \citenamefont
  {Bromberg}}]{Gottlieb:2020raq}%
  \BibitemOpen
  \bibfield  {author} {\bibinfo {author} {\bibfnamefont {O.}~\bibnamefont
  {Gottlieb}}, \bibinfo {author} {\bibfnamefont {E.}~\bibnamefont {Nakar}},\
  and\ \bibinfo {author} {\bibfnamefont {O.}~\bibnamefont {Bromberg}},\ }\href
  {https://doi.org/10.1093/mnras/staa3501} {\bibfield  {journal} {\bibinfo
  {journal} {Mon. Not. Roy. Astron. Soc.}\ }\textbf {\bibinfo {volume} {500}},\
  \bibinfo {pages} {3511} (\bibinfo {year} {2020})},\ \Eprint
  {https://arxiv.org/abs/2006.02466} {arXiv:2006.02466 [astro-ph.HE]}
  \BibitemShut {NoStop}%
\bibitem [{\citenamefont {Aharonian}\ \emph {et~al.}(2010)\citenamefont
  {Aharonian}, \citenamefont {Kelner},\ and\ \citenamefont
  {Prosekin}}]{Aharonian:2010va}%
  \BibitemOpen
  \bibfield  {author} {\bibinfo {author} {\bibfnamefont {F.~A.}\ \bibnamefont
  {Aharonian}}, \bibinfo {author} {\bibfnamefont {S.~R.}\ \bibnamefont
  {Kelner}},\ and\ \bibinfo {author} {\bibfnamefont {A.~Y.}\ \bibnamefont
  {Prosekin}},\ }\href {https://doi.org/10.1103/PhysRevD.82.043002} {\bibfield
  {journal} {\bibinfo  {journal} {Phys. Rev. D}\ }\textbf {\bibinfo {volume}
  {82}},\ \bibinfo {pages} {043002} (\bibinfo {year} {2010})},\ \Eprint
  {https://arxiv.org/abs/1006.1045} {arXiv:1006.1045 [astro-ph.HE]}
  \BibitemShut {NoStop}%
\bibitem [{\citenamefont {Alford}\ \emph {et~al.}(2023)\citenamefont {Alford},
  \citenamefont {Brodie}, \citenamefont {Haber},\ and\ \citenamefont
  {Tews}}]{Alford:2023rgp}%
  \BibitemOpen
  \bibfield  {author} {\bibinfo {author} {\bibfnamefont {M.~G.}\ \bibnamefont
  {Alford}}, \bibinfo {author} {\bibfnamefont {L.}~\bibnamefont {Brodie}},
  \bibinfo {author} {\bibfnamefont {A.}~\bibnamefont {Haber}},\ and\ \bibinfo
  {author} {\bibfnamefont {I.}~\bibnamefont {Tews}},\ }\href
  {https://doi.org/10.1088/1402-4896/ad03c8} {\bibfield  {journal} {\bibinfo
  {journal} {Phys. Scripta}\ }\textbf {\bibinfo {volume} {98}},\ \bibinfo
  {pages} {125302} (\bibinfo {year} {2023})},\ \Eprint
  {https://arxiv.org/abs/2304.07836} {arXiv:2304.07836 [nucl-th]} \BibitemShut
  {NoStop}%
\bibitem [{\citenamefont {Tichy}(2009)}]{Tichy:2009yr}%
  \BibitemOpen
  \bibfield  {author} {\bibinfo {author} {\bibfnamefont {W.}~\bibnamefont
  {Tichy}},\ }\href {https://doi.org/10.1088/0264-9381/26/17/175018} {\bibfield
   {journal} {\bibinfo  {journal} {Class. Quant. Grav.}\ }\textbf {\bibinfo
  {volume} {26}},\ \bibinfo {pages} {175018} (\bibinfo {year} {2009})},\
  \Eprint {https://arxiv.org/abs/0908.0620} {arXiv:0908.0620 [gr-qc]}
  \BibitemShut {NoStop}%
\bibitem [{\citenamefont {Tichy}(2012)}]{Tichy:2012rp}%
  \BibitemOpen
  \bibfield  {author} {\bibinfo {author} {\bibfnamefont {W.}~\bibnamefont
  {Tichy}},\ }\href {https://doi.org/10.1103/PhysRevD.86.064024} {\bibfield
  {journal} {\bibinfo  {journal} {Phys. Rev. D}\ }\textbf {\bibinfo {volume}
  {86}},\ \bibinfo {pages} {064024} (\bibinfo {year} {2012})},\ \Eprint
  {https://arxiv.org/abs/1209.5336} {arXiv:1209.5336 [gr-qc]} \BibitemShut
  {NoStop}%
\bibitem [{\citenamefont {Dietrich}\ \emph
  {et~al.}(2015{\natexlab{b}})\citenamefont {Dietrich}, \citenamefont
  {Moldenhauer}, \citenamefont {Johnson-McDaniel}, \citenamefont {Bernuzzi},
  \citenamefont {Markakis}, \citenamefont {Br\"ugmann},\ and\ \citenamefont
  {Tichy}}]{Dietrich:2015pxa}%
  \BibitemOpen
  \bibfield  {author} {\bibinfo {author} {\bibfnamefont {T.}~\bibnamefont
  {Dietrich}}, \bibinfo {author} {\bibfnamefont {N.}~\bibnamefont
  {Moldenhauer}}, \bibinfo {author} {\bibfnamefont {N.~K.}\ \bibnamefont
  {Johnson-McDaniel}}, \bibinfo {author} {\bibfnamefont {S.}~\bibnamefont
  {Bernuzzi}}, \bibinfo {author} {\bibfnamefont {C.~M.}\ \bibnamefont
  {Markakis}}, \bibinfo {author} {\bibfnamefont {B.}~\bibnamefont
  {Br\"ugmann}},\ and\ \bibinfo {author} {\bibfnamefont {W.}~\bibnamefont
  {Tichy}},\ }\href {https://doi.org/10.1103/PhysRevD.92.124007} {\bibfield
  {journal} {\bibinfo  {journal} {Phys. Rev. D}\ }\textbf {\bibinfo {volume}
  {92}},\ \bibinfo {pages} {124007} (\bibinfo {year} {2015}{\natexlab{b}})},\
  \Eprint {https://arxiv.org/abs/1507.07100} {arXiv:1507.07100 [gr-qc]}
  \BibitemShut {NoStop}%
\bibitem [{\citenamefont {Tichy}\ \emph {et~al.}(2019)\citenamefont {Tichy},
  \citenamefont {Rashti}, \citenamefont {Dietrich}, \citenamefont {Dudi},\ and\
  \citenamefont {Br\"ugmann}}]{Tichy:2019ouu}%
  \BibitemOpen
  \bibfield  {author} {\bibinfo {author} {\bibfnamefont {W.}~\bibnamefont
  {Tichy}}, \bibinfo {author} {\bibfnamefont {A.}~\bibnamefont {Rashti}},
  \bibinfo {author} {\bibfnamefont {T.}~\bibnamefont {Dietrich}}, \bibinfo
  {author} {\bibfnamefont {R.}~\bibnamefont {Dudi}},\ and\ \bibinfo {author}
  {\bibfnamefont {B.}~\bibnamefont {Br\"ugmann}},\ }\href
  {https://doi.org/10.1103/PhysRevD.100.124046} {\bibfield  {journal} {\bibinfo
   {journal} {Phys. Rev. D}\ }\textbf {\bibinfo {volume} {100}},\ \bibinfo
  {pages} {124046} (\bibinfo {year} {2019})},\ \Eprint
  {https://arxiv.org/abs/1910.09690} {arXiv:1910.09690 [gr-qc]} \BibitemShut
  {NoStop}%
\bibitem [{\citenamefont {York}(1999)}]{York:1998hy}%
  \BibitemOpen
  \bibfield  {author} {\bibinfo {author} {\bibfnamefont {J.~W.}\ \bibnamefont
  {York}, \bibfnamefont {Jr.}},\ }\href
  {https://doi.org/10.1103/PhysRevLett.82.1350} {\bibfield  {journal} {\bibinfo
   {journal} {Phys. Rev. Lett.}\ }\textbf {\bibinfo {volume} {82}},\ \bibinfo
  {pages} {1350} (\bibinfo {year} {1999})},\ \Eprint
  {https://arxiv.org/abs/gr-qc/9810051} {arXiv:gr-qc/9810051} \BibitemShut
  {NoStop}%
\bibitem [{\citenamefont {Pfeiffer}\ and\ \citenamefont
  {York}(2003)}]{Pfeiffer:2002iy}%
  \BibitemOpen
  \bibfield  {author} {\bibinfo {author} {\bibfnamefont {H.~P.}\ \bibnamefont
  {Pfeiffer}}\ and\ \bibinfo {author} {\bibfnamefont {J.~W.}\ \bibnamefont
  {York}, \bibfnamefont {Jr.}},\ }\href
  {https://doi.org/10.1103/PhysRevD.67.044022} {\bibfield  {journal} {\bibinfo
  {journal} {Phys. Rev. D}\ }\textbf {\bibinfo {volume} {67}},\ \bibinfo
  {pages} {044022} (\bibinfo {year} {2003})},\ \Eprint
  {https://arxiv.org/abs/gr-qc/0207095} {arXiv:gr-qc/0207095} \BibitemShut
  {NoStop}%
\bibitem [{\citenamefont {Lorimer}(2008)}]{Lorimer:2008se}%
  \BibitemOpen
  \bibfield  {author} {\bibinfo {author} {\bibfnamefont {D.~R.}\ \bibnamefont
  {Lorimer}},\ }\href {https://doi.org/10.12942/lrr-2008-8} {\bibfield
  {journal} {\bibinfo  {journal} {Living Rev. Rel.}\ }\textbf {\bibinfo
  {volume} {11}},\ \bibinfo {pages} {8} (\bibinfo {year} {2008})},\ \Eprint
  {https://arxiv.org/abs/0811.0762} {arXiv:0811.0762 [astro-ph]} \BibitemShut
  {NoStop}%
\bibitem [{\citenamefont {Tauris}\ \emph {et~al.}(2017)\citenamefont {Tauris}
  \emph {et~al.}}]{Tauris:2017omb}%
  \BibitemOpen
  \bibfield  {author} {\bibinfo {author} {\bibfnamefont {T.~M.}\ \bibnamefont
  {Tauris}} \emph {et~al.},\ }\href {https://doi.org/10.3847/1538-4357/aa7e89}
  {\bibfield  {journal} {\bibinfo  {journal} {Astrophys. J.}\ }\textbf
  {\bibinfo {volume} {846}},\ \bibinfo {pages} {170} (\bibinfo {year}
  {2017})},\ \Eprint {https://arxiv.org/abs/1706.09438} {arXiv:1706.09438
  [astro-ph.HE]} \BibitemShut {NoStop}%
\bibitem [{\citenamefont {Kiuchi}\ \emph {et~al.}(2023)\citenamefont {Kiuchi},
  \citenamefont {Fujibayashi}, \citenamefont {Hayashi}, \citenamefont
  {Kyutoku}, \citenamefont {Sekiguchi},\ and\ \citenamefont
  {Shibata}}]{Kiuchi:2022ninc}%
  \BibitemOpen
  \bibfield  {author} {\bibinfo {author} {\bibfnamefont {K.}~\bibnamefont
  {Kiuchi}}, \bibinfo {author} {\bibfnamefont {S.}~\bibnamefont {Fujibayashi}},
  \bibinfo {author} {\bibfnamefont {K.}~\bibnamefont {Hayashi}}, \bibinfo
  {author} {\bibfnamefont {K.}~\bibnamefont {Kyutoku}}, \bibinfo {author}
  {\bibfnamefont {Y.}~\bibnamefont {Sekiguchi}},\ and\ \bibinfo {author}
  {\bibfnamefont {M.}~\bibnamefont {Shibata}},\ }\href
  {https://doi.org/10.1103/PhysRevLett.131.011401} {\bibfield  {journal}
  {\bibinfo  {journal} {Phys. Rev. Lett.}\ }\textbf {\bibinfo {volume} {131}},\
  \bibinfo {pages} {011401} (\bibinfo {year} {2023})},\ \Eprint
  {https://arxiv.org/abs/2211.07637} {arXiv:2211.07637 [astro-ph.HE]}
  \BibitemShut {NoStop}%
\bibitem [{\citenamefont {Most}\ and\ \citenamefont
  {Quataert}(2023)}]{Most:2023sft}%
  \BibitemOpen
  \bibfield  {author} {\bibinfo {author} {\bibfnamefont {E.~R.}\ \bibnamefont
  {Most}}\ and\ \bibinfo {author} {\bibfnamefont {E.}~\bibnamefont
  {Quataert}},\ }\href {https://doi.org/10.3847/2041-8213/acca84} {\bibfield
  {journal} {\bibinfo  {journal} {Astrophys. J. Lett.}\ }\textbf {\bibinfo
  {volume} {947}},\ \bibinfo {pages} {L15} (\bibinfo {year} {2023})},\ \Eprint
  {https://arxiv.org/abs/2303.08062} {arXiv:2303.08062 [astro-ph.HE]}
  \BibitemShut {NoStop}%
\bibitem [{\citenamefont {Combi}\ and\ \citenamefont
  {Siegel}(2023{\natexlab{b}})}]{Combi:2022nhg}%
  \BibitemOpen
  \bibfield  {author} {\bibinfo {author} {\bibfnamefont {L.}~\bibnamefont
  {Combi}}\ and\ \bibinfo {author} {\bibfnamefont {D.~M.}\ \bibnamefont
  {Siegel}},\ }\href {https://doi.org/10.3847/1538-4357/acac29} {\bibfield
  {journal} {\bibinfo  {journal} {Astrophys. J.}\ }\textbf {\bibinfo {volume}
  {944}},\ \bibinfo {pages} {28} (\bibinfo {year} {2023}{\natexlab{b}})},\
  \Eprint {https://arxiv.org/abs/2206.03618} {arXiv:2206.03618 [astro-ph.HE]}
  \BibitemShut {NoStop}%
\bibitem [{\citenamefont {Ruiz}\ \emph {et~al.}(2020)\citenamefont {Ruiz},
  \citenamefont {Tsokaros},\ and\ \citenamefont {Shapiro}}]{Ruiz:2020via}%
  \BibitemOpen
  \bibfield  {author} {\bibinfo {author} {\bibfnamefont {M.}~\bibnamefont
  {Ruiz}}, \bibinfo {author} {\bibfnamefont {A.}~\bibnamefont {Tsokaros}},\
  and\ \bibinfo {author} {\bibfnamefont {S.~L.}\ \bibnamefont {Shapiro}},\
  }\href {https://doi.org/10.1103/PhysRevD.101.064042} {\bibfield  {journal}
  {\bibinfo  {journal} {Phys. Rev. D}\ }\textbf {\bibinfo {volume} {101}},\
  \bibinfo {pages} {064042} (\bibinfo {year} {2020})},\ \Eprint
  {https://arxiv.org/abs/2001.09153} {arXiv:2001.09153 [astro-ph.HE]}
  \BibitemShut {NoStop}%
\bibitem [{\citenamefont {Alford}\ \emph {et~al.}(2022)\citenamefont {Alford},
  \citenamefont {Brodie}, \citenamefont {Haber},\ and\ \citenamefont
  {Tews}}]{Alford:2022bpp}%
  \BibitemOpen
  \bibfield  {author} {\bibinfo {author} {\bibfnamefont {M.~G.}\ \bibnamefont
  {Alford}}, \bibinfo {author} {\bibfnamefont {L.}~\bibnamefont {Brodie}},
  \bibinfo {author} {\bibfnamefont {A.}~\bibnamefont {Haber}},\ and\ \bibinfo
  {author} {\bibfnamefont {I.}~\bibnamefont {Tews}},\ }\href
  {https://doi.org/10.1103/PhysRevC.106.055804} {\bibfield  {journal} {\bibinfo
   {journal} {Phys. Rev. C}\ }\textbf {\bibinfo {volume} {106}},\ \bibinfo
  {pages} {055804} (\bibinfo {year} {2022})},\ \Eprint
  {https://arxiv.org/abs/2205.10283} {arXiv:2205.10283 [nucl-th]} \BibitemShut
  {NoStop}%
\bibitem [{\citenamefont {Chatterjee}\ \emph {et~al.}(2025)\citenamefont
  {Chatterjee} \emph {et~al.}}]{qmc_rmf3}%
  \BibitemOpen
  \bibfield  {author} {\bibinfo {author} {\bibfnamefont {D.}~\bibnamefont
  {Chatterjee}} \emph {et~al.},\ }\href
  {https://doi.org/10.5281/zenodo.14809193} {\bibinfo {title} {Data table for
  eos abht(qmc-rmf3) nparam=3}} (\bibinfo {year} {2025}),\ \Eprint
  {https://arxiv.org/abs/https://doi.org/10.5281/zenodo.14809193}
  {https://doi.org/10.5281/zenodo.14809193} \BibitemShut {NoStop}%
\bibitem [{\citenamefont {Tews}\ \emph {et~al.}(2018)\citenamefont {Tews},
  \citenamefont {Carlson}, \citenamefont {Gandolfi},\ and\ \citenamefont
  {Reddy}}]{Tews:2018kmu}%
  \BibitemOpen
  \bibfield  {author} {\bibinfo {author} {\bibfnamefont {I.}~\bibnamefont
  {Tews}}, \bibinfo {author} {\bibfnamefont {J.}~\bibnamefont {Carlson}},
  \bibinfo {author} {\bibfnamefont {S.}~\bibnamefont {Gandolfi}},\ and\
  \bibinfo {author} {\bibfnamefont {S.}~\bibnamefont {Reddy}},\ }\href
  {https://doi.org/10.3847/1538-4357/aac267} {\bibfield  {journal} {\bibinfo
  {journal} {Astrophys. J.}\ }\textbf {\bibinfo {volume} {860}},\ \bibinfo
  {pages} {149} (\bibinfo {year} {2018})},\ \Eprint
  {https://arxiv.org/abs/1801.01923} {arXiv:1801.01923 [nucl-th]} \BibitemShut
  {NoStop}%
\bibitem [{\citenamefont {Hempel}\ and\ \citenamefont
  {Schaffner-Bielich}(2010)}]{Hempel:2009mc}%
  \BibitemOpen
  \bibfield  {author} {\bibinfo {author} {\bibfnamefont {M.}~\bibnamefont
  {Hempel}}\ and\ \bibinfo {author} {\bibfnamefont {J.}~\bibnamefont
  {Schaffner-Bielich}},\ }\href
  {https://doi.org/10.1016/j.nuclphysa.2010.02.010} {\bibfield  {journal}
  {\bibinfo  {journal} {Nucl. Phys. A}\ }\textbf {\bibinfo {volume} {837}},\
  \bibinfo {pages} {210} (\bibinfo {year} {2010})},\ \Eprint
  {https://arxiv.org/abs/0911.4073} {arXiv:0911.4073 [nucl-th]} \BibitemShut
  {NoStop}%
\bibitem [{\citenamefont {Fattoyev}\ \emph {et~al.}(2010)\citenamefont
  {Fattoyev}, \citenamefont {Horowitz}, \citenamefont {Piekarewicz},\ and\
  \citenamefont {Shen}}]{Fattoyev:2010mx}%
  \BibitemOpen
  \bibfield  {author} {\bibinfo {author} {\bibfnamefont {F.~J.}\ \bibnamefont
  {Fattoyev}}, \bibinfo {author} {\bibfnamefont {C.~J.}\ \bibnamefont
  {Horowitz}}, \bibinfo {author} {\bibfnamefont {J.}~\bibnamefont
  {Piekarewicz}},\ and\ \bibinfo {author} {\bibfnamefont {G.}~\bibnamefont
  {Shen}},\ }\href {https://doi.org/10.1103/PhysRevC.82.055803} {\bibfield
  {journal} {\bibinfo  {journal} {Phys. Rev. C}\ }\textbf {\bibinfo {volume}
  {82}},\ \bibinfo {pages} {055803} (\bibinfo {year} {2010})},\ \Eprint
  {https://arxiv.org/abs/1008.3030} {arXiv:1008.3030 [nucl-th]} \BibitemShut
  {NoStop}%
\bibitem [{\citenamefont {Drischler}\ \emph {et~al.}(2020)\citenamefont
  {Drischler}, \citenamefont {Furnstahl}, \citenamefont {Melendez},\ and\
  \citenamefont {Phillips}}]{Drischler:2020hwi}%
  \BibitemOpen
  \bibfield  {author} {\bibinfo {author} {\bibfnamefont {C.}~\bibnamefont
  {Drischler}}, \bibinfo {author} {\bibfnamefont {R.~J.}\ \bibnamefont
  {Furnstahl}}, \bibinfo {author} {\bibfnamefont {J.~A.}\ \bibnamefont
  {Melendez}},\ and\ \bibinfo {author} {\bibfnamefont {D.~R.}\ \bibnamefont
  {Phillips}},\ }\href {https://doi.org/10.1103/PhysRevLett.125.202702}
  {\bibfield  {journal} {\bibinfo  {journal} {Phys. Rev. Lett.}\ }\textbf
  {\bibinfo {volume} {125}},\ \bibinfo {pages} {202702} (\bibinfo {year}
  {2020})},\ \Eprint {https://arxiv.org/abs/2004.07232} {arXiv:2004.07232
  [nucl-th]} \BibitemShut {NoStop}%
\bibitem [{\citenamefont {Shlomo}\ \emph {et~al.}(2006)\citenamefont {Shlomo},
  \citenamefont {Kolomietz},\ and\ \citenamefont {Col{\`o}}}]{Shlomo:2006ole}%
  \BibitemOpen
  \bibfield  {author} {\bibinfo {author} {\bibfnamefont {S.}~\bibnamefont
  {Shlomo}}, \bibinfo {author} {\bibfnamefont {V.~M.}\ \bibnamefont
  {Kolomietz}},\ and\ \bibinfo {author} {\bibfnamefont {G.}~\bibnamefont
  {Col{\`o}}},\ }\href {https://doi.org/10.1140/epja/i2006-10100-3} {\bibfield
  {journal} {\bibinfo  {journal} {Eur. Phys. J. A}\ }\textbf {\bibinfo {volume}
  {30}},\ \bibinfo {pages} {23} (\bibinfo {year} {2006})}\BibitemShut {NoStop}%
\bibitem [{\citenamefont {Horowitz}\ \emph {et~al.}(2020)\citenamefont
  {Horowitz}, \citenamefont {Piekarewicz},\ and\ \citenamefont
  {Reed}}]{Horowitz:2020evx}%
  \BibitemOpen
  \bibfield  {author} {\bibinfo {author} {\bibfnamefont {C.~J.}\ \bibnamefont
  {Horowitz}}, \bibinfo {author} {\bibfnamefont {J.}~\bibnamefont
  {Piekarewicz}},\ and\ \bibinfo {author} {\bibfnamefont {B.}~\bibnamefont
  {Reed}},\ }\href {https://doi.org/10.1103/PhysRevC.102.044321} {\bibfield
  {journal} {\bibinfo  {journal} {Phys. Rev. C}\ }\textbf {\bibinfo {volume}
  {102}},\ \bibinfo {pages} {044321} (\bibinfo {year} {2020})},\ \Eprint
  {https://arxiv.org/abs/2007.07117} {arXiv:2007.07117 [nucl-th]} \BibitemShut
  {NoStop}%
\bibitem [{\citenamefont {Keller}\ \emph {et~al.}(2021)\citenamefont {Keller},
  \citenamefont {Wellenhofer}, \citenamefont {Hebeler},\ and\ \citenamefont
  {Schwenk}}]{Keller:2020qhx}%
  \BibitemOpen
  \bibfield  {author} {\bibinfo {author} {\bibfnamefont {J.}~\bibnamefont
  {Keller}}, \bibinfo {author} {\bibfnamefont {C.}~\bibnamefont {Wellenhofer}},
  \bibinfo {author} {\bibfnamefont {K.}~\bibnamefont {Hebeler}},\ and\ \bibinfo
  {author} {\bibfnamefont {A.}~\bibnamefont {Schwenk}},\ }\href
  {https://doi.org/10.1103/PhysRevC.103.055806} {\bibfield  {journal} {\bibinfo
   {journal} {Phys. Rev. C}\ }\textbf {\bibinfo {volume} {103}},\ \bibinfo
  {pages} {055806} (\bibinfo {year} {2021})},\ \Eprint
  {https://arxiv.org/abs/2011.05855} {arXiv:2011.05855 [nucl-th]} \BibitemShut
  {NoStop}%
\bibitem [{\citenamefont {Miller}\ \emph {et~al.}(2019)\citenamefont {Miller}
  \emph {et~al.}}]{Miller:2019cac}%
  \BibitemOpen
  \bibfield  {author} {\bibinfo {author} {\bibfnamefont {M.~C.}\ \bibnamefont
  {Miller}} \emph {et~al.},\ }\href {https://doi.org/10.3847/2041-8213/ab50c5}
  {\bibfield  {journal} {\bibinfo  {journal} {Astrophys. J. Lett.}\ }\textbf
  {\bibinfo {volume} {887}},\ \bibinfo {pages} {L24} (\bibinfo {year}
  {2019})},\ \Eprint {https://arxiv.org/abs/1912.05705} {arXiv:1912.05705
  [astro-ph.HE]} \BibitemShut {NoStop}%
\bibitem [{\citenamefont {Riley}\ \emph {et~al.}(2019)\citenamefont {Riley}
  \emph {et~al.}}]{Riley:2019yda}%
  \BibitemOpen
  \bibfield  {author} {\bibinfo {author} {\bibfnamefont {T.~E.}\ \bibnamefont
  {Riley}} \emph {et~al.},\ }\href {https://doi.org/10.3847/2041-8213/ab481c}
  {\bibfield  {journal} {\bibinfo  {journal} {Astrophys. J. Lett.}\ }\textbf
  {\bibinfo {volume} {887}},\ \bibinfo {pages} {L21} (\bibinfo {year}
  {2019})},\ \Eprint {https://arxiv.org/abs/1912.05702} {arXiv:1912.05702
  [astro-ph.HE]} \BibitemShut {NoStop}%
\bibitem [{\citenamefont {Miller}\ \emph {et~al.}(2021)\citenamefont {Miller}
  \emph {et~al.}}]{Miller:2021qha}%
  \BibitemOpen
  \bibfield  {author} {\bibinfo {author} {\bibfnamefont {M.~C.}\ \bibnamefont
  {Miller}} \emph {et~al.},\ }\href {https://doi.org/10.3847/2041-8213/ac089b}
  {\bibfield  {journal} {\bibinfo  {journal} {Astrophys. J. Lett.}\ }\textbf
  {\bibinfo {volume} {918}},\ \bibinfo {pages} {L28} (\bibinfo {year}
  {2021})},\ \Eprint {https://arxiv.org/abs/2105.06979} {arXiv:2105.06979
  [astro-ph.HE]} \BibitemShut {NoStop}%
\bibitem [{\citenamefont {Riley}\ \emph {et~al.}(2021)\citenamefont {Riley}
  \emph {et~al.}}]{Riley:2021pdl}%
  \BibitemOpen
  \bibfield  {author} {\bibinfo {author} {\bibfnamefont {T.~E.}\ \bibnamefont
  {Riley}} \emph {et~al.},\ }\href {https://doi.org/10.3847/2041-8213/ac0a81}
  {\bibfield  {journal} {\bibinfo  {journal} {Astrophys. J. Lett.}\ }\textbf
  {\bibinfo {volume} {918}},\ \bibinfo {pages} {L27} (\bibinfo {year}
  {2021})},\ \Eprint {https://arxiv.org/abs/2105.06980} {arXiv:2105.06980
  [astro-ph.HE]} \BibitemShut {NoStop}%
\bibitem [{\citenamefont {Choudhury}\ \emph {et~al.}(2024)\citenamefont
  {Choudhury} \emph {et~al.}}]{Choudhury:2024xbk}%
  \BibitemOpen
  \bibfield  {author} {\bibinfo {author} {\bibfnamefont {D.}~\bibnamefont
  {Choudhury}} \emph {et~al.},\ }\href
  {https://doi.org/10.3847/2041-8213/ad5a6f} {\bibfield  {journal} {\bibinfo
  {journal} {Astrophys. J. Lett.}\ }\textbf {\bibinfo {volume} {971}},\
  \bibinfo {pages} {L20} (\bibinfo {year} {2024})},\ \Eprint
  {https://arxiv.org/abs/2407.06789} {arXiv:2407.06789 [astro-ph.HE]}
  \BibitemShut {NoStop}%
\bibitem [{\citenamefont {Salmi}\ \emph {et~al.}(2022)\citenamefont {Salmi}
  \emph {et~al.}}]{Salmi:2022cgy}%
  \BibitemOpen
  \bibfield  {author} {\bibinfo {author} {\bibfnamefont {T.}~\bibnamefont
  {Salmi}} \emph {et~al.},\ }\href {https://doi.org/10.3847/1538-4357/ac983d}
  {\bibfield  {journal} {\bibinfo  {journal} {Astrophys. J.}\ }\textbf
  {\bibinfo {volume} {941}},\ \bibinfo {pages} {150} (\bibinfo {year}
  {2022})},\ \Eprint {https://arxiv.org/abs/2209.12840} {arXiv:2209.12840
  [astro-ph.HE]} \BibitemShut {NoStop}%
\bibitem [{\citenamefont {Salmi}\ \emph {et~al.}(2024)\citenamefont {Salmi}
  \emph {et~al.}}]{Salmi:2024aum}%
  \BibitemOpen
  \bibfield  {author} {\bibinfo {author} {\bibfnamefont {T.}~\bibnamefont
  {Salmi}} \emph {et~al.},\ }\href {https://doi.org/10.3847/1538-4357/ad5f1f}
  {\bibfield  {journal} {\bibinfo  {journal} {Astrophys. J.}\ }\textbf
  {\bibinfo {volume} {974}},\ \bibinfo {pages} {294} (\bibinfo {year}
  {2024})},\ \Eprint {https://arxiv.org/abs/2406.14466} {arXiv:2406.14466
  [astro-ph.HE]} \BibitemShut {NoStop}%
\bibitem [{\citenamefont {Dittmann}\ \emph {et~al.}(2024)\citenamefont
  {Dittmann} \emph {et~al.}}]{Dittmann:2024mbo}%
  \BibitemOpen
  \bibfield  {author} {\bibinfo {author} {\bibfnamefont {A.~J.}\ \bibnamefont
  {Dittmann}} \emph {et~al.},\ }\href
  {https://doi.org/10.3847/1538-4357/ad5f1e} {\bibfield  {journal} {\bibinfo
  {journal} {Astrophys. J.}\ }\textbf {\bibinfo {volume} {974}},\ \bibinfo
  {pages} {295} (\bibinfo {year} {2024})},\ \Eprint
  {https://arxiv.org/abs/2406.14467} {arXiv:2406.14467 [astro-ph.HE]}
  \BibitemShut {NoStop}%
\bibitem [{\citenamefont {Fonseca}\ \emph {et~al.}(2021)\citenamefont {Fonseca}
  \emph {et~al.}}]{Fonseca:2021wxt}%
  \BibitemOpen
  \bibfield  {author} {\bibinfo {author} {\bibfnamefont {E.}~\bibnamefont
  {Fonseca}} \emph {et~al.},\ }\href {https://doi.org/10.3847/2041-8213/ac03b8}
  {\bibfield  {journal} {\bibinfo  {journal} {Astrophys. J. Lett.}\ }\textbf
  {\bibinfo {volume} {915}},\ \bibinfo {pages} {L12} (\bibinfo {year}
  {2021})},\ \Eprint {https://arxiv.org/abs/2104.00880} {arXiv:2104.00880
  [astro-ph.HE]} \BibitemShut {NoStop}%
\bibitem [{\citenamefont {Vinciguerra}\ \emph {et~al.}(2024)\citenamefont
  {Vinciguerra} \emph {et~al.}}]{Vinciguerra:2023qxq}%
  \BibitemOpen
  \bibfield  {author} {\bibinfo {author} {\bibfnamefont {S.}~\bibnamefont
  {Vinciguerra}} \emph {et~al.},\ }\href
  {https://doi.org/10.3847/1538-4357/acfb83} {\bibfield  {journal} {\bibinfo
  {journal} {Astrophys. J.}\ }\textbf {\bibinfo {volume} {961}},\ \bibinfo
  {pages} {62} (\bibinfo {year} {2024})},\ \Eprint
  {https://arxiv.org/abs/2308.09469} {arXiv:2308.09469 [astro-ph.HE]}
  \BibitemShut {NoStop}%
\bibitem [{\citenamefont {Campanelli}\ \emph {et~al.}(2006)\citenamefont
  {Campanelli}, \citenamefont {Lousto},\ and\ \citenamefont
  {Zlochower}}]{Campanelli:2006uy}%
  \BibitemOpen
  \bibfield  {author} {\bibinfo {author} {\bibfnamefont {M.}~\bibnamefont
  {Campanelli}}, \bibinfo {author} {\bibfnamefont {C.~O.}\ \bibnamefont
  {Lousto}},\ and\ \bibinfo {author} {\bibfnamefont {Y.}~\bibnamefont
  {Zlochower}},\ }\href {https://doi.org/10.1103/PhysRevD.74.041501} {\bibfield
   {journal} {\bibinfo  {journal} {Phys. Rev. D}\ }\textbf {\bibinfo {volume}
  {74}},\ \bibinfo {pages} {041501} (\bibinfo {year} {2006})},\ \Eprint
  {https://arxiv.org/abs/gr-qc/0604012} {arXiv:gr-qc/0604012} \BibitemShut
  {NoStop}%
\bibitem [{\citenamefont {Dudi}\ \emph {et~al.}(2022)\citenamefont {Dudi},
  \citenamefont {Dietrich}, \citenamefont {Rashti}, \citenamefont {Bruegmann},
  \citenamefont {Steinhoff},\ and\ \citenamefont {Tichy}}]{Dudi:2021wcf}%
  \BibitemOpen
  \bibfield  {author} {\bibinfo {author} {\bibfnamefont {R.}~\bibnamefont
  {Dudi}}, \bibinfo {author} {\bibfnamefont {T.}~\bibnamefont {Dietrich}},
  \bibinfo {author} {\bibfnamefont {A.}~\bibnamefont {Rashti}}, \bibinfo
  {author} {\bibfnamefont {B.}~\bibnamefont {Bruegmann}}, \bibinfo {author}
  {\bibfnamefont {J.}~\bibnamefont {Steinhoff}},\ and\ \bibinfo {author}
  {\bibfnamefont {W.}~\bibnamefont {Tichy}},\ }\href
  {https://doi.org/10.1103/PhysRevD.105.064050} {\bibfield  {journal} {\bibinfo
   {journal} {Phys. Rev. D}\ }\textbf {\bibinfo {volume} {105}},\ \bibinfo
  {pages} {064050} (\bibinfo {year} {2022})},\ \Eprint
  {https://arxiv.org/abs/2108.10429} {arXiv:2108.10429 [gr-qc]} \BibitemShut
  {NoStop}%
\bibitem [{\citenamefont {Hotokezaka}\ \emph {et~al.}(2013)\citenamefont
  {Hotokezaka}, \citenamefont {Kiuchi}, \citenamefont {Kyutoku}, \citenamefont
  {Okawa}, \citenamefont {Sekiguchi}, \citenamefont {Shibata},\ and\
  \citenamefont {Taniguchi}}]{Hotokezaka:2012ze}%
  \BibitemOpen
  \bibfield  {author} {\bibinfo {author} {\bibfnamefont {K.}~\bibnamefont
  {Hotokezaka}}, \bibinfo {author} {\bibfnamefont {K.}~\bibnamefont {Kiuchi}},
  \bibinfo {author} {\bibfnamefont {K.}~\bibnamefont {Kyutoku}}, \bibinfo
  {author} {\bibfnamefont {H.}~\bibnamefont {Okawa}}, \bibinfo {author}
  {\bibfnamefont {Y.-i.}\ \bibnamefont {Sekiguchi}}, \bibinfo {author}
  {\bibfnamefont {M.}~\bibnamefont {Shibata}},\ and\ \bibinfo {author}
  {\bibfnamefont {K.}~\bibnamefont {Taniguchi}},\ }\href
  {https://doi.org/10.1103/PhysRevD.87.024001} {\bibfield  {journal} {\bibinfo
  {journal} {Phys. Rev. D}\ }\textbf {\bibinfo {volume} {87}},\ \bibinfo
  {pages} {024001} (\bibinfo {year} {2013})},\ \Eprint
  {https://arxiv.org/abs/1212.0905} {arXiv:1212.0905 [astro-ph.HE]}
  \BibitemShut {NoStop}%
\bibitem [{\citenamefont {Shibata}\ \emph {et~al.}(2017)\citenamefont
  {Shibata}, \citenamefont {Fujibayashi}, \citenamefont {Hotokezaka},
  \citenamefont {Kiuchi}, \citenamefont {Kyutoku}, \citenamefont {Sekiguchi},\
  and\ \citenamefont {Tanaka}}]{Shibata:2017xdx}%
  \BibitemOpen
  \bibfield  {author} {\bibinfo {author} {\bibfnamefont {M.}~\bibnamefont
  {Shibata}}, \bibinfo {author} {\bibfnamefont {S.}~\bibnamefont
  {Fujibayashi}}, \bibinfo {author} {\bibfnamefont {K.}~\bibnamefont
  {Hotokezaka}}, \bibinfo {author} {\bibfnamefont {K.}~\bibnamefont {Kiuchi}},
  \bibinfo {author} {\bibfnamefont {K.}~\bibnamefont {Kyutoku}}, \bibinfo
  {author} {\bibfnamefont {Y.}~\bibnamefont {Sekiguchi}},\ and\ \bibinfo
  {author} {\bibfnamefont {M.}~\bibnamefont {Tanaka}},\ }\href
  {https://doi.org/10.1103/PhysRevD.96.123012} {\bibfield  {journal} {\bibinfo
  {journal} {Phys. Rev. D}\ }\textbf {\bibinfo {volume} {96}},\ \bibinfo
  {pages} {123012} (\bibinfo {year} {2017})},\ \Eprint
  {https://arxiv.org/abs/1710.07579} {arXiv:1710.07579 [astro-ph.HE]}
  \BibitemShut {NoStop}%
\bibitem [{\citenamefont {Kiuchi}\ \emph {et~al.}(2019)\citenamefont {Kiuchi},
  \citenamefont {Kyutoku}, \citenamefont {Shibata},\ and\ \citenamefont
  {Taniguchi}}]{Kiuchi:2019lls}%
  \BibitemOpen
  \bibfield  {author} {\bibinfo {author} {\bibfnamefont {K.}~\bibnamefont
  {Kiuchi}}, \bibinfo {author} {\bibfnamefont {K.}~\bibnamefont {Kyutoku}},
  \bibinfo {author} {\bibfnamefont {M.}~\bibnamefont {Shibata}},\ and\ \bibinfo
  {author} {\bibfnamefont {K.}~\bibnamefont {Taniguchi}},\ }\href
  {https://doi.org/10.3847/2041-8213/ab1e45} {\bibfield  {journal} {\bibinfo
  {journal} {Astrophys. J. Lett.}\ }\textbf {\bibinfo {volume} {876}},\
  \bibinfo {pages} {L31} (\bibinfo {year} {2019})},\ \Eprint
  {https://arxiv.org/abs/1903.01466} {arXiv:1903.01466 [astro-ph.HE]}
  \BibitemShut {NoStop}%
\bibitem [{\citenamefont {East}\ \emph
  {et~al.}(2016{\natexlab{a}})\citenamefont {East}, \citenamefont
  {Paschalidis}, \citenamefont {Pretorius},\ and\ \citenamefont
  {Shapiro}}]{East:2015vix}%
  \BibitemOpen
  \bibfield  {author} {\bibinfo {author} {\bibfnamefont {W.~E.}\ \bibnamefont
  {East}}, \bibinfo {author} {\bibfnamefont {V.}~\bibnamefont {Paschalidis}},
  \bibinfo {author} {\bibfnamefont {F.}~\bibnamefont {Pretorius}},\ and\
  \bibinfo {author} {\bibfnamefont {S.~L.}\ \bibnamefont {Shapiro}},\ }\href
  {https://doi.org/10.1103/PhysRevD.93.024011} {\bibfield  {journal} {\bibinfo
  {journal} {Phys. Rev. D}\ }\textbf {\bibinfo {volume} {93}},\ \bibinfo
  {pages} {024011} (\bibinfo {year} {2016}{\natexlab{a}})},\ \Eprint
  {https://arxiv.org/abs/1511.01093} {arXiv:1511.01093 [astro-ph.HE]}
  \BibitemShut {NoStop}%
\bibitem [{\citenamefont {Rosswog}\ \emph {et~al.}(2025)\citenamefont
  {Rosswog}, \citenamefont {Sarin}, \citenamefont {Nakhar},\ and\ \citenamefont
  {Diener}}]{Rosswog:2024vfe}%
  \BibitemOpen
  \bibfield  {author} {\bibinfo {author} {\bibfnamefont {S.}~\bibnamefont
  {Rosswog}}, \bibinfo {author} {\bibfnamefont {N.}~\bibnamefont {Sarin}},
  \bibinfo {author} {\bibfnamefont {E.}~\bibnamefont {Nakhar}},\ and\ \bibinfo
  {author} {\bibfnamefont {P.}~\bibnamefont {Diener}},\ }\href
  {https://doi.org/10.1093/mnras/staf324} {\bibfield  {journal} {\bibinfo
  {journal} {Mon. Not. Roy. Astron. Soc.}\ }\textbf {\bibinfo {volume} {538}},\
  \bibinfo {pages} {907} (\bibinfo {year} {2025})},\ \Eprint
  {https://arxiv.org/abs/2411.18813} {arXiv:2411.18813 [astro-ph.HE]}
  \BibitemShut {NoStop}%
\bibitem [{\citenamefont {Price}\ and\ \citenamefont
  {Rosswog}(2006)}]{Price:2006fi}%
  \BibitemOpen
  \bibfield  {author} {\bibinfo {author} {\bibfnamefont {D.}~\bibnamefont
  {Price}}\ and\ \bibinfo {author} {\bibfnamefont {S.}~\bibnamefont
  {Rosswog}},\ }\href {https://doi.org/10.1126/science.1125201} {\bibfield
  {journal} {\bibinfo  {journal} {Science}\ }\textbf {\bibinfo {volume}
  {312}},\ \bibinfo {pages} {719} (\bibinfo {year} {2006})},\ \Eprint
  {https://arxiv.org/abs/astro-ph/0603845} {arXiv:astro-ph/0603845}
  \BibitemShut {NoStop}%
\bibitem [{\citenamefont {Kiuchi}\ \emph {et~al.}(2015)\citenamefont {Kiuchi},
  \citenamefont {Cerd\'a-Dur\'an}, \citenamefont {Kyutoku}, \citenamefont
  {Sekiguchi},\ and\ \citenamefont {Shibata}}]{Kiuchi:2015sga}%
  \BibitemOpen
  \bibfield  {author} {\bibinfo {author} {\bibfnamefont {K.}~\bibnamefont
  {Kiuchi}}, \bibinfo {author} {\bibfnamefont {P.}~\bibnamefont
  {Cerd\'a-Dur\'an}}, \bibinfo {author} {\bibfnamefont {K.}~\bibnamefont
  {Kyutoku}}, \bibinfo {author} {\bibfnamefont {Y.}~\bibnamefont {Sekiguchi}},\
  and\ \bibinfo {author} {\bibfnamefont {M.}~\bibnamefont {Shibata}},\ }\href
  {https://doi.org/10.1103/PhysRevD.92.124034} {\bibfield  {journal} {\bibinfo
  {journal} {Phys. Rev. D}\ }\textbf {\bibinfo {volume} {92}},\ \bibinfo
  {pages} {124034} (\bibinfo {year} {2015})},\ \Eprint
  {https://arxiv.org/abs/1509.09205} {arXiv:1509.09205 [astro-ph.HE]}
  \BibitemShut {NoStop}%
\bibitem [{\citenamefont {Giacomazzo}\ \emph {et~al.}(2015)\citenamefont
  {Giacomazzo}, \citenamefont {Zrake}, \citenamefont {Duffell}, \citenamefont
  {MacFadyen},\ and\ \citenamefont {Perna}}]{Giacomazzo:2014qba}%
  \BibitemOpen
  \bibfield  {author} {\bibinfo {author} {\bibfnamefont {B.}~\bibnamefont
  {Giacomazzo}}, \bibinfo {author} {\bibfnamefont {J.}~\bibnamefont {Zrake}},
  \bibinfo {author} {\bibfnamefont {P.}~\bibnamefont {Duffell}}, \bibinfo
  {author} {\bibfnamefont {A.~I.}\ \bibnamefont {MacFadyen}},\ and\ \bibinfo
  {author} {\bibfnamefont {R.}~\bibnamefont {Perna}},\ }\href
  {https://doi.org/10.1088/0004-637X/809/1/39} {\bibfield  {journal} {\bibinfo
  {journal} {Astrophys. J.}\ }\textbf {\bibinfo {volume} {809}},\ \bibinfo
  {pages} {39} (\bibinfo {year} {2015})},\ \Eprint
  {https://arxiv.org/abs/1410.0013} {arXiv:1410.0013 [astro-ph.HE]}
  \BibitemShut {NoStop}%
\bibitem [{\citenamefont {Aguilera-Miret}\ \emph {et~al.}(2025)\citenamefont
  {Aguilera-Miret}, \citenamefont {Christian}, \citenamefont {Rosswog},\ and\
  \citenamefont {Palenzuela}}]{Aguilera-Miret:2025nts}%
  \BibitemOpen
  \bibfield  {author} {\bibinfo {author} {\bibfnamefont {R.}~\bibnamefont
  {Aguilera-Miret}}, \bibinfo {author} {\bibfnamefont {J.-E.}\ \bibnamefont
  {Christian}}, \bibinfo {author} {\bibfnamefont {S.}~\bibnamefont {Rosswog}},\
  and\ \bibinfo {author} {\bibfnamefont {C.}~\bibnamefont {Palenzuela}},\
  }\bibfield  {journal} {\bibinfo  {journal} {Mon. Not. Roy. Astron. Soc.}\
  }\href {https://doi.org/10.1093/mnras/staf1291} {10.1093/mnras/staf1291}
  (\bibinfo {year} {2025}),\ \Eprint {https://arxiv.org/abs/2504.10604}
  {2504.10604 [astro-ph.HE]} \BibitemShut {NoStop}%
\bibitem [{\citenamefont {Duez}\ \emph
  {et~al.}(2006{\natexlab{a}})\citenamefont {Duez}, \citenamefont {Liu},
  \citenamefont {Shapiro},\ and\ \citenamefont {Shibata}}]{Duez:2006qe}%
  \BibitemOpen
  \bibfield  {author} {\bibinfo {author} {\bibfnamefont {M.~D.}\ \bibnamefont
  {Duez}}, \bibinfo {author} {\bibfnamefont {Y.~T.}\ \bibnamefont {Liu}},
  \bibinfo {author} {\bibfnamefont {S.~L.}\ \bibnamefont {Shapiro}},\ and\
  \bibinfo {author} {\bibfnamefont {M.}~\bibnamefont {Shibata}},\ }\href
  {https://doi.org/10.1103/PhysRevD.73.104015} {\bibfield  {journal} {\bibinfo
  {journal} {Phys. Rev. D}\ }\textbf {\bibinfo {volume} {73}},\ \bibinfo
  {pages} {104015} (\bibinfo {year} {2006}{\natexlab{a}})},\ \Eprint
  {https://arxiv.org/abs/astro-ph/0605331} {arXiv:astro-ph/0605331}
  \BibitemShut {NoStop}%
\bibitem [{\citenamefont {Sun}\ \emph {et~al.}(2019)\citenamefont {Sun},
  \citenamefont {Ruiz},\ and\ \citenamefont {Shapiro}}]{Sun:2018gcl}%
  \BibitemOpen
  \bibfield  {author} {\bibinfo {author} {\bibfnamefont {L.}~\bibnamefont
  {Sun}}, \bibinfo {author} {\bibfnamefont {M.}~\bibnamefont {Ruiz}},\ and\
  \bibinfo {author} {\bibfnamefont {S.~L.}\ \bibnamefont {Shapiro}},\ }\href
  {https://doi.org/10.1103/PhysRevD.99.064057} {\bibfield  {journal} {\bibinfo
  {journal} {Phys. Rev. D}\ }\textbf {\bibinfo {volume} {99}},\ \bibinfo
  {pages} {064057} (\bibinfo {year} {2019})},\ \Eprint
  {https://arxiv.org/abs/1812.03176} {arXiv:1812.03176 [astro-ph.HE]}
  \BibitemShut {NoStop}%
\bibitem [{\citenamefont {Kiuchi}\ \emph {et~al.}(2018)\citenamefont {Kiuchi},
  \citenamefont {Kyutoku}, \citenamefont {Sekiguchi},\ and\ \citenamefont
  {Shibata}}]{Kiuchi:2017zzg}%
  \BibitemOpen
  \bibfield  {author} {\bibinfo {author} {\bibfnamefont {K.}~\bibnamefont
  {Kiuchi}}, \bibinfo {author} {\bibfnamefont {K.}~\bibnamefont {Kyutoku}},
  \bibinfo {author} {\bibfnamefont {Y.}~\bibnamefont {Sekiguchi}},\ and\
  \bibinfo {author} {\bibfnamefont {M.}~\bibnamefont {Shibata}},\ }\href
  {https://doi.org/10.1103/PhysRevD.97.124039} {\bibfield  {journal} {\bibinfo
  {journal} {Phys. Rev. D}\ }\textbf {\bibinfo {volume} {97}},\ \bibinfo
  {pages} {124039} (\bibinfo {year} {2018})},\ \Eprint
  {https://arxiv.org/abs/1710.01311} {arXiv:1710.01311 [astro-ph.HE]}
  \BibitemShut {NoStop}%
\bibitem [{\citenamefont {Duez}\ \emph
  {et~al.}(2006{\natexlab{b}})\citenamefont {Duez}, \citenamefont {Liu},
  \citenamefont {Shapiro}, \citenamefont {Shibata},\ and\ \citenamefont
  {Stephens}}]{Duez:2005cj}%
  \BibitemOpen
  \bibfield  {author} {\bibinfo {author} {\bibfnamefont {M.~D.}\ \bibnamefont
  {Duez}}, \bibinfo {author} {\bibfnamefont {Y.~T.}\ \bibnamefont {Liu}},
  \bibinfo {author} {\bibfnamefont {S.~L.}\ \bibnamefont {Shapiro}}, \bibinfo
  {author} {\bibfnamefont {M.}~\bibnamefont {Shibata}},\ and\ \bibinfo {author}
  {\bibfnamefont {B.~C.}\ \bibnamefont {Stephens}},\ }\href
  {https://doi.org/10.1103/PhysRevLett.96.031101} {\bibfield  {journal}
  {\bibinfo  {journal} {Phys. Rev. Lett.}\ }\textbf {\bibinfo {volume} {96}},\
  \bibinfo {pages} {031101} (\bibinfo {year} {2006}{\natexlab{b}})},\ \Eprint
  {https://arxiv.org/abs/astro-ph/0510653} {arXiv:astro-ph/0510653}
  \BibitemShut {NoStop}%
\bibitem [{\citenamefont {Siegel}\ \emph {et~al.}(2013)\citenamefont {Siegel},
  \citenamefont {Ciolfi}, \citenamefont {Harte},\ and\ \citenamefont
  {Rezzolla}}]{Siegel:2013nrw}%
  \BibitemOpen
  \bibfield  {author} {\bibinfo {author} {\bibfnamefont {D.~M.}\ \bibnamefont
  {Siegel}}, \bibinfo {author} {\bibfnamefont {R.}~\bibnamefont {Ciolfi}},
  \bibinfo {author} {\bibfnamefont {A.~I.}\ \bibnamefont {Harte}},\ and\
  \bibinfo {author} {\bibfnamefont {L.}~\bibnamefont {Rezzolla}},\ }\href
  {https://doi.org/10.1103/PhysRevD.87.121302} {\bibfield  {journal} {\bibinfo
  {journal} {Phys. Rev. D}\ }\textbf {\bibinfo {volume} {87}},\ \bibinfo
  {pages} {121302} (\bibinfo {year} {2013})},\ \Eprint
  {https://arxiv.org/abs/1302.4368} {arXiv:1302.4368 [gr-qc]} \BibitemShut
  {NoStop}%
\bibitem [{\citenamefont {Most}(2023)}]{Most:2023sme}%
  \BibitemOpen
  \bibfield  {author} {\bibinfo {author} {\bibfnamefont {E.~R.}\ \bibnamefont
  {Most}},\ }\href {https://doi.org/10.1103/PhysRevD.108.123012} {\bibfield
  {journal} {\bibinfo  {journal} {Phys. Rev. D}\ }\textbf {\bibinfo {volume}
  {108}},\ \bibinfo {pages} {123012} (\bibinfo {year} {2023})},\ \Eprint
  {https://arxiv.org/abs/2311.03333} {arXiv:2311.03333 [astro-ph.HE]}
  \BibitemShut {NoStop}%
\bibitem [{\citenamefont {Reboul-Salze}\ \emph {et~al.}(2025)\citenamefont
  {Reboul-Salze}, \citenamefont {Barr{\`e}re}, \citenamefont {Kiuchi},
  \citenamefont {Guilet}, \citenamefont {Raynaud}, \citenamefont
  {Fujibayashi},\ and\ \citenamefont {Shibata}}]{Reboul-Salze:2024jst}%
  \BibitemOpen
  \bibfield  {author} {\bibinfo {author} {\bibfnamefont {A.}~\bibnamefont
  {Reboul-Salze}}, \bibinfo {author} {\bibfnamefont {P.}~\bibnamefont
  {Barr{\`e}re}}, \bibinfo {author} {\bibfnamefont {K.}~\bibnamefont {Kiuchi}},
  \bibinfo {author} {\bibfnamefont {J.}~\bibnamefont {Guilet}}, \bibinfo
  {author} {\bibfnamefont {R.}~\bibnamefont {Raynaud}}, \bibinfo {author}
  {\bibfnamefont {S.}~\bibnamefont {Fujibayashi}},\ and\ \bibinfo {author}
  {\bibfnamefont {M.}~\bibnamefont {Shibata}},\ }\href
  {https://doi.org/10.1051/0004-6361/202453126} {\bibfield  {journal} {\bibinfo
   {journal} {Astron. Astrophys.}\ }\textbf {\bibinfo {volume} {699}},\
  \bibinfo {pages} {A4} (\bibinfo {year} {2025})},\ \Eprint
  {https://arxiv.org/abs/2411.19328} {arXiv:2411.19328 [astro-ph.HE]}
  \BibitemShut {NoStop}%
\bibitem [{\citenamefont {Aguilera-Miret}\ \emph {et~al.}(2020)\citenamefont
  {Aguilera-Miret}, \citenamefont {Vigan{\`o}}, \citenamefont {Carrasco},
  \citenamefont {Mi{\~n}ano},\ and\ \citenamefont
  {Palenzuela}}]{Aguilera-Miret:2020dhz}%
  \BibitemOpen
  \bibfield  {author} {\bibinfo {author} {\bibfnamefont {R.}~\bibnamefont
  {Aguilera-Miret}}, \bibinfo {author} {\bibfnamefont {D.}~\bibnamefont
  {Vigan{\`o}}}, \bibinfo {author} {\bibfnamefont {F.}~\bibnamefont
  {Carrasco}}, \bibinfo {author} {\bibfnamefont {B.}~\bibnamefont
  {Mi{\~n}ano}},\ and\ \bibinfo {author} {\bibfnamefont {C.}~\bibnamefont
  {Palenzuela}},\ }\href {https://doi.org/10.1103/PhysRevD.102.103006}
  {\bibfield  {journal} {\bibinfo  {journal} {Phys. Rev. D}\ }\textbf {\bibinfo
  {volume} {102}},\ \bibinfo {pages} {103006} (\bibinfo {year} {2020})},\
  \Eprint {https://arxiv.org/abs/2009.06669} {arXiv:2009.06669 [gr-qc]}
  \BibitemShut {NoStop}%
\bibitem [{\citenamefont {{Kolmogorov}}(1991)}]{Kolmogorov:1991}%
  \BibitemOpen
  \bibfield  {author} {\bibinfo {author} {\bibfnamefont {A.~N.}\ \bibnamefont
  {{Kolmogorov}}},\ }\href {https://doi.org/10.1098/rspa.1991.0075} {\bibfield
  {journal} {\bibinfo  {journal} {Proceedings of the Royal Society of London
  Series A}\ }\textbf {\bibinfo {volume} {434}},\ \bibinfo {pages} {9}
  (\bibinfo {year} {1991})}\BibitemShut {NoStop}%
\bibitem [{\citenamefont {{Kazantsev}}(1968)}]{Kazantsev:1968}%
  \BibitemOpen
  \bibfield  {author} {\bibinfo {author} {\bibfnamefont {A.~P.}\ \bibnamefont
  {{Kazantsev}}},\ }\href@noop {} {\bibfield  {journal} {\bibinfo  {journal}
  {Soviet Journal of Experimental and Theoretical Physics}\ }\textbf {\bibinfo
  {volume} {26}},\ \bibinfo {pages} {1031} (\bibinfo {year}
  {1968})}\BibitemShut {NoStop}%
\bibitem [{\citenamefont {Zrake}\ and\ \citenamefont
  {MacFadyen}(2013)}]{Zrake:2013mra}%
  \BibitemOpen
  \bibfield  {author} {\bibinfo {author} {\bibfnamefont {J.}~\bibnamefont
  {Zrake}}\ and\ \bibinfo {author} {\bibfnamefont {A.~I.}\ \bibnamefont
  {MacFadyen}},\ }\href {https://doi.org/10.1088/2041-8205/769/2/L29}
  {\bibfield  {journal} {\bibinfo  {journal} {Astrophys. J. Lett.}\ }\textbf
  {\bibinfo {volume} {769}},\ \bibinfo {pages} {L29} (\bibinfo {year}
  {2013})},\ \Eprint {https://arxiv.org/abs/1303.1450} {arXiv:1303.1450
  [astro-ph.HE]} \BibitemShut {NoStop}%
\bibitem [{\citenamefont {Balbus}\ and\ \citenamefont
  {Hawley}(1998)}]{Balbus:1998ja}%
  \BibitemOpen
  \bibfield  {author} {\bibinfo {author} {\bibfnamefont {S.~A.}\ \bibnamefont
  {Balbus}}\ and\ \bibinfo {author} {\bibfnamefont {J.~F.}\ \bibnamefont
  {Hawley}},\ }\href {https://doi.org/10.1103/RevModPhys.70.1} {\bibfield
  {journal} {\bibinfo  {journal} {Rev. Mod. Phys.}\ }\textbf {\bibinfo {volume}
  {70}},\ \bibinfo {pages} {1} (\bibinfo {year} {1998})}\BibitemShut {NoStop}%
\bibitem [{\citenamefont {Celora}\ \emph {et~al.}(2025)\citenamefont {Celora},
  \citenamefont {Palenzuela}, \citenamefont {Vigan{\`o}},\ and\ \citenamefont
  {Aguilera-Miret}}]{Celora:2025fxm}%
  \BibitemOpen
  \bibfield  {author} {\bibinfo {author} {\bibfnamefont {T.}~\bibnamefont
  {Celora}}, \bibinfo {author} {\bibfnamefont {C.}~\bibnamefont {Palenzuela}},
  \bibinfo {author} {\bibfnamefont {D.}~\bibnamefont {Vigan{\`o}}},\ and\
  \bibinfo {author} {\bibfnamefont {R.}~\bibnamefont {Aguilera-Miret}},\
  }\Eprint {https://arxiv.org/abs/2505.01208} {arXiv:2505.01208 [astro-ph.HE]}
  (\bibinfo {year} {2025})\BibitemShut {NoStop}%
\bibitem [{\citenamefont {Fern{\'a}ndez}\ and\ \citenamefont
  {Metzger}(2013)}]{Fernandez:2013tya}%
  \BibitemOpen
  \bibfield  {author} {\bibinfo {author} {\bibfnamefont {R.}~\bibnamefont
  {Fern{\'a}ndez}}\ and\ \bibinfo {author} {\bibfnamefont {B.~D.}\ \bibnamefont
  {Metzger}},\ }\href {https://doi.org/10.1093/mnras/stt1312} {\bibfield
  {journal} {\bibinfo  {journal} {Mon. Not. Roy. Astron. Soc.}\ }\textbf
  {\bibinfo {volume} {435}},\ \bibinfo {pages} {502} (\bibinfo {year}
  {2013})},\ \Eprint {https://arxiv.org/abs/1304.6720} {arXiv:1304.6720
  [astro-ph.HE]} \BibitemShut {NoStop}%
\bibitem [{\citenamefont {Just}\ \emph {et~al.}(2015)\citenamefont {Just},
  \citenamefont {Bauswein}, \citenamefont {Pulpillo}, \citenamefont {Goriely},\
  and\ \citenamefont {Janka}}]{Just:2014fka}%
  \BibitemOpen
  \bibfield  {author} {\bibinfo {author} {\bibfnamefont {O.}~\bibnamefont
  {Just}}, \bibinfo {author} {\bibfnamefont {A.}~\bibnamefont {Bauswein}},
  \bibinfo {author} {\bibfnamefont {R.~A.}\ \bibnamefont {Pulpillo}}, \bibinfo
  {author} {\bibfnamefont {S.}~\bibnamefont {Goriely}},\ and\ \bibinfo {author}
  {\bibfnamefont {H.~T.}\ \bibnamefont {Janka}},\ }\href
  {https://doi.org/10.1093/mnras/stv009} {\bibfield  {journal} {\bibinfo
  {journal} {Mon. Not. Roy. Astron. Soc.}\ }\textbf {\bibinfo {volume} {448}},\
  \bibinfo {pages} {541} (\bibinfo {year} {2015})},\ \Eprint
  {https://arxiv.org/abs/1406.2687} {arXiv:1406.2687 [astro-ph.SR]}
  \BibitemShut {NoStop}%
\bibitem [{\citenamefont {Fujibayashi}\ \emph {et~al.}(2018)\citenamefont
  {Fujibayashi}, \citenamefont {Kiuchi}, \citenamefont {Nishimura},
  \citenamefont {Sekiguchi},\ and\ \citenamefont
  {Shibata}}]{Fujibayashi:2017puw}%
  \BibitemOpen
  \bibfield  {author} {\bibinfo {author} {\bibfnamefont {S.}~\bibnamefont
  {Fujibayashi}}, \bibinfo {author} {\bibfnamefont {K.}~\bibnamefont {Kiuchi}},
  \bibinfo {author} {\bibfnamefont {N.}~\bibnamefont {Nishimura}}, \bibinfo
  {author} {\bibfnamefont {Y.}~\bibnamefont {Sekiguchi}},\ and\ \bibinfo
  {author} {\bibfnamefont {M.}~\bibnamefont {Shibata}},\ }\href
  {https://doi.org/10.3847/1538-4357/aabafd} {\bibfield  {journal} {\bibinfo
  {journal} {Astrophys. J.}\ }\textbf {\bibinfo {volume} {860}},\ \bibinfo
  {pages} {64} (\bibinfo {year} {2018})},\ \Eprint
  {https://arxiv.org/abs/1711.02093} {arXiv:1711.02093 [astro-ph.HE]}
  \BibitemShut {NoStop}%
\bibitem [{\citenamefont {Nedora}\ \emph {et~al.}(2019)\citenamefont {Nedora},
  \citenamefont {Bernuzzi}, \citenamefont {Radice}, \citenamefont {Perego},
  \citenamefont {Endrizzi},\ and\ \citenamefont {Ortiz}}]{Nedora:2019jhl}%
  \BibitemOpen
  \bibfield  {author} {\bibinfo {author} {\bibfnamefont {V.}~\bibnamefont
  {Nedora}}, \bibinfo {author} {\bibfnamefont {S.}~\bibnamefont {Bernuzzi}},
  \bibinfo {author} {\bibfnamefont {D.}~\bibnamefont {Radice}}, \bibinfo
  {author} {\bibfnamefont {A.}~\bibnamefont {Perego}}, \bibinfo {author}
  {\bibfnamefont {A.}~\bibnamefont {Endrizzi}},\ and\ \bibinfo {author}
  {\bibfnamefont {N.}~\bibnamefont {Ortiz}},\ }\href
  {https://doi.org/10.3847/2041-8213/ab5794} {\bibfield  {journal} {\bibinfo
  {journal} {Astrophys. J. Lett.}\ }\textbf {\bibinfo {volume} {886}},\
  \bibinfo {pages} {L30} (\bibinfo {year} {2019})},\ \Eprint
  {https://arxiv.org/abs/1907.04872} {arXiv:1907.04872 [astro-ph.HE]}
  \BibitemShut {NoStop}%
\bibitem [{\citenamefont {Shibata}\ and\ \citenamefont
  {Hotokezaka}(2019)}]{Shibata:2019wef}%
  \BibitemOpen
  \bibfield  {author} {\bibinfo {author} {\bibfnamefont {M.}~\bibnamefont
  {Shibata}}\ and\ \bibinfo {author} {\bibfnamefont {K.}~\bibnamefont
  {Hotokezaka}},\ }\href {https://doi.org/10.1146/annurev-nucl-101918-023625}
  {\bibfield  {journal} {\bibinfo  {journal} {Ann. Rev. Nucl. Part. Sci.}\
  }\textbf {\bibinfo {volume} {69}},\ \bibinfo {pages} {41} (\bibinfo {year}
  {2019})},\ \Eprint {https://arxiv.org/abs/1908.02350} {arXiv:1908.02350
  [astro-ph.HE]} \BibitemShut {NoStop}%
\bibitem [{\citenamefont {Endrizzi}\ \emph {et~al.}(2020)\citenamefont
  {Endrizzi}, \citenamefont {Perego}, \citenamefont {Fabbri}, \citenamefont
  {Branca}, \citenamefont {Radice}, \citenamefont {Bernuzzi}, \citenamefont
  {Giacomazzo}, \citenamefont {Pederiva},\ and\ \citenamefont
  {Lovato}}]{Endrizzi:2019trv}%
  \BibitemOpen
  \bibfield  {author} {\bibinfo {author} {\bibfnamefont {A.}~\bibnamefont
  {Endrizzi}}, \bibinfo {author} {\bibfnamefont {A.}~\bibnamefont {Perego}},
  \bibinfo {author} {\bibfnamefont {F.~M.}\ \bibnamefont {Fabbri}}, \bibinfo
  {author} {\bibfnamefont {L.}~\bibnamefont {Branca}}, \bibinfo {author}
  {\bibfnamefont {D.}~\bibnamefont {Radice}}, \bibinfo {author} {\bibfnamefont
  {S.}~\bibnamefont {Bernuzzi}}, \bibinfo {author} {\bibfnamefont
  {B.}~\bibnamefont {Giacomazzo}}, \bibinfo {author} {\bibfnamefont
  {F.}~\bibnamefont {Pederiva}},\ and\ \bibinfo {author} {\bibfnamefont
  {A.}~\bibnamefont {Lovato}},\ }\href
  {https://doi.org/10.1140/epja/s10050-019-00018-6} {\bibfield  {journal}
  {\bibinfo  {journal} {Eur. Phys. J. A}\ }\textbf {\bibinfo {volume} {56}},\
  \bibinfo {pages} {15} (\bibinfo {year} {2020})},\ \Eprint
  {https://arxiv.org/abs/1908.04952} {arXiv:1908.04952 [astro-ph.HE]}
  \BibitemShut {NoStop}%
\bibitem [{\citenamefont {{Lodders}}\ \emph {et~al.}(2025)\citenamefont
  {{Lodders}}, \citenamefont {{Bergemann}},\ and\ \citenamefont
  {{Palme}}}]{Lodders:2025}%
  \BibitemOpen
  \bibfield  {author} {\bibinfo {author} {\bibfnamefont {K.}~\bibnamefont
  {{Lodders}}}, \bibinfo {author} {\bibfnamefont {M.}~\bibnamefont
  {{Bergemann}}},\ and\ \bibinfo {author} {\bibfnamefont {H.}~\bibnamefont
  {{Palme}}},\ }\href {https://doi.org/10.1007/s11214-025-01146-w} {\bibfield
  {journal} {\bibinfo  {journal} {Space Sci. Rev.}\ }\textbf {\bibinfo {volume}
  {221}},\ \bibinfo {eid} {23} (\bibinfo {year} {2025})},\ \Eprint
  {https://arxiv.org/abs/2502.10575} {arXiv:2502.10575 [astro-ph.SR]}
  \BibitemShut {NoStop}%
\bibitem [{\citenamefont {{Prantzos}}\ \emph {et~al.}(2020)\citenamefont
  {{Prantzos}}, \citenamefont {{Abia}}, \citenamefont {{Cristallo}},
  \citenamefont {{Limongi}},\ and\ \citenamefont {{Chieffi}}}]{Prantzos:2020}%
  \BibitemOpen
  \bibfield  {author} {\bibinfo {author} {\bibfnamefont {N.}~\bibnamefont
  {{Prantzos}}}, \bibinfo {author} {\bibfnamefont {C.}~\bibnamefont {{Abia}}},
  \bibinfo {author} {\bibfnamefont {S.}~\bibnamefont {{Cristallo}}}, \bibinfo
  {author} {\bibfnamefont {M.}~\bibnamefont {{Limongi}}},\ and\ \bibinfo
  {author} {\bibfnamefont {A.}~\bibnamefont {{Chieffi}}},\ }\href
  {https://doi.org/10.1093/mnras/stz3154} {\bibfield  {journal} {\bibinfo
  {journal} {Mon. Not. Roy. Astron. Soc.}\ }\textbf {\bibinfo {volume} {491}},\
  \bibinfo {pages} {1832} (\bibinfo {year} {2020})},\ \Eprint
  {https://arxiv.org/abs/1911.02545} {arXiv:1911.02545 [astro-ph.GA]}
  \BibitemShut {NoStop}%
\bibitem [{\citenamefont {Bernuzzi}\ \emph {et~al.}(2025)\citenamefont
  {Bernuzzi}, \citenamefont {Magistrelli}, \citenamefont {Jacobi},
  \citenamefont {Logoteta}, \citenamefont {Perego},\ and\ \citenamefont
  {Radice}}]{Bernuzzi:2024mfx}%
  \BibitemOpen
  \bibfield  {author} {\bibinfo {author} {\bibfnamefont {S.}~\bibnamefont
  {Bernuzzi}}, \bibinfo {author} {\bibfnamefont {F.}~\bibnamefont
  {Magistrelli}}, \bibinfo {author} {\bibfnamefont {M.}~\bibnamefont {Jacobi}},
  \bibinfo {author} {\bibfnamefont {D.}~\bibnamefont {Logoteta}}, \bibinfo
  {author} {\bibfnamefont {A.}~\bibnamefont {Perego}},\ and\ \bibinfo {author}
  {\bibfnamefont {D.}~\bibnamefont {Radice}},\ }\href
  {https://doi.org/10.1093/mnras/staf1147} {\bibfield  {journal} {\bibinfo
  {journal} {Mon. Not. Roy. Astron. Soc.}\ }\textbf {\bibinfo {volume} {256}},\
  \bibinfo {pages} {271} (\bibinfo {year} {2025})},\ \Eprint
  {https://arxiv.org/abs/2409.18185} {arXiv:2409.18185 [astro-ph.HE]}
  \BibitemShut {NoStop}%
\bibitem [{\citenamefont {Ricigliano}\ \emph {et~al.}(2024)\citenamefont
  {Ricigliano}, \citenamefont {Jacobi},\ and\ \citenamefont
  {Arcones}}]{Ricigliano:2024lwf}%
  \BibitemOpen
  \bibfield  {author} {\bibinfo {author} {\bibfnamefont {G.}~\bibnamefont
  {Ricigliano}}, \bibinfo {author} {\bibfnamefont {M.}~\bibnamefont {Jacobi}},\
  and\ \bibinfo {author} {\bibfnamefont {A.}~\bibnamefont {Arcones}},\ }\href
  {https://doi.org/10.1093/mnras/stae1979} {\bibfield  {journal} {\bibinfo
  {journal} {Mon. Not. Roy. Astron. Soc.}\ }\textbf {\bibinfo {volume} {533}},\
  \bibinfo {pages} {2096} (\bibinfo {year} {2024})},\ \Eprint
  {https://arxiv.org/abs/2406.03649} {arXiv:2406.03649 [astro-ph.HE]}
  \BibitemShut {NoStop}%
\bibitem [{\citenamefont {Lippuner}\ and\ \citenamefont
  {Roberts}(2017)}]{Lippuner:2017tyn}%
  \BibitemOpen
  \bibfield  {author} {\bibinfo {author} {\bibfnamefont {J.}~\bibnamefont
  {Lippuner}}\ and\ \bibinfo {author} {\bibfnamefont {L.~F.}\ \bibnamefont
  {Roberts}},\ }\href {https://doi.org/10.3847/1538-4365/aa94cb} {\bibfield
  {journal} {\bibinfo  {journal} {Astrophys. J. Suppl.}\ }\textbf {\bibinfo
  {volume} {233}},\ \bibinfo {pages} {18} (\bibinfo {year} {2017})},\ \Eprint
  {https://arxiv.org/abs/1706.06198} {arXiv:1706.06198 [astro-ph.HE]}
  \BibitemShut {NoStop}%
\bibitem [{\citenamefont {Marketin}\ \emph {et~al.}(2016)\citenamefont
  {Marketin}, \citenamefont {Huther},\ and\ \citenamefont
  {Mart{\'\i}nez-Pinedo}}]{Marketin:2015gya}%
  \BibitemOpen
  \bibfield  {author} {\bibinfo {author} {\bibfnamefont {T.}~\bibnamefont
  {Marketin}}, \bibinfo {author} {\bibfnamefont {L.}~\bibnamefont {Huther}},\
  and\ \bibinfo {author} {\bibfnamefont {G.}~\bibnamefont
  {Mart{\'\i}nez-Pinedo}},\ }\href {https://doi.org/10.1103/PhysRevC.93.025805}
  {\bibfield  {journal} {\bibinfo  {journal} {Phys. Rev. C}\ }\textbf {\bibinfo
  {volume} {93}},\ \bibinfo {pages} {025805} (\bibinfo {year} {2016})},\
  \Eprint {https://arxiv.org/abs/1507.07442} {arXiv:1507.07442 [nucl-th]}
  \BibitemShut {NoStop}%
\bibitem [{\citenamefont {{Pfeiffer}}\ \emph {et~al.}(1997)\citenamefont
  {{Pfeiffer}}, \citenamefont {{Kratz}},\ and\ \citenamefont
  {{Thielemann}}}]{Pfeiffer:1997}%
  \BibitemOpen
  \bibfield  {author} {\bibinfo {author} {\bibfnamefont {B.}~\bibnamefont
  {{Pfeiffer}}}, \bibinfo {author} {\bibfnamefont {K.~L.}\ \bibnamefont
  {{Kratz}}},\ and\ \bibinfo {author} {\bibfnamefont {F.~K.}\ \bibnamefont
  {{Thielemann}}},\ }\href {https://doi.org/10.1007/s002180050237} {\bibfield
  {journal} {\bibinfo  {journal} {Zeitschrift für Physik A Hadrons and
  Nuclei}\ }\textbf {\bibinfo {volume} {357}},\ \bibinfo {pages} {235}
  (\bibinfo {year} {1997})}\BibitemShut {NoStop}%
\bibitem [{\citenamefont {Arcones}\ and\ \citenamefont
  {Martinez-Pinedo}(2011)}]{Arcones:2010dz}%
  \BibitemOpen
  \bibfield  {author} {\bibinfo {author} {\bibfnamefont {A.}~\bibnamefont
  {Arcones}}\ and\ \bibinfo {author} {\bibfnamefont {G.}~\bibnamefont
  {Martinez-Pinedo}},\ }\href {https://doi.org/10.1103/PhysRevC.83.045809}
  {\bibfield  {journal} {\bibinfo  {journal} {Phys. Rev. C}\ }\textbf {\bibinfo
  {volume} {83}},\ \bibinfo {pages} {045809} (\bibinfo {year} {2011})},\
  \Eprint {https://arxiv.org/abs/1008.3890} {arXiv:1008.3890 [astro-ph.SR]}
  \BibitemShut {NoStop}%
\bibitem [{\citenamefont {Eichler}\ \emph {et~al.}(2015)\citenamefont {Eichler}
  \emph {et~al.}}]{Eichler:2014kma}%
  \BibitemOpen
  \bibfield  {author} {\bibinfo {author} {\bibfnamefont {M.}~\bibnamefont
  {Eichler}} \emph {et~al.},\ }\href
  {https://doi.org/10.1088/0004-637X/808/1/30} {\bibfield  {journal} {\bibinfo
  {journal} {Astrophys. J.}\ }\textbf {\bibinfo {volume} {808}},\ \bibinfo
  {pages} {30} (\bibinfo {year} {2015})},\ \Eprint
  {https://arxiv.org/abs/1411.0974} {arXiv:1411.0974 [astro-ph.HE]}
  \BibitemShut {NoStop}%
\bibitem [{\citenamefont {Roederer}\ \emph {et~al.}(2022)\citenamefont
  {Roederer} \emph {et~al.}}]{Roederer:2022exr}%
  \BibitemOpen
  \bibfield  {author} {\bibinfo {author} {\bibfnamefont {I.~U.}\ \bibnamefont
  {Roederer}} \emph {et~al.},\ }\href
  {https://doi.org/10.3847/1538-4357/ac85bc} {\bibfield  {journal} {\bibinfo
  {journal} {Astrophys. J.}\ }\textbf {\bibinfo {volume} {936}},\ \bibinfo
  {pages} {84} (\bibinfo {year} {2022})},\ \Eprint
  {https://arxiv.org/abs/2210.15105} {arXiv:2210.15105 [astro-ph.SR]}
  \BibitemShut {NoStop}%
\bibitem [{\citenamefont {Kuske}\ \emph {et~al.}(2025)\citenamefont {Kuske},
  \citenamefont {Arcones},\ and\ \citenamefont {Reichert}}]{Kuske:2025ffp}%
  \BibitemOpen
  \bibfield  {author} {\bibinfo {author} {\bibfnamefont {J.}~\bibnamefont
  {Kuske}}, \bibinfo {author} {\bibfnamefont {A.}~\bibnamefont {Arcones}},\
  and\ \bibinfo {author} {\bibfnamefont {M.}~\bibnamefont {Reichert}},\ }\href
  {https://doi.org/10.3847/1538-4357/adf0f7} {\bibfield  {journal} {\bibinfo
  {journal} {Astrophys. J.}\ }\textbf {\bibinfo {volume} {990}},\ \bibinfo
  {pages} {37} (\bibinfo {year} {2025})},\ \Eprint
  {https://arxiv.org/abs/2506.00092} {arXiv:2506.00092 [astro-ph.HE]}
  \BibitemShut {NoStop}%
\bibitem [{\citenamefont {de~Jes{\'u}s Mendoza-Temis}\ \emph
  {et~al.}(2015)\citenamefont {de~Jes{\'u}s Mendoza-Temis}, \citenamefont {Wu},
  \citenamefont {Mart{\'\i}nez-Pinedo}, \citenamefont {Langanke}, \citenamefont
  {Bauswein},\ and\ \citenamefont {Janka}}]{deJesusMendoza-Temis:2014owk}%
  \BibitemOpen
  \bibfield  {author} {\bibinfo {author} {\bibfnamefont {J.}~\bibnamefont
  {de~Jes{\'u}s Mendoza-Temis}}, \bibinfo {author} {\bibfnamefont {M.-R.}\
  \bibnamefont {Wu}}, \bibinfo {author} {\bibfnamefont {G.}~\bibnamefont
  {Mart{\'\i}nez-Pinedo}}, \bibinfo {author} {\bibfnamefont {K.}~\bibnamefont
  {Langanke}}, \bibinfo {author} {\bibfnamefont {A.}~\bibnamefont {Bauswein}},\
  and\ \bibinfo {author} {\bibfnamefont {H.-T.}\ \bibnamefont {Janka}},\ }\href
  {https://doi.org/10.1103/PhysRevC.92.055805} {\bibfield  {journal} {\bibinfo
  {journal} {Phys. Rev. C}\ }\textbf {\bibinfo {volume} {92}},\ \bibinfo
  {pages} {055805} (\bibinfo {year} {2015})},\ \Eprint
  {https://arxiv.org/abs/1409.6135} {arXiv:1409.6135 [astro-ph.HE]}
  \BibitemShut {NoStop}%
\bibitem [{\citenamefont {Woosley}\ and\ \citenamefont
  {Hoffman}(1992)}]{Woosley:1992ek}%
  \BibitemOpen
  \bibfield  {author} {\bibinfo {author} {\bibfnamefont {S.~E.}\ \bibnamefont
  {Woosley}}\ and\ \bibinfo {author} {\bibfnamefont {R.~D.}\ \bibnamefont
  {Hoffman}},\ }\href {https://doi.org/10.1086/171644} {\bibfield  {journal}
  {\bibinfo  {journal} {Astrophys. J.}\ }\textbf {\bibinfo {volume} {395}},\
  \bibinfo {pages} {202} (\bibinfo {year} {1992})}\BibitemShut {NoStop}%
\bibitem [{\citenamefont {Metzger}\ \emph {et~al.}(2015)\citenamefont
  {Metzger}, \citenamefont {Bauswein}, \citenamefont {Goriely},\ and\
  \citenamefont {Kasen}}]{Metzger:2014yda}%
  \BibitemOpen
  \bibfield  {author} {\bibinfo {author} {\bibfnamefont {B.~D.}\ \bibnamefont
  {Metzger}}, \bibinfo {author} {\bibfnamefont {A.}~\bibnamefont {Bauswein}},
  \bibinfo {author} {\bibfnamefont {S.}~\bibnamefont {Goriely}},\ and\ \bibinfo
  {author} {\bibfnamefont {D.}~\bibnamefont {Kasen}},\ }\href
  {https://doi.org/10.1093/mnras/stu2225} {\bibfield  {journal} {\bibinfo
  {journal} {Mon. Not. Roy. Astron. Soc.}\ }\textbf {\bibinfo {volume} {446}},\
  \bibinfo {pages} {1115} (\bibinfo {year} {2015})},\ \Eprint
  {https://arxiv.org/abs/1409.0544} {arXiv:1409.0544 [astro-ph.HE]}
  \BibitemShut {NoStop}%
\bibitem [{\citenamefont {Abac}\ \emph {et~al.}(2024)\citenamefont {Abac},
  \citenamefont {Dietrich}, \citenamefont {Buonanno}, \citenamefont
  {Steinhoff},\ and\ \citenamefont {Ujevic}}]{Abac:2023ujg}%
  \BibitemOpen
  \bibfield  {author} {\bibinfo {author} {\bibfnamefont {A.}~\bibnamefont
  {Abac}}, \bibinfo {author} {\bibfnamefont {T.}~\bibnamefont {Dietrich}},
  \bibinfo {author} {\bibfnamefont {A.}~\bibnamefont {Buonanno}}, \bibinfo
  {author} {\bibfnamefont {J.}~\bibnamefont {Steinhoff}},\ and\ \bibinfo
  {author} {\bibfnamefont {M.}~\bibnamefont {Ujevic}},\ }\href
  {https://doi.org/10.1103/PhysRevD.109.024062} {\bibfield  {journal} {\bibinfo
   {journal} {Phys. Rev. D}\ }\textbf {\bibinfo {volume} {109}},\ \bibinfo
  {pages} {024062} (\bibinfo {year} {2024})},\ \Eprint
  {https://arxiv.org/abs/2311.07456} {arXiv:2311.07456 [gr-qc]} \BibitemShut
  {NoStop}%
\bibitem [{\citenamefont {De~Pietri}\ \emph {et~al.}(2018)\citenamefont
  {De~Pietri}, \citenamefont {Feo}, \citenamefont {Font}, \citenamefont
  {L{\"o}ffler}, \citenamefont {Maione}, \citenamefont {Pasquali},\ and\
  \citenamefont {Stergioulas}}]{DePietri:2018tpx}%
  \BibitemOpen
  \bibfield  {author} {\bibinfo {author} {\bibfnamefont {R.}~\bibnamefont
  {De~Pietri}}, \bibinfo {author} {\bibfnamefont {A.}~\bibnamefont {Feo}},
  \bibinfo {author} {\bibfnamefont {J.~A.}\ \bibnamefont {Font}}, \bibinfo
  {author} {\bibfnamefont {F.}~\bibnamefont {L{\"o}ffler}}, \bibinfo {author}
  {\bibfnamefont {F.}~\bibnamefont {Maione}}, \bibinfo {author} {\bibfnamefont
  {M.}~\bibnamefont {Pasquali}},\ and\ \bibinfo {author} {\bibfnamefont
  {N.}~\bibnamefont {Stergioulas}},\ }\href
  {https://doi.org/10.1103/PhysRevLett.120.221101} {\bibfield  {journal}
  {\bibinfo  {journal} {Phys. Rev. Lett.}\ }\textbf {\bibinfo {volume} {120}},\
  \bibinfo {pages} {221101} (\bibinfo {year} {2018})},\ \Eprint
  {https://arxiv.org/abs/1802.03288} {arXiv:1802.03288 [gr-qc]} \BibitemShut
  {NoStop}%
\bibitem [{\citenamefont {De~Pietri}\ \emph {et~al.}(2020)\citenamefont
  {De~Pietri}, \citenamefont {Feo}, \citenamefont {Font}, \citenamefont
  {L{\"o}ffler}, \citenamefont {Pasquali},\ and\ \citenamefont
  {Stergioulas}}]{DePietri:2019mti}%
  \BibitemOpen
  \bibfield  {author} {\bibinfo {author} {\bibfnamefont {R.}~\bibnamefont
  {De~Pietri}}, \bibinfo {author} {\bibfnamefont {A.}~\bibnamefont {Feo}},
  \bibinfo {author} {\bibfnamefont {J.~A.}\ \bibnamefont {Font}}, \bibinfo
  {author} {\bibfnamefont {F.}~\bibnamefont {L{\"o}ffler}}, \bibinfo {author}
  {\bibfnamefont {M.}~\bibnamefont {Pasquali}},\ and\ \bibinfo {author}
  {\bibfnamefont {N.}~\bibnamefont {Stergioulas}},\ }\href
  {https://doi.org/10.1103/PhysRevD.101.064052} {\bibfield  {journal} {\bibinfo
   {journal} {Phys. Rev. D}\ }\textbf {\bibinfo {volume} {101}},\ \bibinfo
  {pages} {064052} (\bibinfo {year} {2020})},\ \Eprint
  {https://arxiv.org/abs/1910.04036} {arXiv:1910.04036 [gr-qc]} \BibitemShut
  {NoStop}%
\bibitem [{\citenamefont {Moore}\ \emph {et~al.}(2015)\citenamefont {Moore},
  \citenamefont {Cole},\ and\ \citenamefont {Berry}}]{Moore:2014lga}%
  \BibitemOpen
  \bibfield  {author} {\bibinfo {author} {\bibfnamefont {C.~J.}\ \bibnamefont
  {Moore}}, \bibinfo {author} {\bibfnamefont {R.~H.}\ \bibnamefont {Cole}},\
  and\ \bibinfo {author} {\bibfnamefont {C.~P.~L.}\ \bibnamefont {Berry}},\
  }\href {https://doi.org/10.1088/0264-9381/32/1/015014} {\bibfield  {journal}
  {\bibinfo  {journal} {Class. Quant. Grav.}\ }\textbf {\bibinfo {volume}
  {32}},\ \bibinfo {pages} {015014} (\bibinfo {year} {2015})},\ \Eprint
  {https://arxiv.org/abs/1408.0740} {arXiv:1408.0740 [gr-qc]} \BibitemShut
  {NoStop}%
\bibitem [{\citenamefont {Welch}(1967)}]{Welch:1967}%
  \BibitemOpen
  \bibfield  {author} {\bibinfo {author} {\bibfnamefont {P.~D.}\ \bibnamefont
  {Welch}},\ }\href {https://doi.org/10.1109/TAU.1967.1161901} {\bibfield
  {journal} {\bibinfo  {journal} {IEEE Transactions on Audio and
  Electroacoustics}\ }\textbf {\bibinfo {volume} {15}},\ \bibinfo {pages} {70}
  (\bibinfo {year} {1967})}\BibitemShut {NoStop}%
\bibitem [{\citenamefont {Stergioulas}\ \emph {et~al.}(2011)\citenamefont
  {Stergioulas}, \citenamefont {Bauswein}, \citenamefont {Zagkouris},\ and\
  \citenamefont {Janka}}]{Stergioulas:2011gd}%
  \BibitemOpen
  \bibfield  {author} {\bibinfo {author} {\bibfnamefont {N.}~\bibnamefont
  {Stergioulas}}, \bibinfo {author} {\bibfnamefont {A.}~\bibnamefont
  {Bauswein}}, \bibinfo {author} {\bibfnamefont {K.}~\bibnamefont
  {Zagkouris}},\ and\ \bibinfo {author} {\bibfnamefont {H.-T.}\ \bibnamefont
  {Janka}},\ }\href {https://doi.org/10.1111/j.1365-2966.2011.19493.x}
  {\bibfield  {journal} {\bibinfo  {journal} {Mon. Not. Roy. Astron. Soc.}\
  }\textbf {\bibinfo {volume} {418}},\ \bibinfo {pages} {427} (\bibinfo {year}
  {2011})},\ \Eprint {https://arxiv.org/abs/1105.0368} {arXiv:1105.0368
  [gr-qc]} \BibitemShut {NoStop}%
\bibitem [{\citenamefont {Vretinaris}\ \emph {et~al.}(2020)\citenamefont
  {Vretinaris}, \citenamefont {Stergioulas},\ and\ \citenamefont
  {Bauswein}}]{Vretinaris:2019spn}%
  \BibitemOpen
  \bibfield  {author} {\bibinfo {author} {\bibfnamefont {S.}~\bibnamefont
  {Vretinaris}}, \bibinfo {author} {\bibfnamefont {N.}~\bibnamefont
  {Stergioulas}},\ and\ \bibinfo {author} {\bibfnamefont {A.}~\bibnamefont
  {Bauswein}},\ }\href {https://doi.org/10.1103/PhysRevD.101.084039} {\bibfield
   {journal} {\bibinfo  {journal} {Phys. Rev. D}\ }\textbf {\bibinfo {volume}
  {101}},\ \bibinfo {pages} {084039} (\bibinfo {year} {2020})},\ \Eprint
  {https://arxiv.org/abs/1910.10856} {arXiv:1910.10856 [gr-qc]} \BibitemShut
  {NoStop}%
\bibitem [{\citenamefont {Radice}\ \emph {et~al.}(2016)\citenamefont {Radice},
  \citenamefont {Bernuzzi},\ and\ \citenamefont {Ott}}]{Radice:2016gym}%
  \BibitemOpen
  \bibfield  {author} {\bibinfo {author} {\bibfnamefont {D.}~\bibnamefont
  {Radice}}, \bibinfo {author} {\bibfnamefont {S.}~\bibnamefont {Bernuzzi}},\
  and\ \bibinfo {author} {\bibfnamefont {C.~D.}\ \bibnamefont {Ott}},\ }\href
  {https://doi.org/10.1103/PhysRevD.94.064011} {\bibfield  {journal} {\bibinfo
  {journal} {Phys. Rev. D}\ }\textbf {\bibinfo {volume} {94}},\ \bibinfo
  {pages} {064011} (\bibinfo {year} {2016})},\ \Eprint
  {https://arxiv.org/abs/1603.05726} {arXiv:1603.05726 [gr-qc]} \BibitemShut
  {NoStop}%
\bibitem [{\citenamefont {East}\ \emph
  {et~al.}(2016{\natexlab{b}})\citenamefont {East}, \citenamefont
  {Paschalidis},\ and\ \citenamefont {Pretorius}}]{East:2016zvv}%
  \BibitemOpen
  \bibfield  {author} {\bibinfo {author} {\bibfnamefont {W.~E.}\ \bibnamefont
  {East}}, \bibinfo {author} {\bibfnamefont {V.}~\bibnamefont {Paschalidis}},\
  and\ \bibinfo {author} {\bibfnamefont {F.}~\bibnamefont {Pretorius}},\ }\href
  {https://doi.org/10.1088/0264-9381/33/24/244004} {\bibfield  {journal}
  {\bibinfo  {journal} {Class. Quant. Grav.}\ }\textbf {\bibinfo {volume}
  {33}},\ \bibinfo {pages} {244004} (\bibinfo {year} {2016}{\natexlab{b}})},\
  \Eprint {https://arxiv.org/abs/1609.00725} {arXiv:1609.00725 [astro-ph.HE]}
  \BibitemShut {NoStop}%
\bibitem [{\citenamefont {Paschalidis}\ \emph {et~al.}(2015)\citenamefont
  {Paschalidis}, \citenamefont {East}, \citenamefont {Pretorius},\ and\
  \citenamefont {Shapiro}}]{Paschalidis:2015mla}%
  \BibitemOpen
  \bibfield  {author} {\bibinfo {author} {\bibfnamefont {V.}~\bibnamefont
  {Paschalidis}}, \bibinfo {author} {\bibfnamefont {W.~E.}\ \bibnamefont
  {East}}, \bibinfo {author} {\bibfnamefont {F.}~\bibnamefont {Pretorius}},\
  and\ \bibinfo {author} {\bibfnamefont {S.~L.}\ \bibnamefont {Shapiro}},\
  }\href {https://doi.org/10.1103/PhysRevD.92.121502} {\bibfield  {journal}
  {\bibinfo  {journal} {Phys. Rev. D}\ }\textbf {\bibinfo {volume} {92}},\
  \bibinfo {pages} {121502} (\bibinfo {year} {2015})},\ \Eprint
  {https://arxiv.org/abs/1510.03432} {arXiv:1510.03432 [astro-ph.HE]}
  \BibitemShut {NoStop}%
\bibitem [{\citenamefont {Lehner}\ \emph {et~al.}(2016)\citenamefont {Lehner},
  \citenamefont {Liebling}, \citenamefont {Palenzuela},\ and\ \citenamefont
  {Motl}}]{Lehner:2016wjg}%
  \BibitemOpen
  \bibfield  {author} {\bibinfo {author} {\bibfnamefont {L.}~\bibnamefont
  {Lehner}}, \bibinfo {author} {\bibfnamefont {S.~L.}\ \bibnamefont
  {Liebling}}, \bibinfo {author} {\bibfnamefont {C.}~\bibnamefont
  {Palenzuela}},\ and\ \bibinfo {author} {\bibfnamefont {P.~M.}\ \bibnamefont
  {Motl}},\ }\href {https://doi.org/10.1103/PhysRevD.94.043003} {\bibfield
  {journal} {\bibinfo  {journal} {Phys. Rev. D}\ }\textbf {\bibinfo {volume}
  {94}},\ \bibinfo {pages} {043003} (\bibinfo {year} {2016})},\ \Eprint
  {https://arxiv.org/abs/1605.02369} {arXiv:1605.02369 [gr-qc]} \BibitemShut
  {NoStop}%
\bibitem [{\citenamefont {Radice}\ and\ \citenamefont
  {Bernuzzi}(2023)}]{Radice:2023zlw}%
  \BibitemOpen
  \bibfield  {author} {\bibinfo {author} {\bibfnamefont {D.}~\bibnamefont
  {Radice}}\ and\ \bibinfo {author} {\bibfnamefont {S.}~\bibnamefont
  {Bernuzzi}},\ }\href {https://doi.org/10.3847/1538-4357/ad0235} {\bibfield
  {journal} {\bibinfo  {journal} {Astrophys. J.}\ }\textbf {\bibinfo {volume}
  {959}},\ \bibinfo {pages} {46} (\bibinfo {year} {2023})},\ \Eprint
  {https://arxiv.org/abs/2306.13709} {arXiv:2306.13709 [astro-ph.HE]}
  \BibitemShut {NoStop}%
\bibitem [{\citenamefont {Takami}\ \emph {et~al.}(2014)\citenamefont {Takami},
  \citenamefont {Rezzolla},\ and\ \citenamefont {Baiotti}}]{Takami:2014zpa}%
  \BibitemOpen
  \bibfield  {author} {\bibinfo {author} {\bibfnamefont {K.}~\bibnamefont
  {Takami}}, \bibinfo {author} {\bibfnamefont {L.}~\bibnamefont {Rezzolla}},\
  and\ \bibinfo {author} {\bibfnamefont {L.}~\bibnamefont {Baiotti}},\ }\href
  {https://doi.org/10.1103/PhysRevLett.113.091104} {\bibfield  {journal}
  {\bibinfo  {journal} {Phys. Rev. Lett.}\ }\textbf {\bibinfo {volume} {113}},\
  \bibinfo {pages} {091104} (\bibinfo {year} {2014})},\ \Eprint
  {https://arxiv.org/abs/1403.5672} {arXiv:1403.5672 [gr-qc]} \BibitemShut
  {NoStop}%
\bibitem [{\citenamefont {Barsotti}\ \emph {et~al.}(2018)\citenamefont
  {Barsotti}, \citenamefont {Mcculler}, \citenamefont {Evans},\ and\
  \citenamefont {Fritschel}}]{Barsotti:2018}%
  \BibitemOpen
  \bibfield  {author} {\bibinfo {author} {\bibfnamefont {L.}~\bibnamefont
  {Barsotti}}, \bibinfo {author} {\bibfnamefont {L.}~\bibnamefont {Mcculler}},
  \bibinfo {author} {\bibfnamefont {M.}~\bibnamefont {Evans}},\ and\ \bibinfo
  {author} {\bibfnamefont {P.}~\bibnamefont {Fritschel}},\ }\href
  {https://dcc.ligo.org/LIGO-T1800042/public} {\emph {\bibinfo {title} {The A+
  design curve}}},\ \bibinfo {type} {Technical Report}\ \bibinfo {number}
  {LIGO-T1800042}\ (\bibinfo  {institution} {LIGO},\ \bibinfo {year}
  {2018})\BibitemShut {NoStop}%
\bibitem [{\citenamefont {Hild}\ \emph {et~al.}(2011)\citenamefont {Hild} \emph
  {et~al.}}]{Hild:2010id}%
  \BibitemOpen
  \bibfield  {author} {\bibinfo {author} {\bibfnamefont {S.}~\bibnamefont
  {Hild}} \emph {et~al.},\ }\href
  {https://doi.org/10.1088/0264-9381/28/9/094013} {\bibfield  {journal}
  {\bibinfo  {journal} {Class. Quant. Grav.}\ }\textbf {\bibinfo {volume}
  {28}},\ \bibinfo {pages} {094013} (\bibinfo {year} {2011})},\ \Eprint
  {https://arxiv.org/abs/1012.0908} {arXiv:1012.0908 [gr-qc]} \BibitemShut
  {NoStop}%
\bibitem [{\citenamefont {Li}\ and\ \citenamefont
  {Paczynski}(1998)}]{Li:1998bw}%
  \BibitemOpen
  \bibfield  {author} {\bibinfo {author} {\bibfnamefont {L.-X.}\ \bibnamefont
  {Li}}\ and\ \bibinfo {author} {\bibfnamefont {B.}~\bibnamefont {Paczynski}},\
  }\href {https://doi.org/10.1086/311680} {\bibfield  {journal} {\bibinfo
  {journal} {Astrophys. J. Lett.}\ }\textbf {\bibinfo {volume} {507}},\
  \bibinfo {pages} {L59} (\bibinfo {year} {1998})},\ \Eprint
  {https://arxiv.org/abs/astro-ph/9807272} {arXiv:astro-ph/9807272}
  \BibitemShut {NoStop}%
\bibitem [{\citenamefont {Darbha}\ and\ \citenamefont
  {Kasen}(2020)}]{Darbha:2020lhz}%
  \BibitemOpen
  \bibfield  {author} {\bibinfo {author} {\bibfnamefont {S.}~\bibnamefont
  {Darbha}}\ and\ \bibinfo {author} {\bibfnamefont {D.}~\bibnamefont {Kasen}},\
  }\href {https://doi.org/10.3847/1538-4357/ab9a34} {\bibfield  {journal}
  {\bibinfo  {journal} {Astrophys. J.}\ }\textbf {\bibinfo {volume} {897}},\
  \bibinfo {eid} {150} (\bibinfo {year} {2020})},\ \Eprint
  {https://arxiv.org/abs/2002.00299} {2002.00299 [astro-ph.HE]} \BibitemShut
  {NoStop}%
\bibitem [{\citenamefont {Kawaguchi}\ \emph {et~al.}(2018)\citenamefont
  {Kawaguchi}, \citenamefont {Shibata},\ and\ \citenamefont
  {Tanaka}}]{Kawaguchi:2018ptg}%
  \BibitemOpen
  \bibfield  {author} {\bibinfo {author} {\bibfnamefont {K.}~\bibnamefont
  {Kawaguchi}}, \bibinfo {author} {\bibfnamefont {M.}~\bibnamefont {Shibata}},\
  and\ \bibinfo {author} {\bibfnamefont {M.}~\bibnamefont {Tanaka}},\ }\href
  {https://doi.org/10.3847/2041-8213/aade02} {\bibfield  {journal} {\bibinfo
  {journal} {Astrophys. J. Lett.}\ }\textbf {\bibinfo {volume} {865}},\
  \bibinfo {pages} {L21} (\bibinfo {year} {2018})},\ \Eprint
  {https://arxiv.org/abs/1806.04088} {arXiv:1806.04088 [astro-ph.HE]}
  \BibitemShut {NoStop}%
\bibitem [{\citenamefont {Korobkin}\ \emph {et~al.}(2021)\citenamefont
  {Korobkin} \emph {et~al.}}]{Korobkin:2020spe}%
  \BibitemOpen
  \bibfield  {author} {\bibinfo {author} {\bibfnamefont {O.}~\bibnamefont
  {Korobkin}} \emph {et~al.},\ }\href
  {https://doi.org/10.3847/1538-4357/abe1b5} {\bibfield  {journal} {\bibinfo
  {journal} {Astrophys. J.}\ }\textbf {\bibinfo {volume} {910}},\ \bibinfo
  {pages} {116} (\bibinfo {year} {2021})},\ \Eprint
  {https://arxiv.org/abs/2004.00102} {arXiv:2004.00102 [astro-ph.HE]}
  \BibitemShut {NoStop}%
\bibitem [{\citenamefont {Collins}\ \emph {et~al.}(2023)\citenamefont
  {Collins}, \citenamefont {Bauswein}, \citenamefont {Sim}, \citenamefont
  {Vijayan}, \citenamefont {Martinez-Pinedo}, \citenamefont {Just},
  \citenamefont {Shingles},\ and\ \citenamefont {Kromer}}]{Collins:2023www}%
  \BibitemOpen
  \bibfield  {author} {\bibinfo {author} {\bibfnamefont {C.}~\bibnamefont
  {Collins}}, \bibinfo {author} {\bibfnamefont {A.}~\bibnamefont {Bauswein}},
  \bibinfo {author} {\bibfnamefont {S.}~\bibnamefont {Sim}}, \bibinfo {author}
  {\bibfnamefont {V.}~\bibnamefont {Vijayan}}, \bibinfo {author} {\bibfnamefont
  {G.}~\bibnamefont {Martinez-Pinedo}}, \bibinfo {author} {\bibfnamefont
  {O.}~\bibnamefont {Just}}, \bibinfo {author} {\bibfnamefont {L.~J.}\
  \bibnamefont {Shingles}},\ and\ \bibinfo {author} {\bibfnamefont
  {M.}~\bibnamefont {Kromer}},\ }\href {https://doi.org/10.22323/1.419.0010}
  {\bibfield  {journal} {\bibinfo  {journal} {PoS}\ }\textbf {\bibinfo {volume}
  {FAIRness2022}},\ \bibinfo {pages} {010} (\bibinfo {year}
  {2023})}\BibitemShut {NoStop}%
\bibitem [{\citenamefont {Groenewegen}\ \emph {et~al.}(2025)\citenamefont
  {Groenewegen}, \citenamefont {Curtis}, \citenamefont {M{\"o}sta},
  \citenamefont {Kasen},\ and\ \citenamefont
  {Brethauer}}]{Groenewegen:2025ezj}%
  \BibitemOpen
  \bibfield  {author} {\bibinfo {author} {\bibfnamefont {L.~S.}\ \bibnamefont
  {Groenewegen}}, \bibinfo {author} {\bibfnamefont {S.}~\bibnamefont {Curtis}},
  \bibinfo {author} {\bibfnamefont {P.}~\bibnamefont {M{\"o}sta}}, \bibinfo
  {author} {\bibfnamefont {D.}~\bibnamefont {Kasen}},\ and\ \bibinfo {author}
  {\bibfnamefont {D.}~\bibnamefont {Brethauer}},\ }\Eprint
  {https://arxiv.org/abs/2508.00062} {arXiv:2508.00062 [astro-ph.HE]}
  (\bibinfo {year} {2025})\BibitemShut {NoStop}%
\bibitem [{\citenamefont {Dekany}\ \emph {et~al.}(2020)\citenamefont {Dekany}
  \emph {et~al.}}]{Dekany:2020tyb}%
  \BibitemOpen
  \bibfield  {author} {\bibinfo {author} {\bibfnamefont {R.}~\bibnamefont
  {Dekany}} \emph {et~al.},\ }\href {https://doi.org/10.1088/1538-3873/ab4ca2}
  {\bibfield  {journal} {\bibinfo  {journal} {Publ. Astron. Soc. Pac.}\
  }\textbf {\bibinfo {volume} {132}},\ \bibinfo {pages} {038001} (\bibinfo
  {year} {2020})},\ \Eprint {https://arxiv.org/abs/2008.04923}
  {arXiv:2008.04923 [astro-ph.IM]} \BibitemShut {NoStop}%
\bibitem [{\citenamefont {Ivezi{\'c}}\ \emph {et~al.}(2018)\citenamefont
  {Ivezi{\'c}} \emph {et~al.}}]{LSST:2017}%
  \BibitemOpen
  \bibfield  {author} {\bibinfo {author} {\bibfnamefont {{\v{Z}}.}~\bibnamefont
  {Ivezi{\'c}}} \emph {et~al.} (\bibinfo {collaboration} {LSST}),\ }\href
  {https://docushare.lsst.org/docushare/dsweb/Get/LPM-17} {\bibinfo {title}
  {{The LSST System Science Requirements Document}}} (\bibinfo {year}
  {2018})\BibitemShut {NoStop}%
\bibitem [{\citenamefont {{Eichler}}\ \emph {et~al.}(1989)\citenamefont
  {{Eichler}}, \citenamefont {{Livio}}, \citenamefont {{Piran}},\ and\
  \citenamefont {{Schramm}}}]{Eichler:1989}%
  \BibitemOpen
  \bibfield  {author} {\bibinfo {author} {\bibfnamefont {D.}~\bibnamefont
  {{Eichler}}}, \bibinfo {author} {\bibfnamefont {M.}~\bibnamefont {{Livio}}},
  \bibinfo {author} {\bibfnamefont {T.}~\bibnamefont {{Piran}}},\ and\ \bibinfo
  {author} {\bibfnamefont {D.~N.}\ \bibnamefont {{Schramm}}},\ }\href
  {https://doi.org/10.1038/340126a0} {\bibfield  {journal} {\bibinfo  {journal}
  {Nature}\ }\textbf {\bibinfo {volume} {340}},\ \bibinfo {pages} {126}
  (\bibinfo {year} {1989})}\BibitemShut {NoStop}%
\bibitem [{\citenamefont {Narayan}\ \emph {et~al.}(1992)\citenamefont
  {Narayan}, \citenamefont {Paczynski},\ and\ \citenamefont
  {Piran}}]{Narayan:1992iy}%
  \BibitemOpen
  \bibfield  {author} {\bibinfo {author} {\bibfnamefont {R.}~\bibnamefont
  {Narayan}}, \bibinfo {author} {\bibfnamefont {B.}~\bibnamefont {Paczynski}},\
  and\ \bibinfo {author} {\bibfnamefont {T.}~\bibnamefont {Piran}},\ }\href
  {https://doi.org/10.1086/186493} {\bibfield  {journal} {\bibinfo  {journal}
  {Astrophys. J. Lett.}\ }\textbf {\bibinfo {volume} {395}},\ \bibinfo {pages}
  {L83} (\bibinfo {year} {1992})},\ \Eprint
  {https://arxiv.org/abs/astro-ph/9204001} {arXiv:astro-ph/9204001}
  \BibitemShut {NoStop}%
\bibitem [{\citenamefont {Berger}(2014)}]{Berger:2013jza}%
  \BibitemOpen
  \bibfield  {author} {\bibinfo {author} {\bibfnamefont {E.}~\bibnamefont
  {Berger}},\ }\href {https://doi.org/10.1146/annurev-astro-081913-035926}
  {\bibfield  {journal} {\bibinfo  {journal} {Ann. Rev. Astron. Astrophys.}\
  }\textbf {\bibinfo {volume} {52}},\ \bibinfo {pages} {43} (\bibinfo {year}
  {2014})},\ \Eprint {https://arxiv.org/abs/1311.2603} {arXiv:1311.2603
  [astro-ph.HE]} \BibitemShut {NoStop}%
\bibitem [{\citenamefont {Koehn}\ \emph {et~al.}(2025)\citenamefont {Koehn},
  \citenamefont {Wouters}, \citenamefont {Pang}, \citenamefont {Bulla},
  \citenamefont {Rose}, \citenamefont {Wichern},\ and\ \citenamefont
  {Dietrich}}]{Koehn:2025zzb}%
  \BibitemOpen
  \bibfield  {author} {\bibinfo {author} {\bibfnamefont {H.}~\bibnamefont
  {Koehn}}, \bibinfo {author} {\bibfnamefont {T.}~\bibnamefont {Wouters}},
  \bibinfo {author} {\bibfnamefont {P.~T.~H.}\ \bibnamefont {Pang}}, \bibinfo
  {author} {\bibfnamefont {M.}~\bibnamefont {Bulla}}, \bibinfo {author}
  {\bibfnamefont {H.}~\bibnamefont {Rose}}, \bibinfo {author} {\bibfnamefont
  {H.}~\bibnamefont {Wichern}},\ and\ \bibinfo {author} {\bibfnamefont
  {T.}~\bibnamefont {Dietrich}},\ }\Eprint {https://arxiv.org/abs/2507.13807}
  {arXiv:2507.13807 [astro-ph.HE]}  (\bibinfo {year} {2025})\BibitemShut
  {NoStop}%
\bibitem [{\citenamefont {Marcowith}\ \emph {et~al.}(2020)\citenamefont
  {Marcowith}, \citenamefont {Ferrand}, \citenamefont {Grech}, \citenamefont
  {Meliani}, \citenamefont {Plotnikov},\ and\ \citenamefont
  {Walder}}]{Marcowith:2020vho}%
  \BibitemOpen
  \bibfield  {author} {\bibinfo {author} {\bibfnamefont {A.}~\bibnamefont
  {Marcowith}}, \bibinfo {author} {\bibfnamefont {G.}~\bibnamefont {Ferrand}},
  \bibinfo {author} {\bibfnamefont {M.}~\bibnamefont {Grech}}, \bibinfo
  {author} {\bibfnamefont {Z.}~\bibnamefont {Meliani}}, \bibinfo {author}
  {\bibfnamefont {I.}~\bibnamefont {Plotnikov}},\ and\ \bibinfo {author}
  {\bibfnamefont {R.}~\bibnamefont {Walder}},\ }\href
  {https://doi.org/10.1007/s41115-020-0007-6} {\bibfield  {journal} {\bibinfo
  {journal} {Liv. Rev. Comput. Astrophys.}\ }\textbf {\bibinfo {volume} {6}},\
  \bibinfo {pages} {1} (\bibinfo {year} {2020})},\ \Eprint
  {https://arxiv.org/abs/2002.09411} {arXiv:2002.09411 [astro-ph.HE]}
  \BibitemShut {NoStop}%
\bibitem [{\citenamefont {Yuan}\ \emph {et~al.}(2025)\citenamefont {Yuan},
  \citenamefont {Chen},\ and\ \citenamefont {Luepker}}]{Yuan:2025war}%
  \BibitemOpen
  \bibfield  {author} {\bibinfo {author} {\bibfnamefont {Y.}~\bibnamefont
  {Yuan}}, \bibinfo {author} {\bibfnamefont {A.~Y.}\ \bibnamefont {Chen}},\
  and\ \bibinfo {author} {\bibfnamefont {M.}~\bibnamefont {Luepker}},\ }\href
  {https://doi.org/10.3847/1538-4357/adce79} {\bibfield  {journal} {\bibinfo
  {journal} {Astrophys. J.}\ }\textbf {\bibinfo {volume} {985}},\ \bibinfo
  {pages} {159} (\bibinfo {year} {2025})},\ \Eprint
  {https://arxiv.org/abs/2503.08487} {arXiv:2503.08487 [astro-ph.HE]}
  \BibitemShut {NoStop}%
\bibitem [{\citenamefont {Vanthieghem}\ \emph {et~al.}(2024)\citenamefont
  {Vanthieghem}, \citenamefont {Tsiolis}, \citenamefont {Fiuza}, \citenamefont
  {Sekiguchi}, \citenamefont {Spitkovsky},\ and\ \citenamefont
  {Todo}}]{Vanthieghem:2024xov}%
  \BibitemOpen
  \bibfield  {author} {\bibinfo {author} {\bibfnamefont {A.}~\bibnamefont
  {Vanthieghem}}, \bibinfo {author} {\bibfnamefont {V.}~\bibnamefont
  {Tsiolis}}, \bibinfo {author} {\bibfnamefont {F.}~\bibnamefont {Fiuza}},
  \bibinfo {author} {\bibfnamefont {K.}~\bibnamefont {Sekiguchi}}, \bibinfo
  {author} {\bibfnamefont {A.}~\bibnamefont {Spitkovsky}},\ and\ \bibinfo
  {author} {\bibfnamefont {Y.}~\bibnamefont {Todo}},\ }\href
  {https://doi.org/10.22323/1.461.0011} {\bibfield  {journal} {\bibinfo
  {journal} {PoS}\ }\textbf {\bibinfo {volume} {HEPROVIII}},\ \bibinfo {pages}
  {011} (\bibinfo {year} {2024})},\ \Eprint {https://arxiv.org/abs/2407.03838}
  {arXiv:2407.03838 [astro-ph.HE]} \BibitemShut {NoStop}%
\bibitem [{\citenamefont {Braun}\ \emph {et~al.}(2019)\citenamefont {Braun},
  \citenamefont {Bonaldi}, \citenamefont {Bourke}, \citenamefont {Keane},\ and\
  \citenamefont {Wagg}}]{Braun:2019gdo}%
  \BibitemOpen
  \bibfield  {author} {\bibinfo {author} {\bibfnamefont {R.}~\bibnamefont
  {Braun}}, \bibinfo {author} {\bibfnamefont {A.}~\bibnamefont {Bonaldi}},
  \bibinfo {author} {\bibfnamefont {T.}~\bibnamefont {Bourke}}, \bibinfo
  {author} {\bibfnamefont {E.}~\bibnamefont {Keane}},\ and\ \bibinfo {author}
  {\bibfnamefont {J.}~\bibnamefont {Wagg}},\ }\Eprint
  {https://arxiv.org/abs/1912.12699} {arXiv:1912.12699 [astro-ph.IM]}
  (\bibinfo {year} {2019})\BibitemShut {NoStop}%
\bibitem [{\citenamefont {Bonaldi}\ \emph {et~al.}(2020)\citenamefont {Bonaldi}
  \emph {et~al.}}]{Bonaldi:2020ukl}%
  \BibitemOpen
  \bibfield  {author} {\bibinfo {author} {\bibfnamefont {A.}~\bibnamefont
  {Bonaldi}} \emph {et~al.},\ }\href {https://doi.org/10.1093/mnras/staa3023}
  {\bibfield  {journal} {\bibinfo  {journal} {Mon. Not. Roy. Astron. Soc.}\
  }\textbf {\bibinfo {volume} {500}},\ \bibinfo {pages} {3821} (\bibinfo {year}
  {2020})},\ \Eprint {https://arxiv.org/abs/2009.13346} {arXiv:2009.13346
  [astro-ph.IM]} \BibitemShut {NoStop}%
\bibitem [{\citenamefont {Bianco}\ \emph {et~al.}(2022)\citenamefont {Bianco}
  \emph {et~al.}}]{Bianco:2021ape}%
  \BibitemOpen
  \bibfield  {author} {\bibinfo {author} {\bibfnamefont {F.~B.}\ \bibnamefont
  {Bianco}} \emph {et~al.},\ }\href {https://doi.org/10.3847/1538-4365/ac3e72}
  {\bibfield  {journal} {\bibinfo  {journal} {Astrophys. J. Supp.}\ }\textbf
  {\bibinfo {volume} {258}},\ \bibinfo {pages} {1} (\bibinfo {year} {2022})},\
  \Eprint {https://arxiv.org/abs/2108.01683} {arXiv:2108.01683 [astro-ph.IM]}
  \BibitemShut {NoStop}%
\bibitem [{\citenamefont {HARMONI}(2025)}]{ELT_HARMONI}%
  \BibitemOpen
  \bibfield  {author} {\bibinfo {author} {\bibnamefont {HARMONI}} (\bibinfo
  {collaboration} {HARMONI for the ELT}),\ }\href
  {https://harmoni-elt.physics.ox.ac.uk/Performances.html} {\bibinfo {title}
  {{HARMONI Performance}}} (\bibinfo {year} {2025})\BibitemShut {NoStop}%
\bibitem [{\citenamefont {Davies}\ \emph {et~al.}(2021)\citenamefont {Davies}
  \emph {et~al.}}]{ELT_MICADO}%
  \BibitemOpen
  \bibfield  {author} {\bibinfo {author} {\bibfnamefont {R.}~\bibnamefont
  {Davies}} \emph {et~al.} (\bibinfo {collaboration} {The MICADO Consortium}),\
  }\href {https://doi.org/10.18727/0722-6691/5217} {\bibfield  {journal}
  {\bibinfo  {journal} {The Messenger}\ }\textbf {\bibinfo {volume} {182}},\
  \bibinfo {pages} {17} (\bibinfo {year} {2021})}\BibitemShut {NoStop}%
\bibitem [{\citenamefont {Laird}\ \emph {et~al.}(2009)\citenamefont {Laird}
  \emph {et~al.}}]{Laird:2008va}%
  \BibitemOpen
  \bibfield  {author} {\bibinfo {author} {\bibfnamefont {E.~S.}\ \bibnamefont
  {Laird}} \emph {et~al.},\ }\href
  {https://doi.org/10.1088/0067-0049/180/1/102} {\bibfield  {journal} {\bibinfo
   {journal} {Astrophys. J. Suppl.}\ }\textbf {\bibinfo {volume} {180}},\
  \bibinfo {pages} {102} (\bibinfo {year} {2009})},\ \Eprint
  {https://arxiv.org/abs/0809.1349} {arXiv:0809.1349 [astro-ph]} \BibitemShut
  {NoStop}%
\bibitem [{\citenamefont {Nandra}\ \emph {et~al.}(2013)\citenamefont {Nandra}
  \emph {et~al.}}]{Nandra:2013jka}%
  \BibitemOpen
  \bibfield  {author} {\bibinfo {author} {\bibfnamefont {K.}~\bibnamefont
  {Nandra}} \emph {et~al.},\ }\Eprint {https://arxiv.org/abs/1306.2307}
  {arXiv:1306.2307 [astro-ph.HE]}  (\bibinfo {year} {2013})\BibitemShut
  {NoStop}%
\bibitem [{\citenamefont {Reynolds}\ \emph {et~al.}(2023)\citenamefont
  {Reynolds} \emph {et~al.}}]{Reynolds:2023vvf}%
  \BibitemOpen
  \bibfield  {author} {\bibinfo {author} {\bibfnamefont {C.~S.}\ \bibnamefont
  {Reynolds}} \emph {et~al.},\ }\href {https://doi.org/10.1117/12.2677468}
  {\bibfield  {journal} {\bibinfo  {journal} {Proc. SPIE Int. Soc. Opt. Eng.}\
  }\textbf {\bibinfo {volume} {12678}},\ \bibinfo {pages} {126781E} (\bibinfo
  {year} {2023})},\ \Eprint {https://arxiv.org/abs/2311.00780}
  {arXiv:2311.00780 [astro-ph.IM]} \BibitemShut {NoStop}%
\bibitem [{\citenamefont {Marchesi}\ \emph {et~al.}(2020)\citenamefont
  {Marchesi} \emph {et~al.}}]{Marchesi:2020smf}%
  \BibitemOpen
  \bibfield  {author} {\bibinfo {author} {\bibfnamefont {S.}~\bibnamefont
  {Marchesi}} \emph {et~al.},\ }\href
  {https://doi.org/10.1051/0004-6361/202038622} {\bibfield  {journal} {\bibinfo
   {journal} {Astron. Astrophys.}\ }\textbf {\bibinfo {volume} {642}},\
  \bibinfo {pages} {A184} (\bibinfo {year} {2020})},\ \Eprint
  {https://arxiv.org/abs/2008.09133} {arXiv:2008.09133 [astro-ph.IM]}
  \BibitemShut {NoStop}%
\bibitem [{\citenamefont {Gieg}\ \emph {et~al.}(2025)\citenamefont {Gieg},
  \citenamefont {Schianchi}, \citenamefont {Ujevic},\ and\ \citenamefont
  {Dietrich}}]{Gieg:2024jxs}%
  \BibitemOpen
  \bibfield  {author} {\bibinfo {author} {\bibfnamefont {H.}~\bibnamefont
  {Gieg}}, \bibinfo {author} {\bibfnamefont {F.}~\bibnamefont {Schianchi}},
  \bibinfo {author} {\bibfnamefont {M.}~\bibnamefont {Ujevic}},\ and\ \bibinfo
  {author} {\bibfnamefont {T.}~\bibnamefont {Dietrich}},\ }\href
  {https://doi.org/10.1103/52ph-53yw} {\bibfield  {journal} {\bibinfo
  {journal} {Phys. Rev. D}\ }\textbf {\bibinfo {volume} {112}},\ \bibinfo
  {pages} {023036} (\bibinfo {year} {2025})},\ \Eprint
  {https://arxiv.org/abs/2409.04420} {arXiv:2409.04420 [gr-qc]} \BibitemShut
  {NoStop}%
\bibitem [{\citenamefont {Ng}\ \emph {et~al.}(2025)\citenamefont {Ng},
  \citenamefont {Musolino}, \citenamefont {Tootle},\ and\ \citenamefont
  {Rezzolla}}]{Ng:2024zve}%
  \BibitemOpen
  \bibfield  {author} {\bibinfo {author} {\bibfnamefont {H.~H.-Y.}\
  \bibnamefont {Ng}}, \bibinfo {author} {\bibfnamefont {C.}~\bibnamefont
  {Musolino}}, \bibinfo {author} {\bibfnamefont {S.~D.}\ \bibnamefont
  {Tootle}},\ and\ \bibinfo {author} {\bibfnamefont {L.}~\bibnamefont
  {Rezzolla}},\ }\href {https://doi.org/10.3847/2041-8213/add324} {\bibfield
  {journal} {\bibinfo  {journal} {Astrophys. J. Lett.}\ }\textbf {\bibinfo
  {volume} {985}},\ \bibinfo {pages} {L36} (\bibinfo {year} {2025})},\ \Eprint
  {https://arxiv.org/abs/2411.19178} {arXiv:2411.19178 [astro-ph.HE]}
  \BibitemShut {NoStop}%
\bibitem [{\citenamefont {Cheong}\ \emph {et~al.}(2025)\citenamefont {Cheong},
  \citenamefont {Tsokaros}, \citenamefont {Ruiz}, \citenamefont {Venturi},
  \citenamefont {Chan}, \citenamefont {Yip},\ and\ \citenamefont
  {Uryu}}]{Cheong:2024stz}%
  \BibitemOpen
  \bibfield  {author} {\bibinfo {author} {\bibfnamefont {P.~C.-K.}\
  \bibnamefont {Cheong}}, \bibinfo {author} {\bibfnamefont {A.}~\bibnamefont
  {Tsokaros}}, \bibinfo {author} {\bibfnamefont {M.}~\bibnamefont {Ruiz}},
  \bibinfo {author} {\bibfnamefont {F.}~\bibnamefont {Venturi}}, \bibinfo
  {author} {\bibfnamefont {J.~C.~L.}\ \bibnamefont {Chan}}, \bibinfo {author}
  {\bibfnamefont {A.~K.~L.}\ \bibnamefont {Yip}},\ and\ \bibinfo {author}
  {\bibfnamefont {K.}~\bibnamefont {Uryu}},\ }\href
  {https://doi.org/10.1103/PhysRevD.111.063030} {\bibfield  {journal} {\bibinfo
   {journal} {Phys. Rev. D}\ }\textbf {\bibinfo {volume} {111}},\ \bibinfo
  {pages} {063030} (\bibinfo {year} {2025})},\ \Eprint
  {https://arxiv.org/abs/2409.10508} {arXiv:2409.10508 [astro-ph.HE]}
  \BibitemShut {NoStop}%
\bibitem [{\citenamefont {Dietrich}\ \emph {et~al.}(2018)\citenamefont
  {Dietrich}, \citenamefont {Radice}, \citenamefont {Bernuzzi}, \citenamefont
  {Zappa}, \citenamefont {Perego}, \citenamefont {Br\"ugmann}, \citenamefont
  {Chaurasia}, \citenamefont {Dudi}, \citenamefont {Tichy},\ and\ \citenamefont
  {Ujevic}}]{Dietrich:2018phi}%
  \BibitemOpen
  \bibfield  {author} {\bibinfo {author} {\bibfnamefont {T.}~\bibnamefont
  {Dietrich}}, \bibinfo {author} {\bibfnamefont {D.}~\bibnamefont {Radice}},
  \bibinfo {author} {\bibfnamefont {S.}~\bibnamefont {Bernuzzi}}, \bibinfo
  {author} {\bibfnamefont {F.}~\bibnamefont {Zappa}}, \bibinfo {author}
  {\bibfnamefont {A.}~\bibnamefont {Perego}}, \bibinfo {author} {\bibfnamefont
  {B.}~\bibnamefont {Br\"ugmann}}, \bibinfo {author} {\bibfnamefont {S.~V.}\
  \bibnamefont {Chaurasia}}, \bibinfo {author} {\bibfnamefont {R.}~\bibnamefont
  {Dudi}}, \bibinfo {author} {\bibfnamefont {W.}~\bibnamefont {Tichy}},\ and\
  \bibinfo {author} {\bibfnamefont {M.}~\bibnamefont {Ujevic}},\ }\href
  {https://doi.org/10.1088/1361-6382/aaebc0} {\bibfield  {journal} {\bibinfo
  {journal} {Class. Quant. Grav.}\ }\textbf {\bibinfo {volume} {35}},\ \bibinfo
  {pages} {24LT01} (\bibinfo {year} {2018})},\ \Eprint
  {https://arxiv.org/abs/1806.01625} {arXiv:1806.01625 [gr-qc]} \BibitemShut
  {NoStop}%
\bibitem [{\citenamefont {Gonzalez}\ \emph {et~al.}(2023)\citenamefont
  {Gonzalez} \emph {et~al.}}]{Gonzalez:2022mgo}%
  \BibitemOpen
  \bibfield  {author} {\bibinfo {author} {\bibfnamefont {A.}~\bibnamefont
  {Gonzalez}} \emph {et~al.},\ }\href
  {https://doi.org/10.1088/1361-6382/acc231} {\bibfield  {journal} {\bibinfo
  {journal} {Class. Quant. Grav.}\ }\textbf {\bibinfo {volume} {40}},\ \bibinfo
  {pages} {085011} (\bibinfo {year} {2023})},\ \Eprint
  {https://arxiv.org/abs/2210.16366} {arXiv:2210.16366 [gr-qc]} \BibitemShut
  {NoStop}%
\bibitem [{\citenamefont {Markin}\ \emph {et~al.}(2025)\citenamefont {Markin},
  \citenamefont {Neuweiler}, \citenamefont {Gieg},\ and\ \citenamefont
  {Dietrich}}]{markin_2025_17303085}%
  \BibitemOpen
  \bibfield  {author} {\bibinfo {author} {\bibfnamefont {I.}~\bibnamefont
  {Markin}}, \bibinfo {author} {\bibfnamefont {A.}~\bibnamefont {Neuweiler}},
  \bibinfo {author} {\bibfnamefont {H.~L.}\ \bibnamefont {Gieg}},\ and\
  \bibinfo {author} {\bibfnamefont {T.}~\bibnamefont {Dietrich}},\ }\href
  {https://doi.org/10.5281/zenodo.17303085} {\bibinfo {title}
  {General-relativistic radiation magnetohydrodynamics simulations of binary
  neutron star mergers: The influence of spin on the multi- messenger picture -
  videos}} (\bibinfo {year} {2025})\BibitemShut {NoStop}%
\bibitem [{\citenamefont {Oreskes}(2004)}]{Oreskes_2004}%
  \BibitemOpen
  \bibfield  {author} {\bibinfo {author} {\bibfnamefont {N.}~\bibnamefont
  {Oreskes}},\ }\href {https://doi.org/10.1126/science.1103618} {\bibfield
  {journal} {\bibinfo  {journal} {Science}\ }\textbf {\bibinfo {volume}
  {306}},\ \bibinfo {pages} {1686} (\bibinfo {year} {2004})}\BibitemShut
  {NoStop}%
\bibitem [{\citenamefont {Doran}\ and\ \citenamefont
  {Zimmerman}(2009)}]{Doran_2009}%
  \BibitemOpen
  \bibfield  {author} {\bibinfo {author} {\bibfnamefont {P.~T.}\ \bibnamefont
  {Doran}}\ and\ \bibinfo {author} {\bibfnamefont {M.~K.}\ \bibnamefont
  {Zimmerman}},\ }\href {https://doi.org/https://doi.org/10.1029/2009EO030002}
  {\bibfield  {journal} {\bibinfo  {journal} {Eos, Transactions American
  Geophysical Union}\ }\textbf {\bibinfo {volume} {90}},\ \bibinfo {pages} {22}
  (\bibinfo {year} {2009})}\BibitemShut {NoStop}%
\bibitem [{\citenamefont {Cook}\ \emph {et~al.}(2013)\citenamefont {Cook},
  \citenamefont {Nuccitelli}, \citenamefont {Green}, \citenamefont
  {Richardson}, \citenamefont {Winkler}, \citenamefont {Painting},
  \citenamefont {Way}, \citenamefont {Jacobs},\ and\ \citenamefont
  {Skuce}}]{Cook_2013}%
  \BibitemOpen
  \bibfield  {author} {\bibinfo {author} {\bibfnamefont {J.}~\bibnamefont
  {Cook}}, \bibinfo {author} {\bibfnamefont {D.}~\bibnamefont {Nuccitelli}},
  \bibinfo {author} {\bibfnamefont {S.~A.}\ \bibnamefont {Green}}, \bibinfo
  {author} {\bibfnamefont {M.}~\bibnamefont {Richardson}}, \bibinfo {author}
  {\bibfnamefont {B.}~\bibnamefont {Winkler}}, \bibinfo {author} {\bibfnamefont
  {R.}~\bibnamefont {Painting}}, \bibinfo {author} {\bibfnamefont
  {R.}~\bibnamefont {Way}}, \bibinfo {author} {\bibfnamefont {P.}~\bibnamefont
  {Jacobs}},\ and\ \bibinfo {author} {\bibfnamefont {A.}~\bibnamefont
  {Skuce}},\ }\href {https://doi.org/10.1088/1748-9326/8/2/024024} {\bibfield
  {journal} {\bibinfo  {journal} {Environmental Research Letters}\ }\textbf
  {\bibinfo {volume} {8}},\ \bibinfo {pages} {024024} (\bibinfo {year}
  {2013})}\BibitemShut {NoStop}%
\bibitem [{\citenamefont {{Cook}}\ \emph {et~al.}(2016)\citenamefont {{Cook}}
  \emph {et~al.}}]{Cook_2016}%
  \BibitemOpen
  \bibfield  {author} {\bibinfo {author} {\bibfnamefont {J.}~\bibnamefont
  {{Cook}}} \emph {et~al.},\ }\href
  {https://doi.org/10.1088/1748-9326/11/4/048002} {\bibfield  {journal}
  {\bibinfo  {journal} {Environmental Research Letters}\ }\textbf {\bibinfo
  {volume} {11}},\ \bibinfo {eid} {048002} (\bibinfo {year}
  {2016})}\BibitemShut {NoStop}%
\bibitem [{\citenamefont {Lynas}\ \emph {et~al.}(2021)\citenamefont {Lynas},
  \citenamefont {Houlton},\ and\ \citenamefont {Perry}}]{Lynas_2021}%
  \BibitemOpen
  \bibfield  {author} {\bibinfo {author} {\bibfnamefont {M.}~\bibnamefont
  {Lynas}}, \bibinfo {author} {\bibfnamefont {B.~Z.}\ \bibnamefont {Houlton}},\
  and\ \bibinfo {author} {\bibfnamefont {S.}~\bibnamefont {Perry}},\ }\href
  {https://doi.org/10.1088/1748-9326/ac2966} {\bibfield  {journal} {\bibinfo
  {journal} {Environmental Research Letters}\ }\textbf {\bibinfo {volume}
  {16}},\ \bibinfo {pages} {114005} (\bibinfo {year} {2021})}\BibitemShut
  {NoStop}%
\bibitem [{\citenamefont {{Myers}}\ \emph {et~al.}(2021)\citenamefont
  {{Myers}}, \citenamefont {{Doran}}, \citenamefont {{Cook}}, \citenamefont
  {{Kotcher}},\ and\ \citenamefont {{Myers}}}]{Myers_2021}%
  \BibitemOpen
  \bibfield  {author} {\bibinfo {author} {\bibfnamefont {K.~F.}\ \bibnamefont
  {{Myers}}}, \bibinfo {author} {\bibfnamefont {P.~T.}\ \bibnamefont
  {{Doran}}}, \bibinfo {author} {\bibfnamefont {J.}~\bibnamefont {{Cook}}},
  \bibinfo {author} {\bibfnamefont {J.~E.}\ \bibnamefont {{Kotcher}}},\ and\
  \bibinfo {author} {\bibfnamefont {T.~A.}\ \bibnamefont {{Myers}}},\ }\href
  {https://doi.org/10.1088/1748-9326/ac2774} {\bibfield  {journal} {\bibinfo
  {journal} {Environmental Research Letters}\ }\textbf {\bibinfo {volume}
  {16}},\ \bibinfo {eid} {104030} (\bibinfo {year} {2021})}\BibitemShut
  {NoStop}%
\bibitem [{\citenamefont {Conrad}\ and\ \citenamefont
  {Lorenz}(2023)}]{HLRS:2023}%
  \BibitemOpen
  \bibfield  {author} {\bibinfo {author} {\bibfnamefont {N.}~\bibnamefont
  {Conrad}}\ and\ \bibinfo {author} {\bibfnamefont {B.}~\bibnamefont
  {Lorenz}},\ }\href
  {https://www.hlrs.de/fileadmin/about/social_responsibility/Sustainability/HLRS_Umwelterklaerung2023_barrierefrei.pdf}
  {\emph {\bibinfo {title} {Umwelterklärung 2023}}},\ \bibinfo {type}
  {Technischer Bericht / Umweltbericht}\ (\bibinfo  {institution}
  {Höchstleistungsrechenzentrum (HLRS), Universität Stuttgart},\ \bibinfo
  {year} {2023})\BibitemShut {NoStop}%
\end{thebibliography}%

\end{document}